\documentclass[unsortedaddress,rmp,aps,reprint, amsmath,amssymb,graphicx,floatfix]{revtex4-1_nosort}

\pdfoutput=1

\usepackage{color}
\usepackage{graphicx}
\setcitestyle{authoryear}

\newcommand\apjl{{ApJ }}%
\newcommand\apjs{{ApJS }}%
\newcommand\aap{{A\&A }}%
\newcommand\mnras{{MNRAS }}%
\newcommand\sovast{{Soviet~Ast. }}%
\newcommand\physrep{{Phys.~Rep. }}%
\newcommand\rpp{{Rep. Prog. Phys.}}%

\begin{document}

\title{Measuring the neutron star equation of state using X-ray timing}

\author{Anna L. Watts}
\affiliation{\mbox{Anton Pannekoek Institute, University of Amsterdam, PO Box 94249, 1090 GE Amsterdam, The Netherlands}}

\author{Nils Andersson}
\affiliation{\mbox{Mathematical Sciences, University of Southampton, Highfield, Southampton SO17 1BJ, UK}}

\author{Deepto Chakrabarty}
\affiliation{\mbox{Kavli Institute for Astrophysics and Space Research, Massachusetts Institute of Technology, Cambridge, MA 02139, USA}}

\author{Marco Feroci}
\affiliation{\mbox{INAF/IASF Roma, via Fosso del Cavaliere 100, I-00133 Roma, Italy}}
\affiliation{\mbox{INFN, Sezione di Roma Tor Vergata, Via della Ricerca Scientifica 1, I-00133 Roma, Italy}}

\author{Kai Hebeler}
\affiliation{\mbox{Institut f\"ur Kernphysik, Technische Universit\"at
Darmstadt, 64289 Darmstadt, Germany}}
\affiliation{\mbox{ExtreMe Matter Institute EMMI, GSI Helmholtzzentrum f\"ur
Schwerionenforschung GmbH, 64291 Darmstadt, Germany}}

\author{Gianluca Israel}
\affiliation{\mbox{INAF/OAR, Via Frascati 33, I-00040, Monte Porzio Catone (Roma), Italy}}

\author{Frederick K. Lamb}
\affiliation{\mbox{Department of Physics, University of Illinois at Urbana-Champaign, 1110 West Green Street, Urbana, IL 61801, USA}}

\author{M. Coleman Miller}
\affiliation{\mbox{Department of Astronomy, University of Maryland, College Park, MD 20742, USA}}

\author{Sharon Morsink}
\affiliation{\mbox{Department of Physics, 4-181 CCIS, University of Alberta, Edmonton, Alberta T6G 2E1, Canada}}

\author{Feryal \"Ozel}
\affiliation{\mbox{Steward Observatory, University of Arizona, 933 N. Cherry Avenue, Tucson, AZ 85721, USA}}

\author{Alessandro Patruno}
\affiliation{\mbox{Leiden Observatory, University of Leiden, P.O. Box 9513, 2300 RA Leiden, The Netherlands}}

\author{Juri Poutanen}
\affiliation{\mbox{Tuorla Observatory, Department of Physics and Astronomy, University of Turku, V\"ais\"al\"antie 20, FIN-21500 Piikki\"o, Finland}}

\author{Dimitrios Psaltis}
\affiliation{\mbox{Steward Observatory, University of Arizona, 933 N. Cherry Avenue, Tucson, AZ 85721, USA}}

\author{Achim Schwenk}
\affiliation{\mbox{Institut f\"ur Kernphysik, Technische Universit\"at
Darmstadt, 64289 Darmstadt, Germany}}
\affiliation{\mbox{ExtreMe Matter Institute EMMI, GSI Helmholtzzentrum f\"ur
Schwerionenforschung GmbH, 64291 Darmstadt, Germany}}

\author{Andrew W. Steiner}
\affiliation{\mbox{Department of Physics and Astronomy, University of Tennessee, Knoxville, Tennessee 37996, USA}}
\affiliation{\mbox{Physics Division, Oak Ridge National Laboratory, Oak Ridge, TN 37831, USA}}

\author{Luigi Stella}
\affiliation{\mbox{INAF/OAR, Via Frascati 33, I-00040, Monte Porzio Catone (Roma), Italy}}

\author{Laura Tolos}
\affiliation{\mbox{Instituto de Ciencias del Espacio (IEEC-CSIC), Campus UAB, Carrer de Can Magrans, s/n, 08193 Cerdanyola del Vall\'es, Spain}}

\author{Michiel van der Klis}
\affiliation{\mbox{Anton Pannekoek Institute, University of Amsterdam, PO Box 94249, 1090 GE Amsterdam, The Netherlands}}

\vspace*{1.5cm}

\begin{abstract}
One of the primary science goals of the next generation of hard X-ray timing instruments is to determine the equation of state of the matter at supranuclear densities inside neutron stars, by measuring the radius of neutron stars with different masses to accuracies of a few percent.  Three main techniques can be used to achieve this goal.  The first involves waveform modelling.  The flux we observe from a hotspot on the neutron star surface offset from the rotational pole will be modulated by the star's rotation, and this periodic modulation at the spin frequency is called a pulsation. As the photons propagate through the curved space-time of the star, information about mass and radius is encoded into the shape of the waveform (pulse profile) via special and general relativistic effects.  Using pulsations from known sources (which have hotspots that develop either during thermonuclear bursts or due to channelled accretion) it is possible to obtain tight constraints on mass and radius.  The second technique involves characterising the spin distribution of accreting neutron stars. A large collecting area enables highly sensitive searches for weak or intermittent pulsations (which yield spin) from the many accreting neutron stars whose spin rates are not yet known. The most rapidly rotating stars provide a very clean constraint, since the limiting spin rate where the equatorial surface velocity is comparable to the local orbital velocity, at which mass-shedding occurs, is a function of mass and radius.   However the overall spin distribution also provides a guide to the torque mechanisms in operation and the moment of inertia, both of which can depend sensitively on dense matter physics. The third technique is to search for quasi-periodic oscillations in X-ray flux associated with global seismic vibrations of magnetars (the most highly magnetized neutron stars), triggered by magnetic explosions.  The vibrational frequencies depend on stellar parameters including the dense matter equation of state, and large area X-ray timing instruments would provide much improved detection capability.  We illustrate how these complementary X-ray timing techniques can be used to constrain the dense matter equation of state, and discuss the results that might be expected from a 10m$^2$ instrument.  We also discuss how the results from such a facility would compare to other astronomical investigations of neutron star properties.   
\end{abstract}

\maketitle

\tableofcontents

\section{Supranuclear density matter}

\subsection{Introduction}  

\begin{figure}
\centering
\includegraphics[width=0.49\textwidth]{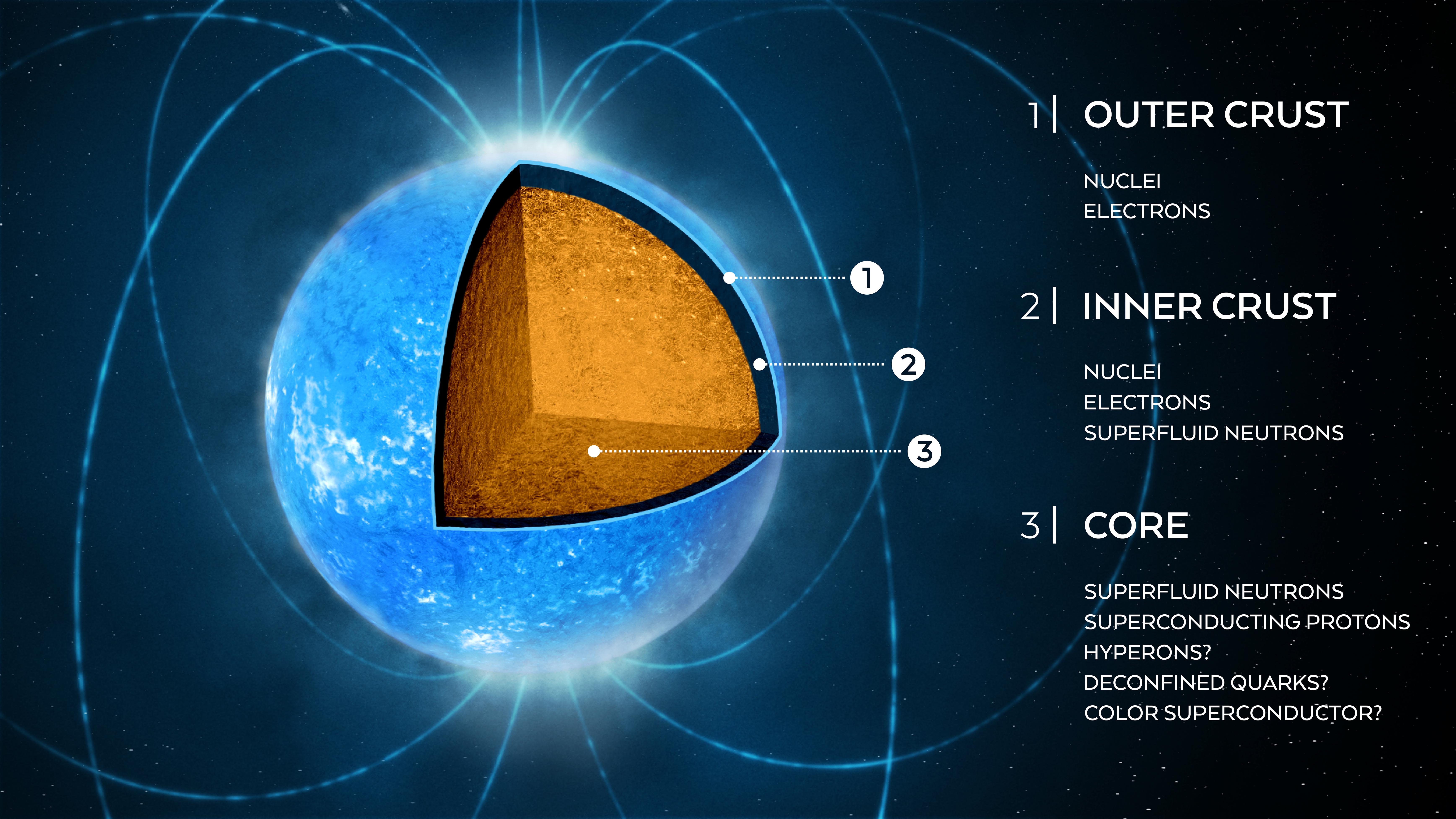}
\caption{Schematic structure of a neutron star. The outer layer is a solid ionic crust supported by electron degeneracy pressure. Neutrons begin to leak out of ions (nuclei) at densities $\sim 4 \times 10^{11} $ g/cm$^3$ (the neutron drip density, which separates inner from outer crust), where neutron degeneracy also starts to play a role.  At densities $\sim 2\times 10^{14}$ g/cm$^3$, the nuclei dissolve completely.  This marks the crust-core boundary. In the core, densities reach several times the nuclear saturation density $\rho_\mathrm{sat} = 2.8\times 10^{14}$ g/cm$^3$ (see text).}
\label{nscut}
\end{figure}

Neutron stars are the densest observable objects in the Universe, attaining physical conditions of matter that cannot be replicated on Earth. Inside neutron stars, the state of matter ranges from ions (nuclei) embedded in a sea of electrons at low densities in the outer crust, through increasingly neutron-rich ions in the inner crust and outer core, to the supranuclear densities reached in the center, where particles are squeezed together more tightly than in atomic nuclei, and theory predicts a host of possible exotic states of matter (Figure \ref{nscut}). The nature of matter at such densities is one of the great unsolved problems in modern science, and this makes neutron stars unparalleled laboratories for nuclear physics and quantum chromodynamics (QCD) under extreme conditions.

The most fundamental macroscopic diagnostic of dense matter is the pressure-density-temperature relation of bulk matter, the equation of state (EOS). The EOS can be used to infer key aspects of the microphysics, such as the role of many-body interactions at nuclear densities or the presence of deconfined quarks at high densities (Section \ref{nmat}).  Measuring the EOS of supranuclear density matter is therefore of major importance to nuclear physics.  However it is also critical to astrophysics.  The dense matter EOS is clearly central to understanding the powerful, violent, and enigmatic objects that are neutron stars. However, neutron star/neutron star and neutron star/black hole binary inspiral and merger, prime sources of gravitational waves and the likely engines of short gamma-ray bursts \citep{Nakar07}, also depend sensitively on the EOS \citep{Shibata11,Faber12,Bauswein12,Lackey12,Takami14}.  The EOS affects merger dynamics, black hole formation timescales, the precise gravitational wave and neutrino signals, any associated mass loss and r-process nucleosynthesis, and the attendant gamma-ray bursts and optical flashes \citep{Metzger10,Hotokezaka11,Rosswog15,Kumar15}.  The EOS of dense matter is also vital to understanding core collapse supernova explosions and their associated gravitational wave and neutrino emission \citep{Janka07}\footnote{Note that whilst most neutron stars, even during the binary inspiral phase, can be described by the cold EOS that is the focus of this Colloquium (see Section \ref{mrtoeos}), temperature corrections must be applied when describing either newborn neutron stars in the immediate aftermath of a supernova, or the hot differentially rotating remnants that may survive for a short period of time following a compact object merger.  The cold and hot EOS must of course connect and be consistent with one another.}.

\subsection{The nature of matter:  major open questions}
\label{nmat}

The properties of neutron stars, like those of atomic nuclei, depend
crucially on the interactions between protons and neutrons (nucleons)
governed by the strong force. This is evident from the seminal work of
Oppenheimer and Volkoff \citep{Oppenheimer39}, which showed
that the maximal mass of neutron stars consisting of non-interacting
neutrons is 0.7 M$_\odot$. To stabilize heavier neutron stars, as
realized in nature, requires repulsive interactions between nucleons,
which set in with increasing density. At low energies, and thus low
densities, the interactions between nucleons are attractive, as they
have to be to bind neutrons and protons into nuclei. However, to
prevent nuclei from collapsing, repulsive two-nucleon and
three-nucleon interactions set in at higher momenta and
densities.   Because neutron stars reach densities exceeding those in atomic nuclei, this makes them particularly sensitive to many-body forces \citep[see, for example,][]{Akmal98}, and recently it was shown that the dominant uncertainty at nuclear densities is due to three-nucleon forces \citep{Hebeler10b,Gandolfi12}

At low energies, effective field theories based on QCD provide a systematic basis for nuclear forces \citep{Epelbaum09}, which make unique predictions for many-body forces \citep{Hammer13} and
neutron-rich matter \citep{Tolos08,Hebeler14,Hebeler15}. While two-nucleon interactions are well constrained,
three-nucleon forces are a frontier in nuclear physics, especially for
neutron-rich nuclei \citep[see, for example,][]{Wienholtz13}. Such exotic nuclei are the focus of present and upcoming
laboratory experiments. Neutron star observations probe the same
nuclear forces at extremes of density and neutron richness. In
addition, to effective field theories, there are nuclear potential
models, such as the Argonne two-nucleon and Urbana/Illinois
three-nucleon potentials, which are fit to two-body scattering data and
light nuclei \citep{Gandolfi14,Carlson14}.

At high densities, neutron stars may be affected by exotic states of
matter. This regime is not accessible to first principle QCD
calculations due to the fermion sign problem \citep[see, for example, the discussions in][]{Hands07,Miller13a}. Therefore, at present, one has to resort to models, and experiment and observation are vital
to test theories and drive progress. In addition, recently,
perturbative QCD calculations have been performed at very high
densities (above 10 GeV/fm$^3$,  $\sim 64 \rho_\mathrm{sat}$), and used to interpolate to the EOS at low densities
\citep{Kurkela14}.

For symmetric matter (with an equal number of neutrons and protons) at the
nuclear saturation density $\rho_\mathrm{sat} = 2.8 \times 10^{14}$ g/cm$^3$ (the central density in very large nuclei when the Coulomb interaction is neglected)
there is a range of experimental constraints. This includes nuclear
masses and charge
radii~\citep[see, for example,][]{Klupfel09,Kortelainen10,Kortelainen14,Niksic15} as well
as giant dipole resonances and dipole
polarizabilities~\citep{Trippa08,Tamii11,Piekarewicz12}. Neutron-rich
matter can be probed by measuring the neutron skin thickness of heavy
nuclei \citep{Horowitz01,RocaMaza11}. However, all of these laboratory
experiments probe only matter at nuclear densities and
below. Low-energy heavy-ion collisions probe hot and dense matter, but
have uncontrolled extrapolations to zero temperature and to extreme
neutron-richness \citep{Tsang09}.  Neutron stars therefore provide a
unique environment for testing our understanding of the physics of the
strong interaction and dense matter.

\begin{figure}
\centering
\includegraphics[width=0.49\textwidth]{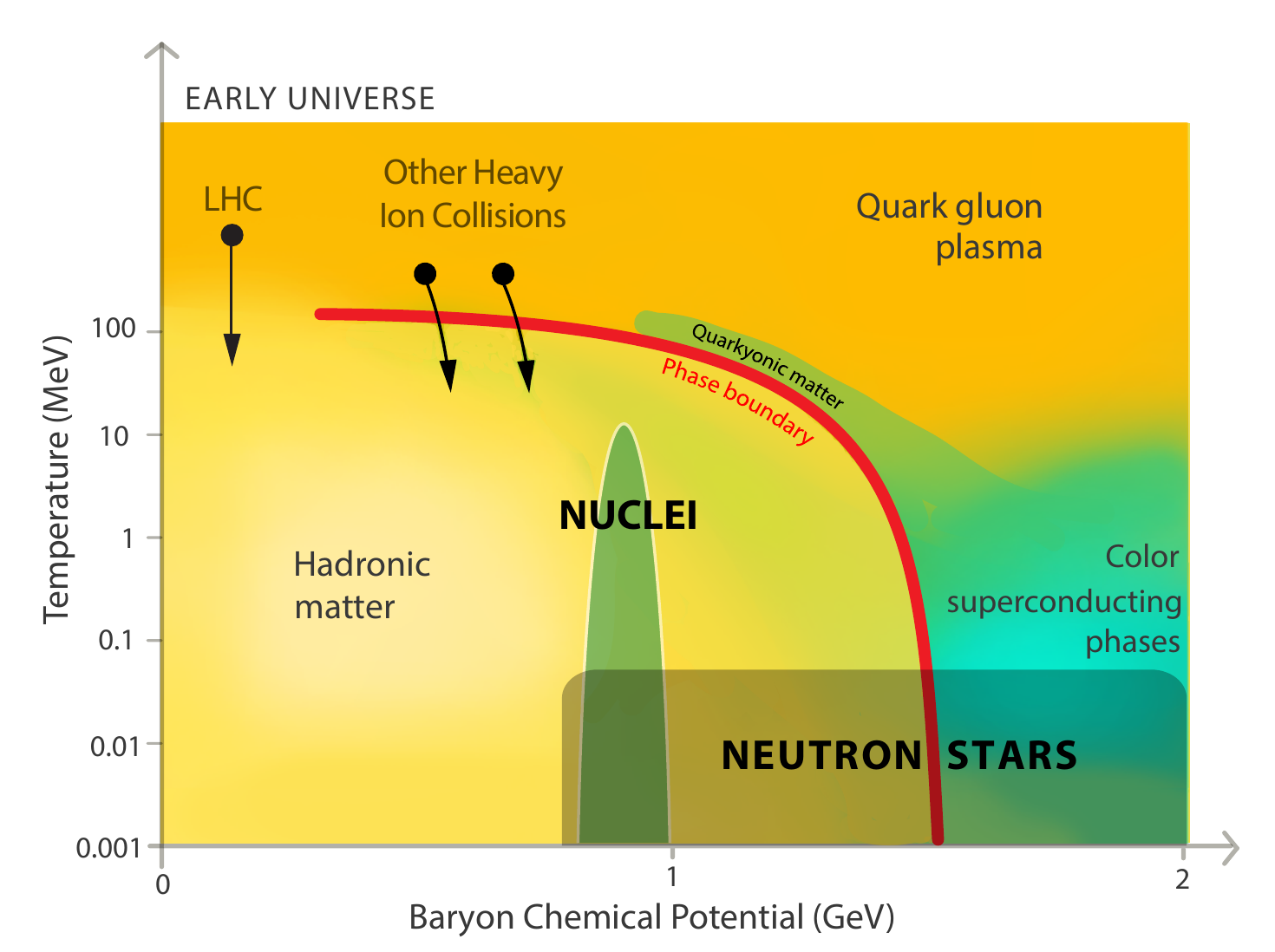}
\caption{Hypothetical states of matter accessed by neutron stars and 
current or planned laboratory experiments (Large Hadron Collider and other heavy ion collision experiments, shown by black arrows), in the parameter space of
temperature against baryon chemical potential (1-2 GeV corresponds to
$\sim$ 1-6 times the density of normal atomic nuclei). Quarkyonic
matter: a hypothesised phase where cold dense quarks experience
confining forces \citep{McLerran07,Fukushima11}. The stabilizing
effect of gravitational confinement in neutron stars permits
long-timescale weak interactions (such as electron captures) to reach
equilibrium, generating matter that is neutron rich (see Figure 2 of
\citealt{Watts14}) and may involve matter with strange quarks. This
means that neutron stars access unique states of matter that can only
be created with extreme difficulty in the laboratory: nuclear
superfluids, strange matter states with hyperons, deconfined quarks,
and color superconducting phases.}
\label{paramspace}
\end{figure}

At very high densities, possibly reached in neutron star cores,
transitions to non-nucleonic states of matter may occur. Some of
the possibilities involve strange quarks: unlike heavy-ion collision
experiments, which always produce very short-lived and hot dense
states, the stable gravitationally confined environment of a neutron
star permits slow-acting weak interactions that can form states of
matter with a high net strangeness. Strange matter possibilities
include the formation of hyperons \citep[strange baryons,][]{Ambartsumyan60,Glendenning82,Balberg99,Vidana15}, deconfined quarks \citep[forming a hybrid star,][]{Collins75}, or color superconducting
phases \citep{Alford08}. It is even possible that the entire star
might convert into a lower energy self-bound state consisting of up,
down and strange quarks, known as a strange quark star
\citep{Bodmer71,Witten84,Haensel86}. Other states that have been hypothesized
include Bose-Einstein condensates of mesons \citep[pions or kaons, the
latter containing a strange quark, see for example][]
{Kaplan86,Kunihiro93}. The densities at which such phases may
appear are highly uncertain.

Figure \ref{paramspace} compares the parameter space that can be
accessed within the laboratory to that which can be explored with
neutron stars. The physical ground state of dense matter is neutron
rich, which develops via weak interactions, and it is unbound so 
gravitational confinement is necessary to realize the ground state
of dense matter in nature. Only neutron stars sample this low
temperature regime of the dense matter EOS. The exotic non-nucleonic
states of matter described in the previous paragraph can be reached
only with extreme difficulty in the laboratory.

\subsection{Methodology: how neutron star mass and radius specify the EOS}
\label{mrtoeos}

The relativistic stellar structure equations relate the EOS to macroscopic observables including the mass $M$ and radius $R$ of the neutron star.  The dependence of the EOS on temperature can be neglected in computing bulk structure for neutron stars older than $\sim 100$ s: by this point the neutron star has cooled far below the Fermi temperature of the particles involved, the matter is degenerate, and hence temperature effects are negligible \citep[see for example][]{Haensel07}.  For non-rotating and non-magnetic stars, the classic Tolman-Oppenheimer Volkoff stellar structure equations would apply \citep{Tolman39,Oppenheimer39}.  However rotation is important, and the equations must be modified accordingly. For neutron stars spinning at a few hundred Hz the slow rotation (to second order) Hartle-Thorne metric is appropriate for most applications \citep{Hartle68}.  One can also compute full GR models for stars spinning at up to break-up speeds using a variety of methods implemented in well-tested codes \citep[for a review see][]{Stergioulas03}.  Codes such as \texttt{rotstar} \citep{Bonazzola98} and \texttt{rns} \citep{Stergioulas95} generate masses and radii for rapidly rotating neutron stars that are accurate to better than one part in $10^{-4}-10^{-5}$. 

\begin{figure*}
\centering
\includegraphics[width=\textwidth]{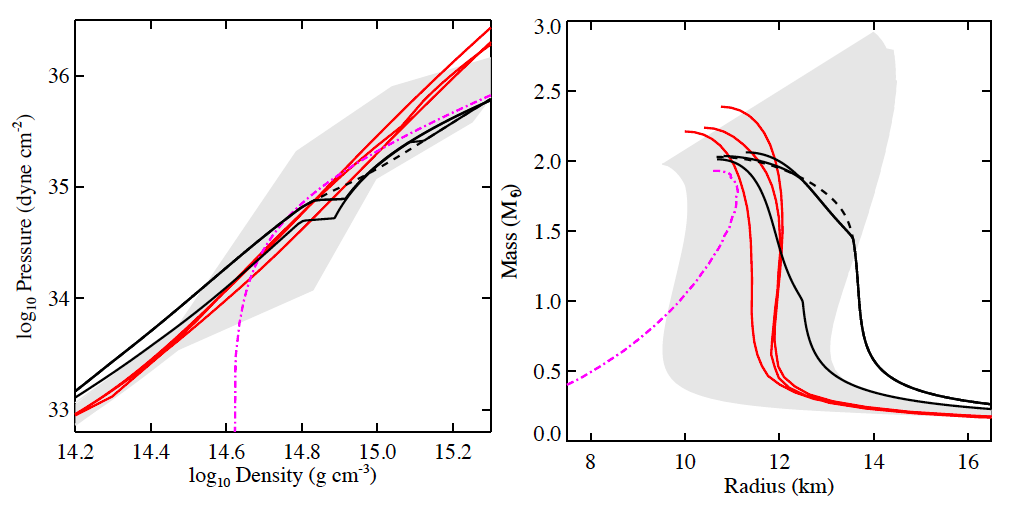}
\caption{The pressure density relation (EOS, left) and the corresponding $M$-$R$ relation (right) based on models
with different microphysics. Red: nucleonic EOS from \citet{Lattimer01}. Black solid: Hybrid models (strange quark core; \citealt{Zdunik13}). Black dashed: Hyperon core models \citep{Bednarek12}. Magenta: A self-bound strange quark star model \citep{Lattimer01}. Grey band: range of a parameterized family of nucleonic EOS based on chiral effective field theory at low densities, which provides
a systematic expansion for nuclear forces that allows one to estimate the theoretical uncertainties involved, combined with using a general extrapolations to high densities \citep[see Figure 12 of][for examples of specific representative EOS lying within this band]{Hebeler13}.}
\label{eos2mr}
\end{figure*}

There is a one to one map from the EOS to the $M$-$R$ relation \citep[see, for example,][]{Lindblom92}. Some examples are shown in Figure \ref{eos2mr}.  A few general features are worthy of note.  For each EOS there is a maximum mass that is a direct consequence of General Relativity \citep[for a review of this topic see][]{Chamel13}, and there are plausible astrophysical mechanisms (formation or accretion) that might lead to this being reached in real neutron stars.  The minimum observable mass, by contrast, is more likely to be set by evolution than by stability.   Radius tends to reduce as mass increases (although for some EOS models, radius increases slightly with increasing mass in the mid-range of masses), and current models suggest radii in the range 8-15 km for masses above 1 M$_\odot$.   

In terms of dependence on the nuclear physics, the maximum mass is determined primarily by the behavior of the cold EOS at the very highest densities \citep[$\sim 5-8 \rho_\mathrm{sat}$,][]{Lattimer05,Ozel09,Read09,Hebeler13}. The presence of non-nucleonic phases (such as hyperons or condensates) softens the EOS, reducing pressure support and leading to a smaller maximum mass. Radius on the other hand, depends more strongly on the behavior of the EOS at $\sim (1-2) \rho_\mathrm{sat}$ \citep{Lattimer01}. The nucleonic EOS at these densities is highly sensitive to three-nucleon forces \citep{Hebeler10,Gandolfi12}, whilst the presence of non-nucleonic phases tends to reduce $R$. The slope of the $M$-$R$ relation (i.e. whether $R$ increases or decreases with $M$), for masses $\gtrsim 1.2$ M$_\odot$ (the observed minimum, consistent with expectations from formation models), depends on the pressure at $\sim 4 \rho_\mathrm{sat}$ \citep{Ozel09}.

In testing EOS models, there are two potential approaches.  One is simply to compute, for a given EOS model, the resulting $M$-$R$ relation, and then determine the likelihood of obtaining the measured values of $M$, $R$ (with uncertainties) if this model is correct.   The other option is to perform the inverse process, and to map from the measured values of $M$-$R$ (with their uncertainties) to the EOS.  The first attempt to address this problem, which made no assumptions about the form of the EOS, was made by \citet{Lindblom92}.  Since then the approach has been refined by several authors.  Newer analyses rely on parameterized representations of the EOS that are a good characterization of many specific EOS models, but contain the lowest possible number of adjustable parameters (e.g. piecewise polytropic fits as employed by \citet{Ozel09,Read09,Steiner10}, or spectral representations as employed by \citet{Lindblom12,Lindblom14}).  \citet{Ozel09} showed that given three measurements of $M$, $R$ with accuracies of $\sim 5$\%, current EOS models could be distinguished at the 3$\sigma$ level using a piecewise-polytropic representation.    Using spectral representations, \citet{Lindblom12} showed that the inversion process itself (reliance on discrete measurements, and the use of a generalized model) introduces errors that are typically less than $\sim 1$ \%\footnote{This was the case for all models tested apart from those with a very strong phase transition, which are not well described by spectral representations with only a few parameters, and are better tested using the alternative method.}.  At this level the anticipated measurement errors on $M$, $R$ (which are at the few percent level) would be the primary determinant of the uncertainty on the inferred EOS.

\subsection{Current observational constraints on the cold dense EOS}
\label{currentobs}

The cleanest constraints on the EOS to date have come from radio pulsar timing, where the mass of neutron stars in compact binaries can be measured very precisely using relativistic effects \citep{Lorimer08}. Since any given EOS has a maximum stable mass (Figure \ref{eos2mr}), high mass stars can rule out particular EOS.   The most massive pulsars have masses $\approx 2$ M$_\odot$ \citep{Demorest10,Antoniadis13}, and that these results would have an impact on EOS studies was immediately clear \citep{Demorest10,Ozel10}.    The requirement to generate neutron stars with masses of at least  2 M$_\odot$ is now an integral part of EOS model development.  \citet{Hebeler13} and \citet{Lattimer14}, for example, have combined insights from  nuclear physics with the new maximum mass requirement in their models. Meanwhile \citet{Kurkela14} have generated models that are constrained to approach the EOS of quark matter at high density computed from state of the art perturbative QCD calculations (which are robust in the very highest density regime).   

One of the largest impacts of the maximum mass measurements has been on studies of hyperons. The presence of hyperons in neutron stars is energetically probable, but should induce a strong softening of the EOS that would lead to maximum masses below 2 M$_\odot$. The solution of this {\it hyperon puzzle} \citep[see for example][]{Chen11,Weissenborn12} requires some additional repulsion to make the EOS stiffer. Possible mechanisms include stiffer hyperon-nucleon and/or hyperon-hyperon interactions, the inclusion of three-body forces with one or more hyperons \citep{Takatsuka08,Logoteta13}, or the appearance of a phase transition to quark matter (see discussion of hybrid stars in Section \ref{nmat}). 

Whilst measuring masses using radio pulsars yields very precise results, measuring radius (or mass and radius simultaneously) is more challenging and, with the methods currently in use, more model-dependent. 
Current efforts to constrain the full $M$-$R$ relation focus primarily on spectroscopic measurements of the surface emission from accreting neutron stars in quiescence \citep{Rutledge99,Heinke03,Heinke06,Webb07,Servillat12,Catuneanu13,Guillot13,Guillot14,Heinke14}, and when they exhibit thermonuclear (Type I) X-ray bursts due to unstable nuclear burning in accreted surface layers \citep{vanParadijs79,vanParadijs87,Ozel06,Ozel09a,Guver10a,Guver10b,Suleimanov11,Ozel12,Guver13}. 

The essence of the method is the fact that the surface spectrum is
close to a diluted blackbody with a color temperature that is larger
than the effective temperature of the star $T_{\rm c}=f_c T_{\rm eff}$
by the color-correction factor $f_{\rm c}\approx 1.3-2$ \citep{London86,Zavlin96,Madej04,Heinke06,Suleimanov11b,Suleimanov12}.  This factor depends on the
atmospheric composition and on the effective surface gravity. For a slowly
spinning neutron star at a known distance $D$, measuring the flux $F$
of surface emission and the color temperature $T_{\rm c}$ gives a relation between $M$ and $R$:
\begin{equation}
R^2\left(1-\frac{2GM}{Rc^2}\right)^{-1}=\frac{F D^2 f_{\rm c}^4}
{\sigma T_{\rm c}^4}\;.
\end{equation}
where $\sigma$ is the Stefan-Boltzmann constant.  Spectroscopic measurements using quiescent neutron stars, for which no additional information is available, result only in broad
constraints between the neutron star masses and radii \citep[see][]{Heinke14,Guillot14}, although ensemble constraints may be tighter \citep{Ozel15}.

Studies of the spectra during thermonuclear X-ray bursts allow 
additional constraints to be obtained that can break the degeneracy
between $M$ and $R$ \citep[see reviews by][]{Lewin93,Ozel06}.  
The radius expansion bursts (during which radiation forces lift
the neutron star photospheres) serve as a good laboratory
because, from the burst flux $F_{\rm Edd}$ at the touchdown point,
one can get a measurement of the Eddington luminosity, which is a
different function of $M$ and $R$
\begin{equation}
L_{\rm Edd} = 4\pi D^2 F_{\rm Edd} = \frac{4\pi GMc}{\kappa_e} \left(
1-\frac{2GM}{Rc^2}\right)^{1/2}\;.
\end{equation}
Here $\kappa_e=0.2(1+X)$~cm$^2$~g$^{-1}$ is the electron scattering
opacity and $X$ is the hydrogen mass function.  Combining this with
the measurement of the surface area based on the thermal flux leads to
a weakly correlated inference of both $M$ and $R$ \citep{Ozel09,Guver10a,Guver10b,Suleimanov11,Ozel12,Guver13,Poutanen14}.

Different spectroscopic methods and burst selection criteria have been used to obtain constraints on the EOS. \citet{Ozel10,Steiner10,Steiner13} find that large ($\ge 13$ km) radii are ruled out, while others \citep{Suleimanov11,Poutanen14} get radii in the range 11-15 km. Clearly, the $\approx 5\%$ accuracy in radius required to pinpoint the value of the pressure beyond the nuclear saturation density has not yet been achieved. 

Both source types are affected by uncertainties in the composition of the atmosphere and (in many cases) lack of prior knowledge of the distance to each source  \citep[see for example][]{Heinke14}\footnote{Note that in principle, spectroscopic measurements rely on the absolute flux calibration of X-ray telescopes.  However, a limited number of studies currently seem to indicate that there is no significant flux calibration bias \citep{Suleimanov11,Guver15}.}.  In the case of quiescent neutron stars, there are additional sources of uncertainty related to the composition of the intervening interstellar medium 
as well as the effects of residual accretion \citep{Heinke14,Guillot14}, but see also \citet{Bahramian15}. In the case of the X-ray
bursters, there are several uncertainties: a systematic spread in the angular size of the source and the Eddington flux \citep{Guver12a,Guver12b} which may reflect some level of non-uniform emission over the stellar surface;  identification of the touchdown point \citep{Galloway08b,Guver12a,Miller13a}; and the role of accretion \citep{Kajava14,Poutanen14}.

Perhaps more importantly, both types of spectroscopic measurements of
neutron star masses and radii rely on the particular astrophysical
interpretations of two types of observations: that the quiescent
emission is powered by deep crustal heating and that the Eddington critical flux can be estimated accurately from the data.
The timing techniques discussed later in this article, on the other
hand, do not rely on these interpretations and, therefore, provide an
independent measurement of masses and radii, with orthogonal
systematic uncertainties and biases. Obtaining multiple measurements
of masses and radii with different techniques, and in many cases for
the same sources, will allow us to address and correct for the
systematic uncertainties of each method.

Timing-based techniques for constraining $M$ and $R$ rely on the presence of surface inhomogeneities, leading to emission that varies periodically as the star rotates. Three types of neutron star systems are suitable for a timing analysis, and several attempts have already been made:  for the accretion-powered pulsars \citep{Poutanen03,Leahy04,Leahy09,Leahy11,Morsink11}; for accreting neutron stars that show oscillations during their thermonuclear X-ray bursts \citep{Bhattacharyya05c}; and for rotation-powered X-ray pulsars \citep{Bogdanov07,Bogdanov08,Bogdanov09,Bogdanov13}.  All of the $M$-$R$ constraints coming from these observations to date have very large error bars. However the method has great promise (see Section \ref{pulse}). 

Laboratory experiments have also provided some constraints on the EOS.  Neutron star radii, for example, have been shown to depend strongly on the density dependence of the nuclear symmetry energy close to $\rho_\mathrm{sat}$ \citep[see, for example,][]{Lattimer14}, and the behaviour of this quantity for densities up to $\rho_\mathrm{sat}$ is now being probed by laboratory experiments \citep{Tsang12}.  Other experimental observables, such as K$^+$ meson production in nuclear collisions at subthreshold energies \citep{Sturm01} and the nuclear elliptic flow in heavy ion collisions \citep{Danielewicz02}, have been used as a sensitive probe for the stiffness of nuclear matter for high densities.  K$^+$ meson production seems to suggest a soft EOS for two to three times saturation density.  Indeed, these results on heavy-ion collisions have been used
to constrain the features of neutron stars \citep{Sagert12}.  Meanwhile results on elliptic flow rule out strongly repulsive nuclear EOS from relativistic mean field theory and weakly repulsive EOS with phase transitions at densities less than three times that of stable nuclei, but not EOS softened at higher densities due to a transformation to quark matter.  However, heavy-ion observables are obtained in highly-symmetric systems at high temperatures, thus the analysis of the EOS for asymmetric zero-temperature systems must be treated with caution. A reliable analysis of the EOS for zero temperature in asymmetric matter can be only achieved by neutron star measurements.

\subsection{Future observational constraints on the cold dense EOS}

The dense matter EOS will be a target science area for a number of different telescopes over the next decade, operating in very different wavebands.  To set the techniques that can be exploited in the hard X-ray band in context, we first review the advances that are expected in other wavebands. 
     
The next decade will see a major expansion in our ability to detect galactic radio pulsars, with the advent of the Square Kilometer Array \citep[SKA,][]{Bourke15} and its pathfinders LOFAR \citep{vanHaarlem13}, ASKAP \citep{Schinckel12} and MeerKAT \citep{Booth12}. The SKA is both more sensitive than current telescopes and will have increased timing precision.  By finding more radio pulsars in binary systems, and being able to determine post-Keplerian orbital parameters more precisely, the SKA should increase the number of measured neutron star masses by a factor of at least $\sim 10$ \citep{Watts14}.  New EOS constraints will result if the maximum mass record is broken, as discussed in Section \ref{currentobs}.   Radio observations can also deliver radius via measurements of the neutron star moment of inertia (determined from spin-orbit coupling).  The moment of inertia of the only known double pulsar system PSR J0737-3039 \citep{Burgay03}, will be determined to within 10\% within the next 20 years \citep{Lattimer05b,Kramer09}, resulting in a constraint on $R$ $\sim 5$\% \citep[see also the discussion in Section 4.2 of][]{Watts14}.  The SKA may discover more systems for which a measurement of moment of inertia is possible, although it will be challenging since the requirements on system geometry are quite restrictive \citep{Watts14}.  

The upgraded gravitational wave telescopes Advanced LIGO \citep{LIGO15} and Advanced VIRGO \citep{Acernese15} begin to enter service from 2015 and will operate well into the next decade. Gravitational waves from the late inspirals of binary neutron stars are sensitive to the EOS, with departures from the point particle waveform constraining $M$ and $R$. Global seismic oscillations excited by coalescence also depend on the EOS \citep{Bauswein12}.  Estimates are that Advanced LIGO and VIRGO could achieve uncertainties of ~10\% (1$\sigma$) in $R$, for the closest detected binaries \citep{Read13}, although event rates are highly uncertain. 

NICER-SEXTANT is a NASA Explorer Mission of Opportunity experiment that is due to be mounted on the International Space Station in late 2016 \citep{Arzoumanian14}.  NICER has an effective area that is 0.2~m$^2$ at 2~keV dropping to 0.06~m$^2$ at 6~keV.  The primary focus of NICER will therefore be on soft X-ray sources, in particular rotation-powered pulsars that emit in both the X-ray and radio bands.   The X-ray emission from rotation-powered pulsars is expected to be steady, which means that long accumulation of data is possible, even though they are dim. Radii will then be inferred using soft X-ray waveform modeling \citep[][and see Section \ref{pulse}]{Bogdanov08}.  If the mass of a neutron star and the pattern of radiation from its surface are known accurately a priori, NICER observations will
achieve an accuracy of $\simeq 2$\% in the measurement of  radius \citep{Gendreau12,Bogdanov13}. In practice, the measurement will limited by uncertainties in these two
requirements.   The uncertainty in the mass measurement of
NICER's primary target, the bright pulsar PSR~J0437$-$4715, is $\sim$5\% \citep{Reardon15}. Other main targets of NICER will be the nearby isolated radio millisecond pulsars PSR J0030+0451 and J2124-3358, however these will produce less stringent constraints since there is no prospect for measuring their masses. The temperature profile on the
neutron star surface, which is believed to be determined by the flux
of return currents circulating in the magnetosphere \cite[see e.g.][]{Bai10,Philippov14} is also highly uncertain, and development of ab initio numerical models that might fully address these questions is still a work in progress.  Waveform modelling for accretion-powered millisecond pulsars may also be possible \citep[see the discussion in][and Section \ref{pulse}]{Gendreau12}.

Athena, the soft X-ray observatory selected for the European Space Agency's L2 launch slot in the late 2020s, has an effective area requirement of 2 m$^2$ at 1 keV, dropping to 0.25 m$^2$ at 6 keV \citep{Nandra13}.  Dense matter is not a primary science goal for Athena.  However it could in principle be used for waveform modelling for the same isolated X-ray pulsars as NICER, and spectral modeling of neutron stars in quiescence or the cooling tails of X-ray bursts \citep{Motch13}.  These techniques have already been discussed in detail both above and in Section \ref{currentobs}.    

ASTROSAT \citep{Singh14}, a multiwavelength Indian astronomy
satellite was launched in September 2015. LAXPC, a 3--80 keV X-ray
timing instrument, has an effective area of 0.8 m$^2$ in the 5--20 keV
range.  SXT, a co-aligned 0.3-8 keV X-ray imager, has an effective
area of 0.01 m$^2$ in the 1--2 keV range.  This combination is
well-suited for spectral modeling of the cooling tails in X-ray
bursts.  LAXPC can also in principle be used for waveform modeling of
accretion-powered millisecond pulsars or X-ray burst oscillations.

\section{Hard X-ray timing techniques that deliver $M$ and $R$}

Hard X-ray timing enables three primary techniques that have the capability of delivering $M$ and $R$:  waveform (or {\it pulse profile}) modeling, spin measurements, and asteroseismology.  These techniques involve different classes of neutron star: accreting neutron stars with thermonuclear bursts, accretion-powered X-ray pulsars, and isolated highly magnetic neutron stars known as magnetars.  The use of multiple techniques and different source types allows cross-calibration of techniques, and independent cross-checks on the EOS.  In this Section we will explore each technique in turn, and discuss the contraints on the EOS that would be achievable with a large area  ($\sim 10$ m$^2$) hard X-ray (2-30 keV) timing telescope.  Much of the work presented here was developed as part of the science case for the proposed Large Observatory for X-ray Timing \citep[LOFT:][and the LOFT ESA M3 Yellow Book, \texttt{http://sci.esa.int/loft/53447-loft-yellow-book}]{Feroci12,Feroci14}.  

\subsection{Waveform modelling}

Millisecond X-ray oscillations are observed from accretion-powered pulsars \citep{Patruno12}, from some thermally-emitting rotation-powered (non-accreting) pulsars \citep{Becker01} and during some thermonuclear bursts on accreting neutron stars \citep[burst oscillations, see][]{Watts12}.  These oscillations are thought to be produced by X-ray emission from a region on the surface of the star that is hotter than the rest of the stellar surface and is offset from the rotational pole of the star.  As the hotter region rotates around the star, the X-ray flux seen by a distant observer is modulated at or near the rotation frequency of the star (Figure \ref{pulse}).   Accretion-powered pulsations are formed as material is channeled onto the magnetic poles of accreting X-ray pulsars.  X-ray emission comes from both a hotspot that forms at the magnetic poles, and from the shock that forms just above the star's surface as the channeled material decelerates abruptly.  Burst oscillations, by contrast, are due to hotspots that form during thermonuclear X-ray bursts on accreting neutron stars \citep{Lewin93,Galloway08}.  Ultimately one would hope to be able to use both types of pulsation when modelling the resulting pulse profile, or waveform, to recover $M$ and $R$: several sources show both phenomena, providing an important cross-check on the results.   

\subsubsection{Factors affecting the waveform}
\label{factorswaveform}

As the photons propagate through the curved space-time of the star, information about $M$ and $R$ is encoded into the shape and energy-dependence of the waveform.  General relativistic light-bending, which depends on compactness $M$/$R$, affects the amplitude of the pulsations.  Special relativistic Doppler boosting, aberration, and the magnitude of time delays depend on the relative orientation of the hot spot and
the line of sight \citep{Braje00}. This introduces a number of effects that lead to a direct measurement of the neutron star
radius. 

In the absence of special relativity, the peak flux occurs when the hot spot is facing the observer and the time for the
flux to rise from minimum to maximum is the same as the time to fall from maximum to minimum. The Doppler boosting effect makes the
blueshifted side of the star appear brighter than the redshifted side, so that the flux maximum occurs earlier in phase than if the special
relativistic effects did not occur. This asymmetry between the rise and fall time is approximately proportional to the projected line-of-sight velocity. Since the angular velocity is known from the pulse frequency, this provides a constraint on $R$ if the inclinations of the
spot and observer are known. The great advantage of the very rapidly rotating neutron stars (with spin frequencies of 400 Hz and larger) is that the Doppler boosting effect is more pronounced in the data which reduces degeneracies. The asymmetry in the rise/fall times is independent of the normalization, so errors in flux calibration are not important.

\begin{figure*}
\centering
\includegraphics[width=\textwidth]{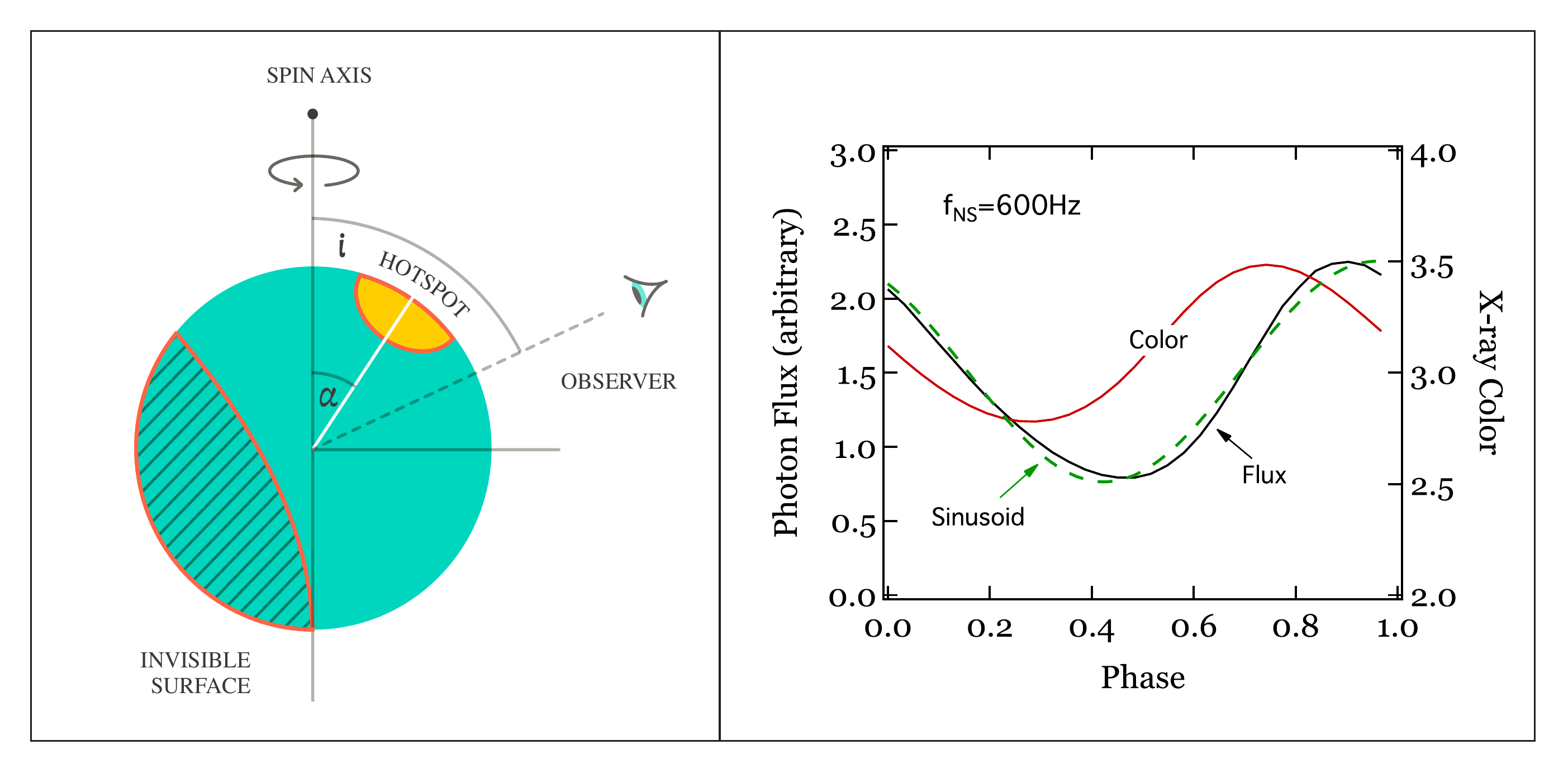}
\caption{Left: As the neutron star rotates, emission from a surface hotspot generates a pulsation. The figure shows observer inclination $i$ and hotspot inclination $\alpha$.  The invisible surface is smaller than a hemisphere due to relativistic light-bending.  Right: A schematic from \citet{Psaltis14b} to illustrate the effects on the waveform arising from relativistic effects: the waveform is modified from a pure sinusoid, and the temperature (indicated by the color, which is the ratio of the number of photons with energies above to those below the blackbody temperature) also varies even though the underlying hotspot has a uniform temperature. The shape of the waveform and its energy dependence can be used to recover $M$ and $R$. See also \citet{Viironen04} for examples of waveforms that include the expected variation in X-ray polarization.}
\label{pulse}
\end{figure*}

There are, of course, other factors that affect the waveforms and these must be taken into account when fitting for $M$ and $R$. The
geometric parameters $\alpha$, the angle between the spin axis and the spot center, and $i$, the observer's inclination angle (see Figure \ref{pulse}) introduce
degeneracies with $M$ and $R$. As the compactness ratio $M/R$ increases, the pulse fraction decreases. However a decrease in the quantity
$\sin(\alpha)\sin(i)$ also decreases the pulse fraction. Similarly an increase in $\sin(\alpha)\sin(i)$ increases the projected line-of-sight
velocity, leading to a degeneracy with the star's equatorial radius. Other geometric parameters such as the hotspot's shape and
size as well as emission from the rest of the star and the disk affect the shape of the light curve. Fortunately, the resulting parameter
dependencies can be resolved when the detailed structure of the waveforms and their dependence on photon energy is taken into account (see below), allowing us to recover $M$ and $R$.

A very important contribution to the parameter degeneracy is the beaming pattern of the radiation. A highly beamed emission pattern could, in some ways mimic the effects of decreased gravitational light bending and special relativistic Doppler boosting.  In the case of the thermal emission from X-ray bursts, this pattern is very well understood from theoretical modelling: it is very close to the limb-darkened pattern of a scattering dominated atmosphere \citep{Madej91,Suleimanov12}, and is not a significant source of uncertainty \citep{Miller13b}.  In contrast, in the case of the accretion-powered pulsations, the theoretical beaming pattern due to Compton scattering introduces a free parameter that is poorly constrained by the observations, leading
to the weak constraints on $M$ and $R$ that result for these stars \citep{Poutanen03, Leahy09, Leahy11, Morsink11}. The beaming pattern for the hydrogen atmosphere models used to compute the light curves of the rotation-powered pulsars \citep{Bogdanov08}  do not formally have free parameters, although there are still some open questions regarding the beaming from such atmospheres that are thought to be heated by relativistic particles from the magnetosphere.  

Additional complications to the waveform modeling from accreting sources come from the fact that the neutron star is surrounded by the accretion disk,  which may block radiation coming from the `southern hemisphere'.  For accreting pulsars, the eclipses by the disk may lead to appearance of strong harmonic structure to the waveform that completely dominates all other sources producing harmonics \citep{Poutanen08}. The evolution of the waveform from the best-studied pulsar SAX J1808.4$-$3658 during the outburst can be explained by varying just one parameter - the disk inner radius \citep{Poutanen09,Ibragimov09,Kajava11}. The way out is to either use data at high accretion rate, when the southern pole  is completely blocked, or to model the eclipse by the disk, but this requires knowledge of the inclination.  In addition the accretion steam may block hotspot emission at some phases,  introducing a dip \citep{Ibragimov09}. It is therefore important to use constraints from different techniques where possible to cross-check.

\subsubsection{Space-time of spinning neutron stars}

As mentioned in Section \ref{factorswaveform}, rapid rotation is desirable for waveform modelling since it can break degeneracies\footnote{The known accretion-powered pulsars and rapidly-rotating burst oscillation sources, the prime targets for this technique, have spin rates in the range 180-620 Hz \citep{Patruno12,Watts12}.  More rapid rotation rates are seen in the radio pulsar population, and rotation rates exceeding 1000 Hz are theoretically possible (see Section \ref{spinm})}.  Extensive work on gravitational lensing in spinning neutron star spacetimes has fully quantified the various levels of approximation  and their effects on the generation of pulse profiles.  The qualitative character of the general-relativistic effects depends on the ratio

 \begin{equation}
\frac{f_{\rm s}}{f_0}=0.24
\left(\frac{f_{\rm s}}{600~{\rm Hz}}\right)
\left(\frac{M}{1.8 ~\mathrm{M}_\odot}\right)^{-1/2}
\left(\frac{R}{10~{\rm km}}\right)^{3/2}\;,
\end{equation}
of the spin frequency of the neutron star, $f_{\rm s}$, and the characteristic frequency $f_0=\sqrt{GM/R^3}/(2\pi)$. To zeroth order in $f_{\rm s}/f_0$, the
neutron star is spherically symmetric and its external spacetime is described by the Schwarzschild metric, which depends only on the mass
of the star. Waveform calculations in this limit were performed in the pioneering work of \citet{Pechenick83}. A simple approximation for
the light-bending integral, introduced by \citet{Beloborodov02}, is also useful in many cases.

To first order in $f_{\rm s}/f_0$, the external spacetime of a slowly rotating neutron star and the exterior of the Kerr metric are the same \citep{Hartle67}.  In this case, the amount of gravitational lensing depends also on the spin angular momentum of the star (because of the effects of
frame dragging), which in turn depends on the density profile inside the star and hence on the equation of state. The effects of frame
dragging on gravitational lensing are negligible \citep{Braje00} and, therefore, the external spacetime (for the purpose of calculating
waveforms) is still well-described by the Schwarzschild metric. However, at the same order, special relativistic effects (Doppler boosts and aberration) as well as time delays become significant and can be calculated in the Schwarzschild + Doppler approximation \citep{Miller98,Poutanen03,Poutanen06}.  This Schwarzschild + Doppler approach has been shown to be an excellent approximation for spin frequencies less than about 300 Hz \citep{Cadeau07} through comparisons with waveforms calculated in exact, numerically generated neutron star spacetimes.

To second order in $f_{\rm s}/f_0$, the neutron star becomes oblate and its external spacetime is described by the Hartle--Thorne metric \citep{Hartle68},
which depends on the mass of the neutron star, on its spin angular momentum, and on its quadrupole mass moment. For spin frequencies
$\lesssim 600$~Hz, the effect of the stellar oblateness on the waveforms can be as large as 10-30\% \citep{Morsink07,Psaltis14a,Miller15} whereas the effect of the spacetime quadrupole is of the order of $1-5$\% \citep{Psaltis14a}. The
effects of the stellar oblateness alone can be incorporated into an approximation that makes use of the light-bending formula arising from
the Schwarzschild metric, and a shape function that only depends on $M/R$ and the spin frequency \citep{Morsink07,Baubock13,AlGendy14}.

Finally, to even higher orders in $f_{\rm s}/f_0$, the external spacetime of the neutron star depends on multipole moments that are of
increasing order. In principle the external spacetime of a rapidly spinning neutron star can be accurately described by the analytic
solution of \citet{Manko00a,Manko00b}, see \citep{Berti04,Berti05}.  However, the form of this metric is impractical for use in ray-tracing applications. On the other hand, spacetimes of rapidly spinning neutron stars can be calculated numerically \citep[see for example][]{Cook94a,Stergioulas95,Bonazzola98}.  Simulations of waveforms for such rapidly rotating neutron stars with particular choices of the equation of state have been performed by \citet{Cadeau07}.

\subsubsection{Inversion: from waveform to $M$ and $R$}

 The key question, when considering how to apply the waveform modelling technique in practice, is how many photons must be accumulated for a given source.  Some insight is given by the study of \citet{Psaltis14b}, which generated simulated waveforms under various assumptions (including that of a small hotspot - angular radius less than 20$^\circ$ - and isotropic emission from the hotspot), and studied the dependence of some key waveform properties on $M$, $R$ and other relevant parameters.  By using information about the shape and energy-dependence of the waveform, the different dependence on $M$ and $R$ of the various observables can in principle be used to break the degeneracies with the geometric parameters to recover $M$ and $R$.   This study also resulted in an order of magnitude estimate for the number of photons from the hotspot that would need to be accumulated in order to reach precisions of a few \% in $R$, $\sim 10^6$ counts.  

\begin{figure*}
\centering
\includegraphics{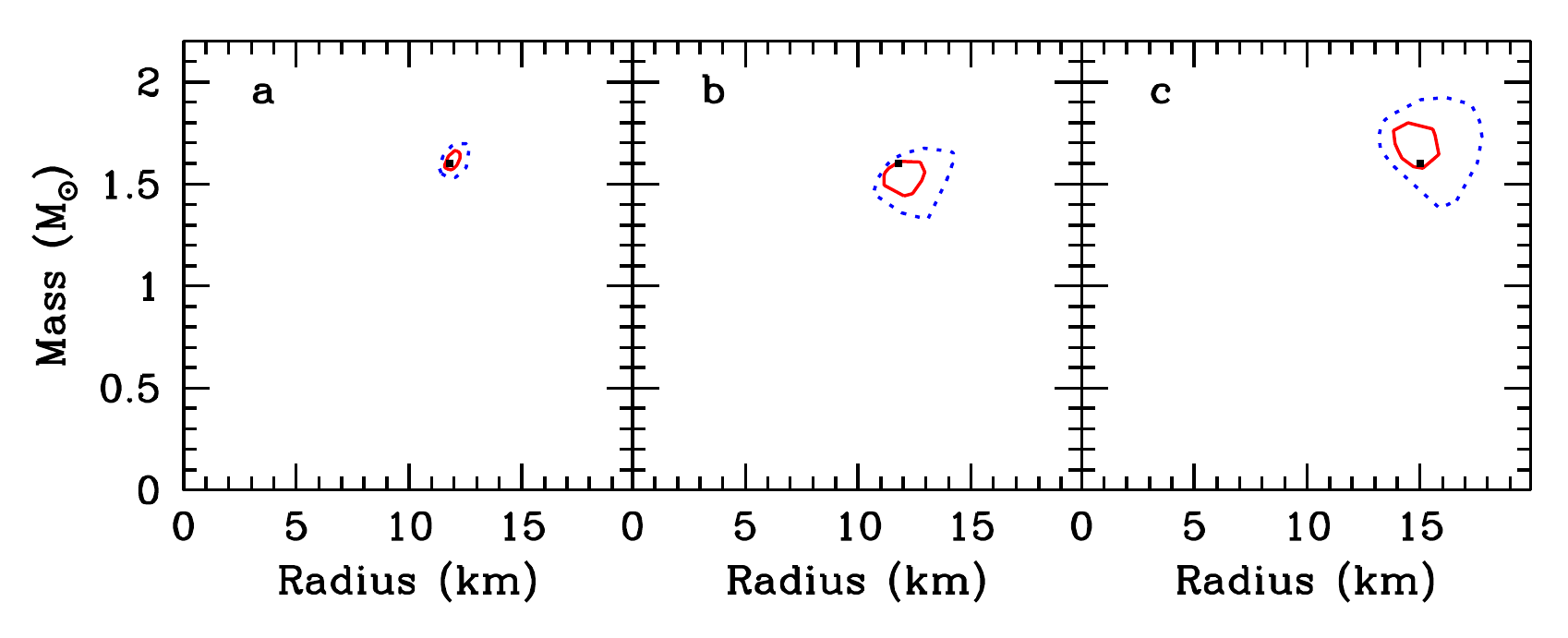}
\caption{Constraints on $M$ and $R$ obtained by fitting a waveform model to synthetic observed waveform data \citep[from][]{Miller15}. The black square indicates the mass and radius used to construct the synthetic data, the solid red lines show the 1$\sigma$ constraints, and the blue dashed lines show the 2$\sigma$ constraints. Here the spin rate $f_s$=600 Hz, hotspot inclination $\alpha$ and observer inclination $i$ are 90$^\circ$, the fractional rms amplitude of the oscillations is 10.8\%, and the total number of counts is $10^7$. $M$ and $R$ are tightly constrained: $\Delta M/M$=2.8 \% and $\Delta R/R$=2.9\%. The parameter values used to generate the waveform data used in Panel (b) are the same as in Panel (a), except the rotation rate, which is much lower (300 Hz), causing the constraints on $M$ and $R$ to be weaker: $\Delta M/M$=5.6\% and $\Delta R/R$=7.6 \%. Panel (c) shows that when the spot is at an intermediate colatitude (here 60$^\circ$), the constraints on $M$ and $R$ are weaker ($\Delta M/M$=6.5\% and $\Delta R/R$=6.7 \%), even if the star has a large radius (here 15 km) and is rapidly rotating (here, at 600 Hz).}
\label{contours}
\end{figure*}

The inversion problem has also been examined by \citet{Lo13} and \citet{Miller15}.  These studies employed a Bayesian approach and Markov Chain Monte Carlo sampling methods in an extensive parameter estimation study that estimated $M$ and $R$ by fitting waveform models to synthetic waveform data, and determined confidence regions in the $M$-$R$ plane. Some example results are shown in Figure \ref{contours}.  The uncertainties in $M$ and $R$ estimates are most sensitive to the stellar rotation rate (rapid rotation leading to smaller uncertainties), the spot inclination, and the observer inclination. They also depend on the background (be that from the accretion disk, astrophysical background, or instrumental background), but much more weakly \citep[see also][]{Psaltis14b}. This technique does not require knowledge of the distance because one fits the fractional amplitude of the pulsations, not the absolute value. \citet{Lo13} showed that for fixed values of the other system properties that affect the waveform, uncertainties in $M$ and $R$ estimates scale as ${\cal R}^{-1}$, where ${\cal R} \equiv N_{\rm osc}/\sqrt{N_{\rm tot}} = 1.4f_{\rm rms}\sqrt{N_{\rm tot}}$. Here $N_{\rm osc}$ is the total number of counts in the oscillating component of the waveform, $N_{\rm tot}$ is the total number of counts collected, and $f_{\rm rms}$ is the fractional rms amplitude of the oscillation. If the stellar rotation rate is $\gtrsim\,$300~Hz and the spot center and the observer's sightline are both within 30$^\circ$ of the star's rotational equator, burst oscillation waveform data with ${\cal R} \gtrsim\,$400 allow $M$ and $R$ to be determined with uncertainties $\lesssim\,$10\%.

A key result from the \citet{Lo13} and \citet{Miller15} studies is their conclusions on the effects of systematic errors. They considered the consequences of differences in the actual spot shape, beaming pattern and energy spectrum from what was assumed in the model.  None of these cases yielded simultaneously (1) a statistically good fit, (2) apparently tight constraints (at the desired few percent level) on $M$ and $R$, and (3) significantly biased masses and radii.  Thus if an analysis yields a good fit with tight constraints, the inferred mass and radius are reliable.  This statement is currently unique among proposed methods to measure neutron star radii.

\subsubsection{Instrument requirements and observing strategy}

It is the necessity of obtaining ${\cal R} \approx 400$ for systems with favorable geometry that drives the requirement for large-area instruments to properly exploit this technique.  The observing times necessary to achieve this have been studied in detail as part of the LOFT assessment process (see LOFT ESA M3 Yellow Book, \texttt{http://sci.esa.int/loft/53447-loft-yellow-book}), with observing strategy focusing on the burst oscillation sources given their well understood spectrum \citep{Suleimanov12} and very low ($\sim$ 1\%) atmospheric model uncertainties \citep{Miller13b}.  Using burst and burst oscillation properties observed with RXTE \citep[burst brightness, burst oscillation fractional amplitude, as summarized in][]{Galloway08} for the known burst oscillation sources (such as 4U 1636-536), it was shown that with a $\sim 10$ m$^2$ instrument one would need to combine data from $\sim 10-30$ bursts with oscillations to meet the required target\footnote{For a more detailed discussion of how one would combine data from different bursts, and how one can account for potential changes in the size and position of the hotspot, see \citet{Lo13}.}.  Given the percentage of bursts that show oscillations (which is not 100\%), and the mean burst rate \citep[again using properties observed with RXTE,][]{Galloway08} this would require observing sources for a few hundred ks, easily feasible within anticipated mission lifetimes.  Burst oscillations do seem to occur preferentially in certain accretion states \citep{Muno04}, and targeting these states with a suitable all-sky monitor would reduce the necessary observing time.  There are at present 27 known sources with burst oscillations and/or accretion-powered pulsations spinning at 100 Hz or faster.  Given that some are transient sources with long periods of quiescence, new discoveries are to be expected.   Having such a large number of sources to choose from will help to select a sample with optimal observational characteristics (such as flux, pulse amplitude and harmonic content), the latter being easily achievable. 

\begin{figure}
\centering
\includegraphics[width=0.49\textwidth]{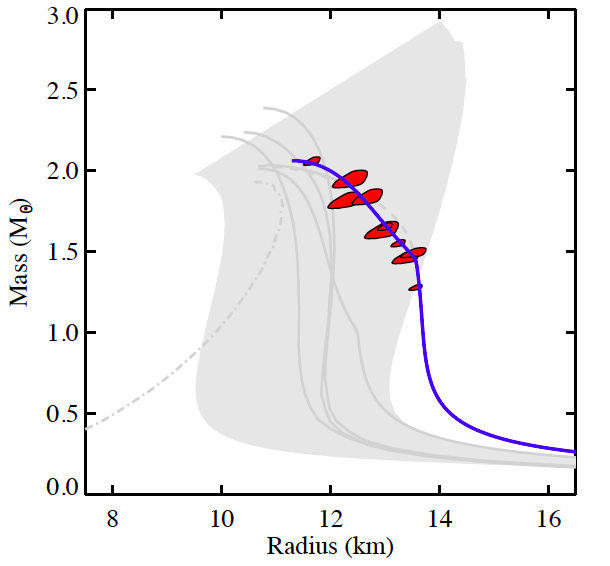}
\caption{$1\sigma$ confidence regions illustrating the constraints on the EOS that would be expected from a $\sim 10$ m$^2$ hard X-ray timing telescope. For the sake of illustration, the regions are assumed to trace the $M$--$R$ curve of a proposed EOS that produces a neutron star with a strange-matter core. The larger regions assume observing time has been dedicated to reach ${\cal R}\approx 400$ for the 50\% of known burst oscillation systems (here 10) that have optimal spot and observer inclinations ($i=60^\circ$--$90^\circ$) if their orientations are randomly distributed. The size and shape of these regions are from the Bayesian analyses of waveform data by \citet{Lo13,Miller15} that assume the hot spots and observers are in the equatorial plane and that there is no independent knowledge of any of the model waveform parameters. The smaller $1\sigma$ confidence regions show the constraints that could be achieved by a deep follow-up to halve the fractional uncertainties in $M$ and $R$ for the 3--4 stars that are most constraining, in terms of their locations in the $M$--$R$ plane, the potential for reducing their uncertainties, and the availability of complementary constraints from other observations. Constraints like those shown will exclude many of the EOS models shown in this figure. In this illustration, the models shown as grey lines are excluded. }
\label{ellipses}
\end{figure}

As indicated above, for favorable geometry that will be reflected in higher harmonic content of the pulsations, ${\cal R} \approx 400$ is sufficient to measure $M$ and $R$ to a few percent precision (Figure \ref{ellipses}).  Less favorable geometries would require more observing time, since errors on $M$ and $R$ scale roughly as the inverse square root of the total number of counts  \citep{Lo13,Psaltis14b}. Since a mix of geometries among sources is a reasonable expectation, observing strategy must be flexible enough to ensure that the goals can be met no matter what system geometries we encounter. Flexibility can also allow responsiveness to preliminary findings:  ideally one would schedule longer observations of the sources that are most constraining in the $M$-$R$ plane, in order to further reduce the size of their error ellipses. By thus tailoring observations one can confirm key findings at a much higher confidence. Figure \ref{ellipses} illustrates the type of constraints on the $M$-$R$ relation that could be delivered by such a strategy.  

Independent knowledge of any of the relevant parameters improves the uncertainties, with the biggest improvement coming from knowledge of the observer inclination.  There are very good prospects in the next $\sim 10$ years for determining the angle of our line of sight to the axis of the binary orbit using Fe line modeling \citep{Cackett10,Egron11}, Doppler shifting of burst oscillation frequencies \citep{Strohmayer02,Casares06}, and burst echo mapping \citep[using the time delay between X-ray burst emission from the neutron star and the optical echo of that emission as it is reprocessed by the surface of the companion star,][note that this requires simultaneous optical observations]{Casares10}.  Since mass transfer is expected to cause the stellar rotation axis to align with the orbital axis in our target systems \citep{Hills83,Bhattacharya91,Guillemot14}, this will yield the observer inclination. 

Using accreting sources has the advantage that it enables independent cross-checks. Several potential targets show both accretion-powered pulsations and burst oscillations, allowing checks using two independent waveform models.   One can also use the continuum spectral modelling constraints outlined in Section \ref{currentobs}, and there are potential new continuum fitting techniques that would be enabled by high quality spectra enabled from a large area detector \citep{Lo13}.  Observation of an identifiable surface atomic line in the hot-spot emission would also provide a tight contraint \citep[see for example][]{Rauch08}.   The rotational broadening of an atomic line depends on $R$, so combining this with its centroid (which depends on $M/R$) yields a measurement of $M$ and $R$ independently, modulo the unknown inclination \citep{Ozel03}.  To avoid the degeneracy with the inclination, one can also use the equivalent width of the line, which depends on the effective gravitational acceleration $g_\mathrm{eff} = GM (1-2GM/Rc^2)^{-1/2}/R^2$ \citep{Chang05}.  Combining these two pieces of information yields separate measurements of $M$ and $R$, that are complementary to and independent of the constraints obtained from waveform fitting.   Note that whilst the effects of frame dragging can be neglected \citep{Bhattacharyya05b}, the effect of stellar oblateness and the space-time quadrupole on the line profile must be taken into account for rotation rates $\gtrsim 300$ Hz \citep{Baubock13}.  \citet{Lo13} explore how the constraints that would result from the detection of an atomic line could be combined with the results of waveform fitting.  

\subsection{Spin measurements}
\label{spinm}

The spin distribution of neutron stars offers another way of constraining the EOS, and a large-area instrument, with a correspondingly high sensitivity to pulsations, offers a unique opportunity to fully characterise this function.  

\subsubsection{Rapid rotation}

\begin{figure}
\centering
\includegraphics[width=0.49\textwidth]{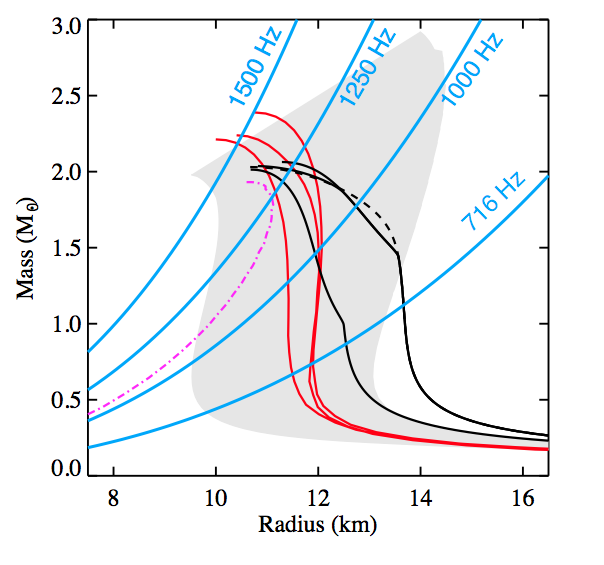}
\caption{Spin limits on the EOS, reproduced from \citet{Watts14}.  Neutron stars of a given spin rate must lie to the left of the relevant limiting line in the $M$-$R$ plane (shown in blue for various spins). The current record holder, which spins at 716 Hz \citep{Hessels06} is not constraining.  However given a high enough spin, individual EOS can be ruled out.  Between 1 kHz and 1.25 kHz, for example, some individual EOS in the grey band \citep[which represents a parameterized family of EOS models from][]{Hebeler13} would be excluded.}
\label{spinlim}
\end{figure}

At the very simplest level, one can obtain constraints from the most rapidly rotating neutron stars.  The limiting spin rate $f_\mathrm{max}$, at which the equatorial surface velocity is comparable to the local orbital velocity and mass-shedding occurs, is a function of $M$ and $R$ and hence fast spins constrain the EOS.  The mass-shedding frequency is given to good approximation \citep{Haensel09}  by the empirical formula 

\begin{equation}
f_\mathrm{max} \approx {\cal C} \left[\frac{M}{M_\odot}\right]^\frac{1}{2} \left[\frac{R}{10 ~\mathrm{km}} \right]^{-\frac{3}{2}} ~\mathrm{kHz}
\end{equation}
where $R$ is the radius of the non-rotating star of mass $M$.  Softer EOS have smaller $R$ for a given $M$ and hence have higher limiting spin rates.  More rapidly spinning neutron stars place increasingly stringent constraints on the EOS.  The deviation of ${\cal C}$ from its Newtonian value of 1.838 depends, in GR, \citep[as computed by][]{Haensel09} on the neutron star interior mass distribution.  For a hadronic EOS (one that consists of baryons or mesons), ${\cal C}$ = 1.08, whilst for a strange star with a crust, ${\cal C} \approx$ 1.15.   This can be recast as a limit on $R$

\begin{equation}
R<10 {\cal C}^\frac{2}{3} \left[\frac{M}{M_\odot}\right]^\frac{1}{3} \left[\frac{f_s}{1 ~\mathrm{kHz}}\right]^{-\frac{2}{3}} ~\mathrm{km} 
\end{equation}
which places a constraint on the EOS (Figure \ref{spinlim}). 

The most rapidly spinning neutron star known, a 716 Hz radio pulsar \citep{Hessels06}, is not spinning rapidly enough to be constraining.  However the discovery of a neutron star with sub-millisecond spin would pose a strong and clean constraint on the EOS.  What then are the prospects therefore for finding more rapidly spinning stars?  Since the standard formation route for the millisecond radio pulsars (MSPs) is via spin-up due to accretion \citep{Alpar82,Radhakrishnan82,Bhattacharya91}, it is clear that we should look in the X-ray as well as the radio, and theory has long suggested that accretion could spin stars up close to the break-up limit \citep{Cook94b}.  

\begin{figure}
\includegraphics[width=0.49\textwidth]{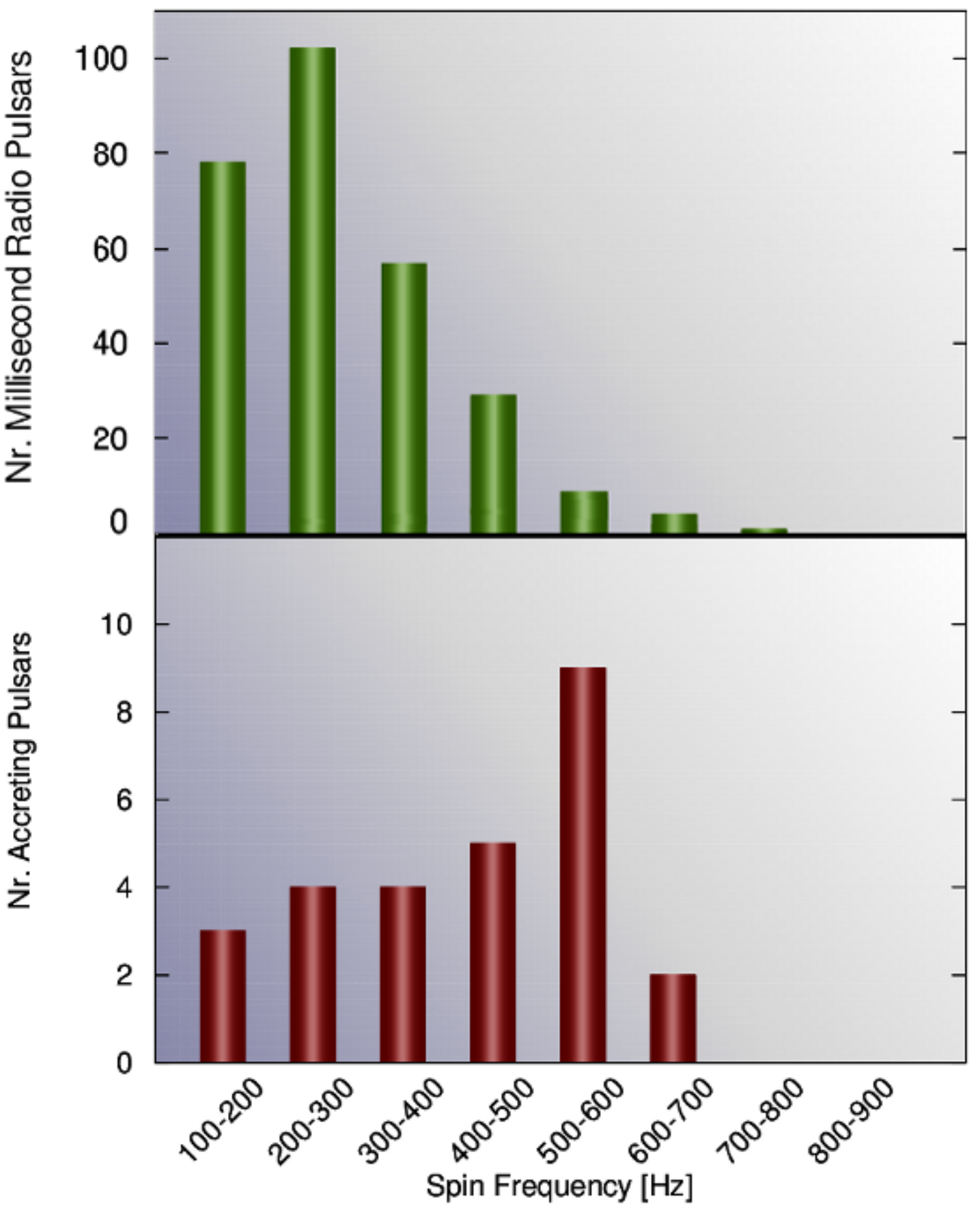}
\caption{Spin distributions of neutron stars with rotation rates above 100 Hz.  Top panel:  radio pulsars.  Lower panel:  accreting neutron stars (accretion powered millisecond pulsars and burst oscillation sources).}
\label{spindist}
\end{figure}

Figure \ref{spindist} shows the current spin distributions for the MSPs and their proposed progenitors, the rapidly spinning accreting neutron stars. For the far more numerous radio pulsars, there is a clear fall off in the distribution at high spin.  Until a few years ago, the sometimes prohibitive computational costs of large surveys sensitive to very rapidly rotating objects limited radio pulsar searches, and this is likely still reflected in the measured distributions \citep[for a more in-depth discussion, see][]{Watts14}.  The improvement in computational capabilities in recent years, and targeted searches for sources from the Fermi catalogues have led to the discovery of several new binary millisecond radio pulsars with spin frequencies above 500 Hz.  Interestingly the most rapidly rotating stars, including the 716 Hz spin record holder, appear to be in eclipsing systems, where matter from the companion obscures the radio pulsations for large fractions of the orbit.  This suggests that there may be an observational bias against finding these systems in the radio \citep[see][for further discussion]{Watts14}.  

For the accreting systems, it is clear that we are still in the regime of small number statistics: however the drop-off at high spin rates seen in the radio is not apparent.  There are moreover physical reasons why it may have been difficult to find the most rapidly-spinning accreting sources.  Rapid spin is most likely in sources that accrete at high rates, for example, and when accretion rate is high episodes of channeled accretion are expected to be intermittent \citep{Romanova08}.   Strong accretion may also drive a star towards alignment, making pulsations weak \citep{Ruderman91,Lamb09b}.   For bursters, the most rapid spins may suppress flame spread \citep{Spitkovsky02,Cavecchi13}, weakening and shortening bursts.  This means that the existence of rapidly spinning stars, that have been out of reach of current detectors, is certainly plausible both on observational and theoretical grounds.  

Accretion-powered pulsations and burst oscillations are strongest in the hard X-ray (2-30 keV) band, and sensitivity scales directly with signal to noise.  As such, a large area hard X-ray timing instrument is necessary to discover more neutron star spins.  Intermittent accretion-powered pulsations have already been detected from a number of rapidly spinning sources \citep{Galloway07,Casella08,Altamirano08}:  to detect more of these events, sensitivity to brief pulsation trains is key.  A 10 m$^2$ instrument with response similar to that of RXTE would be able to detect 100-s duration pulse trains down to an amplitude of $\sim$ 0.4\% rms for a 100 mCrab source (5$\sigma$), $\sim 100$ s being the duration of intermittent pulsations observed from Aql X-1 by \citet{Casella08}.   RXTE would have needed 15 times as long to reach the same sensitivity, so that 100 s pulse trains were severely diluted; longer pulse trains suffered from Doppler smearing.   Searches for weak (rather than intermittent) accretion-powered pulsations using the sophisticated techniques being used for the Fermi pulsar surveys \citep{Atwood06,Abdo09,Messenger11,Pletsch12}, which compensate for orbital Doppler smearing, would yield 5 $\sigma$ multi-trial pulsation sensitivities of $\sim 0.03-0.003$ \% (rms) in bright neutron stars ($>$ 100 mCrab) and $\sim 0.2-0.04$ \% in faint neutron stars (10-100 mCrab), taking into account expected observing times and prior knowledge of the orbits.   A 10 m$^2$ instrument would also be able to detect oscillations in individual Type I X-ray bursts down to amplitudes of 0.6\% rms (for a 1s exposure and a typical burst of brightness of 4 Crab); by stacking bursts sensitivity would improve.  

\subsubsection{Spin distribution and evolution}

Mapping the spin distribution more fully, so that the accreting neutron star sample is no longer limited by small number statistics, is also extremely valuable.  One of the big open questions in stellar evolution is how precisely the recycling scenario progresses - and whether it does indeed account for the formation of the entire MSP population.  The discovery of the first accreting millisecond X-ray pulsar by \citet{Wijnands98}, and the recent detection of transitional pulsars, that switch from radio pulsars to accreting X-ray sources \citep{Archibald09,Papitto13,Bassa14,Patruno14,Stappers14,Bogdanov14,Bogdanov15}, seems to confirm the basic picture.  However key details of the evolutionary process, in particular the specifics of mass transfer and magnetic field decay, remain to be resolved \citep[see for example the discussion in][]{Tauris12}. Comparison of the spin distributions of the MSPs and the accreting neutron stars is a vital part of that effort.  

The torques that operate on rapidly spinning accreting neutron stars also remain an important topic of investigation.  Accretion torques, mediated by the interaction between the star's magnetic field and the accretion flow \citep[first explored in detail by][]{Ghosh78,Ghosh79a,Ghosh79b}, clearly play a very large role \cite[see][for a review of more recent work]{Patruno12}.  There are also several mechanisms, such as core r-modes \citep[a global oscillation of the fluid, restored by the Coriolis force, see][for a recent review]{Haskell15} and crustal mountains \citep[see][for a recent review]{Chamel13}, that may generate gravitational waves and hence a spin-down torque.  These mechanisms are expected to depend in part on the EOS \citep[see, for example,][]{Ho11,Moustakidis15}.  In addition there are potential interactions between internal magnetic fields and an unstable r-mode that may be important \citep[see for example][]{Mendell01}, and the physics of the weak interaction at high densities also becomes relevant, since weak interactions control the viscous processes that are an integral part of the gravitational wave torque mechanisms \citep{Alford12}.  

Torque mechanisms can be probed in two ways:  firstly, by examining the maximum spin reached, which may be below theoretical break-up rates, since both magnetic torques and gravitational wave torques may act to halt spin-up \citep{Bildsten98,Lamb05,Andersson05}; and secondly by high precision tracking of spin evolution, enabled by increased sensitivity to pulsations.  Whilst extracting EOS information from the spin distribution and spin evolution will clearly be more challenging than the clean constraint that would come from the detection of a single rapid spin, it is nonetheless an important part of the models and one that can be tested.  Ultimately, more and better quality timing data are needed to confirm if it is, indeed, the magnetic field that regulates the spin of the fastest observed accreting neutron stars or if additional torques are needed. On the one hand, it has been argued that the spin-evolution during and following an accretion outburst of IGR~J00291+5934 is consistent with the `standard' magnetic accretion model \cite{Falanga05,Patruno10,Hartman11}. On the other hand, the results are not quite consistent and there is still room for refinements and/or additional torques \cite{Andersson14,Ho14}. Whether this means that there is scope for a gravitational-wave element or not remains unclear \cite{Ho11,Haskell12}, but a large area X-ray instrument should take us much closer to the answer.

Precision ephemerides from X-ray timing are very important enablers for simultaneous gravitational wave searches, since one has to fold long periods of data to detect the weak signals, and the gravitational wave frequency depends on spin rate in both mountain and r-mode mechanisms. Without such ephemerides, the number of templates that must be searched makes detection of continuous wave emission from sources like Sco X-1 very difficult \citep{Watts08b}. This is very clear when one compares the limits currently obtained for continuous wave gravitational wave searches where ephemerides are known \citep[the radio pulsars,][]{Abbott07,Abbott08} to those obtained for systems where the spin is not known (non-pulsing systems like Cas A, \citealt{Abadie10} and Sco X-1, \citealt{Aasi14}). A direct detection of gravitational waves from such a system would of course have immediate consequences for potential gravitational wave emission mechanisms, and any EOS dependence.

\subsection{Asteroseismology}

Asteroseismology is now firmly established as a precision technique for the study of the interiors of normal stars.  As such the detection of seismic vibrations in neutron stars was one of RXTE's most exciting discoveries.  They were found in magnetars, young, highly magnetized neutron stars that emit bursts of hard X-ray/gamma-rays powered by decay of the strong magnetic field \citep[see][for a review]{Woods06}.  What triggers the flares remains unknown, but most likely involves either starquakes or magnetospheric instabilities.  Rapid reconfiguration/reconnection powers the electromagnetic burst:  however the events are so powerful that it had already been suggested that they might set the star vibrating \citep{Duncan98}.   These vibrations, which manifest as Quasi-Periodic Oscillations (QPOs) in hard X-ray emission, were first detected in the several hundred second long tails of the most energetic giant flares from two magnetars \citep{Israel05,Strohmayer05,Strohmayer06a,Watts06}.  Similar QPOs have since been discovered during storms of short, low fluence bursts from several magnetars \citep{Huppenkothen13,Huppenkothen14a,Huppenkothen14b}.   The QPOs have frequencies that range from 18 to 1800 Hz.  

Seismic vibrations offer us a unique way to explore the interiors of neutron stars.  The QPOs were initially tentatively identified with torsional shear modes of the neutron star crust, and torsional Alfv\'en modes of the highly magnetized fluid core. These identifications were based on the expected mode frequencies, which are set by both the size of the resonant volume (determined by the star's radius) and the relevant wave speed. The fact that the oscillations must be computed in a relativistic framework introduces additional dependences, and for this reason they can be used to diagnose $M$ and $R$ (see for example \citealt{Samuelsson07} for relativistic crust modes, and \citealt{Sotani08} for relativistic core Alfv\'en modes). Seismic vibrations also take us beyond the simple $M$-$R$ relation, constraining the non-isotropic components of the stress tensor of supranuclear density material.

In fact, for a star with a magnetar strength magnetic field, crustal vibrations and core vibrations should couple together on very short timescales \citep{Levin07}. The current viewpoint is that the QPOs are associated with global magneto-elastic axial (torsional) oscillations of the star \citep{Glampedakis06,Lee08,Andersson09,Steiner09,vanHoven11,vanHoven12,Colaiuda11,Colaiuda12,Gabler12,Gabler13a, Passamonti13,Passamonti14,Asai14,Glampedakis14}. Since coupled oscillations depend on the same physics, they have frequencies in the same range as the natural frequencies of the isolated elements. 

Current magneto-torsional oscillation models can in principle easily explain the presence of oscillations at 155 Hz and below. Until recently there was a significant problem with the higher frequency QPOs, which appeared to persist much longer than the models predicted, but this has now been resolved \citep{Huppenkothen14c}. Issues currently being addressed include questions of emission \citep{Timokhin08,DAngelo12,Gabler14}, excitation \citep{Link14}, coupling to polar Alfv\'en modes \citep{Lander10,Lander11,Colaiuda12}, and resonances between the crust and core that might develop as a result of superfluid effects \citep{Gabler13b,Passamonti14}. The latter in particular can have a large effect on the characteristics of the mode spectrum, and since superfluidity is certainly present in neutron stars, mode models must start to take this into account properly before we can make firm mode identifications. What is now clear is that mode frequencies depend not only on $M$ and $R$, but also on magnetic field strength/geometry, superfluidity, and crust composition.

\begin{figure}
\centering
\includegraphics[width=0.49\textwidth]{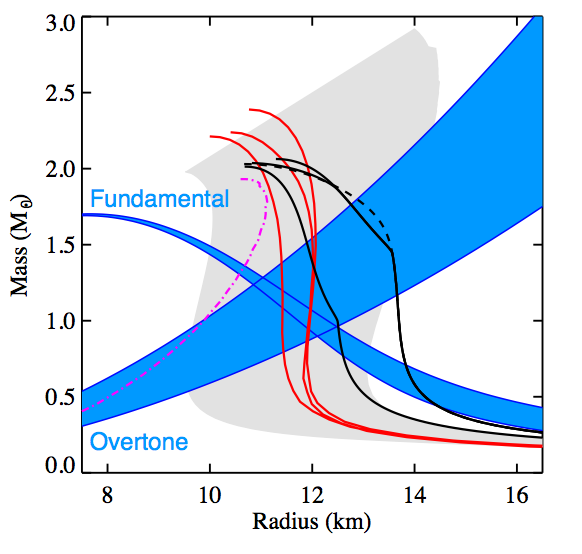}
\caption{$M$-$R$ diagram showing the seismological constraints for the soft gamma-ray repeater SGR 1806--20 using the relativistic torsional crust oscillation model of \citet{Samuelsson07}, in which the 29 Hz QPO is identified as the fundamental and the 625 Hz QPO as the first radial overtone. The neutron star lies in the box where the constraints from the two frequency bands overlap. Once QPOs are detected, frequency measurement errors are negligible for this purpose.  This model is very simple (it does not include) crust-core coupling, but it gives some idea of the type of constraints that might result from the detection of a harmonic sequence of seismic vibrations.  More sophisticated models that take into account coupling and the other relevant physical effects are under development.}
\label{seis}
\end{figure}

Several papers have specifically explored EOS dependencies in neutron star asteroseismology \citep{Strohmayer05,Strohmayer06a,Watts07,Samuelsson07,Sotani08,Steiner09,Gabler12}.  Figure \ref{seis} illustrates the constraints that result when one models the QPOs detected in the SGR 1806--20 hyperflare as torsional shear oscillations of the neutron star crust, \citep{Samuelsson07}.   This model is simple, in that it does not include magnetic coupling between crust and core.  However it gives some idea of the types of constraints on $M$ and $R$ that can result from the detection of several frequencies in a single event, where having multiple simultaneous frequencies assists mode identification (the burst storm identifications discussed above involve combining data from multiple bursts, so are less useful in this regard).  

Sadly giant flares are rare, occurring only every $\sim$ 10 years.   Ideally therefore we would like the ability to make similar detections in the more frequent but less bright events.    Intermediate flares, which are detected roughly once per year, have similar peak fluxes and spectra to the tails of the giant flares, but are too brief ($\sim$ 1 s) to permit detection of similar QPOs with current instrumentation. A $\sim 10$ m$^2$ hard X-ray timing instrument would be sensitive to QPOs in intermediate flares with similar fractional amplitudes as those observed in the tails of giant flares, provided that the collimator permitted the transmission of higher energy (above 30 keV) photons.  The latter is important since intermediate flares are unpredictable and likely to be observed off-axis, although one can increase the odds of capturing them by scheduling pointed observations during periods of high burst activity \citep{Israel08}.  Theoretically the expectation of similar fractional amplitudes is justified: mode excitation at substantial amplitude even by events releasing energies typical of intermediate flares is feasible \citep{Duncan98}.  Empirically, QPOs in giant flares tend to appear rather late in the tails, when luminosities are similar to those in intermediate flares, and given that they appear and disappear multiple time in these tails, may be triggered by magnetic starquakes at these ‘low’ fluxes \citep{Strohmayer06a}.  The development of similar fractional amplitude QPOs in intermediate flares is thus considered plausible.  This idea has also been given a boost by the discovery of QPOs in short burst storms from two different magnetars \citep{Huppenkothen13,Huppenkothen14a,Huppenkothen14b} including one that had also shown QPOs in a giant flare, since individually these bursts are much less energetic than the intermediate flares. The amplitudes at which the oscillations were detected in the burst storms are comparable to those of the detections in the giant flares. Upper limits on the presence of QPOs in the intermediate flares observed by current instruments, however, are above this level.

\section{Summary}

Neutron stars are unique testing grounds for fundamental nuclear physics, the only place where one can study the  equation of state of cold matter in equilibrium, at up to ten times normal nuclear densities.  The stable gravitational confinement permits the formation of matter which is extremely neutron rich, and which may involve matter with strange quarks.  The relativistic stellar structure equations show that there is a one to one mapping between the bulk properties of neutron stars, in particular their mass and radius, and the dense matter EOS.   Efforts to measure neutron star properties for this purpose are being made by both electromagnetic and gravitational wave astronomers.  In this Colloquium, we explored the techniques available using hard X-ray timing instruments:  waveform fitting, spin measurements, and asteroseismology. Hard X-ray timing offers unique advantages in terms of the numbers of known sources, and the potential for cross-checks using independent techniques and source classes.   

The previous generation of hard X-ray timing telescopes, in particular RXTE (a 0.6 m$^2$ telescope which operated from 1995 to 2012), uncovered many of the phenomena described in this Colloquium.  To exploit them to measure the EOS, however, requires larger area instruments, and various mission concepts are now being proposed.  These have included the 3 m$^2$ Advanced X-ray Timing Array \citep[AXTAR:][]{Ray10}, and the 8.5-10 m$^2$ Large Observatory for X-ray Timing, LOFT \citep[][see also the LOFT ESA M3 Yellow Book \texttt{http://sci.esa.int/loft/53447-loft-yellow-book}]{Feroci12,Feroci14}.   None have yet been successful in securing a launch slot.   However the advantages that such a telescope would offer in terms of measuring the dense matter equation of state are sufficiently highly compelling that mission concept development continues apace.  

\begin{acknowledgments}
The authors would like to thank all of the members of the LOFT Consortium, in particular the members of the LOFT Dense Matter Working Group, for useful discussions.  ALW acknowledges support from NWO Vidi Grant 639.042.916, NWO Vrije Competitie Grant 614.001.201, and ERC Starting Grant 639217 CSINEUTRONSTAR.  The work of KH and AS is supported by ERC Grant No. 307986 STRONGINT and the DFG through Grant SFB 634.  MF, GI, and LS acknowledge support from the Italian Space Agency (ASI) under contract I/021/12/0.  LT acknowledges support from the Ramon y Cajal Research Programme and from Contracts No. FPA2010-16963 and No. FPA2013-43425-P of Ministerio de Economia y Competitividad, from FP7-PEOPLE-2011-CIG under Contract No. PCIG09-GA-2011-291679 as well as NewCompStar (COST Action MP1304).  SMM acknowledges support from NSERC. JP acknowledges the Academy of Finland grant 268740. AP acknowledges support from NWO Vidi Grant 639.042.319.  AWS was supported by the U.S. Department of Energy Office of Nuclear Physics.
\end{acknowledgments}

\bibliography{eos_rmp}

\begin{thebibliography}{273}%
\makeatletter
\providecommand \@ifxundefined [1]{%
 \@ifx{#1\undefined}
}%
\providecommand \@ifnum [1]{%
 \ifnum #1\expandafter \@firstoftwo
 \else \expandafter \@secondoftwo
 \fi
}%
\providecommand \@ifx [1]{%
 \ifx #1\expandafter \@firstoftwo
 \else \expandafter \@secondoftwo
 \fi
}%
\providecommand \natexlab [1]{#1}%
\providecommand \enquote  [1]{``#1''}%
\providecommand \bibnamefont  [1]{#1}%
\providecommand \bibfnamefont [1]{#1}%
\providecommand \citenamefont [1]{#1}%
\providecommand \href@noop [0]{\@secondoftwo}%
\providecommand \href [0]{\begingroup \@sanitize@url \@href}%
\providecommand \@href[1]{\@@startlink{#1}\@@href}%
\providecommand \@@href[1]{\endgroup#1\@@endlink}%
\providecommand \@sanitize@url [0]{\catcode `\\12\catcode `\$12\catcode
  `\&12\catcode `\#12\catcode `\^12\catcode `\_12\catcode `\%12\relax}%
\providecommand \@@startlink[1]{}%
\providecommand \@@endlink[0]{}%
\providecommand \url  [0]{\begingroup\@sanitize@url \@url }%
\providecommand \@url [1]{\endgroup\@href {#1}{\urlprefix }}%
\providecommand \urlprefix  [0]{URL }%
\providecommand \Eprint [0]{\href }%
\providecommand \doibase [0]{http://dx.doi.org/}%
\providecommand \selectlanguage [0]{\@gobble}%
\providecommand \bibinfo  [0]{\@secondoftwo}%
\providecommand \bibfield  [0]{\@secondoftwo}%
\providecommand \translation [1]{[#1]}%
\providecommand \BibitemOpen [0]{}%
\providecommand \bibitemStop [0]{}%
\providecommand \bibitemNoStop [0]{.\EOS\space}%
\providecommand \EOS [0]{\spacefactor3000\relax}%
\providecommand \BibitemShut  [1]{\csname bibitem#1\endcsname}%
\let\auto@bib@innerbib\@empty
\bibitem [{\citenamefont {{Aasi}}\ \emph {et~al.}(2014)\citenamefont {{Aasi}},
  \citenamefont {{Abbott}}, \citenamefont {{Abbott}}, \citenamefont {{Abbott}},
  \citenamefont {{Abernathy}}, \citenamefont {{Accadia}}, \citenamefont
  {{Acernese}}, \citenamefont {{Ackley}}, \citenamefont {{Adams}},
  \citenamefont {{Adams}} \emph {et~al.}}]{Aasi14}%
  \BibitemOpen
  \bibfield  {author} {\bibinfo {author} {\bibnamefont {{Aasi}}, \bibfnamefont
  {J.}}, \bibinfo {author} {\bibfnamefont {B.~P.}\ \bibnamefont {{Abbott}}},
  \bibinfo {author} {\bibfnamefont {R.}~\bibnamefont {{Abbott}}}, \bibinfo
  {author} {\bibfnamefont {T.}~\bibnamefont {{Abbott}}}, \bibinfo {author}
  {\bibfnamefont {M.~R.}\ \bibnamefont {{Abernathy}}}, \bibinfo {author}
  {\bibfnamefont {T.}~\bibnamefont {{Accadia}}}, \bibinfo {author}
  {\bibfnamefont {F.}~\bibnamefont {{Acernese}}}, \bibinfo {author}
  {\bibfnamefont {K.}~\bibnamefont {{Ackley}}}, \bibinfo {author}
  {\bibfnamefont {C.}~\bibnamefont {{Adams}}}, \bibinfo {author} {\bibfnamefont
  {T.}~\bibnamefont {{Adams}}},  \emph {et~al.}} (\bibinfo {year} {2014}),\
  \href {\doibase 10.1103/PhysRevD.90.062010} {\bibfield  {journal} {\bibinfo
  {journal} {\prd}\ }\textbf {\bibinfo {volume} {90}}~(\bibinfo {number} {6}),\
  \bibinfo {eid} {062010}},\ \Eprint {http://arxiv.org/abs/1405.7904}
  {arXiv:1405.7904 [gr-qc]} \BibitemShut {NoStop}%
\bibitem [{\citenamefont {{Abadie}}\ \emph {et~al.}(2010)\citenamefont
  {{Abadie}}, \citenamefont {{Abbott}}, \citenamefont {{Abbott}}, \citenamefont
  {{Abernathy}}, \citenamefont {{Adams}}, \citenamefont {{Adhikari}},
  \citenamefont {{Ajith}}, \citenamefont {{Allen}}, \citenamefont {{Allen}},
  \citenamefont {{Amador Ceron}} \emph {et~al.}}]{Abadie10}%
  \BibitemOpen
  \bibfield  {author} {\bibinfo {author} {\bibnamefont {{Abadie}},
  \bibfnamefont {J.}}, \bibinfo {author} {\bibfnamefont {B.~P.}\ \bibnamefont
  {{Abbott}}}, \bibinfo {author} {\bibfnamefont {R.}~\bibnamefont {{Abbott}}},
  \bibinfo {author} {\bibfnamefont {M.}~\bibnamefont {{Abernathy}}}, \bibinfo
  {author} {\bibfnamefont {C.}~\bibnamefont {{Adams}}}, \bibinfo {author}
  {\bibfnamefont {R.}~\bibnamefont {{Adhikari}}}, \bibinfo {author}
  {\bibfnamefont {P.}~\bibnamefont {{Ajith}}}, \bibinfo {author} {\bibfnamefont
  {B.}~\bibnamefont {{Allen}}}, \bibinfo {author} {\bibfnamefont
  {G.}~\bibnamefont {{Allen}}}, \bibinfo {author} {\bibfnamefont
  {E.}~\bibnamefont {{Amador Ceron}}},  \emph {et~al.}} (\bibinfo {year}
  {2010}),\ \href {\doibase 10.1088/0004-637X/722/2/1504} {\bibfield  {journal}
  {\bibinfo  {journal} {\apj}\ }\textbf {\bibinfo {volume} {722}},\ \bibinfo
  {pages} {1504}},\ \Eprint {http://arxiv.org/abs/1006.2535} {arXiv:1006.2535
  [gr-qc]} \BibitemShut {NoStop}%
\bibitem [{\citenamefont {{Abbott}}\ \emph {et~al.}(2007)\citenamefont
  {{Abbott}}, \citenamefont {{Abbott}}, \citenamefont {{Adhikari}},
  \citenamefont {{Agresti}}, \citenamefont {{Ajith}}, \citenamefont {{Allen}},
  \citenamefont {{Amin}}, \citenamefont {{Anderson}}, \citenamefont
  {{Anderson}}, \citenamefont {{Arain}} \emph {et~al.}}]{Abbott07}%
  \BibitemOpen
  \bibfield  {author} {\bibinfo {author} {\bibnamefont {{Abbott}},
  \bibfnamefont {B.}}, \bibinfo {author} {\bibfnamefont {R.}~\bibnamefont
  {{Abbott}}}, \bibinfo {author} {\bibfnamefont {R.}~\bibnamefont
  {{Adhikari}}}, \bibinfo {author} {\bibfnamefont {J.}~\bibnamefont
  {{Agresti}}}, \bibinfo {author} {\bibfnamefont {P.}~\bibnamefont {{Ajith}}},
  \bibinfo {author} {\bibfnamefont {B.}~\bibnamefont {{Allen}}}, \bibinfo
  {author} {\bibfnamefont {R.}~\bibnamefont {{Amin}}}, \bibinfo {author}
  {\bibfnamefont {S.~B.}\ \bibnamefont {{Anderson}}}, \bibinfo {author}
  {\bibfnamefont {W.~G.}\ \bibnamefont {{Anderson}}}, \bibinfo {author}
  {\bibfnamefont {M.}~\bibnamefont {{Arain}}},  \emph {et~al.}} (\bibinfo
  {year} {2007}),\ \href {\doibase 10.1103/PhysRevD.76.042001} {\bibfield
  {journal} {\bibinfo  {journal} {\prd}\ }\textbf {\bibinfo {volume}
  {76}}~(\bibinfo {number} {4}),\ \bibinfo {eid} {042001}},\ \Eprint
  {http://arxiv.org/abs/gr-qc/0702039} {gr-qc/0702039} \BibitemShut {NoStop}%
\bibitem [{\citenamefont {{Abbott}}\ \emph {et~al.}(2008)\citenamefont
  {{Abbott}}, \citenamefont {{Abbott}}, \citenamefont {{Adhikari}},
  \citenamefont {{Agresti}}, \citenamefont {{Ajith}}, \citenamefont {{Allen}},
  \citenamefont {{Amin}}, \citenamefont {{Anderson}}, \citenamefont
  {{Anderson}}, \citenamefont {{Arain}} \emph {et~al.}}]{Abbott08}%
  \BibitemOpen
  \bibfield  {author} {\bibinfo {author} {\bibnamefont {{Abbott}},
  \bibfnamefont {B.}}, \bibinfo {author} {\bibfnamefont {R.}~\bibnamefont
  {{Abbott}}}, \bibinfo {author} {\bibfnamefont {R.}~\bibnamefont
  {{Adhikari}}}, \bibinfo {author} {\bibfnamefont {J.}~\bibnamefont
  {{Agresti}}}, \bibinfo {author} {\bibfnamefont {P.}~\bibnamefont {{Ajith}}},
  \bibinfo {author} {\bibfnamefont {B.}~\bibnamefont {{Allen}}}, \bibinfo
  {author} {\bibfnamefont {R.}~\bibnamefont {{Amin}}}, \bibinfo {author}
  {\bibfnamefont {S.~B.}\ \bibnamefont {{Anderson}}}, \bibinfo {author}
  {\bibfnamefont {W.~G.}\ \bibnamefont {{Anderson}}}, \bibinfo {author}
  {\bibfnamefont {M.}~\bibnamefont {{Arain}}},  \emph {et~al.}} (\bibinfo
  {year} {2008}),\ \href {\doibase 10.1103/PhysRevD.77.022001} {\bibfield
  {journal} {\bibinfo  {journal} {\prd}\ }\textbf {\bibinfo {volume}
  {77}}~(\bibinfo {number} {2}),\ \bibinfo {eid} {022001}},\ \Eprint
  {http://arxiv.org/abs/0708.3818} {arXiv:0708.3818 [gr-qc]} \BibitemShut
  {NoStop}%
\bibitem [{\citenamefont {{Abdo}}\ \emph {et~al.}(2009)\citenamefont {{Abdo}},
  \citenamefont {{Ackermann}}, \citenamefont {{Ajello}}, \citenamefont
  {{Anderson}}, \citenamefont {{Atwood}}, \citenamefont {{Axelsson}},
  \citenamefont {{Baldini}}, \citenamefont {{Ballet}}, \citenamefont
  {{Barbiellini}}, \citenamefont {{Baring}} \emph {et~al.}}]{Abdo09}%
  \BibitemOpen
  \bibfield  {author} {\bibinfo {author} {\bibnamefont {{Abdo}}, \bibfnamefont
  {A.~A.}}, \bibinfo {author} {\bibfnamefont {M.}~\bibnamefont {{Ackermann}}},
  \bibinfo {author} {\bibfnamefont {M.}~\bibnamefont {{Ajello}}}, \bibinfo
  {author} {\bibfnamefont {B.}~\bibnamefont {{Anderson}}}, \bibinfo {author}
  {\bibfnamefont {W.~B.}\ \bibnamefont {{Atwood}}}, \bibinfo {author}
  {\bibfnamefont {M.}~\bibnamefont {{Axelsson}}}, \bibinfo {author}
  {\bibfnamefont {L.}~\bibnamefont {{Baldini}}}, \bibinfo {author}
  {\bibfnamefont {J.}~\bibnamefont {{Ballet}}}, \bibinfo {author}
  {\bibfnamefont {G.}~\bibnamefont {{Barbiellini}}}, \bibinfo {author}
  {\bibfnamefont {M.~G.}\ \bibnamefont {{Baring}}},  \emph {et~al.}} (\bibinfo
  {year} {2009}),\ \href {\doibase 10.1126/science.1175558} {\bibfield
  {journal} {\bibinfo  {journal} {Science}\ }\textbf {\bibinfo {volume}
  {325}},\ \bibinfo {pages} {840}},\ \Eprint {http://arxiv.org/abs/1009.0748}
  {arXiv:1009.0748 [astro-ph.GA]} \BibitemShut {NoStop}%
\bibitem [{\citenamefont {{Acernese}}\ \emph {et~al.}(2015)\citenamefont
  {{Acernese}}, \citenamefont {{Agathos}}, \citenamefont {{Agatsuma}},
  \citenamefont {{Aisa}}, \citenamefont {{Allemandou}}, \citenamefont
  {{Allocca}}, \citenamefont {{Amarni}}, \citenamefont {{Astone}},
  \citenamefont {{Balestri}}, \citenamefont {{Ballardin}},\ and\ \citenamefont
  {et~al.}}]{Acernese15}%
  \BibitemOpen
  \bibfield  {author} {\bibinfo {author} {\bibnamefont {{Acernese}},
  \bibfnamefont {F.}}, \bibinfo {author} {\bibfnamefont {M.}~\bibnamefont
  {{Agathos}}}, \bibinfo {author} {\bibfnamefont {K.}~\bibnamefont
  {{Agatsuma}}}, \bibinfo {author} {\bibfnamefont {D.}~\bibnamefont {{Aisa}}},
  \bibinfo {author} {\bibfnamefont {N.}~\bibnamefont {{Allemandou}}}, \bibinfo
  {author} {\bibfnamefont {A.}~\bibnamefont {{Allocca}}}, \bibinfo {author}
  {\bibfnamefont {J.}~\bibnamefont {{Amarni}}}, \bibinfo {author}
  {\bibfnamefont {P.}~\bibnamefont {{Astone}}}, \bibinfo {author}
  {\bibfnamefont {G.}~\bibnamefont {{Balestri}}}, \bibinfo {author}
  {\bibfnamefont {G.}~\bibnamefont {{Ballardin}}}, \ and\ \bibinfo {author}
  {\bibnamefont {et~al.}}} (\bibinfo {year} {2015}),\ \href {\doibase
  10.1088/0264-9381/32/2/024001} {\bibfield  {journal} {\bibinfo  {journal}
  {Classical and Quantum Gravity}\ }\textbf {\bibinfo {volume} {32}}~(\bibinfo
  {number} {2}),\ \bibinfo {eid} {024001}},\ \Eprint
  {http://arxiv.org/abs/1408.3978} {arXiv:1408.3978 [gr-qc]} \BibitemShut
  {NoStop}%
\bibitem [{\citenamefont {{Akmal}}\ \emph {et~al.}(1998)\citenamefont
  {{Akmal}}, \citenamefont {{Pandharipande}},\ and\ \citenamefont
  {{Ravenhall}}}]{Akmal98}%
  \BibitemOpen
  \bibfield  {author} {\bibinfo {author} {\bibnamefont {{Akmal}}, \bibfnamefont
  {A.}}, \bibinfo {author} {\bibfnamefont {V.~R.}\ \bibnamefont
  {{Pandharipande}}}, \ and\ \bibinfo {author} {\bibfnamefont {D.~G.}\
  \bibnamefont {{Ravenhall}}}} (\bibinfo {year} {1998}),\ \href {\doibase
  10.1103/PhysRevC.58.1804} {\bibfield  {journal} {\bibinfo  {journal} {\prc}\
  }\textbf {\bibinfo {volume} {58}},\ \bibinfo {pages} {1804}},\ \Eprint
  {http://arxiv.org/abs/nucl-th/9804027} {nucl-th/9804027} \BibitemShut
  {NoStop}%
\bibitem [{\citenamefont {{Alford}}\ \emph {et~al.}(2012)\citenamefont
  {{Alford}}, \citenamefont {{Mahmoodifar}},\ and\ \citenamefont
  {{Schwenzer}}}]{Alford12}%
  \BibitemOpen
  \bibfield  {author} {\bibinfo {author} {\bibnamefont {{Alford}},
  \bibfnamefont {M.~G.}}, \bibinfo {author} {\bibfnamefont {S.}~\bibnamefont
  {{Mahmoodifar}}}, \ and\ \bibinfo {author} {\bibfnamefont {K.}~\bibnamefont
  {{Schwenzer}}}} (\bibinfo {year} {2012}),\ \href {\doibase
  10.1103/PhysRevD.85.024007} {\bibfield  {journal} {\bibinfo  {journal}
  {\prd}\ }\textbf {\bibinfo {volume} {85}}~(\bibinfo {number} {2}),\ \bibinfo
  {eid} {024007}},\ \Eprint {http://arxiv.org/abs/1012.4883} {arXiv:1012.4883
  [astro-ph.HE]} \BibitemShut {NoStop}%
\bibitem [{\citenamefont {{Alford}}\ \emph {et~al.}(2008)\citenamefont
  {{Alford}}, \citenamefont {{Schmitt}}, \citenamefont {{Rajagopal}},\ and\
  \citenamefont {{Sch{\"a}fer}}}]{Alford08}%
  \BibitemOpen
  \bibfield  {author} {\bibinfo {author} {\bibnamefont {{Alford}},
  \bibfnamefont {M.~G.}}, \bibinfo {author} {\bibfnamefont {A.}~\bibnamefont
  {{Schmitt}}}, \bibinfo {author} {\bibfnamefont {K.}~\bibnamefont
  {{Rajagopal}}}, \ and\ \bibinfo {author} {\bibfnamefont {T.}~\bibnamefont
  {{Sch{\"a}fer}}}} (\bibinfo {year} {2008}),\ \href {\doibase
  10.1103/RevModPhys.80.1455} {\bibfield  {journal} {\bibinfo  {journal}
  {Reviews of Modern Physics}\ }\textbf {\bibinfo {volume} {80}},\ \bibinfo
  {pages} {1455}},\ \Eprint {http://arxiv.org/abs/0709.4635} {arXiv:0709.4635
  [hep-ph]} \BibitemShut {NoStop}%
\bibitem [{\citenamefont {{AlGendy}}\ and\ \citenamefont
  {{Morsink}}(2014)}]{AlGendy14}%
  \BibitemOpen
  \bibfield  {author} {\bibinfo {author} {\bibnamefont {{AlGendy}},
  \bibfnamefont {M.}}, \ and\ \bibinfo {author} {\bibfnamefont {S.~M.}\
  \bibnamefont {{Morsink}}}} (\bibinfo {year} {2014}),\ \href {\doibase
  10.1088/0004-637X/791/2/78} {\bibfield  {journal} {\bibinfo  {journal}
  {\apj}\ }\textbf {\bibinfo {volume} {791}},\ \bibinfo {eid} {78}},\ \Eprint
  {http://arxiv.org/abs/1404.0609} {arXiv:1404.0609 [astro-ph.HE]} \BibitemShut
  {NoStop}%
\bibitem [{\citenamefont {{Alpar}}\ \emph {et~al.}(1982)\citenamefont
  {{Alpar}}, \citenamefont {{Cheng}}, \citenamefont {{Ruderman}},\ and\
  \citenamefont {{Shaham}}}]{Alpar82}%
  \BibitemOpen
  \bibfield  {author} {\bibinfo {author} {\bibnamefont {{Alpar}}, \bibfnamefont
  {M.~A.}}, \bibinfo {author} {\bibfnamefont {A.~F.}\ \bibnamefont {{Cheng}}},
  \bibinfo {author} {\bibfnamefont {M.~A.}\ \bibnamefont {{Ruderman}}}, \ and\
  \bibinfo {author} {\bibfnamefont {J.}~\bibnamefont {{Shaham}}}} (\bibinfo
  {year} {1982}),\ \href {\doibase 10.1038/300728a0} {\bibfield  {journal}
  {\bibinfo  {journal} {\nat}\ }\textbf {\bibinfo {volume} {300}},\ \bibinfo
  {pages} {728}}\BibitemShut {NoStop}%
\bibitem [{\citenamefont {{Altamirano}}\ \emph {et~al.}(2008)\citenamefont
  {{Altamirano}}, \citenamefont {{Casella}}, \citenamefont {{Patruno}},
  \citenamefont {{Wijnands}},\ and\ \citenamefont {{van der
  Klis}}}]{Altamirano08}%
  \BibitemOpen
  \bibfield  {author} {\bibinfo {author} {\bibnamefont {{Altamirano}},
  \bibfnamefont {D.}}, \bibinfo {author} {\bibfnamefont {P.}~\bibnamefont
  {{Casella}}}, \bibinfo {author} {\bibfnamefont {A.}~\bibnamefont
  {{Patruno}}}, \bibinfo {author} {\bibfnamefont {R.}~\bibnamefont
  {{Wijnands}}}, \ and\ \bibinfo {author} {\bibfnamefont {M.}~\bibnamefont
  {{van der Klis}}}} (\bibinfo {year} {2008}),\ \href {\doibase 10.1086/528983}
  {\bibfield  {journal} {\bibinfo  {journal} {\apjl}\ }\textbf {\bibinfo
  {volume} {674}},\ \bibinfo {pages} {L45}},\ \Eprint
  {http://arxiv.org/abs/0708.1316} {arXiv:0708.1316} \BibitemShut {NoStop}%
\bibitem [{\citenamefont {{Ambartsumyan}}\ and\ \citenamefont
  {{Saakyan}}(1960)}]{Ambartsumyan60}%
  \BibitemOpen
  \bibfield  {author} {\bibinfo {author} {\bibnamefont {{Ambartsumyan}},
  \bibfnamefont {V.~A.}}, \ and\ \bibinfo {author} {\bibfnamefont {G.~S.}\
  \bibnamefont {{Saakyan}}}} (\bibinfo {year} {1960}),\ \href@noop {}
  {\bibfield  {journal} {\bibinfo  {journal} {\sovast}\ }\textbf {\bibinfo
  {volume} {4}},\ \bibinfo {pages} {187}}\BibitemShut {NoStop}%
\bibitem [{\citenamefont {{Andersson}}\ \emph {et~al.}(2005)\citenamefont
  {{Andersson}}, \citenamefont {{Glampedakis}}, \citenamefont {{Haskell}},\
  and\ \citenamefont {{Watts}}}]{Andersson05}%
  \BibitemOpen
  \bibfield  {author} {\bibinfo {author} {\bibnamefont {{Andersson}},
  \bibfnamefont {N.}}, \bibinfo {author} {\bibfnamefont {K.}~\bibnamefont
  {{Glampedakis}}}, \bibinfo {author} {\bibfnamefont {B.}~\bibnamefont
  {{Haskell}}}, \ and\ \bibinfo {author} {\bibfnamefont {A.~L.}\ \bibnamefont
  {{Watts}}}} (\bibinfo {year} {2005}),\ \href {\doibase
  10.1111/j.1365-2966.2005.09167.x} {\bibfield  {journal} {\bibinfo  {journal}
  {\mnras}\ }\textbf {\bibinfo {volume} {361}},\ \bibinfo {pages} {1153}},\
  \Eprint {http://arxiv.org/abs/astro-ph/0411747} {astro-ph/0411747}
  \BibitemShut {NoStop}%
\bibitem [{\citenamefont {{Andersson}}\ \emph {et~al.}(2009)\citenamefont
  {{Andersson}}, \citenamefont {{Glampedakis}},\ and\ \citenamefont
  {{Samuelsson}}}]{Andersson09}%
  \BibitemOpen
  \bibfield  {author} {\bibinfo {author} {\bibnamefont {{Andersson}},
  \bibfnamefont {N.}}, \bibinfo {author} {\bibfnamefont {K.}~\bibnamefont
  {{Glampedakis}}}, \ and\ \bibinfo {author} {\bibfnamefont {L.}~\bibnamefont
  {{Samuelsson}}}} (\bibinfo {year} {2009}),\ \href {\doibase
  10.1111/j.1365-2966.2009.14734.x} {\bibfield  {journal} {\bibinfo  {journal}
  {\mnras}\ }\textbf {\bibinfo {volume} {396}},\ \bibinfo {pages} {894}},\
  \Eprint {http://arxiv.org/abs/0812.2417} {arXiv:0812.2417} \BibitemShut
  {NoStop}%
\bibitem [{\citenamefont {{Andersson}}\ \emph {et~al.}(2014)\citenamefont
  {{Andersson}}, \citenamefont {{Jones}},\ and\ \citenamefont
  {{Ho}}}]{Andersson14}%
  \BibitemOpen
  \bibfield  {author} {\bibinfo {author} {\bibnamefont {{Andersson}},
  \bibfnamefont {N.}}, \bibinfo {author} {\bibfnamefont {D.~I.}\ \bibnamefont
  {{Jones}}}, \ and\ \bibinfo {author} {\bibfnamefont {W.~C.~G.}\ \bibnamefont
  {{Ho}}}} (\bibinfo {year} {2014}),\ \href {\doibase 10.1093/mnras/stu870}
  {\bibfield  {journal} {\bibinfo  {journal} {\mnras}\ }\textbf {\bibinfo
  {volume} {442}},\ \bibinfo {pages} {1786}},\ \Eprint
  {http://arxiv.org/abs/1403.0860} {arXiv:1403.0860 [astro-ph.SR]} \BibitemShut
  {NoStop}%
\bibitem [{\citenamefont {{Antoniadis}}\ \emph {et~al.}(2013)\citenamefont
  {{Antoniadis}}, \citenamefont {{Freire}}, \citenamefont {{Wex}},
  \citenamefont {{Tauris}}, \citenamefont {{Lynch}}, \citenamefont {{van
  Kerkwijk}}, \citenamefont {{Kramer}}, \citenamefont {{Bassa}}, \citenamefont
  {{Dhillon}}, \citenamefont {{Driebe}} \emph {et~al.}}]{Antoniadis13}%
  \BibitemOpen
  \bibfield  {author} {\bibinfo {author} {\bibnamefont {{Antoniadis}},
  \bibfnamefont {J.}}, \bibinfo {author} {\bibfnamefont {P.~C.~C.}\
  \bibnamefont {{Freire}}}, \bibinfo {author} {\bibfnamefont {N.}~\bibnamefont
  {{Wex}}}, \bibinfo {author} {\bibfnamefont {T.~M.}\ \bibnamefont {{Tauris}}},
  \bibinfo {author} {\bibfnamefont {R.~S.}\ \bibnamefont {{Lynch}}}, \bibinfo
  {author} {\bibfnamefont {M.~H.}\ \bibnamefont {{van Kerkwijk}}}, \bibinfo
  {author} {\bibfnamefont {M.}~\bibnamefont {{Kramer}}}, \bibinfo {author}
  {\bibfnamefont {C.}~\bibnamefont {{Bassa}}}, \bibinfo {author} {\bibfnamefont
  {V.~S.}\ \bibnamefont {{Dhillon}}}, \bibinfo {author} {\bibfnamefont
  {T.}~\bibnamefont {{Driebe}}},  \emph {et~al.}} (\bibinfo {year} {2013}),\
  \href {\doibase 10.1126/science.1233232} {\bibfield  {journal} {\bibinfo
  {journal} {Science}\ }\textbf {\bibinfo {volume} {340}},\ \bibinfo {pages}
  {448}},\ \Eprint {http://arxiv.org/abs/1304.6875} {arXiv:1304.6875
  [astro-ph.HE]} \BibitemShut {NoStop}%
\bibitem [{\citenamefont {{Archibald}}\ \emph {et~al.}(2009)\citenamefont
  {{Archibald}}, \citenamefont {{Stairs}}, \citenamefont {{Ransom}},
  \citenamefont {{Kaspi}}, \citenamefont {{Kondratiev}}, \citenamefont
  {{Lorimer}}, \citenamefont {{McLaughlin}}, \citenamefont {{Boyles}},
  \citenamefont {{Hessels}}, \citenamefont {{Lynch}} \emph
  {et~al.}}]{Archibald09}%
  \BibitemOpen
  \bibfield  {author} {\bibinfo {author} {\bibnamefont {{Archibald}},
  \bibfnamefont {A.~M.}}, \bibinfo {author} {\bibfnamefont {I.~H.}\
  \bibnamefont {{Stairs}}}, \bibinfo {author} {\bibfnamefont {S.~M.}\
  \bibnamefont {{Ransom}}}, \bibinfo {author} {\bibfnamefont {V.~M.}\
  \bibnamefont {{Kaspi}}}, \bibinfo {author} {\bibfnamefont {V.~I.}\
  \bibnamefont {{Kondratiev}}}, \bibinfo {author} {\bibfnamefont {D.~R.}\
  \bibnamefont {{Lorimer}}}, \bibinfo {author} {\bibfnamefont {M.~A.}\
  \bibnamefont {{McLaughlin}}}, \bibinfo {author} {\bibfnamefont
  {J.}~\bibnamefont {{Boyles}}}, \bibinfo {author} {\bibfnamefont {J.~W.~T.}\
  \bibnamefont {{Hessels}}}, \bibinfo {author} {\bibfnamefont {R.}~\bibnamefont
  {{Lynch}}},  \emph {et~al.}} (\bibinfo {year} {2009}),\ \href {\doibase
  10.1126/science.1172740} {\bibfield  {journal} {\bibinfo  {journal}
  {Science}\ }\textbf {\bibinfo {volume} {324}},\ \bibinfo {pages} {1411}},\
  \Eprint {http://arxiv.org/abs/0905.3397} {arXiv:0905.3397 [astro-ph.HE]}
  \BibitemShut {NoStop}%
\bibitem [{\citenamefont {{Arzoumanian}}\ \emph {et~al.}(2014)\citenamefont
  {{Arzoumanian}}, \citenamefont {{Gendreau}}, \citenamefont {{Baker}},
  \citenamefont {{Cazeau}}, \citenamefont {{Hestnes}}, \citenamefont
  {{Kellogg}}, \citenamefont {{Kenyon}}, \citenamefont {{Kozon}}, \citenamefont
  {{Liu}}, \citenamefont {{Manthripragada}}, \citenamefont {{Markwardt}},
  \citenamefont {{Mitchell}}, \citenamefont {{Mitchell}}, \citenamefont
  {{Monroe}}, \citenamefont {{Okajima}}, \citenamefont {{Pollard}},
  \citenamefont {{Powers}}, \citenamefont {{Savadkin}}, \citenamefont
  {{Winternitz}}, \citenamefont {{Chen}}, \citenamefont {{Wright}},
  \citenamefont {{Foster}}, \citenamefont {{Prigozhin}}, \citenamefont
  {{Remillard}},\ and\ \citenamefont {{Doty}}}]{Arzoumanian14}%
  \BibitemOpen
  \bibfield  {author} {\bibinfo {author} {\bibnamefont {{Arzoumanian}},
  \bibfnamefont {Z.}}, \bibinfo {author} {\bibfnamefont {K.~C.}\ \bibnamefont
  {{Gendreau}}}, \bibinfo {author} {\bibfnamefont {C.~L.}\ \bibnamefont
  {{Baker}}}, \bibinfo {author} {\bibfnamefont {T.}~\bibnamefont {{Cazeau}}},
  \bibinfo {author} {\bibfnamefont {P.}~\bibnamefont {{Hestnes}}}, \bibinfo
  {author} {\bibfnamefont {J.~W.}\ \bibnamefont {{Kellogg}}}, \bibinfo {author}
  {\bibfnamefont {S.~J.}\ \bibnamefont {{Kenyon}}}, \bibinfo {author}
  {\bibfnamefont {R.~P.}\ \bibnamefont {{Kozon}}}, \bibinfo {author}
  {\bibfnamefont {K.-C.}\ \bibnamefont {{Liu}}}, \bibinfo {author}
  {\bibfnamefont {S.~S.}\ \bibnamefont {{Manthripragada}}}, \bibinfo {author}
  {\bibfnamefont {C.~B.}\ \bibnamefont {{Markwardt}}}, \bibinfo {author}
  {\bibfnamefont {A.~L.}\ \bibnamefont {{Mitchell}}}, \bibinfo {author}
  {\bibfnamefont {J.~W.}\ \bibnamefont {{Mitchell}}}, \bibinfo {author}
  {\bibfnamefont {C.~A.}\ \bibnamefont {{Monroe}}}, \bibinfo {author}
  {\bibfnamefont {T.}~\bibnamefont {{Okajima}}}, \bibinfo {author}
  {\bibfnamefont {S.~E.}\ \bibnamefont {{Pollard}}}, \bibinfo {author}
  {\bibfnamefont {D.~F.}\ \bibnamefont {{Powers}}}, \bibinfo {author}
  {\bibfnamefont {B.~J.}\ \bibnamefont {{Savadkin}}}, \bibinfo {author}
  {\bibfnamefont {L.~B.}\ \bibnamefont {{Winternitz}}}, \bibinfo {author}
  {\bibfnamefont {P.~T.}\ \bibnamefont {{Chen}}}, \bibinfo {author}
  {\bibfnamefont {M.~R.}\ \bibnamefont {{Wright}}}, \bibinfo {author}
  {\bibfnamefont {R.}~\bibnamefont {{Foster}}}, \bibinfo {author}
  {\bibfnamefont {G.}~\bibnamefont {{Prigozhin}}}, \bibinfo {author}
  {\bibfnamefont {R.}~\bibnamefont {{Remillard}}}, \ and\ \bibinfo {author}
  {\bibfnamefont {J.}~\bibnamefont {{Doty}}}} (\bibinfo {year} {2014}),\ in\
  \href {\doibase 10.1117/12.2056811} {\emph {\bibinfo {booktitle} {Society of
  Photo-Optical Instrumentation Engineers (SPIE) Conference Series}}},\
  \bibinfo {series} {Society of Photo-Optical Instrumentation Engineers (SPIE)
  Conference Series}, Vol.\ \bibinfo {volume} {9144},\ p.~\bibinfo {pages}
  {20}\BibitemShut {NoStop}%
\bibitem [{\citenamefont {{Asai}}\ and\ \citenamefont {{Lee}}(2014)}]{Asai14}%
  \BibitemOpen
  \bibfield  {author} {\bibinfo {author} {\bibnamefont {{Asai}}, \bibfnamefont
  {H.}}, \ and\ \bibinfo {author} {\bibfnamefont {U.}~\bibnamefont {{Lee}}}}
  (\bibinfo {year} {2014}),\ \href {\doibase 10.1088/0004-637X/790/1/66}
  {\bibfield  {journal} {\bibinfo  {journal} {\apj}\ }\textbf {\bibinfo
  {volume} {790}},\ \bibinfo {eid} {66}},\ \Eprint
  {http://arxiv.org/abs/1406.2786} {arXiv:1406.2786 [astro-ph.SR]} \BibitemShut
  {NoStop}%
\bibitem [{\citenamefont {{Atwood}}\ \emph {et~al.}(2006)\citenamefont
  {{Atwood}}, \citenamefont {{Ziegler}}, \citenamefont {{Johnson}},\ and\
  \citenamefont {{Baughman}}}]{Atwood06}%
  \BibitemOpen
  \bibfield  {author} {\bibinfo {author} {\bibnamefont {{Atwood}},
  \bibfnamefont {W.~B.}}, \bibinfo {author} {\bibfnamefont {M.}~\bibnamefont
  {{Ziegler}}}, \bibinfo {author} {\bibfnamefont {R.~P.}\ \bibnamefont
  {{Johnson}}}, \ and\ \bibinfo {author} {\bibfnamefont {B.~M.}\ \bibnamefont
  {{Baughman}}}} (\bibinfo {year} {2006}),\ \href {\doibase 10.1086/510018}
  {\bibfield  {journal} {\bibinfo  {journal} {\apjl}\ }\textbf {\bibinfo
  {volume} {652}},\ \bibinfo {pages} {L49}}\BibitemShut {NoStop}%
\bibitem [{\citenamefont {{Bahramian}}\ \emph {et~al.}(2015)\citenamefont
  {{Bahramian}}, \citenamefont {{Heinke}}, \citenamefont {{Degenaar}},
  \citenamefont {{Chomiuk}}, \citenamefont {{Wijnands}}, \citenamefont
  {{Strader}}, \citenamefont {{Ho}},\ and\ \citenamefont
  {{Pooley}}}]{Bahramian15}%
  \BibitemOpen
  \bibfield  {author} {\bibinfo {author} {\bibnamefont {{Bahramian}},
  \bibfnamefont {A.}}, \bibinfo {author} {\bibfnamefont {C.~O.}\ \bibnamefont
  {{Heinke}}}, \bibinfo {author} {\bibfnamefont {N.}~\bibnamefont
  {{Degenaar}}}, \bibinfo {author} {\bibfnamefont {L.}~\bibnamefont
  {{Chomiuk}}}, \bibinfo {author} {\bibfnamefont {R.}~\bibnamefont
  {{Wijnands}}}, \bibinfo {author} {\bibfnamefont {J.}~\bibnamefont
  {{Strader}}}, \bibinfo {author} {\bibfnamefont {W.~C.~G.}\ \bibnamefont
  {{Ho}}}, \ and\ \bibinfo {author} {\bibfnamefont {D.}~\bibnamefont
  {{Pooley}}}} (\bibinfo {year} {2015}),\ \href {\doibase
  10.1093/mnras/stv1585} {\bibfield  {journal} {\bibinfo  {journal} {\mnras}\
  }\textbf {\bibinfo {volume} {452}},\ \bibinfo {pages} {3475}},\ \Eprint
  {http://arxiv.org/abs/1507.03994} {arXiv:1507.03994 [astro-ph.HE]}
  \BibitemShut {NoStop}%
\bibitem [{\citenamefont {{Bai}}\ and\ \citenamefont
  {{Spitkovsky}}(2010)}]{Bai10}%
  \BibitemOpen
  \bibfield  {author} {\bibinfo {author} {\bibnamefont {{Bai}}, \bibfnamefont
  {X.-N.}}, \ and\ \bibinfo {author} {\bibfnamefont {A.}~\bibnamefont
  {{Spitkovsky}}}} (\bibinfo {year} {2010}),\ \href {\doibase
  10.1088/0004-637X/715/2/1270} {\bibfield  {journal} {\bibinfo  {journal}
  {\apj}\ }\textbf {\bibinfo {volume} {715}},\ \bibinfo {pages} {1270}},\
  \Eprint {http://arxiv.org/abs/0910.5740} {arXiv:0910.5740 [astro-ph.HE]}
  \BibitemShut {NoStop}%
\bibitem [{\citenamefont {{Balberg}}\ \emph {et~al.}(1999)\citenamefont
  {{Balberg}}, \citenamefont {{Lichtenstadt}},\ and\ \citenamefont
  {{Cook}}}]{Balberg99}%
  \BibitemOpen
  \bibfield  {author} {\bibinfo {author} {\bibnamefont {{Balberg}},
  \bibfnamefont {S.}}, \bibinfo {author} {\bibfnamefont {I.}~\bibnamefont
  {{Lichtenstadt}}}, \ and\ \bibinfo {author} {\bibfnamefont {G.~B.}\
  \bibnamefont {{Cook}}}} (\bibinfo {year} {1999}),\ \href@noop {} {\bibfield
  {journal} {\bibinfo  {journal} {\apjs}\ }\textbf {\bibinfo {volume} {121}},\
  \bibinfo {pages} {515}},\ \Eprint {http://arxiv.org/abs/astro-ph/9810361}
  {astro-ph/9810361} \BibitemShut {NoStop}%
\bibitem [{\citenamefont {{Bassa}}\ \emph {et~al.}(2014)\citenamefont
  {{Bassa}}, \citenamefont {{Patruno}}, \citenamefont {{Hessels}},
  \citenamefont {{Keane}}, \citenamefont {{Monard}}, \citenamefont {{Mahony}},
  \citenamefont {{Bogdanov}}, \citenamefont {{Corbel}}, \citenamefont
  {{Edwards}}, \citenamefont {{Archibald}}, \citenamefont {{Janssen}},
  \citenamefont {{Stappers}},\ and\ \citenamefont {{Tendulkar}}}]{Bassa14}%
  \BibitemOpen
  \bibfield  {author} {\bibinfo {author} {\bibnamefont {{Bassa}}, \bibfnamefont
  {C.~G.}}, \bibinfo {author} {\bibfnamefont {A.}~\bibnamefont {{Patruno}}},
  \bibinfo {author} {\bibfnamefont {J.~W.~T.}\ \bibnamefont {{Hessels}}},
  \bibinfo {author} {\bibfnamefont {E.~F.}\ \bibnamefont {{Keane}}}, \bibinfo
  {author} {\bibfnamefont {B.}~\bibnamefont {{Monard}}}, \bibinfo {author}
  {\bibfnamefont {E.~K.}\ \bibnamefont {{Mahony}}}, \bibinfo {author}
  {\bibfnamefont {S.}~\bibnamefont {{Bogdanov}}}, \bibinfo {author}
  {\bibfnamefont {S.}~\bibnamefont {{Corbel}}}, \bibinfo {author}
  {\bibfnamefont {P.~G.}\ \bibnamefont {{Edwards}}}, \bibinfo {author}
  {\bibfnamefont {A.~M.}\ \bibnamefont {{Archibald}}}, \bibinfo {author}
  {\bibfnamefont {G.~H.}\ \bibnamefont {{Janssen}}}, \bibinfo {author}
  {\bibfnamefont {B.~W.}\ \bibnamefont {{Stappers}}}, \ and\ \bibinfo {author}
  {\bibfnamefont {S.}~\bibnamefont {{Tendulkar}}}} (\bibinfo {year} {2014}),\
  \href {\doibase 10.1093/mnras/stu708} {\bibfield  {journal} {\bibinfo
  {journal} {\mnras}\ }\textbf {\bibinfo {volume} {441}},\ \bibinfo {pages}
  {1825}},\ \Eprint {http://arxiv.org/abs/1402.0765} {arXiv:1402.0765
  [astro-ph.HE]} \BibitemShut {NoStop}%
\bibitem [{\citenamefont {{Baub{\"o}ck}}\ \emph {et~al.}(2013)\citenamefont
  {{Baub{\"o}ck}}, \citenamefont {{Berti}}, \citenamefont {{Psaltis}},\ and\
  \citenamefont {{{\"O}zel}}}]{Baubock13}%
  \BibitemOpen
  \bibfield  {author} {\bibinfo {author} {\bibnamefont {{Baub{\"o}ck}},
  \bibfnamefont {M.}}, \bibinfo {author} {\bibfnamefont {E.}~\bibnamefont
  {{Berti}}}, \bibinfo {author} {\bibfnamefont {D.}~\bibnamefont {{Psaltis}}},
  \ and\ \bibinfo {author} {\bibfnamefont {F.}~\bibnamefont {{{\"O}zel}}}}
  (\bibinfo {year} {2013}),\ \href {\doibase 10.1088/0004-637X/777/1/68}
  {\bibfield  {journal} {\bibinfo  {journal} {\apj}\ }\textbf {\bibinfo
  {volume} {777}},\ \bibinfo {eid} {68}},\ \Eprint
  {http://arxiv.org/abs/1306.0569} {arXiv:1306.0569 [astro-ph.HE]} \BibitemShut
  {NoStop}%
\bibitem [{\citenamefont {{Bauswein}}\ \emph {et~al.}(2012)\citenamefont
  {{Bauswein}}, \citenamefont {{Janka}}, \citenamefont {{Hebeler}},\ and\
  \citenamefont {{Schwenk}}}]{Bauswein12}%
  \BibitemOpen
  \bibfield  {author} {\bibinfo {author} {\bibnamefont {{Bauswein}},
  \bibfnamefont {A.}}, \bibinfo {author} {\bibfnamefont {H.-T.}\ \bibnamefont
  {{Janka}}}, \bibinfo {author} {\bibfnamefont {K.}~\bibnamefont {{Hebeler}}},
  \ and\ \bibinfo {author} {\bibfnamefont {A.}~\bibnamefont {{Schwenk}}}}
  (\bibinfo {year} {2012}),\ \href {\doibase 10.1103/PhysRevD.86.063001}
  {\bibfield  {journal} {\bibinfo  {journal} {\prd}\ }\textbf {\bibinfo
  {volume} {86}}~(\bibinfo {number} {6}),\ \bibinfo {eid} {063001}},\ \Eprint
  {http://arxiv.org/abs/1204.1888} {arXiv:1204.1888 [astro-ph.SR]} \BibitemShut
  {NoStop}%
\bibitem [{\citenamefont {{Becker}}(2001)}]{Becker01}%
  \BibitemOpen
  \bibfield  {author} {\bibinfo {author} {\bibnamefont {{Becker}},
  \bibfnamefont {W.}}} (\bibinfo {year} {2001}),\ \href {\doibase
  10.1063/1.1434615} {\bibfield  {journal} {\bibinfo  {journal} {X-ray
  Astronomy: Stellar Endpoints, AGN, and the Diffuse X-ray Background}\
  }\textbf {\bibinfo {volume} {599}},\ \bibinfo {pages} {13}}\BibitemShut
  {NoStop}%
\bibitem [{\citenamefont {{Bednarek}}\ \emph {et~al.}(2012)\citenamefont
  {{Bednarek}}, \citenamefont {{Haensel}}, \citenamefont {{Zdunik}},
  \citenamefont {{Bejger}},\ and\ \citenamefont {{Ma{\'n}ka}}}]{Bednarek12}%
  \BibitemOpen
  \bibfield  {author} {\bibinfo {author} {\bibnamefont {{Bednarek}},
  \bibfnamefont {I.}}, \bibinfo {author} {\bibfnamefont {P.}~\bibnamefont
  {{Haensel}}}, \bibinfo {author} {\bibfnamefont {J.~L.}\ \bibnamefont
  {{Zdunik}}}, \bibinfo {author} {\bibfnamefont {M.}~\bibnamefont {{Bejger}}},
  \ and\ \bibinfo {author} {\bibfnamefont {R.}~\bibnamefont {{Ma{\'n}ka}}}}
  (\bibinfo {year} {2012}),\ \href {\doibase 10.1051/0004-6361/201118560}
  {\bibfield  {journal} {\bibinfo  {journal} {\aap}\ }\textbf {\bibinfo
  {volume} {543}},\ \bibinfo {eid} {A157}},\ \Eprint
  {http://arxiv.org/abs/1111.6942} {arXiv:1111.6942 [astro-ph.SR]} \BibitemShut
  {NoStop}%
\bibitem [{\citenamefont {{Beloborodov}}(2002)}]{Beloborodov02}%
  \BibitemOpen
  \bibfield  {author} {\bibinfo {author} {\bibnamefont {{Beloborodov}},
  \bibfnamefont {A.~M.}}} (\bibinfo {year} {2002}),\ \href {\doibase
  10.1086/339511} {\bibfield  {journal} {\bibinfo  {journal} {\apjl}\ }\textbf
  {\bibinfo {volume} {566}},\ \bibinfo {pages} {L85}},\ \Eprint
  {http://arxiv.org/abs/astro-ph/0201117} {astro-ph/0201117} \BibitemShut
  {NoStop}%
\bibitem [{\citenamefont {{Berti}}\ and\ \citenamefont
  {{Stergioulas}}(2004)}]{Berti04}%
  \BibitemOpen
  \bibfield  {author} {\bibinfo {author} {\bibnamefont {{Berti}}, \bibfnamefont
  {E.}}, \ and\ \bibinfo {author} {\bibfnamefont {N.}~\bibnamefont
  {{Stergioulas}}}} (\bibinfo {year} {2004}),\ \href {\doibase
  10.1111/j.1365-2966.2004.07740.x} {\bibfield  {journal} {\bibinfo  {journal}
  {\mnras}\ }\textbf {\bibinfo {volume} {350}},\ \bibinfo {pages} {1416}},\
  \Eprint {http://arxiv.org/abs/gr-qc/0310061} {gr-qc/0310061} \BibitemShut
  {NoStop}%
\bibitem [{\citenamefont {{Berti}}\ \emph {et~al.}(2005)\citenamefont
  {{Berti}}, \citenamefont {{White}}, \citenamefont {{Maniopoulou}},\ and\
  \citenamefont {{Bruni}}}]{Berti05}%
  \BibitemOpen
  \bibfield  {author} {\bibinfo {author} {\bibnamefont {{Berti}}, \bibfnamefont
  {E.}}, \bibinfo {author} {\bibfnamefont {F.}~\bibnamefont {{White}}},
  \bibinfo {author} {\bibfnamefont {A.}~\bibnamefont {{Maniopoulou}}}, \ and\
  \bibinfo {author} {\bibfnamefont {M.}~\bibnamefont {{Bruni}}}} (\bibinfo
  {year} {2005}),\ \href {\doibase 10.1111/j.1365-2966.2005.08812.x} {\bibfield
   {journal} {\bibinfo  {journal} {\mnras}\ }\textbf {\bibinfo {volume}
  {358}},\ \bibinfo {pages} {923}},\ \Eprint
  {http://arxiv.org/abs/gr-qc/0405146} {gr-qc/0405146} \BibitemShut {NoStop}%
\bibitem [{\citenamefont {{Bhattacharya}}\ and\ \citenamefont {{van den
  Heuvel}}(1991)}]{Bhattacharya91}%
  \BibitemOpen
  \bibfield  {author} {\bibinfo {author} {\bibnamefont {{Bhattacharya}},
  \bibfnamefont {D.}}, \ and\ \bibinfo {author} {\bibfnamefont {E.~P.~J.}\
  \bibnamefont {{van den Heuvel}}}} (\bibinfo {year} {1991}),\ \href@noop {}
  {\bibfield  {journal} {\bibinfo  {journal} {\physrep}\ }\textbf {\bibinfo
  {volume} {203}},\ \bibinfo {pages} {1}}\BibitemShut {NoStop}%
\bibitem [{\citenamefont {{Bhattacharyya}}\ \emph
  {et~al.}(2005{\natexlab{a}})\citenamefont {{Bhattacharyya}}, \citenamefont
  {{Miller}},\ and\ \citenamefont {{Lamb}}}]{Bhattacharyya05b}%
  \BibitemOpen
  \bibfield  {author} {\bibinfo {author} {\bibnamefont {{Bhattacharyya}},
  \bibfnamefont {S.}}, \bibinfo {author} {\bibfnamefont {M.~C.}\ \bibnamefont
  {{Miller}}}, \ and\ \bibinfo {author} {\bibfnamefont {F.~K.}\ \bibnamefont
  {{Lamb}}}} (\bibinfo {year} {2005}{\natexlab{a}}),\ in\ \href {\doibase
  10.1063/1.1960941} {\emph {\bibinfo {booktitle} {X-ray Diagnostics of
  Astrophysical Plasmas: Theory, Experiment, and Observation}}},\ \bibinfo
  {series} {American Institute of Physics Conference Series}, Vol.\ \bibinfo
  {volume} {774},\ \bibinfo {editor} {edited by\ \bibinfo {editor}
  {\bibfnamefont {R.}~\bibnamefont {{Smith}}}},\ pp.\ \bibinfo {pages}
  {291--293}\BibitemShut {NoStop}%
\bibitem [{\citenamefont {{Bhattacharyya}}\ \emph
  {et~al.}(2005{\natexlab{b}})\citenamefont {{Bhattacharyya}}, \citenamefont
  {{Strohmayer}}, \citenamefont {{Miller}},\ and\ \citenamefont
  {{Markwardt}}}]{Bhattacharyya05c}%
  \BibitemOpen
  \bibfield  {author} {\bibinfo {author} {\bibnamefont {{Bhattacharyya}},
  \bibfnamefont {S.}}, \bibinfo {author} {\bibfnamefont {T.~E.}\ \bibnamefont
  {{Strohmayer}}}, \bibinfo {author} {\bibfnamefont {M.~C.}\ \bibnamefont
  {{Miller}}}, \ and\ \bibinfo {author} {\bibfnamefont {C.~B.}\ \bibnamefont
  {{Markwardt}}}} (\bibinfo {year} {2005}{\natexlab{b}}),\ \href {\doibase
  10.1086/426383} {\bibfield  {journal} {\bibinfo  {journal} {\apj}\ }\textbf
  {\bibinfo {volume} {619}},\ \bibinfo {pages} {483}},\ \Eprint
  {http://arxiv.org/abs/astro-ph/0402534} {astro-ph/0402534} \BibitemShut
  {NoStop}%
\bibitem [{\citenamefont {{Bildsten}}(1998)}]{Bildsten98}%
  \BibitemOpen
  \bibfield  {author} {\bibinfo {author} {\bibnamefont {{Bildsten}},
  \bibfnamefont {L.}}} (\bibinfo {year} {1998}),\ \href {\doibase
  10.1086/311440} {\bibfield  {journal} {\bibinfo  {journal} {\apjl}\ }\textbf
  {\bibinfo {volume} {501}},\ \bibinfo {pages} {L89}},\ \Eprint
  {http://arxiv.org/abs/astro-ph/9804325} {astro-ph/9804325} \BibitemShut
  {NoStop}%
\bibitem [{\citenamefont {{Bodmer}}(1971)}]{Bodmer71}%
  \BibitemOpen
  \bibfield  {author} {\bibinfo {author} {\bibnamefont {{Bodmer}},
  \bibfnamefont {A.~R.}}} (\bibinfo {year} {1971}),\ \href@noop {} {\bibfield
  {journal} {\bibinfo  {journal} {\prd}\ }\textbf {\bibinfo {volume} {4}},\
  \bibinfo {pages} {1601}}\BibitemShut {NoStop}%
\bibitem [{\citenamefont {{Bogdanov}}(2013)}]{Bogdanov13}%
  \BibitemOpen
  \bibfield  {author} {\bibinfo {author} {\bibnamefont {{Bogdanov}},
  \bibfnamefont {S.}}} (\bibinfo {year} {2013}),\ \href {\doibase
  10.1088/0004-637X/762/2/96} {\bibfield  {journal} {\bibinfo  {journal}
  {\apj}\ }\textbf {\bibinfo {volume} {762}},\ \bibinfo {eid} {96}},\ \Eprint
  {http://arxiv.org/abs/1211.6113} {arXiv:1211.6113 [astro-ph.HE]} \BibitemShut
  {NoStop}%
\bibitem [{\citenamefont {{Bogdanov}}\ and\ \citenamefont
  {{Grindlay}}(2009)}]{Bogdanov09}%
  \BibitemOpen
  \bibfield  {author} {\bibinfo {author} {\bibnamefont {{Bogdanov}},
  \bibfnamefont {S.}}, \ and\ \bibinfo {author} {\bibfnamefont {J.~E.}\
  \bibnamefont {{Grindlay}}}} (\bibinfo {year} {2009}),\ \href {\doibase
  10.1088/0004-637X/703/2/1557} {\bibfield  {journal} {\bibinfo  {journal}
  {\apj}\ }\textbf {\bibinfo {volume} {703}},\ \bibinfo {pages} {1557}},\
  \Eprint {http://arxiv.org/abs/0908.1971} {arXiv:0908.1971 [astro-ph.HE]}
  \BibitemShut {NoStop}%
\bibitem [{\citenamefont {{Bogdanov}}\ \emph {et~al.}(2008)\citenamefont
  {{Bogdanov}}, \citenamefont {{Grindlay}},\ and\ \citenamefont
  {{Rybicki}}}]{Bogdanov08}%
  \BibitemOpen
  \bibfield  {author} {\bibinfo {author} {\bibnamefont {{Bogdanov}},
  \bibfnamefont {S.}}, \bibinfo {author} {\bibfnamefont {J.~E.}\ \bibnamefont
  {{Grindlay}}}, \ and\ \bibinfo {author} {\bibfnamefont {G.~B.}\ \bibnamefont
  {{Rybicki}}}} (\bibinfo {year} {2008}),\ \href {\doibase 10.1086/592341}
  {\bibfield  {journal} {\bibinfo  {journal} {\apj}\ }\textbf {\bibinfo
  {volume} {689}},\ \bibinfo {pages} {407}},\ \Eprint
  {http://arxiv.org/abs/0801.4030} {arXiv:0801.4030} \BibitemShut {NoStop}%
\bibitem [{\citenamefont {{Bogdanov}}\ and\ \citenamefont
  {{Halpern}}(2015)}]{Bogdanov15}%
  \BibitemOpen
  \bibfield  {author} {\bibinfo {author} {\bibnamefont {{Bogdanov}},
  \bibfnamefont {S.}}, \ and\ \bibinfo {author} {\bibfnamefont {J.~P.}\
  \bibnamefont {{Halpern}}}} (\bibinfo {year} {2015}),\ \href {\doibase
  10.1088/2041-8205/803/2/L27} {\bibfield  {journal} {\bibinfo  {journal}
  {\apjl}\ }\textbf {\bibinfo {volume} {803}},\ \bibinfo {eid} {L27}},\ \Eprint
  {http://arxiv.org/abs/1503.01698} {arXiv:1503.01698 [astro-ph.HE]}
  \BibitemShut {NoStop}%
\bibitem [{\citenamefont {{Bogdanov}}\ \emph {et~al.}(2014)\citenamefont
  {{Bogdanov}}, \citenamefont {{Patruno}}, \citenamefont {{Archibald}},
  \citenamefont {{Bassa}}, \citenamefont {{Hessels}}, \citenamefont
  {{Janssen}},\ and\ \citenamefont {{Stappers}}}]{Bogdanov14}%
  \BibitemOpen
  \bibfield  {author} {\bibinfo {author} {\bibnamefont {{Bogdanov}},
  \bibfnamefont {S.}}, \bibinfo {author} {\bibfnamefont {A.}~\bibnamefont
  {{Patruno}}}, \bibinfo {author} {\bibfnamefont {A.~M.}\ \bibnamefont
  {{Archibald}}}, \bibinfo {author} {\bibfnamefont {C.}~\bibnamefont
  {{Bassa}}}, \bibinfo {author} {\bibfnamefont {J.~W.~T.}\ \bibnamefont
  {{Hessels}}}, \bibinfo {author} {\bibfnamefont {G.~H.}\ \bibnamefont
  {{Janssen}}}, \ and\ \bibinfo {author} {\bibfnamefont {B.~W.}\ \bibnamefont
  {{Stappers}}}} (\bibinfo {year} {2014}),\ \href {\doibase
  10.1088/0004-637X/789/1/40} {\bibfield  {journal} {\bibinfo  {journal}
  {\apj}\ }\textbf {\bibinfo {volume} {789}},\ \bibinfo {eid} {40}},\ \Eprint
  {http://arxiv.org/abs/1402.6324} {arXiv:1402.6324 [astro-ph.HE]} \BibitemShut
  {NoStop}%
\bibitem [{\citenamefont {{Bogdanov}}\ \emph {et~al.}(2007)\citenamefont
  {{Bogdanov}}, \citenamefont {{Rybicki}},\ and\ \citenamefont
  {{Grindlay}}}]{Bogdanov07}%
  \BibitemOpen
  \bibfield  {author} {\bibinfo {author} {\bibnamefont {{Bogdanov}},
  \bibfnamefont {S.}}, \bibinfo {author} {\bibfnamefont {G.~B.}\ \bibnamefont
  {{Rybicki}}}, \ and\ \bibinfo {author} {\bibfnamefont {J.~E.}\ \bibnamefont
  {{Grindlay}}}} (\bibinfo {year} {2007}),\ \href {\doibase 10.1086/520793}
  {\bibfield  {journal} {\bibinfo  {journal} {\apj}\ }\textbf {\bibinfo
  {volume} {670}},\ \bibinfo {pages} {668}},\ \Eprint
  {http://arxiv.org/abs/astro-ph/0612791} {astro-ph/0612791} \BibitemShut
  {NoStop}%
\bibitem [{\citenamefont {{Bonazzola}}\ \emph {et~al.}(1998)\citenamefont
  {{Bonazzola}}, \citenamefont {{Gourgoulhon}},\ and\ \citenamefont
  {{Marck}}}]{Bonazzola98}%
  \BibitemOpen
  \bibfield  {author} {\bibinfo {author} {\bibnamefont {{Bonazzola}},
  \bibfnamefont {S.}}, \bibinfo {author} {\bibfnamefont {E.}~\bibnamefont
  {{Gourgoulhon}}}, \ and\ \bibinfo {author} {\bibfnamefont {J.-A.}\
  \bibnamefont {{Marck}}}} (\bibinfo {year} {1998}),\ \href {\doibase
  10.1103/PhysRevD.58.104020} {\bibfield  {journal} {\bibinfo  {journal}
  {\prd}\ }\textbf {\bibinfo {volume} {58}}~(\bibinfo {number} {10}),\ \bibinfo
  {eid} {104020}},\ \Eprint {http://arxiv.org/abs/astro-ph/9803086}
  {astro-ph/9803086} \BibitemShut {NoStop}%
\bibitem [{\citenamefont {{Booth}}\ and\ \citenamefont
  {{Jonas}}(2012)}]{Booth12}%
  \BibitemOpen
  \bibfield  {author} {\bibinfo {author} {\bibnamefont {{Booth}}, \bibfnamefont
  {R.~S.}}, \ and\ \bibinfo {author} {\bibfnamefont {J.~L.}\ \bibnamefont
  {{Jonas}}}} (\bibinfo {year} {2012}),\ \href@noop {} {\bibfield  {journal}
  {\bibinfo  {journal} {African Skies}\ }\textbf {\bibinfo {volume} {16}},\
  \bibinfo {pages} {101}}\BibitemShut {NoStop}%
\bibitem [{\citenamefont {Bourke}\ \emph {et~al.}(2015)\citenamefont {Bourke},
  \citenamefont {Braun}, \citenamefont {Fender}, \citenamefont {Govoni},
  \citenamefont {Green}, \citenamefont {Hoare}, \citenamefont {Jarvis},
  \citenamefont {Johnston-Hollitt}, \citenamefont {Keane}, \citenamefont
  {Koopmans} \emph {et~al.}}]{Bourke15}%
  \BibitemOpen
  \bibinfo {editor} {\bibnamefont {Bourke}, \bibfnamefont {T.~L.}}, \bibinfo
  {editor} {\bibfnamefont {R.}~\bibnamefont {Braun}}, \bibinfo {editor}
  {\bibfnamefont {R.}~\bibnamefont {Fender}}, \bibinfo {editor} {\bibfnamefont
  {F.}~\bibnamefont {Govoni}}, \bibinfo {editor} {\bibfnamefont
  {J.}~\bibnamefont {Green}}, \bibinfo {editor} {\bibfnamefont
  {M.}~\bibnamefont {Hoare}}, \bibinfo {editor} {\bibfnamefont
  {M.}~\bibnamefont {Jarvis}}, \bibinfo {editor} {\bibfnamefont
  {M.}~\bibnamefont {Johnston-Hollitt}}, \bibinfo {editor} {\bibfnamefont
  {E.}~\bibnamefont {Keane}}, \bibinfo {editor} {\bibfnamefont
  {L.}~\bibnamefont {Koopmans}},  \emph {et~al.},\ Eds. (\bibinfo {year}
  {2015}),\ \href {http://pos.sissa.it/cgi-bin/reader/conf.cgi?confid=215}
  {\emph {\bibinfo {title} {{Proceedings, Advancing Astrophysics with the
  Square Kilometre Array (AASKA14)}}}},\ Vol.\ \bibinfo {volume} {AASKA14},\
  \bibinfo {organization} {SISSA}\ (\bibinfo  {publisher} {SISSA})\BibitemShut
  {NoStop}%
\bibitem [{\citenamefont {{Braje}}\ \emph {et~al.}(2000)\citenamefont
  {{Braje}}, \citenamefont {{Romani}},\ and\ \citenamefont
  {{Rauch}}}]{Braje00}%
  \BibitemOpen
  \bibfield  {author} {\bibinfo {author} {\bibnamefont {{Braje}}, \bibfnamefont
  {T.~M.}}, \bibinfo {author} {\bibfnamefont {R.~W.}\ \bibnamefont {{Romani}}},
  \ and\ \bibinfo {author} {\bibfnamefont {K.~P.}\ \bibnamefont {{Rauch}}}}
  (\bibinfo {year} {2000}),\ \href {\doibase 10.1086/308448} {\bibfield
  {journal} {\bibinfo  {journal} {\apj}\ }\textbf {\bibinfo {volume} {531}},\
  \bibinfo {pages} {447}},\ \Eprint {http://arxiv.org/abs/astro-ph/0004411}
  {astro-ph/0004411} \BibitemShut {NoStop}%
\bibitem [{\citenamefont {{Burgay}}\ \emph {et~al.}(2003)\citenamefont
  {{Burgay}}, \citenamefont {{D'Amico}}, \citenamefont {{Possenti}},
  \citenamefont {{Manchester}}, \citenamefont {{Lyne}}, \citenamefont
  {{Joshi}}, \citenamefont {{McLaughlin}}, \citenamefont {{Kramer}},
  \citenamefont {{Sarkissian}}, \citenamefont {{Camilo}}, \citenamefont
  {{Kalogera}}, \citenamefont {{Kim}},\ and\ \citenamefont
  {{Lorimer}}}]{Burgay03}%
  \BibitemOpen
  \bibfield  {author} {\bibinfo {author} {\bibnamefont {{Burgay}},
  \bibfnamefont {M.}}, \bibinfo {author} {\bibfnamefont {N.}~\bibnamefont
  {{D'Amico}}}, \bibinfo {author} {\bibfnamefont {A.}~\bibnamefont
  {{Possenti}}}, \bibinfo {author} {\bibfnamefont {R.~N.}\ \bibnamefont
  {{Manchester}}}, \bibinfo {author} {\bibfnamefont {A.~G.}\ \bibnamefont
  {{Lyne}}}, \bibinfo {author} {\bibfnamefont {B.~C.}\ \bibnamefont {{Joshi}}},
  \bibinfo {author} {\bibfnamefont {M.~A.}\ \bibnamefont {{McLaughlin}}},
  \bibinfo {author} {\bibfnamefont {M.}~\bibnamefont {{Kramer}}}, \bibinfo
  {author} {\bibfnamefont {J.~M.}\ \bibnamefont {{Sarkissian}}}, \bibinfo
  {author} {\bibfnamefont {F.}~\bibnamefont {{Camilo}}}, \bibinfo {author}
  {\bibfnamefont {V.}~\bibnamefont {{Kalogera}}}, \bibinfo {author}
  {\bibfnamefont {C.}~\bibnamefont {{Kim}}}, \ and\ \bibinfo {author}
  {\bibfnamefont {D.~R.}\ \bibnamefont {{Lorimer}}}} (\bibinfo {year} {2003}),\
  \href {\doibase 10.1038/nature02124} {\bibfield  {journal} {\bibinfo
  {journal} {\nat}\ }\textbf {\bibinfo {volume} {426}},\ \bibinfo {pages}
  {531}},\ \Eprint {http://arxiv.org/abs/astro-ph/0312071} {astro-ph/0312071}
  \BibitemShut {NoStop}%
\bibitem [{\citenamefont {{Cackett}}\ \emph {et~al.}(2010)\citenamefont
  {{Cackett}}, \citenamefont {{Miller}}, \citenamefont {{Ballantyne}},
  \citenamefont {{Barret}}, \citenamefont {{Bhattacharyya}}, \citenamefont
  {{Boutelier}}, \citenamefont {{Miller}}, \citenamefont {{Strohmayer}},\ and\
  \citenamefont {{Wijnands}}}]{Cackett10}%
  \BibitemOpen
  \bibfield  {author} {\bibinfo {author} {\bibnamefont {{Cackett}},
  \bibfnamefont {E.~M.}}, \bibinfo {author} {\bibfnamefont {J.~M.}\
  \bibnamefont {{Miller}}}, \bibinfo {author} {\bibfnamefont {D.~R.}\
  \bibnamefont {{Ballantyne}}}, \bibinfo {author} {\bibfnamefont
  {D.}~\bibnamefont {{Barret}}}, \bibinfo {author} {\bibfnamefont
  {S.}~\bibnamefont {{Bhattacharyya}}}, \bibinfo {author} {\bibfnamefont
  {M.}~\bibnamefont {{Boutelier}}}, \bibinfo {author} {\bibfnamefont {M.~C.}\
  \bibnamefont {{Miller}}}, \bibinfo {author} {\bibfnamefont {T.~E.}\
  \bibnamefont {{Strohmayer}}}, \ and\ \bibinfo {author} {\bibfnamefont
  {R.}~\bibnamefont {{Wijnands}}}} (\bibinfo {year} {2010}),\ \href {\doibase
  10.1088/0004-637X/720/1/205} {\bibfield  {journal} {\bibinfo  {journal}
  {\apj}\ }\textbf {\bibinfo {volume} {720}},\ \bibinfo {pages} {205}},\
  \Eprint {http://arxiv.org/abs/0908.1098} {arXiv:0908.1098 [astro-ph.HE]}
  \BibitemShut {NoStop}%
\bibitem [{\citenamefont {{Cadeau}}\ \emph {et~al.}(2007)\citenamefont
  {{Cadeau}}, \citenamefont {{Morsink}}, \citenamefont {{Leahy}},\ and\
  \citenamefont {{Campbell}}}]{Cadeau07}%
  \BibitemOpen
  \bibfield  {author} {\bibinfo {author} {\bibnamefont {{Cadeau}},
  \bibfnamefont {C.}}, \bibinfo {author} {\bibfnamefont {S.~M.}\ \bibnamefont
  {{Morsink}}}, \bibinfo {author} {\bibfnamefont {D.}~\bibnamefont {{Leahy}}},
  \ and\ \bibinfo {author} {\bibfnamefont {S.~S.}\ \bibnamefont {{Campbell}}}}
  (\bibinfo {year} {2007}),\ \href {\doibase 10.1086/509103} {\bibfield
  {journal} {\bibinfo  {journal} {\apj}\ }\textbf {\bibinfo {volume} {654}},\
  \bibinfo {pages} {458}},\ \Eprint {http://arxiv.org/abs/astro-ph/0609325}
  {astro-ph/0609325} \BibitemShut {NoStop}%
\bibitem [{\citenamefont {{Carlson}}\ \emph {et~al.}(2014)\citenamefont
  {{Carlson}}, \citenamefont {{Gandolfi}}, \citenamefont {{Pederiva}},
  \citenamefont {{Pieper}}, \citenamefont {{Schiavilla}}, \citenamefont
  {{Schmidt}},\ and\ \citenamefont {{Wiringa}}}]{Carlson14}%
  \BibitemOpen
  \bibfield  {author} {\bibinfo {author} {\bibnamefont {{Carlson}},
  \bibfnamefont {J.}}, \bibinfo {author} {\bibfnamefont {S.}~\bibnamefont
  {{Gandolfi}}}, \bibinfo {author} {\bibfnamefont {F.}~\bibnamefont
  {{Pederiva}}}, \bibinfo {author} {\bibfnamefont {S.~C.}\ \bibnamefont
  {{Pieper}}}, \bibinfo {author} {\bibfnamefont {R.}~\bibnamefont
  {{Schiavilla}}}, \bibinfo {author} {\bibfnamefont {K.~E.}\ \bibnamefont
  {{Schmidt}}}, \ and\ \bibinfo {author} {\bibfnamefont {R.~B.}\ \bibnamefont
  {{Wiringa}}}} (\bibinfo {year} {2014}),\ \href@noop {} {\bibfield  {journal}
  {\bibinfo  {journal} {ArXiv e-prints}\ }}\Eprint
  {http://arxiv.org/abs/1412.3081} {arXiv:1412.3081 [nucl-th]} \BibitemShut
  {NoStop}%
\bibitem [{\citenamefont {{Casares}}(2010)}]{Casares10}%
  \BibitemOpen
  \bibfield  {author} {\bibinfo {author} {\bibnamefont {{Casares}},
  \bibfnamefont {J.}}} (\bibinfo {year} {2010}),\ in\ \href {\doibase
  10.1007/978-3-642-11250-8_1} {\emph {\bibinfo {booktitle} {Highlights of
  Spanish Astrophysics V}}},\ \bibinfo {editor} {edited by\ \bibinfo {editor}
  {\bibfnamefont {J.~M.}\ \bibnamefont {{Diego}}}, \bibinfo {editor}
  {\bibfnamefont {L.~J.}\ \bibnamefont {{Goicoechea}}}, \bibinfo {editor}
  {\bibfnamefont {J.~I.}\ \bibnamefont {{Gonz{\'a}lez-Serrano}}}, \ and\
  \bibinfo {editor} {\bibfnamefont {J.}~\bibnamefont {{Gorgas}}}},\ pp.\
  \bibinfo {pages} {3--642},\ \Eprint {http://arxiv.org/abs/0904.1116}
  {arXiv:0904.1116 [astro-ph.GA]} \BibitemShut {NoStop}%
\bibitem [{\citenamefont {{Casares}}\ \emph {et~al.}(2006)\citenamefont
  {{Casares}}, \citenamefont {{Cornelisse}}, \citenamefont {{Steeghs}},
  \citenamefont {{Charles}}, \citenamefont {{Hynes}}, \citenamefont
  {{O'Brien}},\ and\ \citenamefont {{Strohmayer}}}]{Casares06}%
  \BibitemOpen
  \bibfield  {author} {\bibinfo {author} {\bibnamefont {{Casares}},
  \bibfnamefont {J.}}, \bibinfo {author} {\bibfnamefont {R.}~\bibnamefont
  {{Cornelisse}}}, \bibinfo {author} {\bibfnamefont {D.}~\bibnamefont
  {{Steeghs}}}, \bibinfo {author} {\bibfnamefont {P.~A.}\ \bibnamefont
  {{Charles}}}, \bibinfo {author} {\bibfnamefont {R.~I.}\ \bibnamefont
  {{Hynes}}}, \bibinfo {author} {\bibfnamefont {K.}~\bibnamefont {{O'Brien}}},
  \ and\ \bibinfo {author} {\bibfnamefont {T.~E.}\ \bibnamefont
  {{Strohmayer}}}} (\bibinfo {year} {2006}),\ \href {\doibase
  10.1111/j.1365-2966.2006.11106.x} {\bibfield  {journal} {\bibinfo  {journal}
  {\mnras}\ }\textbf {\bibinfo {volume} {373}},\ \bibinfo {pages} {1235}},\
  \Eprint {http://arxiv.org/abs/astro-ph/0610086} {astro-ph/0610086}
  \BibitemShut {NoStop}%
\bibitem [{\citenamefont {{Casella}}\ \emph {et~al.}(2008)\citenamefont
  {{Casella}}, \citenamefont {{Altamirano}}, \citenamefont {{Patruno}},
  \citenamefont {{Wijnands}},\ and\ \citenamefont {{van der
  Klis}}}]{Casella08}%
  \BibitemOpen
  \bibfield  {author} {\bibinfo {author} {\bibnamefont {{Casella}},
  \bibfnamefont {P.}}, \bibinfo {author} {\bibfnamefont {D.}~\bibnamefont
  {{Altamirano}}}, \bibinfo {author} {\bibfnamefont {A.}~\bibnamefont
  {{Patruno}}}, \bibinfo {author} {\bibfnamefont {R.}~\bibnamefont
  {{Wijnands}}}, \ and\ \bibinfo {author} {\bibfnamefont {M.}~\bibnamefont
  {{van der Klis}}}} (\bibinfo {year} {2008}),\ \href {\doibase 10.1086/528982}
  {\bibfield  {journal} {\bibinfo  {journal} {\apjl}\ }\textbf {\bibinfo
  {volume} {674}},\ \bibinfo {pages} {L41}},\ \Eprint
  {http://arxiv.org/abs/0708.1110} {arXiv:0708.1110} \BibitemShut {NoStop}%
\bibitem [{\citenamefont {{Catuneanu}}\ \emph {et~al.}(2013)\citenamefont
  {{Catuneanu}}, \citenamefont {{Heinke}}, \citenamefont {{Sivakoff}},
  \citenamefont {{Ho}},\ and\ \citenamefont {{Servillat}}}]{Catuneanu13}%
  \BibitemOpen
  \bibfield  {author} {\bibinfo {author} {\bibnamefont {{Catuneanu}},
  \bibfnamefont {A.}}, \bibinfo {author} {\bibfnamefont {C.~O.}\ \bibnamefont
  {{Heinke}}}, \bibinfo {author} {\bibfnamefont {G.~R.}\ \bibnamefont
  {{Sivakoff}}}, \bibinfo {author} {\bibfnamefont {W.~C.~G.}\ \bibnamefont
  {{Ho}}}, \ and\ \bibinfo {author} {\bibfnamefont {M.}~\bibnamefont
  {{Servillat}}}} (\bibinfo {year} {2013}),\ \href {\doibase
  10.1088/0004-637X/764/2/145} {\bibfield  {journal} {\bibinfo  {journal}
  {\apj}\ }\textbf {\bibinfo {volume} {764}},\ \bibinfo {eid} {145}},\ \Eprint
  {http://arxiv.org/abs/1301.3768} {arXiv:1301.3768 [astro-ph.HE]} \BibitemShut
  {NoStop}%
\bibitem [{\citenamefont {{Cavecchi}}\ \emph {et~al.}(2013)\citenamefont
  {{Cavecchi}}, \citenamefont {{Watts}}, \citenamefont {{Braithwaite}},\ and\
  \citenamefont {{Levin}}}]{Cavecchi13}%
  \BibitemOpen
  \bibfield  {author} {\bibinfo {author} {\bibnamefont {{Cavecchi}},
  \bibfnamefont {Y.}}, \bibinfo {author} {\bibfnamefont {A.~L.}\ \bibnamefont
  {{Watts}}}, \bibinfo {author} {\bibfnamefont {J.}~\bibnamefont
  {{Braithwaite}}}, \ and\ \bibinfo {author} {\bibfnamefont {Y.}~\bibnamefont
  {{Levin}}}} (\bibinfo {year} {2013}),\ \href {\doibase 10.1093/mnras/stt1273}
  {\bibfield  {journal} {\bibinfo  {journal} {\mnras}\ }\textbf {\bibinfo
  {volume} {434}},\ \bibinfo {pages} {3526}},\ \Eprint
  {http://arxiv.org/abs/1212.2872} {arXiv:1212.2872 [astro-ph.HE]} \BibitemShut
  {NoStop}%
\bibitem [{\citenamefont {{Chamel}}\ \emph {et~al.}(2013)\citenamefont
  {{Chamel}}, \citenamefont {{Haensel}}, \citenamefont {{Zdunik}},\ and\
  \citenamefont {{Fantina}}}]{Chamel13}%
  \BibitemOpen
  \bibfield  {author} {\bibinfo {author} {\bibnamefont {{Chamel}},
  \bibfnamefont {N.}}, \bibinfo {author} {\bibfnamefont {P.}~\bibnamefont
  {{Haensel}}}, \bibinfo {author} {\bibfnamefont {J.~L.}\ \bibnamefont
  {{Zdunik}}}, \ and\ \bibinfo {author} {\bibfnamefont {A.~F.}\ \bibnamefont
  {{Fantina}}}} (\bibinfo {year} {2013}),\ \href {\doibase
  10.1142/S021830131330018X} {\bibfield  {journal} {\bibinfo  {journal}
  {International Journal of Modern Physics E}\ }\textbf {\bibinfo {volume}
  {22}},\ \bibinfo {eid} {1330018}},\ \Eprint {http://arxiv.org/abs/1307.3995}
  {arXiv:1307.3995 [astro-ph.HE]} \BibitemShut {NoStop}%
\bibitem [{\citenamefont {{Chang}}\ \emph {et~al.}(2005)\citenamefont
  {{Chang}}, \citenamefont {{Bildsten}},\ and\ \citenamefont
  {{Wasserman}}}]{Chang05}%
  \BibitemOpen
  \bibfield  {author} {\bibinfo {author} {\bibnamefont {{Chang}}, \bibfnamefont
  {P.}}, \bibinfo {author} {\bibfnamefont {L.}~\bibnamefont {{Bildsten}}}, \
  and\ \bibinfo {author} {\bibfnamefont {I.}~\bibnamefont {{Wasserman}}}}
  (\bibinfo {year} {2005}),\ \href {\doibase 10.1086/431730} {\bibfield
  {journal} {\bibinfo  {journal} {\apj}\ }\textbf {\bibinfo {volume} {629}},\
  \bibinfo {pages} {998}},\ \Eprint {http://arxiv.org/abs/astro-ph/0505062}
  {astro-ph/0505062} \BibitemShut {NoStop}%
\bibitem [{\citenamefont {{Chen}}\ \emph {et~al.}(2011)\citenamefont {{Chen}},
  \citenamefont {{Baldo}}, \citenamefont {{Burgio}},\ and\ \citenamefont
  {{Schulze}}}]{Chen11}%
  \BibitemOpen
  \bibfield  {author} {\bibinfo {author} {\bibnamefont {{Chen}}, \bibfnamefont
  {H.}}, \bibinfo {author} {\bibfnamefont {M.}~\bibnamefont {{Baldo}}},
  \bibinfo {author} {\bibfnamefont {G.~F.}\ \bibnamefont {{Burgio}}}, \ and\
  \bibinfo {author} {\bibfnamefont {H.-J.}\ \bibnamefont {{Schulze}}}}
  (\bibinfo {year} {2011}),\ \href {\doibase 10.1103/PhysRevD.84.105023}
  {\bibfield  {journal} {\bibinfo  {journal} {\prd}\ }\textbf {\bibinfo
  {volume} {84}}~(\bibinfo {number} {10}),\ \bibinfo {eid} {105023}},\ \Eprint
  {http://arxiv.org/abs/1107.2497} {arXiv:1107.2497 [nucl-th]} \BibitemShut
  {NoStop}%
\bibitem [{\citenamefont {{Colaiuda}}\ and\ \citenamefont
  {{Kokkotas}}(2011)}]{Colaiuda11}%
  \BibitemOpen
  \bibfield  {author} {\bibinfo {author} {\bibnamefont {{Colaiuda}},
  \bibfnamefont {A.}}, \ and\ \bibinfo {author} {\bibfnamefont {K.~D.}\
  \bibnamefont {{Kokkotas}}}} (\bibinfo {year} {2011}),\ \href {\doibase
  10.1111/j.1365-2966.2011.18602.x} {\bibfield  {journal} {\bibinfo  {journal}
  {\mnras}\ }\textbf {\bibinfo {volume} {414}},\ \bibinfo {pages} {3014}},\
  \Eprint {http://arxiv.org/abs/1012.3103} {arXiv:1012.3103 [gr-qc]}
  \BibitemShut {NoStop}%
\bibitem [{\citenamefont {{Colaiuda}}\ and\ \citenamefont
  {{Kokkotas}}(2012)}]{Colaiuda12}%
  \BibitemOpen
  \bibfield  {author} {\bibinfo {author} {\bibnamefont {{Colaiuda}},
  \bibfnamefont {A.}}, \ and\ \bibinfo {author} {\bibfnamefont {K.~D.}\
  \bibnamefont {{Kokkotas}}}} (\bibinfo {year} {2012}),\ \href {\doibase
  10.1111/j.1365-2966.2012.20919.x} {\bibfield  {journal} {\bibinfo  {journal}
  {\mnras}\ }\textbf {\bibinfo {volume} {423}},\ \bibinfo {pages} {811}},\
  \Eprint {http://arxiv.org/abs/1112.3561} {arXiv:1112.3561 [astro-ph.HE]}
  \BibitemShut {NoStop}%
\bibitem [{\citenamefont {{Collins}}\ and\ \citenamefont
  {{Perry}}(1975)}]{Collins75}%
  \BibitemOpen
  \bibfield  {author} {\bibinfo {author} {\bibnamefont {{Collins}},
  \bibfnamefont {J.~C.}}, \ and\ \bibinfo {author} {\bibfnamefont {M.~J.}\
  \bibnamefont {{Perry}}}} (\bibinfo {year} {1975}),\ \href@noop {} {\bibfield
  {journal} {\bibinfo  {journal} {Physical Review Letters}\ }\textbf {\bibinfo
  {volume} {34}},\ \bibinfo {pages} {1353}}\BibitemShut {NoStop}%
\bibitem [{\citenamefont {{Cook}}\ \emph
  {et~al.}(1994{\natexlab{a}})\citenamefont {{Cook}}, \citenamefont
  {{Shapiro}},\ and\ \citenamefont {{Teukolsky}}}]{Cook94a}%
  \BibitemOpen
  \bibfield  {author} {\bibinfo {author} {\bibnamefont {{Cook}}, \bibfnamefont
  {G.~B.}}, \bibinfo {author} {\bibfnamefont {S.~L.}\ \bibnamefont
  {{Shapiro}}}, \ and\ \bibinfo {author} {\bibfnamefont {S.~A.}\ \bibnamefont
  {{Teukolsky}}}} (\bibinfo {year} {1994}{\natexlab{a}}),\ \href {\doibase
  10.1086/173934} {\bibfield  {journal} {\bibinfo  {journal} {\apj}\ }\textbf
  {\bibinfo {volume} {424}},\ \bibinfo {pages} {823}}\BibitemShut {NoStop}%
\bibitem [{\citenamefont {{Cook}}\ \emph
  {et~al.}(1994{\natexlab{b}})\citenamefont {{Cook}}, \citenamefont
  {{Shapiro}},\ and\ \citenamefont {{Teukolsky}}}]{Cook94b}%
  \BibitemOpen
  \bibfield  {author} {\bibinfo {author} {\bibnamefont {{Cook}}, \bibfnamefont
  {G.~B.}}, \bibinfo {author} {\bibfnamefont {S.~L.}\ \bibnamefont
  {{Shapiro}}}, \ and\ \bibinfo {author} {\bibfnamefont {S.~A.}\ \bibnamefont
  {{Teukolsky}}}} (\bibinfo {year} {1994}{\natexlab{b}}),\ \href {\doibase
  10.1086/187250} {\bibfield  {journal} {\bibinfo  {journal} {\apjl}\ }\textbf
  {\bibinfo {volume} {423}},\ \bibinfo {pages} {L117}}\BibitemShut {NoStop}%
\bibitem [{\citenamefont {{D'Angelo}}\ and\ \citenamefont
  {{Watts}}(2012)}]{DAngelo12}%
  \BibitemOpen
  \bibfield  {author} {\bibinfo {author} {\bibnamefont {{D'Angelo}},
  \bibfnamefont {C.~R.}}, \ and\ \bibinfo {author} {\bibfnamefont {A.~L.}\
  \bibnamefont {{Watts}}}} (\bibinfo {year} {2012}),\ \href {\doibase
  10.1088/2041-8205/751/2/L41} {\bibfield  {journal} {\bibinfo  {journal}
  {\apjl}\ }\textbf {\bibinfo {volume} {751}},\ \bibinfo {eid} {L41}},\ \Eprint
  {http://arxiv.org/abs/1204.3053} {arXiv:1204.3053 [astro-ph.SR]} \BibitemShut
  {NoStop}%
\bibitem [{\citenamefont {{Danielewicz}}\ \emph {et~al.}(2002)\citenamefont
  {{Danielewicz}}, \citenamefont {{Lacey}},\ and\ \citenamefont
  {{Lynch}}}]{Danielewicz02}%
  \BibitemOpen
  \bibfield  {author} {\bibinfo {author} {\bibnamefont {{Danielewicz}},
  \bibfnamefont {P.}}, \bibinfo {author} {\bibfnamefont {R.}~\bibnamefont
  {{Lacey}}}, \ and\ \bibinfo {author} {\bibfnamefont {W.~G.}\ \bibnamefont
  {{Lynch}}}} (\bibinfo {year} {2002}),\ \href@noop {} {\bibfield  {journal}
  {\bibinfo  {journal} {Science}\ }\textbf {\bibinfo {volume} {298}},\ \bibinfo
  {pages} {1592}},\ \Eprint {http://arxiv.org/abs/nucl-th/0208016}
  {nucl-th/0208016} \BibitemShut {NoStop}%
\bibitem [{\citenamefont {{Demorest}}\ \emph {et~al.}(2010)\citenamefont
  {{Demorest}}, \citenamefont {{Pennucci}}, \citenamefont {{Ransom}},
  \citenamefont {{Roberts}},\ and\ \citenamefont {{Hessels}}}]{Demorest10}%
  \BibitemOpen
  \bibfield  {author} {\bibinfo {author} {\bibnamefont {{Demorest}},
  \bibfnamefont {P.~B.}}, \bibinfo {author} {\bibfnamefont {T.}~\bibnamefont
  {{Pennucci}}}, \bibinfo {author} {\bibfnamefont {S.~M.}\ \bibnamefont
  {{Ransom}}}, \bibinfo {author} {\bibfnamefont {M.~S.~E.}\ \bibnamefont
  {{Roberts}}}, \ and\ \bibinfo {author} {\bibfnamefont {J.~W.~T.}\
  \bibnamefont {{Hessels}}}} (\bibinfo {year} {2010}),\ \href {\doibase
  10.1038/nature09466} {\bibfield  {journal} {\bibinfo  {journal} {\nat}\
  }\textbf {\bibinfo {volume} {467}},\ \bibinfo {pages} {1081}},\ \Eprint
  {http://arxiv.org/abs/1010.5788} {arXiv:1010.5788 [astro-ph.HE]} \BibitemShut
  {NoStop}%
\bibitem [{\citenamefont {{Duncan}}(1998)}]{Duncan98}%
  \BibitemOpen
  \bibfield  {author} {\bibinfo {author} {\bibnamefont {{Duncan}},
  \bibfnamefont {R.~C.}}} (\bibinfo {year} {1998}),\ \href {\doibase
  10.1086/311303} {\bibfield  {journal} {\bibinfo  {journal} {\apjl}\ }\textbf
  {\bibinfo {volume} {498}},\ \bibinfo {pages} {L45}},\ \Eprint
  {http://arxiv.org/abs/astro-ph/9803060} {astro-ph/9803060} \BibitemShut
  {NoStop}%
\bibitem [{\citenamefont {{Egron}}\ \emph {et~al.}(2011)\citenamefont
  {{Egron}}, \citenamefont {{di Salvo}}, \citenamefont {{Burderi}},
  \citenamefont {{Papitto}}, \citenamefont {{Barrag{\'a}n}}, \citenamefont
  {{Dauser}}, \citenamefont {{Wilms}}, \citenamefont {{D'A{\`i}}},
  \citenamefont {{Riggio}}, \citenamefont {{Iaria}},\ and\ \citenamefont
  {{Robba}}}]{Egron11}%
  \BibitemOpen
  \bibfield  {author} {\bibinfo {author} {\bibnamefont {{Egron}}, \bibfnamefont
  {E.}}, \bibinfo {author} {\bibfnamefont {T.}~\bibnamefont {{di Salvo}}},
  \bibinfo {author} {\bibfnamefont {L.}~\bibnamefont {{Burderi}}}, \bibinfo
  {author} {\bibfnamefont {A.}~\bibnamefont {{Papitto}}}, \bibinfo {author}
  {\bibfnamefont {L.}~\bibnamefont {{Barrag{\'a}n}}}, \bibinfo {author}
  {\bibfnamefont {T.}~\bibnamefont {{Dauser}}}, \bibinfo {author}
  {\bibfnamefont {J.}~\bibnamefont {{Wilms}}}, \bibinfo {author} {\bibfnamefont
  {A.}~\bibnamefont {{D'A{\`i}}}}, \bibinfo {author} {\bibfnamefont
  {A.}~\bibnamefont {{Riggio}}}, \bibinfo {author} {\bibfnamefont
  {R.}~\bibnamefont {{Iaria}}}, \ and\ \bibinfo {author} {\bibfnamefont
  {N.~R.}\ \bibnamefont {{Robba}}}} (\bibinfo {year} {2011}),\ \href@noop {}
  {\bibfield  {journal} {\bibinfo  {journal} {\aap}\ }\textbf {\bibinfo
  {volume} {530}},\ \bibinfo {eid} {A99}},\ \Eprint
  {http://arxiv.org/abs/1104.0566} {arXiv:1104.0566 [astro-ph.HE]} \BibitemShut
  {NoStop}%
\bibitem [{\citenamefont {{Epelbaum}}\ \emph {et~al.}(2009)\citenamefont
  {{Epelbaum}}, \citenamefont {{Hammer}},\ and\ \citenamefont
  {{Mei{\ss}ner}}}]{Epelbaum09}%
  \BibitemOpen
  \bibfield  {author} {\bibinfo {author} {\bibnamefont {{Epelbaum}},
  \bibfnamefont {E.}}, \bibinfo {author} {\bibfnamefont {H.-W.}\ \bibnamefont
  {{Hammer}}}, \ and\ \bibinfo {author} {\bibfnamefont {U.-G.}\ \bibnamefont
  {{Mei{\ss}ner}}}} (\bibinfo {year} {2009}),\ \href {\doibase
  10.1103/RevModPhys.81.1773} {\bibfield  {journal} {\bibinfo  {journal}
  {Reviews of Modern Physics}\ }\textbf {\bibinfo {volume} {81}},\ \bibinfo
  {pages} {1773}},\ \Eprint {http://arxiv.org/abs/0811.1338} {arXiv:0811.1338
  [nucl-th]} \BibitemShut {NoStop}%
\bibitem [{\citenamefont {{Faber}}\ and\ \citenamefont
  {{Rasio}}(2012)}]{Faber12}%
  \BibitemOpen
  \bibfield  {author} {\bibinfo {author} {\bibnamefont {{Faber}}, \bibfnamefont
  {J.~A.}}, \ and\ \bibinfo {author} {\bibfnamefont {F.~A.}\ \bibnamefont
  {{Rasio}}}} (\bibinfo {year} {2012}),\ \href {\doibase 10.12942/lrr-2012-8}
  {\bibfield  {journal} {\bibinfo  {journal} {Living Reviews in Relativity}\
  }\textbf {\bibinfo {volume} {15}},\ \bibinfo {pages} {8}},\ \Eprint
  {http://arxiv.org/abs/1204.3858} {arXiv:1204.3858 [gr-qc]} \BibitemShut
  {NoStop}%
\bibitem [{\citenamefont {{Falanga}}\ \emph {et~al.}(2005)\citenamefont
  {{Falanga}}, \citenamefont {{Kuiper}}, \citenamefont {{Poutanen}},
  \citenamefont {{Bonning}}, \citenamefont {{Hermsen}}, \citenamefont {{di
  Salvo}}, \citenamefont {{Goldoni}}, \citenamefont {{Goldwurm}}, \citenamefont
  {{Shaw}},\ and\ \citenamefont {{Stella}}}]{Falanga05}%
  \BibitemOpen
  \bibfield  {author} {\bibinfo {author} {\bibnamefont {{Falanga}},
  \bibfnamefont {M.}}, \bibinfo {author} {\bibfnamefont {L.}~\bibnamefont
  {{Kuiper}}}, \bibinfo {author} {\bibfnamefont {J.}~\bibnamefont
  {{Poutanen}}}, \bibinfo {author} {\bibfnamefont {E.~W.}\ \bibnamefont
  {{Bonning}}}, \bibinfo {author} {\bibfnamefont {W.}~\bibnamefont
  {{Hermsen}}}, \bibinfo {author} {\bibfnamefont {T.}~\bibnamefont {{di
  Salvo}}}, \bibinfo {author} {\bibfnamefont {P.}~\bibnamefont {{Goldoni}}},
  \bibinfo {author} {\bibfnamefont {A.}~\bibnamefont {{Goldwurm}}}, \bibinfo
  {author} {\bibfnamefont {S.~E.}\ \bibnamefont {{Shaw}}}, \ and\ \bibinfo
  {author} {\bibfnamefont {L.}~\bibnamefont {{Stella}}}} (\bibinfo {year}
  {2005}),\ \href {\doibase 10.1051/0004-6361:20053472} {\bibfield  {journal}
  {\bibinfo  {journal} {\aap}\ }\textbf {\bibinfo {volume} {444}},\ \bibinfo
  {pages} {15}},\ \Eprint {http://arxiv.org/abs/astro-ph/0508613}
  {astro-ph/0508613} \BibitemShut {NoStop}%
\bibitem [{\citenamefont {{Feroci}}\ \emph {et~al.}(2014)\citenamefont
  {{Feroci}}, \citenamefont {{den Herder}}, \citenamefont {{Bozzo}},
  \citenamefont {{Barret}}, \citenamefont {{Brandt}}, \citenamefont
  {{Hernanz}}, \citenamefont {{van der Klis}}, \citenamefont {{Pohl}},
  \citenamefont {{Santangelo}}, \citenamefont {{Stella}} \emph
  {et~al.}}]{Feroci14}%
  \BibitemOpen
  \bibfield  {author} {\bibinfo {author} {\bibnamefont {{Feroci}},
  \bibfnamefont {M.}}, \bibinfo {author} {\bibfnamefont {J.~W.}\ \bibnamefont
  {{den Herder}}}, \bibinfo {author} {\bibfnamefont {E.}~\bibnamefont
  {{Bozzo}}}, \bibinfo {author} {\bibfnamefont {D.}~\bibnamefont {{Barret}}},
  \bibinfo {author} {\bibfnamefont {S.}~\bibnamefont {{Brandt}}}, \bibinfo
  {author} {\bibfnamefont {M.}~\bibnamefont {{Hernanz}}}, \bibinfo {author}
  {\bibfnamefont {M.}~\bibnamefont {{van der Klis}}}, \bibinfo {author}
  {\bibfnamefont {M.}~\bibnamefont {{Pohl}}}, \bibinfo {author} {\bibfnamefont
  {A.}~\bibnamefont {{Santangelo}}}, \bibinfo {author} {\bibfnamefont
  {L.}~\bibnamefont {{Stella}}},  \emph {et~al.}} (\bibinfo {year} {2014}),\
  in\ \href {\doibase 10.1117/12.2055913} {\emph {\bibinfo {booktitle} {Society
  of Photo-Optical Instrumentation Engineers (SPIE) Conference Series}}},\
  \bibinfo {series} {Society of Photo-Optical Instrumentation Engineers (SPIE)
  Conference Series}, Vol.\ \bibinfo {volume} {9144},\ p.~\bibinfo {pages}
  {2},\ \Eprint {http://arxiv.org/abs/1408.6526} {arXiv:1408.6526
  [astro-ph.IM]} \BibitemShut {NoStop}%
\bibitem [{\citenamefont {{Feroci}}\ \emph {et~al.}(2012)\citenamefont
  {{Feroci}}, \citenamefont {{Stella}}, \citenamefont {{van der Klis}},
  \citenamefont {{Courvoisier}}, \citenamefont {{Hernanz}}, \citenamefont
  {{Hudec}}, \citenamefont {{Santangelo}}, \citenamefont {{Walton}},
  \citenamefont {{Zdziarski}}, \citenamefont {{Barret}} \emph
  {et~al.}}]{Feroci12}%
  \BibitemOpen
  \bibfield  {author} {\bibinfo {author} {\bibnamefont {{Feroci}},
  \bibfnamefont {M.}}, \bibinfo {author} {\bibfnamefont {L.}~\bibnamefont
  {{Stella}}}, \bibinfo {author} {\bibfnamefont {M.}~\bibnamefont {{van der
  Klis}}}, \bibinfo {author} {\bibfnamefont {T.~J.-L.}\ \bibnamefont
  {{Courvoisier}}}, \bibinfo {author} {\bibfnamefont {M.}~\bibnamefont
  {{Hernanz}}}, \bibinfo {author} {\bibfnamefont {R.}~\bibnamefont {{Hudec}}},
  \bibinfo {author} {\bibfnamefont {A.}~\bibnamefont {{Santangelo}}}, \bibinfo
  {author} {\bibfnamefont {D.}~\bibnamefont {{Walton}}}, \bibinfo {author}
  {\bibfnamefont {A.}~\bibnamefont {{Zdziarski}}}, \bibinfo {author}
  {\bibfnamefont {D.}~\bibnamefont {{Barret}}},  \emph {et~al.}} (\bibinfo
  {year} {2012}),\ \href {\doibase 10.1007/s10686-011-9237-2} {\bibfield
  {journal} {\bibinfo  {journal} {Experimental Astronomy}\ }\textbf {\bibinfo
  {volume} {34}},\ \bibinfo {pages} {415}},\ \Eprint
  {http://arxiv.org/abs/1107.0436} {arXiv:1107.0436 [astro-ph.IM]} \BibitemShut
  {NoStop}%
\bibitem [{\citenamefont {{Fukushima}}\ and\ \citenamefont
  {{Hatsuda}}(2011)}]{Fukushima11}%
  \BibitemOpen
  \bibfield  {author} {\bibinfo {author} {\bibnamefont {{Fukushima}},
  \bibfnamefont {K.}}, \ and\ \bibinfo {author} {\bibfnamefont
  {T.}~\bibnamefont {{Hatsuda}}}} (\bibinfo {year} {2011}),\ \href@noop {}
  {\bibfield  {journal} {\bibinfo  {journal} {\rpp}\ }\textbf {\bibinfo
  {volume} {74}},\ \bibinfo {eid} {1}}\BibitemShut {NoStop}%
\bibitem [{\citenamefont {{Gabler}}\ \emph
  {et~al.}(2013{\natexlab{a}})\citenamefont {{Gabler}}, \citenamefont
  {{Cerd{\'a}-Dur{\'a}n}}, \citenamefont {{Font}}, \citenamefont
  {{M{\"u}ller}},\ and\ \citenamefont {{Stergioulas}}}]{Gabler13a}%
  \BibitemOpen
  \bibfield  {author} {\bibinfo {author} {\bibnamefont {{Gabler}},
  \bibfnamefont {M.}}, \bibinfo {author} {\bibfnamefont {P.}~\bibnamefont
  {{Cerd{\'a}-Dur{\'a}n}}}, \bibinfo {author} {\bibfnamefont {J.~A.}\
  \bibnamefont {{Font}}}, \bibinfo {author} {\bibfnamefont {E.}~\bibnamefont
  {{M{\"u}ller}}}, \ and\ \bibinfo {author} {\bibfnamefont {N.}~\bibnamefont
  {{Stergioulas}}}} (\bibinfo {year} {2013}{\natexlab{a}}),\ \href {\doibase
  10.1093/mnras/sts721} {\bibfield  {journal} {\bibinfo  {journal} {\mnras}\
  }\textbf {\bibinfo {volume} {430}},\ \bibinfo {pages} {1811}},\ \Eprint
  {http://arxiv.org/abs/1208.6443} {arXiv:1208.6443 [astro-ph.SR]} \BibitemShut
  {NoStop}%
\bibitem [{\citenamefont {{Gabler}}\ \emph {et~al.}(2012)\citenamefont
  {{Gabler}}, \citenamefont {{Cerd{\'a}-Dur{\'a}n}}, \citenamefont
  {{Stergioulas}}, \citenamefont {{Font}},\ and\ \citenamefont
  {{M{\"u}ller}}}]{Gabler12}%
  \BibitemOpen
  \bibfield  {author} {\bibinfo {author} {\bibnamefont {{Gabler}},
  \bibfnamefont {M.}}, \bibinfo {author} {\bibfnamefont {P.}~\bibnamefont
  {{Cerd{\'a}-Dur{\'a}n}}}, \bibinfo {author} {\bibfnamefont {N.}~\bibnamefont
  {{Stergioulas}}}, \bibinfo {author} {\bibfnamefont {J.~A.}\ \bibnamefont
  {{Font}}}, \ and\ \bibinfo {author} {\bibfnamefont {E.}~\bibnamefont
  {{M{\"u}ller}}}} (\bibinfo {year} {2012}),\ \href {\doibase
  10.1111/j.1365-2966.2012.20454.x} {\bibfield  {journal} {\bibinfo  {journal}
  {\mnras}\ }\textbf {\bibinfo {volume} {421}},\ \bibinfo {pages} {2054}},\
  \Eprint {http://arxiv.org/abs/1109.6233} {arXiv:1109.6233 [astro-ph.HE]}
  \BibitemShut {NoStop}%
\bibitem [{\citenamefont {{Gabler}}\ \emph
  {et~al.}(2013{\natexlab{b}})\citenamefont {{Gabler}}, \citenamefont
  {{Cerd{\'a}-Dur{\'a}n}}, \citenamefont {{Stergioulas}}, \citenamefont
  {{Font}},\ and\ \citenamefont {{M{\"u}ller}}}]{Gabler13b}%
  \BibitemOpen
  \bibfield  {author} {\bibinfo {author} {\bibnamefont {{Gabler}},
  \bibfnamefont {M.}}, \bibinfo {author} {\bibfnamefont {P.}~\bibnamefont
  {{Cerd{\'a}-Dur{\'a}n}}}, \bibinfo {author} {\bibfnamefont {N.}~\bibnamefont
  {{Stergioulas}}}, \bibinfo {author} {\bibfnamefont {J.~A.}\ \bibnamefont
  {{Font}}}, \ and\ \bibinfo {author} {\bibfnamefont {E.}~\bibnamefont
  {{M{\"u}ller}}}} (\bibinfo {year} {2013}{\natexlab{b}}),\ \href {\doibase
  10.1103/PhysRevLett.111.211102} {\bibfield  {journal} {\bibinfo  {journal}
  {Physical Review Letters}\ }\textbf {\bibinfo {volume} {111}}~(\bibinfo
  {number} {21}),\ \bibinfo {eid} {211102}},\ \Eprint
  {http://arxiv.org/abs/1304.3566} {arXiv:1304.3566 [astro-ph.HE]} \BibitemShut
  {NoStop}%
\bibitem [{\citenamefont {{Gabler}}\ \emph {et~al.}(2014)\citenamefont
  {{Gabler}}, \citenamefont {{Cerd{\'a}-Dur{\'a}n}}, \citenamefont
  {{Stergioulas}}, \citenamefont {{Font}},\ and\ \citenamefont
  {{M{\"u}ller}}}]{Gabler14}%
  \BibitemOpen
  \bibfield  {author} {\bibinfo {author} {\bibnamefont {{Gabler}},
  \bibfnamefont {M.}}, \bibinfo {author} {\bibfnamefont {P.}~\bibnamefont
  {{Cerd{\'a}-Dur{\'a}n}}}, \bibinfo {author} {\bibfnamefont {N.}~\bibnamefont
  {{Stergioulas}}}, \bibinfo {author} {\bibfnamefont {J.~A.}\ \bibnamefont
  {{Font}}}, \ and\ \bibinfo {author} {\bibfnamefont {E.}~\bibnamefont
  {{M{\"u}ller}}}} (\bibinfo {year} {2014}),\ \href {\doibase
  10.1093/mnras/stu1263} {\bibfield  {journal} {\bibinfo  {journal} {\mnras}\
  }\textbf {\bibinfo {volume} {443}},\ \bibinfo {pages} {1416}},\ \Eprint
  {http://arxiv.org/abs/1407.7672} {arXiv:1407.7672 [astro-ph.HE]} \BibitemShut
  {NoStop}%
\bibitem [{\citenamefont {{Galloway}}\ \emph {et~al.}(2007)\citenamefont
  {{Galloway}}, \citenamefont {{Morgan}}, \citenamefont {{Krauss}},
  \citenamefont {{Kaaret}},\ and\ \citenamefont {{Chakrabarty}}}]{Galloway07}%
  \BibitemOpen
  \bibfield  {author} {\bibinfo {author} {\bibnamefont {{Galloway}},
  \bibfnamefont {D.~K.}}, \bibinfo {author} {\bibfnamefont {E.~H.}\
  \bibnamefont {{Morgan}}}, \bibinfo {author} {\bibfnamefont {M.~I.}\
  \bibnamefont {{Krauss}}}, \bibinfo {author} {\bibfnamefont {P.}~\bibnamefont
  {{Kaaret}}}, \ and\ \bibinfo {author} {\bibfnamefont {D.}~\bibnamefont
  {{Chakrabarty}}}} (\bibinfo {year} {2007}),\ \href {\doibase 10.1086/510741}
  {\bibfield  {journal} {\bibinfo  {journal} {\apjl}\ }\textbf {\bibinfo
  {volume} {654}},\ \bibinfo {pages} {L73}},\ \Eprint
  {http://arxiv.org/abs/astro-ph/0609693} {astro-ph/0609693} \BibitemShut
  {NoStop}%
\bibitem [{\citenamefont {{Galloway}}\ \emph
  {et~al.}(2008{\natexlab{a}})\citenamefont {{Galloway}}, \citenamefont
  {{Muno}}, \citenamefont {{Hartman}}, \citenamefont {{Psaltis}},\ and\
  \citenamefont {{Chakrabarty}}}]{Galloway08}%
  \BibitemOpen
  \bibfield  {author} {\bibinfo {author} {\bibnamefont {{Galloway}},
  \bibfnamefont {D.~K.}}, \bibinfo {author} {\bibfnamefont {M.~P.}\
  \bibnamefont {{Muno}}}, \bibinfo {author} {\bibfnamefont {J.~M.}\
  \bibnamefont {{Hartman}}}, \bibinfo {author} {\bibfnamefont {D.}~\bibnamefont
  {{Psaltis}}}, \ and\ \bibinfo {author} {\bibfnamefont {D.}~\bibnamefont
  {{Chakrabarty}}}} (\bibinfo {year} {2008}{\natexlab{a}}),\ \href {\doibase
  10.1086/592044} {\bibfield  {journal} {\bibinfo  {journal} {\apjs}\ }\textbf
  {\bibinfo {volume} {179}},\ \bibinfo {pages} {360}},\ \Eprint
  {http://arxiv.org/abs/astro-ph/0608259} {astro-ph/0608259} \BibitemShut
  {NoStop}%
\bibitem [{\citenamefont {{Galloway}}\ \emph
  {et~al.}(2008{\natexlab{b}})\citenamefont {{Galloway}}, \citenamefont
  {{{\"O}zel}},\ and\ \citenamefont {{Psaltis}}}]{Galloway08b}%
  \BibitemOpen
  \bibfield  {author} {\bibinfo {author} {\bibnamefont {{Galloway}},
  \bibfnamefont {D.~K.}}, \bibinfo {author} {\bibfnamefont {F.}~\bibnamefont
  {{{\"O}zel}}}, \ and\ \bibinfo {author} {\bibfnamefont {D.}~\bibnamefont
  {{Psaltis}}}} (\bibinfo {year} {2008}{\natexlab{b}}),\ \href {\doibase
  10.1111/j.1365-2966.2008.13219.x} {\bibfield  {journal} {\bibinfo  {journal}
  {\mnras}\ }\textbf {\bibinfo {volume} {387}},\ \bibinfo {pages} {268}},\
  \Eprint {http://arxiv.org/abs/0712.0412} {arXiv:0712.0412} \BibitemShut
  {NoStop}%
\bibitem [{\citenamefont {{Gandolfi}}\ \emph {et~al.}(2012)\citenamefont
  {{Gandolfi}}, \citenamefont {{Carlson}},\ and\ \citenamefont
  {{Reddy}}}]{Gandolfi12}%
  \BibitemOpen
  \bibfield  {author} {\bibinfo {author} {\bibnamefont {{Gandolfi}},
  \bibfnamefont {S.}}, \bibinfo {author} {\bibfnamefont {J.}~\bibnamefont
  {{Carlson}}}, \ and\ \bibinfo {author} {\bibfnamefont {S.}~\bibnamefont
  {{Reddy}}}} (\bibinfo {year} {2012}),\ \href {\doibase
  10.1103/PhysRevC.85.032801} {\bibfield  {journal} {\bibinfo  {journal}
  {\prc}\ }\textbf {\bibinfo {volume} {85}}~(\bibinfo {number} {3}),\ \bibinfo
  {eid} {032801}},\ \Eprint {http://arxiv.org/abs/1101.1921} {arXiv:1101.1921
  [nucl-th]} \BibitemShut {NoStop}%
\bibitem [{\citenamefont {{Gandolfi}}\ \emph {et~al.}(2014)\citenamefont
  {{Gandolfi}}, \citenamefont {{Carlson}}, \citenamefont {{Reddy}},
  \citenamefont {{Steiner}},\ and\ \citenamefont {{Wiringa}}}]{Gandolfi14}%
  \BibitemOpen
  \bibfield  {author} {\bibinfo {author} {\bibnamefont {{Gandolfi}},
  \bibfnamefont {S.}}, \bibinfo {author} {\bibfnamefont {J.}~\bibnamefont
  {{Carlson}}}, \bibinfo {author} {\bibfnamefont {S.}~\bibnamefont {{Reddy}}},
  \bibinfo {author} {\bibfnamefont {A.~W.}\ \bibnamefont {{Steiner}}}, \ and\
  \bibinfo {author} {\bibfnamefont {R.~B.}\ \bibnamefont {{Wiringa}}}}
  (\bibinfo {year} {2014}),\ \href {\doibase 10.1140/epja/i2014-14010-5}
  {\bibfield  {journal} {\bibinfo  {journal} {European Physical Journal A}\
  }\textbf {\bibinfo {volume} {50}},\ \bibinfo {eid} {10}},\ \Eprint
  {http://arxiv.org/abs/1307.5815} {arXiv:1307.5815 [nucl-th]} \BibitemShut
  {NoStop}%
\bibitem [{\citenamefont {{Gendreau}}\ \emph {et~al.}(2012)\citenamefont
  {{Gendreau}}, \citenamefont {{Arzoumanian}},\ and\ \citenamefont
  {{Okajima}}}]{Gendreau12}%
  \BibitemOpen
  \bibfield  {author} {\bibinfo {author} {\bibnamefont {{Gendreau}},
  \bibfnamefont {K.~C.}}, \bibinfo {author} {\bibfnamefont {Z.}~\bibnamefont
  {{Arzoumanian}}}, \ and\ \bibinfo {author} {\bibfnamefont {T.}~\bibnamefont
  {{Okajima}}}} (\bibinfo {year} {2012}),\ in\ \href {\doibase
  10.1117/12.926396} {\emph {\bibinfo {booktitle} {Society of Photo-Optical
  Instrumentation Engineers (SPIE) Conference Series}}},\ \bibinfo {series}
  {Society of Photo-Optical Instrumentation Engineers (SPIE) Conference
  Series}, Vol.\ \bibinfo {volume} {8443},\ p.~\bibinfo {pages}
  {13}\BibitemShut {NoStop}%
\bibitem [{\citenamefont {{Ghosh}}\ and\ \citenamefont
  {{Lamb}}(1978)}]{Ghosh78}%
  \BibitemOpen
  \bibfield  {author} {\bibinfo {author} {\bibnamefont {{Ghosh}}, \bibfnamefont
  {P.}}, \ and\ \bibinfo {author} {\bibfnamefont {F.~K.}\ \bibnamefont
  {{Lamb}}}} (\bibinfo {year} {1978}),\ \href {\doibase 10.1086/182734}
  {\bibfield  {journal} {\bibinfo  {journal} {\apjl}\ }\textbf {\bibinfo
  {volume} {223}},\ \bibinfo {pages} {L83}}\BibitemShut {NoStop}%
\bibitem [{\citenamefont {{Ghosh}}\ and\ \citenamefont
  {{Lamb}}(1979{\natexlab{a}})}]{Ghosh79a}%
  \BibitemOpen
  \bibfield  {author} {\bibinfo {author} {\bibnamefont {{Ghosh}}, \bibfnamefont
  {P.}}, \ and\ \bibinfo {author} {\bibfnamefont {F.~K.}\ \bibnamefont
  {{Lamb}}}} (\bibinfo {year} {1979}{\natexlab{a}}),\ \href {\doibase
  10.1086/157285} {\bibfield  {journal} {\bibinfo  {journal} {\apj}\ }\textbf
  {\bibinfo {volume} {232}},\ \bibinfo {pages} {259}}\BibitemShut {NoStop}%
\bibitem [{\citenamefont {{Ghosh}}\ and\ \citenamefont
  {{Lamb}}(1979{\natexlab{b}})}]{Ghosh79b}%
  \BibitemOpen
  \bibfield  {author} {\bibinfo {author} {\bibnamefont {{Ghosh}}, \bibfnamefont
  {P.}}, \ and\ \bibinfo {author} {\bibfnamefont {F.~K.}\ \bibnamefont
  {{Lamb}}}} (\bibinfo {year} {1979}{\natexlab{b}}),\ \href {\doibase
  10.1086/157498} {\bibfield  {journal} {\bibinfo  {journal} {\apj}\ }\textbf
  {\bibinfo {volume} {234}},\ \bibinfo {pages} {296}}\BibitemShut {NoStop}%
\bibitem [{\citenamefont {{Glampedakis}}\ and\ \citenamefont
  {{Jones}}(2014)}]{Glampedakis14}%
  \BibitemOpen
  \bibfield  {author} {\bibinfo {author} {\bibnamefont {{Glampedakis}},
  \bibfnamefont {K.}}, \ and\ \bibinfo {author} {\bibfnamefont {D.~I.}\
  \bibnamefont {{Jones}}}} (\bibinfo {year} {2014}),\ \href {\doibase
  10.1093/mnras/stu017} {\bibfield  {journal} {\bibinfo  {journal} {\mnras}\
  }\textbf {\bibinfo {volume} {439}},\ \bibinfo {pages} {1522}},\ \Eprint
  {http://arxiv.org/abs/1307.7078} {arXiv:1307.7078 [astro-ph.SR]} \BibitemShut
  {NoStop}%
\bibitem [{\citenamefont {{Glampedakis}}\ \emph {et~al.}(2006)\citenamefont
  {{Glampedakis}}, \citenamefont {{Samuelsson}},\ and\ \citenamefont
  {{Andersson}}}]{Glampedakis06}%
  \BibitemOpen
  \bibfield  {author} {\bibinfo {author} {\bibnamefont {{Glampedakis}},
  \bibfnamefont {K.}}, \bibinfo {author} {\bibfnamefont {L.}~\bibnamefont
  {{Samuelsson}}}, \ and\ \bibinfo {author} {\bibfnamefont {N.}~\bibnamefont
  {{Andersson}}}} (\bibinfo {year} {2006}),\ \href {\doibase
  10.1111/j.1745-3933.2006.00211.x} {\bibfield  {journal} {\bibinfo  {journal}
  {\mnras}\ }\textbf {\bibinfo {volume} {371}},\ \bibinfo {pages} {L74}},\
  \Eprint {http://arxiv.org/abs/astro-ph/0605461} {astro-ph/0605461}
  \BibitemShut {NoStop}%
\bibitem [{\citenamefont {{Glendenning}}(1982)}]{Glendenning82}%
  \BibitemOpen
  \bibfield  {author} {\bibinfo {author} {\bibnamefont {{Glendenning}},
  \bibfnamefont {N.~K.}}} (\bibinfo {year} {1982}),\ \href {\doibase
  10.1016/0370-2693(82)90078-8} {\bibfield  {journal} {\bibinfo  {journal}
  {Physics Letters B}\ }\textbf {\bibinfo {volume} {114}},\ \bibinfo {pages}
  {392}}\BibitemShut {NoStop}%
\bibitem [{\citenamefont {{Guillemot}}\ and\ \citenamefont
  {{Tauris}}(2014)}]{Guillemot14}%
  \BibitemOpen
  \bibfield  {author} {\bibinfo {author} {\bibnamefont {{Guillemot}},
  \bibfnamefont {L.}}, \ and\ \bibinfo {author} {\bibfnamefont {T.~M.}\
  \bibnamefont {{Tauris}}}} (\bibinfo {year} {2014}),\ \href {\doibase
  10.1093/mnras/stu082} {\bibfield  {journal} {\bibinfo  {journal} {\mnras}\
  }\textbf {\bibinfo {volume} {439}},\ \bibinfo {pages} {2033}},\ \Eprint
  {http://arxiv.org/abs/1401.2773} {arXiv:1401.2773 [astro-ph.HE]} \BibitemShut
  {NoStop}%
\bibitem [{\citenamefont {{Guillot}}\ and\ \citenamefont
  {{Rutledge}}(2014)}]{Guillot14}%
  \BibitemOpen
  \bibfield  {author} {\bibinfo {author} {\bibnamefont {{Guillot}},
  \bibfnamefont {S.}}, \ and\ \bibinfo {author} {\bibfnamefont {R.~E.}\
  \bibnamefont {{Rutledge}}}} (\bibinfo {year} {2014}),\ \href@noop {}
  {\bibfield  {journal} {\bibinfo  {journal} {\apjl}\ }\textbf {\bibinfo
  {volume} {796}},\ \bibinfo {eid} {L3}}\BibitemShut {NoStop}%
\bibitem [{\citenamefont {{Guillot}}\ \emph {et~al.}(2013)\citenamefont
  {{Guillot}}, \citenamefont {{Servillat}}, \citenamefont {{Webb}},\ and\
  \citenamefont {{Rutledge}}}]{Guillot13}%
  \BibitemOpen
  \bibfield  {author} {\bibinfo {author} {\bibnamefont {{Guillot}},
  \bibfnamefont {S.}}, \bibinfo {author} {\bibfnamefont {M.}~\bibnamefont
  {{Servillat}}}, \bibinfo {author} {\bibfnamefont {N.~A.}\ \bibnamefont
  {{Webb}}}, \ and\ \bibinfo {author} {\bibfnamefont {R.~E.}\ \bibnamefont
  {{Rutledge}}}} (\bibinfo {year} {2013}),\ \href {\doibase
  10.1088/0004-637X/772/1/7} {\bibfield  {journal} {\bibinfo  {journal} {\apj}\
  }\textbf {\bibinfo {volume} {772}},\ \bibinfo {eid} {7}},\ \Eprint
  {http://arxiv.org/abs/1302.0023} {arXiv:1302.0023 [astro-ph.HE]} \BibitemShut
  {NoStop}%
\bibitem [{\citenamefont {{G{\"u}ver}}\ and\ \citenamefont
  {{{\"O}zel}}(2013)}]{Guver13}%
  \BibitemOpen
  \bibfield  {author} {\bibinfo {author} {\bibnamefont {{G{\"u}ver}},
  \bibfnamefont {T.}}, \ and\ \bibinfo {author} {\bibfnamefont
  {F.}~\bibnamefont {{{\"O}zel}}}} (\bibinfo {year} {2013}),\ \href {\doibase
  10.1088/2041-8205/765/1/L1} {\bibfield  {journal} {\bibinfo  {journal}
  {\apjl}\ }\textbf {\bibinfo {volume} {765}},\ \bibinfo {eid} {L1}},\ \Eprint
  {http://arxiv.org/abs/1301.0831} {arXiv:1301.0831 [astro-ph.HE]} \BibitemShut
  {NoStop}%
\bibitem [{\citenamefont {{G{\"u}ver}}\ \emph
  {et~al.}(2010{\natexlab{a}})\citenamefont {{G{\"u}ver}}, \citenamefont
  {{{\"O}zel}}, \citenamefont {{Cabrera-Lavers}},\ and\ \citenamefont
  {{Wroblewski}}}]{Guver10a}%
  \BibitemOpen
  \bibfield  {author} {\bibinfo {author} {\bibnamefont {{G{\"u}ver}},
  \bibfnamefont {T.}}, \bibinfo {author} {\bibfnamefont {F.}~\bibnamefont
  {{{\"O}zel}}}, \bibinfo {author} {\bibfnamefont {A.}~\bibnamefont
  {{Cabrera-Lavers}}}, \ and\ \bibinfo {author} {\bibfnamefont
  {P.}~\bibnamefont {{Wroblewski}}}} (\bibinfo {year} {2010}{\natexlab{a}}),\
  \href {\doibase 10.1088/0004-637X/712/2/964} {\bibfield  {journal} {\bibinfo
  {journal} {\apj}\ }\textbf {\bibinfo {volume} {712}},\ \bibinfo {pages}
  {964}},\ \Eprint {http://arxiv.org/abs/0811.3979} {arXiv:0811.3979}
  \BibitemShut {NoStop}%
\bibitem [{\citenamefont {{G\"uver}}\ \emph {et~al.}(2015)\citenamefont
  {{G\"uver}}, \citenamefont {{\"Ozel}}, \citenamefont {{Marshall}},
  \citenamefont {{Psaltis}}, \citenamefont {{Guainazzi}},\ and\ \citenamefont
  {{Diaz-Trigo}}}]{Guver15}%
  \BibitemOpen
  \bibfield  {author} {\bibinfo {author} {\bibnamefont {{G\"uver}},
  \bibfnamefont {T.}}, \bibinfo {author} {\bibfnamefont {F.}~\bibnamefont
  {{\"Ozel}}}, \bibinfo {author} {\bibfnamefont {H.}~\bibnamefont
  {{Marshall}}}, \bibinfo {author} {\bibfnamefont {D.}~\bibnamefont
  {{Psaltis}}}, \bibinfo {author} {\bibfnamefont {M.}~\bibnamefont
  {{Guainazzi}}}, \ and\ \bibinfo {author} {\bibfnamefont {M.}~\bibnamefont
  {{Diaz-Trigo}}}} (\bibinfo {year} {2015}),\ \href@noop {} {\bibfield
  {journal} {\bibinfo  {journal} {ArXiv e-prints}\ }}\Eprint
  {http://arxiv.org/abs/1501.05330} {arXiv:1501.05330 [astro-ph.HE]}
  \BibitemShut {NoStop}%
\bibitem [{\citenamefont {{G{\"u}ver}}\ \emph
  {et~al.}(2012{\natexlab{a}})\citenamefont {{G{\"u}ver}}, \citenamefont
  {{{\"O}zel}},\ and\ \citenamefont {{Psaltis}}}]{Guver12b}%
  \BibitemOpen
  \bibfield  {author} {\bibinfo {author} {\bibnamefont {{G{\"u}ver}},
  \bibfnamefont {T.}}, \bibinfo {author} {\bibfnamefont {F.}~\bibnamefont
  {{{\"O}zel}}}, \ and\ \bibinfo {author} {\bibfnamefont {D.}~\bibnamefont
  {{Psaltis}}}} (\bibinfo {year} {2012}{\natexlab{a}}),\ \href {\doibase
  10.1088/0004-637X/747/1/77} {\bibfield  {journal} {\bibinfo  {journal}
  {\apj}\ }\textbf {\bibinfo {volume} {747}},\ \bibinfo {eid} {77}},\ \Eprint
  {http://arxiv.org/abs/1104.2602} {arXiv:1104.2602 [astro-ph.HE]} \BibitemShut
  {NoStop}%
\bibitem [{\citenamefont {{G{\"u}ver}}\ \emph
  {et~al.}(2012{\natexlab{b}})\citenamefont {{G{\"u}ver}}, \citenamefont
  {{Psaltis}},\ and\ \citenamefont {{{\"O}zel}}}]{Guver12a}%
  \BibitemOpen
  \bibfield  {author} {\bibinfo {author} {\bibnamefont {{G{\"u}ver}},
  \bibfnamefont {T.}}, \bibinfo {author} {\bibfnamefont {D.}~\bibnamefont
  {{Psaltis}}}, \ and\ \bibinfo {author} {\bibfnamefont {F.}~\bibnamefont
  {{{\"O}zel}}}} (\bibinfo {year} {2012}{\natexlab{b}}),\ \href {\doibase
  10.1088/0004-637X/747/1/76} {\bibfield  {journal} {\bibinfo  {journal}
  {\apj}\ }\textbf {\bibinfo {volume} {747}},\ \bibinfo {eid} {76}},\ \Eprint
  {http://arxiv.org/abs/1103.5767} {arXiv:1103.5767 [astro-ph.HE]} \BibitemShut
  {NoStop}%
\bibitem [{\citenamefont {{G{\"u}ver}}\ \emph
  {et~al.}(2010{\natexlab{b}})\citenamefont {{G{\"u}ver}}, \citenamefont
  {{Wroblewski}}, \citenamefont {{Camarota}},\ and\ \citenamefont
  {{{\"O}zel}}}]{Guver10b}%
  \BibitemOpen
  \bibfield  {author} {\bibinfo {author} {\bibnamefont {{G{\"u}ver}},
  \bibfnamefont {T.}}, \bibinfo {author} {\bibfnamefont {P.}~\bibnamefont
  {{Wroblewski}}}, \bibinfo {author} {\bibfnamefont {L.}~\bibnamefont
  {{Camarota}}}, \ and\ \bibinfo {author} {\bibfnamefont {F.}~\bibnamefont
  {{{\"O}zel}}}} (\bibinfo {year} {2010}{\natexlab{b}}),\ \href {\doibase
  10.1088/0004-637X/719/2/1807} {\bibfield  {journal} {\bibinfo  {journal}
  {\apj}\ }\textbf {\bibinfo {volume} {719}},\ \bibinfo {pages} {1807}},\
  \Eprint {http://arxiv.org/abs/1002.3825} {arXiv:1002.3825 [astro-ph.HE]}
  \BibitemShut {NoStop}%
\bibitem [{\citenamefont {{Haensel}}\ \emph {et~al.}(2007)\citenamefont
  {{Haensel}}, \citenamefont {{Potekhin}},\ and\ \citenamefont
  {{Yakovlev}}}]{Haensel07}%
  \BibitemOpen
  \bibinfo {editor} {\bibnamefont {{Haensel}}, \bibfnamefont {P.}}, \bibinfo
  {editor} {\bibfnamefont {A.~Y.}\ \bibnamefont {{Potekhin}}}, \ and\ \bibinfo
  {editor} {\bibfnamefont {D.~G.}\ \bibnamefont {{Yakovlev}}},\ Eds. (\bibinfo
  {year} {2007}),\ \href@noop {} {\emph {\bibinfo {title} {Astrophysics and
  Space Science Library}}},\ \bibinfo {series} {Astrophysics and Space Science
  Library}, Vol.\ \bibinfo {volume} {326}\BibitemShut {NoStop}%
\bibitem [{\citenamefont {{Haensel}}\ \emph {et~al.}(2009)\citenamefont
  {{Haensel}}, \citenamefont {{Zdunik}}, \citenamefont {{Bejger}},\ and\
  \citenamefont {{Lattimer}}}]{Haensel09}%
  \BibitemOpen
  \bibfield  {author} {\bibinfo {author} {\bibnamefont {{Haensel}},
  \bibfnamefont {P.}}, \bibinfo {author} {\bibfnamefont {J.~L.}\ \bibnamefont
  {{Zdunik}}}, \bibinfo {author} {\bibfnamefont {M.}~\bibnamefont {{Bejger}}},
  \ and\ \bibinfo {author} {\bibfnamefont {J.~M.}\ \bibnamefont {{Lattimer}}}}
  (\bibinfo {year} {2009}),\ \href {\doibase 10.1051/0004-6361/200811605}
  {\bibfield  {journal} {\bibinfo  {journal} {\aap}\ }\textbf {\bibinfo
  {volume} {502}},\ \bibinfo {pages} {605}},\ \Eprint
  {http://arxiv.org/abs/0901.1268} {arXiv:0901.1268 [astro-ph.SR]} \BibitemShut
  {NoStop}%
\bibitem [{\citenamefont {{Haensel}}\ \emph {et~al.}(1986)\citenamefont
  {{Haensel}}, \citenamefont {{Zdunik}},\ and\ \citenamefont
  {{Schaefer}}}]{Haensel86}%
  \BibitemOpen
  \bibfield  {author} {\bibinfo {author} {\bibnamefont {{Haensel}},
  \bibfnamefont {P.}}, \bibinfo {author} {\bibfnamefont {J.~L.}\ \bibnamefont
  {{Zdunik}}}, \ and\ \bibinfo {author} {\bibfnamefont {R.}~\bibnamefont
  {{Schaefer}}}} (\bibinfo {year} {1986}),\ \href@noop {} {\bibfield  {journal}
  {\bibinfo  {journal} {\aap}\ }\textbf {\bibinfo {volume} {160}},\ \bibinfo
  {pages} {121}}\BibitemShut {NoStop}%
\bibitem [{\citenamefont {{Hammer}}\ \emph {et~al.}(2013)\citenamefont
  {{Hammer}}, \citenamefont {{Nogga}},\ and\ \citenamefont
  {{Schwenk}}}]{Hammer13}%
  \BibitemOpen
  \bibfield  {author} {\bibinfo {author} {\bibnamefont {{Hammer}},
  \bibfnamefont {H.-W.}}, \bibinfo {author} {\bibfnamefont {A.}~\bibnamefont
  {{Nogga}}}, \ and\ \bibinfo {author} {\bibfnamefont {A.}~\bibnamefont
  {{Schwenk}}}} (\bibinfo {year} {2013}),\ \href {\doibase
  10.1103/RevModPhys.85.197} {\bibfield  {journal} {\bibinfo  {journal}
  {Reviews of Modern Physics}\ }\textbf {\bibinfo {volume} {85}},\ \bibinfo
  {pages} {197}},\ \Eprint {http://arxiv.org/abs/1210.4273} {arXiv:1210.4273
  [nucl-th]} \BibitemShut {NoStop}%
\bibitem [{\citenamefont {{Hands}}(2007)}]{Hands07}%
  \BibitemOpen
  \bibfield  {author} {\bibinfo {author} {\bibnamefont {{Hands}}, \bibfnamefont
  {S.}}} (\bibinfo {year} {2007}),\ \href {\doibase 10.1143/PTPS.168.253}
  {\bibfield  {journal} {\bibinfo  {journal} {Progress of Theoretical Physics
  Supplement}\ }\textbf {\bibinfo {volume} {168}},\ \bibinfo {pages} {253}},\
  \Eprint {http://arxiv.org/abs/hep-lat/0703017} {hep-lat/0703017} \BibitemShut
  {NoStop}%
\bibitem [{\citenamefont {{Hartle}}(1967)}]{Hartle67}%
  \BibitemOpen
  \bibfield  {author} {\bibinfo {author} {\bibnamefont {{Hartle}},
  \bibfnamefont {J.~B.}}} (\bibinfo {year} {1967}),\ \href {\doibase
  10.1086/149400} {\bibfield  {journal} {\bibinfo  {journal} {\apj}\ }\textbf
  {\bibinfo {volume} {150}},\ \bibinfo {pages} {1005}}\BibitemShut {NoStop}%
\bibitem [{\citenamefont {{Hartle}}\ and\ \citenamefont
  {{Thorne}}(1968)}]{Hartle68}%
  \BibitemOpen
  \bibfield  {author} {\bibinfo {author} {\bibnamefont {{Hartle}},
  \bibfnamefont {J.~B.}}, \ and\ \bibinfo {author} {\bibfnamefont {K.~S.}\
  \bibnamefont {{Thorne}}}} (\bibinfo {year} {1968}),\ \href {\doibase
  10.1086/149707} {\bibfield  {journal} {\bibinfo  {journal} {\apj}\ }\textbf
  {\bibinfo {volume} {153}},\ \bibinfo {pages} {807}}\BibitemShut {NoStop}%
\bibitem [{\citenamefont {{Hartman}}\ \emph {et~al.}(2011)\citenamefont
  {{Hartman}}, \citenamefont {{Galloway}},\ and\ \citenamefont
  {{Chakrabarty}}}]{Hartman11}%
  \BibitemOpen
  \bibfield  {author} {\bibinfo {author} {\bibnamefont {{Hartman}},
  \bibfnamefont {J.~M.}}, \bibinfo {author} {\bibfnamefont {D.~K.}\
  \bibnamefont {{Galloway}}}, \ and\ \bibinfo {author} {\bibfnamefont
  {D.}~\bibnamefont {{Chakrabarty}}}} (\bibinfo {year} {2011}),\ \href
  {\doibase 10.1088/0004-637X/726/1/26} {\bibfield  {journal} {\bibinfo
  {journal} {\apj}\ }\textbf {\bibinfo {volume} {726}},\ \bibinfo {eid} {26}},\
  \Eprint {http://arxiv.org/abs/1006.1908} {arXiv:1006.1908 [astro-ph.HE]}
  \BibitemShut {NoStop}%
\bibitem [{\citenamefont {{Haskell}}(2015)}]{Haskell15}%
  \BibitemOpen
  \bibfield  {author} {\bibinfo {author} {\bibnamefont {{Haskell}},
  \bibfnamefont {B.}}} (\bibinfo {year} {2015}),\ \href {\doibase
  10.1142/S0218301315410074} {\bibfield  {journal} {\bibinfo  {journal}
  {International Journal of Modern Physics E}\ }\textbf {\bibinfo {volume}
  {24}},\ \bibinfo {eid} {1541007}},\ \Eprint {http://arxiv.org/abs/1509.04370}
  {arXiv:1509.04370 [astro-ph.HE]} \BibitemShut {NoStop}%
\bibitem [{\citenamefont {{Haskell}}\ \emph {et~al.}(2012)\citenamefont
  {{Haskell}}, \citenamefont {{Degenaar}},\ and\ \citenamefont
  {{Ho}}}]{Haskell12}%
  \BibitemOpen
  \bibfield  {author} {\bibinfo {author} {\bibnamefont {{Haskell}},
  \bibfnamefont {B.}}, \bibinfo {author} {\bibfnamefont {N.}~\bibnamefont
  {{Degenaar}}}, \ and\ \bibinfo {author} {\bibfnamefont {W.~C.~G.}\
  \bibnamefont {{Ho}}}} (\bibinfo {year} {2012}),\ \href {\doibase
  10.1111/j.1365-2966.2012.21171.x} {\bibfield  {journal} {\bibinfo  {journal}
  {\mnras}\ }\textbf {\bibinfo {volume} {424}},\ \bibinfo {pages} {93}},\
  \Eprint {http://arxiv.org/abs/1201.2101} {arXiv:1201.2101 [astro-ph.SR]}
  \BibitemShut {NoStop}%
\bibitem [{\citenamefont {{Hebeler}}\ \emph {et~al.}(2015)\citenamefont
  {{Hebeler}}, \citenamefont {{Holt}}, \citenamefont {{Men{\'e}ndez}},\ and\
  \citenamefont {{Schwenk}}}]{Hebeler15}%
  \BibitemOpen
  \bibfield  {author} {\bibinfo {author} {\bibnamefont {{Hebeler}},
  \bibfnamefont {K.}}, \bibinfo {author} {\bibfnamefont {J.~D.}\ \bibnamefont
  {{Holt}}}, \bibinfo {author} {\bibfnamefont {J.}~\bibnamefont
  {{Men{\'e}ndez}}}, \ and\ \bibinfo {author} {\bibfnamefont {A.}~\bibnamefont
  {{Schwenk}}}} (\bibinfo {year} {2015}),\ \href {\doibase
  10.1146/annurev-nucl-102313-025446} {\bibfield  {journal} {\bibinfo
  {journal} {Annual Review of Nuclear and Particle Science}\ }\textbf {\bibinfo
  {volume} {65}},\ \bibinfo {pages} {19008}},\ \Eprint
  {http://arxiv.org/abs/1508.06893} {arXiv:1508.06893 [nucl-th]} \BibitemShut
  {NoStop}%
\bibitem [{\citenamefont {{Hebeler}}\ \emph {et~al.}(2010)\citenamefont
  {{Hebeler}}, \citenamefont {{Lattimer}}, \citenamefont {{Pethick}},\ and\
  \citenamefont {{Schwenk}}}]{Hebeler10b}%
  \BibitemOpen
  \bibfield  {author} {\bibinfo {author} {\bibnamefont {{Hebeler}},
  \bibfnamefont {K.}}, \bibinfo {author} {\bibfnamefont {J.~M.}\ \bibnamefont
  {{Lattimer}}}, \bibinfo {author} {\bibfnamefont {C.~J.}\ \bibnamefont
  {{Pethick}}}, \ and\ \bibinfo {author} {\bibfnamefont {A.}~\bibnamefont
  {{Schwenk}}}} (\bibinfo {year} {2010}),\ \href {\doibase
  10.1103/PhysRevLett.105.161102} {\bibfield  {journal} {\bibinfo  {journal}
  {Physical Review Letters}\ }\textbf {\bibinfo {volume} {105}}~(\bibinfo
  {number} {16}),\ \bibinfo {eid} {161102}},\ \Eprint
  {http://arxiv.org/abs/1007.1746} {arXiv:1007.1746 [nucl-th]} \BibitemShut
  {NoStop}%
\bibitem [{\citenamefont {{Hebeler}}\ \emph {et~al.}(2013)\citenamefont
  {{Hebeler}}, \citenamefont {{Lattimer}}, \citenamefont {{Pethick}},\ and\
  \citenamefont {{Schwenk}}}]{Hebeler13}%
  \BibitemOpen
  \bibfield  {author} {\bibinfo {author} {\bibnamefont {{Hebeler}},
  \bibfnamefont {K.}}, \bibinfo {author} {\bibfnamefont {J.~M.}\ \bibnamefont
  {{Lattimer}}}, \bibinfo {author} {\bibfnamefont {C.~J.}\ \bibnamefont
  {{Pethick}}}, \ and\ \bibinfo {author} {\bibfnamefont {A.}~\bibnamefont
  {{Schwenk}}}} (\bibinfo {year} {2013}),\ \href {\doibase
  10.1088/0004-637X/773/1/11} {\bibfield  {journal} {\bibinfo  {journal}
  {\apj}\ }\textbf {\bibinfo {volume} {773}},\ \bibinfo {eid} {11}},\ \Eprint
  {http://arxiv.org/abs/1303.4662} {arXiv:1303.4662 [astro-ph.SR]} \BibitemShut
  {NoStop}%
\bibitem [{\citenamefont {{Hebeler}}\ and\ \citenamefont
  {{Schwenk}}(2010)}]{Hebeler10}%
  \BibitemOpen
  \bibfield  {author} {\bibinfo {author} {\bibnamefont {{Hebeler}},
  \bibfnamefont {K.}}, \ and\ \bibinfo {author} {\bibfnamefont
  {A.}~\bibnamefont {{Schwenk}}}} (\bibinfo {year} {2010}),\ \href {\doibase
  10.1103/PhysRevC.82.014314} {\bibfield  {journal} {\bibinfo  {journal}
  {\prc}\ }\textbf {\bibinfo {volume} {82}}~(\bibinfo {number} {1}),\ \bibinfo
  {eid} {014314}},\ \Eprint {http://arxiv.org/abs/0911.0483} {arXiv:0911.0483
  [nucl-th]} \BibitemShut {NoStop}%
\bibitem [{\citenamefont {{Hebeler}}\ and\ \citenamefont
  {{Schwenk}}(2014)}]{Hebeler14}%
  \BibitemOpen
  \bibfield  {author} {\bibinfo {author} {\bibnamefont {{Hebeler}},
  \bibfnamefont {K.}}, \ and\ \bibinfo {author} {\bibfnamefont
  {A.}~\bibnamefont {{Schwenk}}}} (\bibinfo {year} {2014}),\ \href {\doibase
  10.1140/epja/i2014-14011-4} {\bibfield  {journal} {\bibinfo  {journal}
  {European Physical Journal A}\ }\textbf {\bibinfo {volume} {50}},\ \bibinfo
  {eid} {11}},\ \Eprint {http://arxiv.org/abs/1401.5822} {arXiv:1401.5822
  [nucl-th]} \BibitemShut {NoStop}%
\bibitem [{\citenamefont {{Heinke}}\ \emph {et~al.}(2014)\citenamefont
  {{Heinke}}, \citenamefont {{Cohn}}, \citenamefont {{Lugger}}, \citenamefont
  {{Webb}}, \citenamefont {{Ho}}, \citenamefont {{Anderson}}, \citenamefont
  {{Campana}}, \citenamefont {{Bogdanov}}, \citenamefont {{Haggard}},
  \citenamefont {{Cool}},\ and\ \citenamefont {{Grindlay}}}]{Heinke14}%
  \BibitemOpen
  \bibfield  {author} {\bibinfo {author} {\bibnamefont {{Heinke}},
  \bibfnamefont {C.~O.}}, \bibinfo {author} {\bibfnamefont {H.~N.}\
  \bibnamefont {{Cohn}}}, \bibinfo {author} {\bibfnamefont {P.~M.}\
  \bibnamefont {{Lugger}}}, \bibinfo {author} {\bibfnamefont {N.~A.}\
  \bibnamefont {{Webb}}}, \bibinfo {author} {\bibfnamefont {W.~C.~G.}\
  \bibnamefont {{Ho}}}, \bibinfo {author} {\bibfnamefont {J.}~\bibnamefont
  {{Anderson}}}, \bibinfo {author} {\bibfnamefont {S.}~\bibnamefont
  {{Campana}}}, \bibinfo {author} {\bibfnamefont {S.}~\bibnamefont
  {{Bogdanov}}}, \bibinfo {author} {\bibfnamefont {D.}~\bibnamefont
  {{Haggard}}}, \bibinfo {author} {\bibfnamefont {A.~M.}\ \bibnamefont
  {{Cool}}}, \ and\ \bibinfo {author} {\bibfnamefont {J.~E.}\ \bibnamefont
  {{Grindlay}}}} (\bibinfo {year} {2014}),\ \href@noop {} {\bibfield  {journal}
  {\bibinfo  {journal} {\mnras}\ }\textbf {\bibinfo {volume} {444}},\ \bibinfo
  {pages} {443}},\ \Eprint {http://arxiv.org/abs/1406.1497} {arXiv:1406.1497
  [astro-ph.HE]} \BibitemShut {NoStop}%
\bibitem [{\citenamefont {{Heinke}}\ \emph {et~al.}(2003)\citenamefont
  {{Heinke}}, \citenamefont {{Grindlay}}, \citenamefont {{Lloyd}},\ and\
  \citenamefont {{Edmonds}}}]{Heinke03}%
  \BibitemOpen
  \bibfield  {author} {\bibinfo {author} {\bibnamefont {{Heinke}},
  \bibfnamefont {C.~O.}}, \bibinfo {author} {\bibfnamefont {J.~E.}\
  \bibnamefont {{Grindlay}}}, \bibinfo {author} {\bibfnamefont {D.~A.}\
  \bibnamefont {{Lloyd}}}, \ and\ \bibinfo {author} {\bibfnamefont {P.~D.}\
  \bibnamefont {{Edmonds}}}} (\bibinfo {year} {2003}),\ \href {\doibase
  10.1086/374039} {\bibfield  {journal} {\bibinfo  {journal} {\apj}\ }\textbf
  {\bibinfo {volume} {588}},\ \bibinfo {pages} {452}},\ \Eprint
  {http://arxiv.org/abs/astro-ph/0301235} {astro-ph/0301235} \BibitemShut
  {NoStop}%
\bibitem [{\citenamefont {{Heinke}}\ \emph {et~al.}(2006)\citenamefont
  {{Heinke}}, \citenamefont {{Rybicki}}, \citenamefont {{Narayan}},\ and\
  \citenamefont {{Grindlay}}}]{Heinke06}%
  \BibitemOpen
  \bibfield  {author} {\bibinfo {author} {\bibnamefont {{Heinke}},
  \bibfnamefont {C.~O.}}, \bibinfo {author} {\bibfnamefont {G.~B.}\
  \bibnamefont {{Rybicki}}}, \bibinfo {author} {\bibfnamefont {R.}~\bibnamefont
  {{Narayan}}}, \ and\ \bibinfo {author} {\bibfnamefont {J.~E.}\ \bibnamefont
  {{Grindlay}}}} (\bibinfo {year} {2006}),\ \href {\doibase 10.1086/503701}
  {\bibfield  {journal} {\bibinfo  {journal} {\apj}\ }\textbf {\bibinfo
  {volume} {644}},\ \bibinfo {pages} {1090}},\ \Eprint
  {http://arxiv.org/abs/astro-ph/0506563} {astro-ph/0506563} \BibitemShut
  {NoStop}%
\bibitem [{\citenamefont {{Hessels}}\ \emph {et~al.}(2006)\citenamefont
  {{Hessels}}, \citenamefont {{Ransom}}, \citenamefont {{Stairs}},
  \citenamefont {{Freire}}, \citenamefont {{Kaspi}},\ and\ \citenamefont
  {{Camilo}}}]{Hessels06}%
  \BibitemOpen
  \bibfield  {author} {\bibinfo {author} {\bibnamefont {{Hessels}},
  \bibfnamefont {J.~W.~T.}}, \bibinfo {author} {\bibfnamefont {S.~M.}\
  \bibnamefont {{Ransom}}}, \bibinfo {author} {\bibfnamefont {I.~H.}\
  \bibnamefont {{Stairs}}}, \bibinfo {author} {\bibfnamefont {P.~C.~C.}\
  \bibnamefont {{Freire}}}, \bibinfo {author} {\bibfnamefont {V.~M.}\
  \bibnamefont {{Kaspi}}}, \ and\ \bibinfo {author} {\bibfnamefont
  {F.}~\bibnamefont {{Camilo}}}} (\bibinfo {year} {2006}),\ \href {\doibase
  10.1126/science.1123430} {\bibfield  {journal} {\bibinfo  {journal}
  {Science}\ }\textbf {\bibinfo {volume} {311}},\ \bibinfo {pages} {1901}},\
  \Eprint {http://arxiv.org/abs/astro-ph/0601337} {astro-ph/0601337}
  \BibitemShut {NoStop}%
\bibitem [{\citenamefont {{Hills}}(1983)}]{Hills83}%
  \BibitemOpen
  \bibfield  {author} {\bibinfo {author} {\bibnamefont {{Hills}}, \bibfnamefont
  {J.~G.}}} (\bibinfo {year} {1983}),\ \href {\doibase 10.1086/160871}
  {\bibfield  {journal} {\bibinfo  {journal} {\apj}\ }\textbf {\bibinfo
  {volume} {267}},\ \bibinfo {pages} {322}}\BibitemShut {NoStop}%
\bibitem [{\citenamefont {{Ho}}\ \emph {et~al.}(2011)\citenamefont {{Ho}},
  \citenamefont {{Andersson}},\ and\ \citenamefont {{Haskell}}}]{Ho11}%
  \BibitemOpen
  \bibfield  {author} {\bibinfo {author} {\bibnamefont {{Ho}}, \bibfnamefont
  {W.~C.~G.}}, \bibinfo {author} {\bibfnamefont {N.}~\bibnamefont
  {{Andersson}}}, \ and\ \bibinfo {author} {\bibfnamefont {B.}~\bibnamefont
  {{Haskell}}}} (\bibinfo {year} {2011}),\ \href {\doibase
  10.1103/PhysRevLett.107.101101} {\bibfield  {journal} {\bibinfo  {journal}
  {Physical Review Letters}\ }\textbf {\bibinfo {volume} {107}}~(\bibinfo
  {number} {10}),\ \bibinfo {eid} {101101}},\ \Eprint
  {http://arxiv.org/abs/1107.5064} {arXiv:1107.5064 [astro-ph.HE]} \BibitemShut
  {NoStop}%
\bibitem [{\citenamefont {{Ho}}\ \emph {et~al.}(2014)\citenamefont {{Ho}},
  \citenamefont {{Klus}}, \citenamefont {{Coe}},\ and\ \citenamefont
  {{Andersson}}}]{Ho14}%
  \BibitemOpen
  \bibfield  {author} {\bibinfo {author} {\bibnamefont {{Ho}}, \bibfnamefont
  {W.~C.~G.}}, \bibinfo {author} {\bibfnamefont {H.}~\bibnamefont {{Klus}}},
  \bibinfo {author} {\bibfnamefont {M.~J.}\ \bibnamefont {{Coe}}}, \ and\
  \bibinfo {author} {\bibfnamefont {N.}~\bibnamefont {{Andersson}}}} (\bibinfo
  {year} {2014}),\ \href {\doibase 10.1093/mnras/stt2193} {\bibfield  {journal}
  {\bibinfo  {journal} {\mnras}\ }\textbf {\bibinfo {volume} {437}},\ \bibinfo
  {pages} {3664}},\ \Eprint {http://arxiv.org/abs/1311.1969} {arXiv:1311.1969
  [astro-ph.SR]} \BibitemShut {NoStop}%
\bibitem [{\citenamefont {Horowitz}\ \emph {et~al.}(2001)\citenamefont
  {Horowitz}, \citenamefont {Pollock}, \citenamefont {Souder},\ and\
  \citenamefont {Michaels}}]{Horowitz01}%
  \BibitemOpen
  \bibfield  {author} {\bibinfo {author} {\bibnamefont {Horowitz},
  \bibfnamefont {C.~J.}}, \bibinfo {author} {\bibfnamefont {S.~J.}\
  \bibnamefont {Pollock}}, \bibinfo {author} {\bibfnamefont {P.~A.}\
  \bibnamefont {Souder}}, \ and\ \bibinfo {author} {\bibfnamefont
  {R.}~\bibnamefont {Michaels}}} (\bibinfo {year} {2001}),\ \href {\doibase
  10.1103/PhysRevC.63.025501} {\bibfield  {journal} {\bibinfo  {journal} {Phys.
  Rev. C}\ }\textbf {\bibinfo {volume} {63}},\ \bibinfo {pages}
  {025501}}\BibitemShut {NoStop}%
\bibitem [{\citenamefont {{Hotokezaka}}\ \emph {et~al.}(2011)\citenamefont
  {{Hotokezaka}}, \citenamefont {{Kyutoku}}, \citenamefont {{Okawa}},
  \citenamefont {{Shibata}},\ and\ \citenamefont {{Kiuchi}}}]{Hotokezaka11}%
  \BibitemOpen
  \bibfield  {author} {\bibinfo {author} {\bibnamefont {{Hotokezaka}},
  \bibfnamefont {K.}}, \bibinfo {author} {\bibfnamefont {K.}~\bibnamefont
  {{Kyutoku}}}, \bibinfo {author} {\bibfnamefont {H.}~\bibnamefont {{Okawa}}},
  \bibinfo {author} {\bibfnamefont {M.}~\bibnamefont {{Shibata}}}, \ and\
  \bibinfo {author} {\bibfnamefont {K.}~\bibnamefont {{Kiuchi}}}} (\bibinfo
  {year} {2011}),\ \href {\doibase 10.1103/PhysRevD.83.124008} {\bibfield
  {journal} {\bibinfo  {journal} {\prd}\ }\textbf {\bibinfo {volume}
  {83}}~(\bibinfo {number} {12}),\ \bibinfo {eid} {124008}},\ \Eprint
  {http://arxiv.org/abs/1105.4370} {arXiv:1105.4370 [astro-ph.HE]} \BibitemShut
  {NoStop}%
\bibitem [{\citenamefont {{Huppenkothen}}\ \emph
  {et~al.}(2014{\natexlab{a}})\citenamefont {{Huppenkothen}}, \citenamefont
  {{D'Angelo}}, \citenamefont {{Watts}}, \citenamefont {{Heil}}, \citenamefont
  {{van der Klis}}, \citenamefont {{van der Horst}}, \citenamefont
  {{Kouveliotou}}, \citenamefont {{Baring}}, \citenamefont {{G{\"o}{\u
  g}{\"u}{\c s}}}, \citenamefont {{Granot}}, \citenamefont {{Kaneko}},
  \citenamefont {{Lin}}, \citenamefont {{von Kienlin}},\ and\ \citenamefont
  {{Younes}}}]{Huppenkothen14b}%
  \BibitemOpen
  \bibfield  {author} {\bibinfo {author} {\bibnamefont {{Huppenkothen}},
  \bibfnamefont {D.}}, \bibinfo {author} {\bibfnamefont {C.}~\bibnamefont
  {{D'Angelo}}}, \bibinfo {author} {\bibfnamefont {A.~L.}\ \bibnamefont
  {{Watts}}}, \bibinfo {author} {\bibfnamefont {L.}~\bibnamefont {{Heil}}},
  \bibinfo {author} {\bibfnamefont {M.}~\bibnamefont {{van der Klis}}},
  \bibinfo {author} {\bibfnamefont {A.~J.}\ \bibnamefont {{van der Horst}}},
  \bibinfo {author} {\bibfnamefont {C.}~\bibnamefont {{Kouveliotou}}}, \bibinfo
  {author} {\bibfnamefont {M.~G.}\ \bibnamefont {{Baring}}}, \bibinfo {author}
  {\bibfnamefont {E.}~\bibnamefont {{G{\"o}{\u g}{\"u}{\c s}}}}, \bibinfo
  {author} {\bibfnamefont {J.}~\bibnamefont {{Granot}}}, \bibinfo {author}
  {\bibfnamefont {Y.}~\bibnamefont {{Kaneko}}}, \bibinfo {author}
  {\bibfnamefont {L.}~\bibnamefont {{Lin}}}, \bibinfo {author} {\bibfnamefont
  {A.}~\bibnamefont {{von Kienlin}}}, \ and\ \bibinfo {author} {\bibfnamefont
  {G.}~\bibnamefont {{Younes}}}} (\bibinfo {year} {2014}{\natexlab{a}}),\ \href
  {\doibase 10.1088/0004-637X/787/2/128} {\bibfield  {journal} {\bibinfo
  {journal} {\apj}\ }\textbf {\bibinfo {volume} {787}},\ \bibinfo {eid}
  {128}},\ \Eprint {http://arxiv.org/abs/1404.2756} {arXiv:1404.2756
  [astro-ph.HE]} \BibitemShut {NoStop}%
\bibitem [{\citenamefont {{Huppenkothen}}\ \emph
  {et~al.}(2014{\natexlab{b}})\citenamefont {{Huppenkothen}}, \citenamefont
  {{Heil}}, \citenamefont {{Watts}},\ and\ \citenamefont {{G{\"o}{\u g}{\"u}{\c
  s}}}}]{Huppenkothen14a}%
  \BibitemOpen
  \bibfield  {author} {\bibinfo {author} {\bibnamefont {{Huppenkothen}},
  \bibfnamefont {D.}}, \bibinfo {author} {\bibfnamefont {L.~M.}\ \bibnamefont
  {{Heil}}}, \bibinfo {author} {\bibfnamefont {A.~L.}\ \bibnamefont {{Watts}}},
  \ and\ \bibinfo {author} {\bibfnamefont {E.}~\bibnamefont {{G{\"o}{\u
  g}{\"u}{\c s}}}}} (\bibinfo {year} {2014}{\natexlab{b}}),\ \href {\doibase
  10.1088/0004-637X/795/2/114} {\bibfield  {journal} {\bibinfo  {journal}
  {\apj}\ }\textbf {\bibinfo {volume} {795}},\ \bibinfo {eid} {114}},\ \Eprint
  {http://arxiv.org/abs/1409.7642} {arXiv:1409.7642 [astro-ph.HE]} \BibitemShut
  {NoStop}%
\bibitem [{\citenamefont {{Huppenkothen}}\ \emph
  {et~al.}(2014{\natexlab{c}})\citenamefont {{Huppenkothen}}, \citenamefont
  {{Watts}},\ and\ \citenamefont {{Levin}}}]{Huppenkothen14c}%
  \BibitemOpen
  \bibfield  {author} {\bibinfo {author} {\bibnamefont {{Huppenkothen}},
  \bibfnamefont {D.}}, \bibinfo {author} {\bibfnamefont {A.~L.}\ \bibnamefont
  {{Watts}}}, \ and\ \bibinfo {author} {\bibfnamefont {Y.}~\bibnamefont
  {{Levin}}}} (\bibinfo {year} {2014}{\natexlab{c}}),\ \href {\doibase
  10.1088/0004-637X/793/2/129} {\bibfield  {journal} {\bibinfo  {journal}
  {\apj}\ }\textbf {\bibinfo {volume} {793}},\ \bibinfo {eid} {129}},\ \Eprint
  {http://arxiv.org/abs/1408.0734} {arXiv:1408.0734 [astro-ph.HE]} \BibitemShut
  {NoStop}%
\bibitem [{\citenamefont {{Huppenkothen}}\ \emph {et~al.}(2013)\citenamefont
  {{Huppenkothen}}, \citenamefont {{Watts}}, \citenamefont {{Uttley}},
  \citenamefont {{van der Horst}}, \citenamefont {{van der Klis}},
  \citenamefont {{Kouveliotou}}, \citenamefont {{G{\"o}{\v g}{\"u}{\c s}}},
  \citenamefont {{Granot}}, \citenamefont {{Vaughan}},\ and\ \citenamefont
  {{Finger}}}]{Huppenkothen13}%
  \BibitemOpen
  \bibfield  {author} {\bibinfo {author} {\bibnamefont {{Huppenkothen}},
  \bibfnamefont {D.}}, \bibinfo {author} {\bibfnamefont {A.~L.}\ \bibnamefont
  {{Watts}}}, \bibinfo {author} {\bibfnamefont {P.}~\bibnamefont {{Uttley}}},
  \bibinfo {author} {\bibfnamefont {A.~J.}\ \bibnamefont {{van der Horst}}},
  \bibinfo {author} {\bibfnamefont {M.}~\bibnamefont {{van der Klis}}},
  \bibinfo {author} {\bibfnamefont {C.}~\bibnamefont {{Kouveliotou}}}, \bibinfo
  {author} {\bibfnamefont {E.}~\bibnamefont {{G{\"o}{\v g}{\"u}{\c s}}}},
  \bibinfo {author} {\bibfnamefont {J.}~\bibnamefont {{Granot}}}, \bibinfo
  {author} {\bibfnamefont {S.}~\bibnamefont {{Vaughan}}}, \ and\ \bibinfo
  {author} {\bibfnamefont {M.~H.}\ \bibnamefont {{Finger}}}} (\bibinfo {year}
  {2013}),\ \href {\doibase 10.1088/0004-637X/768/1/87} {\bibfield  {journal}
  {\bibinfo  {journal} {\apj}\ }\textbf {\bibinfo {volume} {768}},\ \bibinfo
  {eid} {87}},\ \Eprint {http://arxiv.org/abs/1212.1011} {arXiv:1212.1011
  [astro-ph.HE]} \BibitemShut {NoStop}%
\bibitem [{\citenamefont {{Ibragimov}}\ and\ \citenamefont
  {{Poutanen}}(2009)}]{Ibragimov09}%
  \BibitemOpen
  \bibfield  {author} {\bibinfo {author} {\bibnamefont {{Ibragimov}},
  \bibfnamefont {A.}}, \ and\ \bibinfo {author} {\bibfnamefont
  {J.}~\bibnamefont {{Poutanen}}}} (\bibinfo {year} {2009}),\ \href {\doibase
  10.1111/j.1365-2966.2009.15477.x} {\bibfield  {journal} {\bibinfo  {journal}
  {\mnras}\ }\textbf {\bibinfo {volume} {400}},\ \bibinfo {pages} {492}},\
  \Eprint {http://arxiv.org/abs/0901.0073} {arXiv:0901.0073 [astro-ph.SR]}
  \BibitemShut {NoStop}%
\bibitem [{\citenamefont {{Israel}}\ \emph {et~al.}(2005)\citenamefont
  {{Israel}}, \citenamefont {{Belloni}}, \citenamefont {{Stella}},
  \citenamefont {{Rephaeli}}, \citenamefont {{Gruber}}, \citenamefont
  {{Casella}}, \citenamefont {{Dall'Osso}}, \citenamefont {{Rea}},
  \citenamefont {{Persic}},\ and\ \citenamefont {{Rothschild}}}]{Israel05}%
  \BibitemOpen
  \bibfield  {author} {\bibinfo {author} {\bibnamefont {{Israel}},
  \bibfnamefont {G.~L.}}, \bibinfo {author} {\bibfnamefont {T.}~\bibnamefont
  {{Belloni}}}, \bibinfo {author} {\bibfnamefont {L.}~\bibnamefont {{Stella}}},
  \bibinfo {author} {\bibfnamefont {Y.}~\bibnamefont {{Rephaeli}}}, \bibinfo
  {author} {\bibfnamefont {D.~E.}\ \bibnamefont {{Gruber}}}, \bibinfo {author}
  {\bibfnamefont {P.}~\bibnamefont {{Casella}}}, \bibinfo {author}
  {\bibfnamefont {S.}~\bibnamefont {{Dall'Osso}}}, \bibinfo {author}
  {\bibfnamefont {N.}~\bibnamefont {{Rea}}}, \bibinfo {author} {\bibfnamefont
  {M.}~\bibnamefont {{Persic}}}, \ and\ \bibinfo {author} {\bibfnamefont
  {R.~E.}\ \bibnamefont {{Rothschild}}}} (\bibinfo {year} {2005}),\ \href
  {\doibase 10.1086/432615} {\bibfield  {journal} {\bibinfo  {journal} {\apjl}\
  }\textbf {\bibinfo {volume} {628}},\ \bibinfo {pages} {L53}},\ \Eprint
  {http://arxiv.org/abs/astro-ph/0505255} {astro-ph/0505255} \BibitemShut
  {NoStop}%
\bibitem [{\citenamefont {{Israel}}\ \emph {et~al.}(2008)\citenamefont
  {{Israel}}, \citenamefont {{Romano}}, \citenamefont {{Mangano}},
  \citenamefont {{Dall'Osso}}, \citenamefont {{Chincarini}}, \citenamefont
  {{Stella}}, \citenamefont {{Campana}}, \citenamefont {{Belloni}},
  \citenamefont {{Tagliaferri}}, \citenamefont {{Blustin}} \emph
  {et~al.}}]{Israel08}%
  \BibitemOpen
  \bibfield  {author} {\bibinfo {author} {\bibnamefont {{Israel}},
  \bibfnamefont {G.~L.}}, \bibinfo {author} {\bibfnamefont {P.}~\bibnamefont
  {{Romano}}}, \bibinfo {author} {\bibfnamefont {V.}~\bibnamefont {{Mangano}}},
  \bibinfo {author} {\bibfnamefont {S.}~\bibnamefont {{Dall'Osso}}}, \bibinfo
  {author} {\bibfnamefont {G.}~\bibnamefont {{Chincarini}}}, \bibinfo {author}
  {\bibfnamefont {L.}~\bibnamefont {{Stella}}}, \bibinfo {author}
  {\bibfnamefont {S.}~\bibnamefont {{Campana}}}, \bibinfo {author}
  {\bibfnamefont {T.}~\bibnamefont {{Belloni}}}, \bibinfo {author}
  {\bibfnamefont {G.}~\bibnamefont {{Tagliaferri}}}, \bibinfo {author}
  {\bibfnamefont {A.~J.}\ \bibnamefont {{Blustin}}},  \emph {et~al.}} (\bibinfo
  {year} {2008}),\ \href {\doibase 10.1086/590486} {\bibfield  {journal}
  {\bibinfo  {journal} {\apj}\ }\textbf {\bibinfo {volume} {685}},\ \bibinfo
  {pages} {1114}},\ \Eprint {http://arxiv.org/abs/0805.3919} {arXiv:0805.3919}
  \BibitemShut {NoStop}%
\bibitem [{\citenamefont {{Janka}}\ \emph {et~al.}(2007)\citenamefont
  {{Janka}}, \citenamefont {{Langanke}}, \citenamefont {{Marek}}, \citenamefont
  {{Mart{\'{\i}}nez-Pinedo}},\ and\ \citenamefont {{M{\"u}ller}}}]{Janka07}%
  \BibitemOpen
  \bibfield  {author} {\bibinfo {author} {\bibnamefont {{Janka}}, \bibfnamefont
  {H.-T.}}, \bibinfo {author} {\bibfnamefont {K.}~\bibnamefont {{Langanke}}},
  \bibinfo {author} {\bibfnamefont {A.}~\bibnamefont {{Marek}}}, \bibinfo
  {author} {\bibfnamefont {G.}~\bibnamefont {{Mart{\'{\i}}nez-Pinedo}}}, \ and\
  \bibinfo {author} {\bibfnamefont {B.}~\bibnamefont {{M{\"u}ller}}}} (\bibinfo
  {year} {2007}),\ \href {\doibase 10.1016/j.physrep.2007.02.002} {\bibfield
  {journal} {\bibinfo  {journal} {\physrep}\ }\textbf {\bibinfo {volume}
  {442}},\ \bibinfo {pages} {38}},\ \Eprint
  {http://arxiv.org/abs/astro-ph/0612072} {astro-ph/0612072} \BibitemShut
  {NoStop}%
\bibitem [{\citenamefont {{Kajava}}\ \emph {et~al.}(2011)\citenamefont
  {{Kajava}}, \citenamefont {{Ibragimov}}, \citenamefont {{Annala}},
  \citenamefont {{Patruno}},\ and\ \citenamefont {{Poutanen}}}]{Kajava11}%
  \BibitemOpen
  \bibfield  {author} {\bibinfo {author} {\bibnamefont {{Kajava}},
  \bibfnamefont {J.~J.~E.}}, \bibinfo {author} {\bibfnamefont {A.}~\bibnamefont
  {{Ibragimov}}}, \bibinfo {author} {\bibfnamefont {M.}~\bibnamefont
  {{Annala}}}, \bibinfo {author} {\bibfnamefont {A.}~\bibnamefont {{Patruno}}},
  \ and\ \bibinfo {author} {\bibfnamefont {J.}~\bibnamefont {{Poutanen}}}}
  (\bibinfo {year} {2011}),\ \href {\doibase 10.1111/j.1365-2966.2011.19360.x}
  {\bibfield  {journal} {\bibinfo  {journal} {\mnras}\ }\textbf {\bibinfo
  {volume} {417}},\ \bibinfo {pages} {1454}},\ \Eprint
  {http://arxiv.org/abs/1107.0180} {arXiv:1107.0180 [astro-ph.HE]} \BibitemShut
  {NoStop}%
\bibitem [{\citenamefont {{Kajava}}\ \emph {et~al.}(2014)\citenamefont
  {{Kajava}}, \citenamefont {{N{\"a}ttil{\"a}}}, \citenamefont {{Latvala}},
  \citenamefont {{Pursiainen}}, \citenamefont {{Poutanen}}, \citenamefont
  {{Suleimanov}}, \citenamefont {{Revnivtsev}}, \citenamefont {{Kuulkers}},\
  and\ \citenamefont {{Galloway}}}]{Kajava14}%
  \BibitemOpen
  \bibfield  {author} {\bibinfo {author} {\bibnamefont {{Kajava}},
  \bibfnamefont {J.~J.~E.}}, \bibinfo {author} {\bibfnamefont {J.}~\bibnamefont
  {{N{\"a}ttil{\"a}}}}, \bibinfo {author} {\bibfnamefont {O.-M.}\ \bibnamefont
  {{Latvala}}}, \bibinfo {author} {\bibfnamefont {M.}~\bibnamefont
  {{Pursiainen}}}, \bibinfo {author} {\bibfnamefont {J.}~\bibnamefont
  {{Poutanen}}}, \bibinfo {author} {\bibfnamefont {V.~F.}\ \bibnamefont
  {{Suleimanov}}}, \bibinfo {author} {\bibfnamefont {M.~G.}\ \bibnamefont
  {{Revnivtsev}}}, \bibinfo {author} {\bibfnamefont {E.}~\bibnamefont
  {{Kuulkers}}}, \ and\ \bibinfo {author} {\bibfnamefont {D.~K.}\ \bibnamefont
  {{Galloway}}}} (\bibinfo {year} {2014}),\ \href {\doibase
  10.1093/mnras/stu2073} {\bibfield  {journal} {\bibinfo  {journal} {\mnras}\
  }\textbf {\bibinfo {volume} {445}},\ \bibinfo {pages} {4218}},\ \Eprint
  {http://arxiv.org/abs/1406.0322} {arXiv:1406.0322 [astro-ph.HE]} \BibitemShut
  {NoStop}%
\bibitem [{\citenamefont {{Kaplan}}\ and\ \citenamefont
  {{Nelson}}(1986)}]{Kaplan86}%
  \BibitemOpen
  \bibfield  {author} {\bibinfo {author} {\bibnamefont {{Kaplan}},
  \bibfnamefont {D.~B.}}, \ and\ \bibinfo {author} {\bibfnamefont {A.~E.}\
  \bibnamefont {{Nelson}}}} (\bibinfo {year} {1986}),\ \href@noop {} {\bibfield
   {journal} {\bibinfo  {journal} {Physics Letters B}\ }\textbf {\bibinfo
  {volume} {175}},\ \bibinfo {pages} {57}}\BibitemShut {NoStop}%
\bibitem [{\citenamefont {{Kl{\"u}pfel}}\ \emph {et~al.}(2009)\citenamefont
  {{Kl{\"u}pfel}}, \citenamefont {{Reinhard}}, \citenamefont
  {{B{\"u}rvenich}},\ and\ \citenamefont {{Maruhn}}}]{Klupfel09}%
  \BibitemOpen
  \bibfield  {author} {\bibinfo {author} {\bibnamefont {{Kl{\"u}pfel}},
  \bibfnamefont {P.}}, \bibinfo {author} {\bibfnamefont {P.-G.}\ \bibnamefont
  {{Reinhard}}}, \bibinfo {author} {\bibfnamefont {T.~J.}\ \bibnamefont
  {{B{\"u}rvenich}}}, \ and\ \bibinfo {author} {\bibfnamefont {J.~A.}\
  \bibnamefont {{Maruhn}}}} (\bibinfo {year} {2009}),\ \href {\doibase
  10.1103/PhysRevC.79.034310} {\bibfield  {journal} {\bibinfo  {journal}
  {\prc}\ }\textbf {\bibinfo {volume} {79}}~(\bibinfo {number} {3}),\ \bibinfo
  {eid} {034310}}\BibitemShut {NoStop}%
\bibitem [{\citenamefont {{Kortelainen}}\ \emph {et~al.}(2010)\citenamefont
  {{Kortelainen}}, \citenamefont {{Lesinski}}, \citenamefont {{Mor{\'e}}},
  \citenamefont {{Nazarewicz}}, \citenamefont {{Sarich}}, \citenamefont
  {{Schunck}}, \citenamefont {{Stoitsov}},\ and\ \citenamefont
  {{Wild}}}]{Kortelainen10}%
  \BibitemOpen
  \bibfield  {author} {\bibinfo {author} {\bibnamefont {{Kortelainen}},
  \bibfnamefont {M.}}, \bibinfo {author} {\bibfnamefont {T.}~\bibnamefont
  {{Lesinski}}}, \bibinfo {author} {\bibfnamefont {J.}~\bibnamefont
  {{Mor{\'e}}}}, \bibinfo {author} {\bibfnamefont {W.}~\bibnamefont
  {{Nazarewicz}}}, \bibinfo {author} {\bibfnamefont {J.}~\bibnamefont
  {{Sarich}}}, \bibinfo {author} {\bibfnamefont {N.}~\bibnamefont {{Schunck}}},
  \bibinfo {author} {\bibfnamefont {M.~V.}\ \bibnamefont {{Stoitsov}}}, \ and\
  \bibinfo {author} {\bibfnamefont {S.}~\bibnamefont {{Wild}}}} (\bibinfo
  {year} {2010}),\ \href {\doibase 10.1103/PhysRevC.82.024313} {\bibfield
  {journal} {\bibinfo  {journal} {\prc}\ }\textbf {\bibinfo {volume}
  {82}}~(\bibinfo {number} {2}),\ \bibinfo {eid} {024313}},\ \Eprint
  {http://arxiv.org/abs/1005.5145} {arXiv:1005.5145 [nucl-th]} \BibitemShut
  {NoStop}%
\bibitem [{\citenamefont {{Kortelainen}}\ \emph {et~al.}(2014)\citenamefont
  {{Kortelainen}}, \citenamefont {{McDonnell}}, \citenamefont {{Nazarewicz}},
  \citenamefont {{Olsen}}, \citenamefont {{Reinhard}}, \citenamefont
  {{Sarich}}, \citenamefont {{Schunck}}, \citenamefont {{Wild}}, \citenamefont
  {{Davesne}}, \citenamefont {{Erler}},\ and\ \citenamefont
  {{Pastore}}}]{Kortelainen14}%
  \BibitemOpen
  \bibfield  {author} {\bibinfo {author} {\bibnamefont {{Kortelainen}},
  \bibfnamefont {M.}}, \bibinfo {author} {\bibfnamefont {J.}~\bibnamefont
  {{McDonnell}}}, \bibinfo {author} {\bibfnamefont {W.}~\bibnamefont
  {{Nazarewicz}}}, \bibinfo {author} {\bibfnamefont {E.}~\bibnamefont
  {{Olsen}}}, \bibinfo {author} {\bibfnamefont {P.-G.}\ \bibnamefont
  {{Reinhard}}}, \bibinfo {author} {\bibfnamefont {J.}~\bibnamefont
  {{Sarich}}}, \bibinfo {author} {\bibfnamefont {N.}~\bibnamefont {{Schunck}}},
  \bibinfo {author} {\bibfnamefont {S.~M.}\ \bibnamefont {{Wild}}}, \bibinfo
  {author} {\bibfnamefont {D.}~\bibnamefont {{Davesne}}}, \bibinfo {author}
  {\bibfnamefont {J.}~\bibnamefont {{Erler}}}, \ and\ \bibinfo {author}
  {\bibfnamefont {A.}~\bibnamefont {{Pastore}}}} (\bibinfo {year} {2014}),\
  \href {\doibase 10.1103/PhysRevC.89.054314} {\bibfield  {journal} {\bibinfo
  {journal} {\prc}\ }\textbf {\bibinfo {volume} {89}}~(\bibinfo {number} {5}),\
  \bibinfo {eid} {054314}},\ \Eprint {http://arxiv.org/abs/1312.1746}
  {arXiv:1312.1746 [nucl-th]} \BibitemShut {NoStop}%
\bibitem [{\citenamefont {{Kramer}}\ and\ \citenamefont
  {{Wex}}(2009)}]{Kramer09}%
  \BibitemOpen
  \bibfield  {author} {\bibinfo {author} {\bibnamefont {{Kramer}},
  \bibfnamefont {M.}}, \ and\ \bibinfo {author} {\bibfnamefont
  {N.}~\bibnamefont {{Wex}}}} (\bibinfo {year} {2009}),\ \href {\doibase
  10.1088/0264-9381/26/7/073001} {\bibfield  {journal} {\bibinfo  {journal}
  {Classical and Quantum Gravity}\ }\textbf {\bibinfo {volume} {26}}~(\bibinfo
  {number} {7}),\ \bibinfo {eid} {073001}}\BibitemShut {NoStop}%
\bibitem [{\citenamefont {{Kumar}}\ and\ \citenamefont
  {{Zhang}}(2015)}]{Kumar15}%
  \BibitemOpen
  \bibfield  {author} {\bibinfo {author} {\bibnamefont {{Kumar}}, \bibfnamefont
  {P.}}, \ and\ \bibinfo {author} {\bibfnamefont {B.}~\bibnamefont {{Zhang}}}}
  (\bibinfo {year} {2015}),\ \href {\doibase 10.1016/j.physrep.2014.09.008}
  {\bibfield  {journal} {\bibinfo  {journal} {\physrep}\ }\textbf {\bibinfo
  {volume} {561}},\ \bibinfo {pages} {1}},\ \Eprint
  {http://arxiv.org/abs/1410.0679} {arXiv:1410.0679 [astro-ph.HE]} \BibitemShut
  {NoStop}%
\bibitem [{\citenamefont {{Kunihiro}}\ \emph {et~al.}(1993)\citenamefont
  {{Kunihiro}}, \citenamefont {{Takatsuka}},\ and\ \citenamefont
  {{Tamagaki}}}]{Kunihiro93}%
  \BibitemOpen
  \bibfield  {author} {\bibinfo {author} {\bibnamefont {{Kunihiro}},
  \bibfnamefont {T.}}, \bibinfo {author} {\bibfnamefont {T.}~\bibnamefont
  {{Takatsuka}}}, \ and\ \bibinfo {author} {\bibfnamefont {R.}~\bibnamefont
  {{Tamagaki}}}} (\bibinfo {year} {1993}),\ \href@noop {} {\bibfield  {journal}
  {\bibinfo  {journal} {Progress of Theoretical Physics Supplement}\ }\textbf
  {\bibinfo {volume} {112}},\ \bibinfo {pages} {197}}\BibitemShut {NoStop}%
\bibitem [{\citenamefont {{Kurkela}}\ \emph {et~al.}(2014)\citenamefont
  {{Kurkela}}, \citenamefont {{Fraga}}, \citenamefont {{Schaffner-Bielich}},\
  and\ \citenamefont {{Vuorinen}}}]{Kurkela14}%
  \BibitemOpen
  \bibfield  {author} {\bibinfo {author} {\bibnamefont {{Kurkela}},
  \bibfnamefont {A.}}, \bibinfo {author} {\bibfnamefont {E.~S.}\ \bibnamefont
  {{Fraga}}}, \bibinfo {author} {\bibfnamefont {J.}~\bibnamefont
  {{Schaffner-Bielich}}}, \ and\ \bibinfo {author} {\bibfnamefont
  {A.}~\bibnamefont {{Vuorinen}}}} (\bibinfo {year} {2014}),\ \href@noop {}
  {\bibfield  {journal} {\bibinfo  {journal} {\apj}\ }\textbf {\bibinfo
  {volume} {789}},\ \bibinfo {eid} {127}}\BibitemShut {NoStop}%
\bibitem [{\citenamefont {{Lackey}}\ \emph {et~al.}(2012)\citenamefont
  {{Lackey}}, \citenamefont {{Kyutoku}}, \citenamefont {{Shibata}},
  \citenamefont {{Brady}},\ and\ \citenamefont {{Friedman}}}]{Lackey12}%
  \BibitemOpen
  \bibfield  {author} {\bibinfo {author} {\bibnamefont {{Lackey}},
  \bibfnamefont {B.~D.}}, \bibinfo {author} {\bibfnamefont {K.}~\bibnamefont
  {{Kyutoku}}}, \bibinfo {author} {\bibfnamefont {M.}~\bibnamefont
  {{Shibata}}}, \bibinfo {author} {\bibfnamefont {P.~R.}\ \bibnamefont
  {{Brady}}}, \ and\ \bibinfo {author} {\bibfnamefont {J.~L.}\ \bibnamefont
  {{Friedman}}}} (\bibinfo {year} {2012}),\ \href {\doibase
  10.1103/PhysRevD.85.044061} {\bibfield  {journal} {\bibinfo  {journal}
  {\prd}\ }\textbf {\bibinfo {volume} {85}}~(\bibinfo {number} {4}),\ \bibinfo
  {eid} {044061}},\ \Eprint {http://arxiv.org/abs/1109.3402} {arXiv:1109.3402
  [astro-ph.HE]} \BibitemShut {NoStop}%
\bibitem [{\citenamefont {{Lamb}}\ and\ \citenamefont {{Yu}}(2005)}]{Lamb05}%
  \BibitemOpen
  \bibfield  {author} {\bibinfo {author} {\bibnamefont {{Lamb}}, \bibfnamefont
  {F.}}, \ and\ \bibinfo {author} {\bibfnamefont {W.}~\bibnamefont {{Yu}}}}
  (\bibinfo {year} {2005}),\ in\ \href@noop {} {\emph {\bibinfo {booktitle}
  {Binary Radio Pulsars}}},\ \bibinfo {series} {Astronomical Society of the
  Pacific Conference Series}, Vol.\ \bibinfo {volume} {328},\ \bibinfo {editor}
  {edited by\ \bibinfo {editor} {\bibfnamefont {F.~A.}\ \bibnamefont
  {{Rasio}}}\ and\ \bibinfo {editor} {\bibfnamefont {I.~H.}\ \bibnamefont
  {{Stairs}}}},\ p.\ \bibinfo {pages} {299},\ \Eprint
  {http://arxiv.org/abs/astro-ph/0408459} {astro-ph/0408459} \BibitemShut
  {NoStop}%
\bibitem [{\citenamefont {{Lamb}}\ \emph {et~al.}(2009)\citenamefont {{Lamb}},
  \citenamefont {{Boutloukos}}, \citenamefont {{Van Wassenhove}}, \citenamefont
  {{Chamberlain}}, \citenamefont {{Lo}},\ and\ \citenamefont
  {{Miller}}}]{Lamb09b}%
  \BibitemOpen
  \bibfield  {author} {\bibinfo {author} {\bibnamefont {{Lamb}}, \bibfnamefont
  {F.~K.}}, \bibinfo {author} {\bibfnamefont {S.}~\bibnamefont {{Boutloukos}}},
  \bibinfo {author} {\bibfnamefont {S.}~\bibnamefont {{Van Wassenhove}}},
  \bibinfo {author} {\bibfnamefont {R.~T.}\ \bibnamefont {{Chamberlain}}},
  \bibinfo {author} {\bibfnamefont {K.~H.}\ \bibnamefont {{Lo}}}, \ and\
  \bibinfo {author} {\bibfnamefont {M.~C.}\ \bibnamefont {{Miller}}}} (\bibinfo
  {year} {2009}),\ \href {\doibase 10.1088/0004-637X/705/1/L36} {\bibfield
  {journal} {\bibinfo  {journal} {\apjl}\ }\textbf {\bibinfo {volume} {705}},\
  \bibinfo {pages} {L36}},\ \Eprint {http://arxiv.org/abs/0809.4016}
  {arXiv:0809.4016} \BibitemShut {NoStop}%
\bibitem [{\citenamefont {{Lander}}\ and\ \citenamefont
  {{Jones}}(2011)}]{Lander11}%
  \BibitemOpen
  \bibfield  {author} {\bibinfo {author} {\bibnamefont {{Lander}},
  \bibfnamefont {S.~K.}}, \ and\ \bibinfo {author} {\bibfnamefont {D.~I.}\
  \bibnamefont {{Jones}}}} (\bibinfo {year} {2011}),\ \href {\doibase
  10.1111/j.1365-2966.2010.18009.x} {\bibfield  {journal} {\bibinfo  {journal}
  {\mnras}\ }\textbf {\bibinfo {volume} {412}},\ \bibinfo {pages} {1730}},\
  \Eprint {http://arxiv.org/abs/1010.0614} {arXiv:1010.0614 [astro-ph.SR]}
  \BibitemShut {NoStop}%
\bibitem [{\citenamefont {{Lander}}\ \emph {et~al.}(2010)\citenamefont
  {{Lander}}, \citenamefont {{Jones}},\ and\ \citenamefont
  {{Passamonti}}}]{Lander10}%
  \BibitemOpen
  \bibfield  {author} {\bibinfo {author} {\bibnamefont {{Lander}},
  \bibfnamefont {S.~K.}}, \bibinfo {author} {\bibfnamefont {D.~I.}\
  \bibnamefont {{Jones}}}, \ and\ \bibinfo {author} {\bibfnamefont
  {A.}~\bibnamefont {{Passamonti}}}} (\bibinfo {year} {2010}),\ \href {\doibase
  10.1111/j.1365-2966.2010.16435.x} {\bibfield  {journal} {\bibinfo  {journal}
  {\mnras}\ }\textbf {\bibinfo {volume} {405}},\ \bibinfo {pages} {318}},\
  \Eprint {http://arxiv.org/abs/0912.3480} {arXiv:0912.3480 [astro-ph.SR]}
  \BibitemShut {NoStop}%
\bibitem [{\citenamefont {{Lattimer}}\ and\ \citenamefont
  {{Prakash}}(2001)}]{Lattimer01}%
  \BibitemOpen
  \bibfield  {author} {\bibinfo {author} {\bibnamefont {{Lattimer}},
  \bibfnamefont {J.~M.}}, \ and\ \bibinfo {author} {\bibfnamefont
  {M.}~\bibnamefont {{Prakash}}}} (\bibinfo {year} {2001}),\ \href {\doibase
  10.1086/319702} {\bibfield  {journal} {\bibinfo  {journal} {\apj}\ }\textbf
  {\bibinfo {volume} {550}},\ \bibinfo {pages} {426}},\ \Eprint
  {http://arxiv.org/abs/astro-ph/0002232} {astro-ph/0002232} \BibitemShut
  {NoStop}%
\bibitem [{\citenamefont {Lattimer}\ and\ \citenamefont
  {Prakash}(2005)}]{Lattimer05}%
  \BibitemOpen
  \bibfield  {author} {\bibinfo {author} {\bibnamefont {Lattimer},
  \bibfnamefont {J.~M.}}, \ and\ \bibinfo {author} {\bibfnamefont
  {M.}~\bibnamefont {Prakash}}} (\bibinfo {year} {2005}),\ \href {\doibase
  10.1103/PhysRevLett.94.111101} {\bibfield  {journal} {\bibinfo  {journal}
  {Phys. Rev. Lett.}\ }\textbf {\bibinfo {volume} {94}},\ \bibinfo {pages}
  {111101}}\BibitemShut {NoStop}%
\bibitem [{\citenamefont {{Lattimer}}\ and\ \citenamefont
  {{Schutz}}(2005)}]{Lattimer05b}%
  \BibitemOpen
  \bibfield  {author} {\bibinfo {author} {\bibnamefont {{Lattimer}},
  \bibfnamefont {J.~M.}}, \ and\ \bibinfo {author} {\bibfnamefont {B.~F.}\
  \bibnamefont {{Schutz}}}} (\bibinfo {year} {2005}),\ \href {\doibase
  10.1086/431543} {\bibfield  {journal} {\bibinfo  {journal} {\apj}\ }\textbf
  {\bibinfo {volume} {629}},\ \bibinfo {pages} {979}},\ \Eprint
  {http://arxiv.org/abs/astro-ph/0411470} {astro-ph/0411470} \BibitemShut
  {NoStop}%
\bibitem [{\citenamefont {{Lattimer}}\ and\ \citenamefont
  {{Steiner}}(2014)}]{Lattimer14}%
  \BibitemOpen
  \bibfield  {author} {\bibinfo {author} {\bibnamefont {{Lattimer}},
  \bibfnamefont {J.~M.}}, \ and\ \bibinfo {author} {\bibfnamefont {A.~W.}\
  \bibnamefont {{Steiner}}}} (\bibinfo {year} {2014}),\ \href {\doibase
  10.1140/epja/i2014-14040-y} {\bibfield  {journal} {\bibinfo  {journal}
  {European Physical Journal A}\ }\textbf {\bibinfo {volume} {50}},\ \bibinfo
  {eid} {40}},\ \Eprint {http://arxiv.org/abs/1403.1186} {arXiv:1403.1186
  [nucl-th]} \BibitemShut {NoStop}%
\bibitem [{\citenamefont {{Leahy}}(2004)}]{Leahy04}%
  \BibitemOpen
  \bibfield  {author} {\bibinfo {author} {\bibnamefont {{Leahy}}, \bibfnamefont
  {D.~A.}}} (\bibinfo {year} {2004}),\ \href {\doibase 10.1086/422905}
  {\bibfield  {journal} {\bibinfo  {journal} {\apj}\ }\textbf {\bibinfo
  {volume} {613}},\ \bibinfo {pages} {517}}\BibitemShut {NoStop}%
\bibitem [{\citenamefont {{Leahy}}\ \emph {et~al.}(2011)\citenamefont
  {{Leahy}}, \citenamefont {{Morsink}},\ and\ \citenamefont
  {{Chou}}}]{Leahy11}%
  \BibitemOpen
  \bibfield  {author} {\bibinfo {author} {\bibnamefont {{Leahy}}, \bibfnamefont
  {D.~A.}}, \bibinfo {author} {\bibfnamefont {S.~M.}\ \bibnamefont
  {{Morsink}}}, \ and\ \bibinfo {author} {\bibfnamefont {Y.}~\bibnamefont
  {{Chou}}}} (\bibinfo {year} {2011}),\ \href {\doibase
  10.1088/0004-637X/742/1/17} {\bibfield  {journal} {\bibinfo  {journal}
  {\apj}\ }\textbf {\bibinfo {volume} {742}},\ \bibinfo {eid} {17}},\ \Eprint
  {http://arxiv.org/abs/1106.3131} {arXiv:1106.3131 [astro-ph.HE]} \BibitemShut
  {NoStop}%
\bibitem [{\citenamefont {{Leahy}}\ \emph {et~al.}(2009)\citenamefont
  {{Leahy}}, \citenamefont {{Morsink}}, \citenamefont {{Chung}},\ and\
  \citenamefont {{Chou}}}]{Leahy09}%
  \BibitemOpen
  \bibfield  {author} {\bibinfo {author} {\bibnamefont {{Leahy}}, \bibfnamefont
  {D.~A.}}, \bibinfo {author} {\bibfnamefont {S.~M.}\ \bibnamefont
  {{Morsink}}}, \bibinfo {author} {\bibfnamefont {Y.-Y.}\ \bibnamefont
  {{Chung}}}, \ and\ \bibinfo {author} {\bibfnamefont {Y.}~\bibnamefont
  {{Chou}}}} (\bibinfo {year} {2009}),\ \href {\doibase
  10.1088/0004-637X/691/2/1235} {\bibfield  {journal} {\bibinfo  {journal}
  {\apj}\ }\textbf {\bibinfo {volume} {691}},\ \bibinfo {pages} {1235}},\
  \Eprint {http://arxiv.org/abs/0806.0824} {arXiv:0806.0824} \BibitemShut
  {NoStop}%
\bibitem [{\citenamefont {{Lee}}(2008)}]{Lee08}%
  \BibitemOpen
  \bibfield  {author} {\bibinfo {author} {\bibnamefont {{Lee}}, \bibfnamefont
  {U.}}} (\bibinfo {year} {2008}),\ \href {\doibase
  10.1111/j.1365-2966.2008.12965.x} {\bibfield  {journal} {\bibinfo  {journal}
  {\mnras}\ }\textbf {\bibinfo {volume} {385}},\ \bibinfo {pages} {2069}},\
  \Eprint {http://arxiv.org/abs/0710.4986} {arXiv:0710.4986} \BibitemShut
  {NoStop}%
\bibitem [{\citenamefont {{Levin}}(2007)}]{Levin07}%
  \BibitemOpen
  \bibfield  {author} {\bibinfo {author} {\bibnamefont {{Levin}}, \bibfnamefont
  {Y.}}} (\bibinfo {year} {2007}),\ \href {\doibase
  10.1111/j.1365-2966.2007.11582.x} {\bibfield  {journal} {\bibinfo  {journal}
  {\mnras}\ }\textbf {\bibinfo {volume} {377}},\ \bibinfo {pages} {159}},\
  \Eprint {http://arxiv.org/abs/astro-ph/0612725} {astro-ph/0612725}
  \BibitemShut {NoStop}%
\bibitem [{\citenamefont {{Lewin}}\ \emph {et~al.}(1993)\citenamefont
  {{Lewin}}, \citenamefont {{van Paradijs}},\ and\ \citenamefont
  {{Taam}}}]{Lewin93}%
  \BibitemOpen
  \bibfield  {author} {\bibinfo {author} {\bibnamefont {{Lewin}}, \bibfnamefont
  {W.~H.~G.}}, \bibinfo {author} {\bibfnamefont {J.}~\bibnamefont {{van
  Paradijs}}}, \ and\ \bibinfo {author} {\bibfnamefont {R.~E.}\ \bibnamefont
  {{Taam}}}} (\bibinfo {year} {1993}),\ \href {\doibase 10.1007/BF00196124}
  {\bibfield  {journal} {\bibinfo  {journal} {Space Science Reviews}\ }\textbf
  {\bibinfo {volume} {62}},\ \bibinfo {pages} {223}}\BibitemShut {NoStop}%
\bibitem [{\citenamefont {{Lindblom}}(1992)}]{Lindblom92}%
  \BibitemOpen
  \bibfield  {author} {\bibinfo {author} {\bibnamefont {{Lindblom}},
  \bibfnamefont {L.}}} (\bibinfo {year} {1992}),\ \href {\doibase
  10.1086/171882} {\bibfield  {journal} {\bibinfo  {journal} {\apj}\ }\textbf
  {\bibinfo {volume} {398}},\ \bibinfo {pages} {569}}\BibitemShut {NoStop}%
\bibitem [{\citenamefont {{Lindblom}}\ and\ \citenamefont
  {{Indik}}(2012)}]{Lindblom12}%
  \BibitemOpen
  \bibfield  {author} {\bibinfo {author} {\bibnamefont {{Lindblom}},
  \bibfnamefont {L.}}, \ and\ \bibinfo {author} {\bibfnamefont {N.~M.}\
  \bibnamefont {{Indik}}}} (\bibinfo {year} {2012}),\ \href {\doibase
  10.1103/PhysRevD.86.084003} {\bibfield  {journal} {\bibinfo  {journal}
  {\prd}\ }\textbf {\bibinfo {volume} {86}}~(\bibinfo {number} {8}),\ \bibinfo
  {eid} {084003}},\ \Eprint {http://arxiv.org/abs/1207.3744} {arXiv:1207.3744
  [astro-ph.HE]} \BibitemShut {NoStop}%
\bibitem [{\citenamefont {{Lindblom}}\ and\ \citenamefont
  {{Indik}}(2014)}]{Lindblom14}%
  \BibitemOpen
  \bibfield  {author} {\bibinfo {author} {\bibnamefont {{Lindblom}},
  \bibfnamefont {L.}}, \ and\ \bibinfo {author} {\bibfnamefont {N.~M.}\
  \bibnamefont {{Indik}}}} (\bibinfo {year} {2014}),\ \href {\doibase
  10.1103/PhysRevD.89.064003} {\bibfield  {journal} {\bibinfo  {journal}
  {\prd}\ }\textbf {\bibinfo {volume} {89}}~(\bibinfo {number} {6}),\ \bibinfo
  {eid} {064003}},\ \Eprint {http://arxiv.org/abs/1310.0803} {arXiv:1310.0803
  [astro-ph.HE]} \BibitemShut {NoStop}%
\bibitem [{\citenamefont {{Link}}(2014)}]{Link14}%
  \BibitemOpen
  \bibfield  {author} {\bibinfo {author} {\bibnamefont {{Link}}, \bibfnamefont
  {B.}}} (\bibinfo {year} {2014}),\ \href {\doibase 10.1093/mnras/stu584}
  {\bibfield  {journal} {\bibinfo  {journal} {\mnras}\ }\textbf {\bibinfo
  {volume} {441}},\ \bibinfo {pages} {2676}},\ \Eprint
  {http://arxiv.org/abs/1312.5144} {arXiv:1312.5144 [astro-ph.SR]} \BibitemShut
  {NoStop}%
\bibitem [{\citenamefont {{Lo}}\ \emph {et~al.}(2013)\citenamefont {{Lo}},
  \citenamefont {{Miller}}, \citenamefont {{Bhattacharyya}},\ and\
  \citenamefont {{Lamb}}}]{Lo13}%
  \BibitemOpen
  \bibfield  {author} {\bibinfo {author} {\bibnamefont {{Lo}}, \bibfnamefont
  {K.~H.}}, \bibinfo {author} {\bibfnamefont {M.~C.}\ \bibnamefont {{Miller}}},
  \bibinfo {author} {\bibfnamefont {S.}~\bibnamefont {{Bhattacharyya}}}, \ and\
  \bibinfo {author} {\bibfnamefont {F.~K.}\ \bibnamefont {{Lamb}}}} (\bibinfo
  {year} {2013}),\ \href {\doibase 10.1088/0004-637X/776/1/19} {\bibfield
  {journal} {\bibinfo  {journal} {\apj}\ }\textbf {\bibinfo {volume} {776}},\
  \bibinfo {eid} {19}},\ \Eprint {http://arxiv.org/abs/1304.2330}
  {arXiv:1304.2330 [astro-ph.HE]} \BibitemShut {NoStop}%
\bibitem [{\citenamefont {{Logoteta}}\ \emph {et~al.}(2013)\citenamefont
  {{Logoteta}}, \citenamefont {{Vida{\~n}a}},\ and\ \citenamefont
  {{Provid{\^e}ncia}}}]{Logoteta13}%
  \BibitemOpen
  \bibfield  {author} {\bibinfo {author} {\bibnamefont {{Logoteta}},
  \bibfnamefont {D.}}, \bibinfo {author} {\bibfnamefont {I.}~\bibnamefont
  {{Vida{\~n}a}}}, \ and\ \bibinfo {author} {\bibfnamefont {C.}~\bibnamefont
  {{Provid{\^e}ncia}}}} (\bibinfo {year} {2013}),\ \href {\doibase
  10.1016/j.nuclphysa.2013.01.022} {\bibfield  {journal} {\bibinfo  {journal}
  {Nuclear Physics A}\ }\textbf {\bibinfo {volume} {914}},\ \bibinfo {pages}
  {433}}\BibitemShut {NoStop}%
\bibitem [{\citenamefont {{London}}\ \emph {et~al.}(1986)\citenamefont
  {{London}}, \citenamefont {{Taam}},\ and\ \citenamefont
  {{Howard}}}]{London86}%
  \BibitemOpen
  \bibfield  {author} {\bibinfo {author} {\bibnamefont {{London}},
  \bibfnamefont {R.~A.}}, \bibinfo {author} {\bibfnamefont {R.~E.}\
  \bibnamefont {{Taam}}}, \ and\ \bibinfo {author} {\bibfnamefont {W.~M.}\
  \bibnamefont {{Howard}}}} (\bibinfo {year} {1986}),\ \href {\doibase
  10.1086/164330} {\bibfield  {journal} {\bibinfo  {journal} {\apj}\ }\textbf
  {\bibinfo {volume} {306}},\ \bibinfo {pages} {170}}\BibitemShut {NoStop}%
\bibitem [{\citenamefont {{Lorimer}}(2008)}]{Lorimer08}%
  \BibitemOpen
  \bibfield  {author} {\bibinfo {author} {\bibnamefont {{Lorimer}},
  \bibfnamefont {D.~R.}}} (\bibinfo {year} {2008}),\ \href {\doibase
  10.12942/lrr-2008-8} {\bibfield  {journal} {\bibinfo  {journal} {Living
  Reviews in Relativity}\ }\textbf {\bibinfo {volume} {11}},\ \bibinfo {pages}
  {8}},\ \Eprint {http://arxiv.org/abs/0811.0762} {arXiv:0811.0762}
  \BibitemShut {NoStop}%
\bibitem [{\citenamefont {{Madej}}(1991)}]{Madej91}%
  \BibitemOpen
  \bibfield  {author} {\bibinfo {author} {\bibnamefont {{Madej}}, \bibfnamefont
  {J.}}} (\bibinfo {year} {1991}),\ \href {\doibase 10.1086/170264} {\bibfield
  {journal} {\bibinfo  {journal} {\apj}\ }\textbf {\bibinfo {volume} {376}},\
  \bibinfo {pages} {161}}\BibitemShut {NoStop}%
\bibitem [{\citenamefont {{Madej}}\ \emph {et~al.}(2004)\citenamefont
  {{Madej}}, \citenamefont {{Joss}},\ and\ \citenamefont
  {{R{\'o}{\.z}a{\'n}ska}}}]{Madej04}%
  \BibitemOpen
  \bibfield  {author} {\bibinfo {author} {\bibnamefont {{Madej}}, \bibfnamefont
  {J.}}, \bibinfo {author} {\bibfnamefont {P.~C.}\ \bibnamefont {{Joss}}}, \
  and\ \bibinfo {author} {\bibfnamefont {A.}~\bibnamefont
  {{R{\'o}{\.z}a{\'n}ska}}}} (\bibinfo {year} {2004}),\ \href {\doibase
  10.1086/379761} {\bibfield  {journal} {\bibinfo  {journal} {\apj}\ }\textbf
  {\bibinfo {volume} {602}},\ \bibinfo {pages} {904}}\BibitemShut {NoStop}%
\bibitem [{\citenamefont {{Manko}}\ \emph
  {et~al.}(2000{\natexlab{a}})\citenamefont {{Manko}}, \citenamefont
  {{Mielke}},\ and\ \citenamefont {{Sanabria-G{\'o}mez}}}]{Manko00a}%
  \BibitemOpen
  \bibfield  {author} {\bibinfo {author} {\bibnamefont {{Manko}}, \bibfnamefont
  {V.~S.}}, \bibinfo {author} {\bibfnamefont {E.~W.}\ \bibnamefont {{Mielke}}},
  \ and\ \bibinfo {author} {\bibfnamefont {J.~D.}\ \bibnamefont
  {{Sanabria-G{\'o}mez}}}} (\bibinfo {year} {2000}{\natexlab{a}}),\ \href
  {\doibase 10.1103/PhysRevD.61.081501} {\bibfield  {journal} {\bibinfo
  {journal} {\prd}\ }\textbf {\bibinfo {volume} {61}}~(\bibinfo {number} {8}),\
  \bibinfo {eid} {081501}},\ \Eprint {http://arxiv.org/abs/gr-qc/0001081}
  {gr-qc/0001081} \BibitemShut {NoStop}%
\bibitem [{\citenamefont {{Manko}}\ \emph
  {et~al.}(2000{\natexlab{b}})\citenamefont {{Manko}}, \citenamefont
  {{Sanabria-G{\'o}mez}},\ and\ \citenamefont {{Manko}}}]{Manko00b}%
  \BibitemOpen
  \bibfield  {author} {\bibinfo {author} {\bibnamefont {{Manko}}, \bibfnamefont
  {V.~S.}}, \bibinfo {author} {\bibfnamefont {J.~D.}\ \bibnamefont
  {{Sanabria-G{\'o}mez}}}, \ and\ \bibinfo {author} {\bibfnamefont {O.~V.}\
  \bibnamefont {{Manko}}}} (\bibinfo {year} {2000}{\natexlab{b}}),\ \href
  {\doibase 10.1103/PhysRevD.62.044048} {\bibfield  {journal} {\bibinfo
  {journal} {\prd}\ }\textbf {\bibinfo {volume} {62}}~(\bibinfo {number} {4}),\
  \bibinfo {eid} {044048}}\BibitemShut {NoStop}%
\bibitem [{\citenamefont {{McLerran}}\ and\ \citenamefont
  {{Pisarski}}(2007)}]{McLerran07}%
  \BibitemOpen
  \bibfield  {author} {\bibinfo {author} {\bibnamefont {{McLerran}},
  \bibfnamefont {L.}}, \ and\ \bibinfo {author} {\bibfnamefont {R.~D.}\
  \bibnamefont {{Pisarski}}}} (\bibinfo {year} {2007}),\ \href@noop {}
  {\bibfield  {journal} {\bibinfo  {journal} {Nucl. Phys. A}\ }\textbf
  {\bibinfo {volume} {796}},\ \bibinfo {pages} {83}}\BibitemShut {NoStop}%
\bibitem [{\citenamefont {{Mendell}}(2001)}]{Mendell01}%
  \BibitemOpen
  \bibfield  {author} {\bibinfo {author} {\bibnamefont {{Mendell}},
  \bibfnamefont {G.}}} (\bibinfo {year} {2001}),\ \href {\doibase
  10.1103/PhysRevD.64.044009} {\bibfield  {journal} {\bibinfo  {journal}
  {\prd}\ }\textbf {\bibinfo {volume} {64}}~(\bibinfo {number} {4}),\ \bibinfo
  {eid} {044009}},\ \Eprint {http://arxiv.org/abs/gr-qc/0102042}
  {gr-qc/0102042} \BibitemShut {NoStop}%
\bibitem [{\citenamefont {{Messenger}}(2011)}]{Messenger11}%
  \BibitemOpen
  \bibfield  {author} {\bibinfo {author} {\bibnamefont {{Messenger}},
  \bibfnamefont {C.}}} (\bibinfo {year} {2011}),\ \href {\doibase
  10.1103/PhysRevD.84.083003} {\bibfield  {journal} {\bibinfo  {journal}
  {\prd}\ }\textbf {\bibinfo {volume} {84}}~(\bibinfo {number} {8}),\ \bibinfo
  {eid} {083003}},\ \Eprint {http://arxiv.org/abs/1109.0501} {arXiv:1109.0501
  [gr-qc]} \BibitemShut {NoStop}%
\bibitem [{\citenamefont {{Metzger}}\ \emph {et~al.}(2010)\citenamefont
  {{Metzger}}, \citenamefont {{Mart{\'{\i}}nez-Pinedo}}, \citenamefont
  {{Darbha}}, \citenamefont {{Quataert}}, \citenamefont {{Arcones}},
  \citenamefont {{Kasen}}, \citenamefont {{Thomas}}, \citenamefont {{Nugent}},
  \citenamefont {{Panov}},\ and\ \citenamefont {{Zinner}}}]{Metzger10}%
  \BibitemOpen
  \bibfield  {author} {\bibinfo {author} {\bibnamefont {{Metzger}},
  \bibfnamefont {B.~D.}}, \bibinfo {author} {\bibfnamefont {G.}~\bibnamefont
  {{Mart{\'{\i}}nez-Pinedo}}}, \bibinfo {author} {\bibfnamefont
  {S.}~\bibnamefont {{Darbha}}}, \bibinfo {author} {\bibfnamefont
  {E.}~\bibnamefont {{Quataert}}}, \bibinfo {author} {\bibfnamefont
  {A.}~\bibnamefont {{Arcones}}}, \bibinfo {author} {\bibfnamefont
  {D.}~\bibnamefont {{Kasen}}}, \bibinfo {author} {\bibfnamefont
  {R.}~\bibnamefont {{Thomas}}}, \bibinfo {author} {\bibfnamefont
  {P.}~\bibnamefont {{Nugent}}}, \bibinfo {author} {\bibfnamefont {I.~V.}\
  \bibnamefont {{Panov}}}, \ and\ \bibinfo {author} {\bibfnamefont {N.~T.}\
  \bibnamefont {{Zinner}}}} (\bibinfo {year} {2010}),\ \href {\doibase
  10.1111/j.1365-2966.2010.16864.x} {\bibfield  {journal} {\bibinfo  {journal}
  {\mnras}\ }\textbf {\bibinfo {volume} {406}},\ \bibinfo {pages} {2650}},\
  \Eprint {http://arxiv.org/abs/1001.5029} {arXiv:1001.5029 [astro-ph.HE]}
  \BibitemShut {NoStop}%
\bibitem [{\citenamefont {{Miller}}(2013)}]{Miller13a}%
  \BibitemOpen
  \bibfield  {author} {\bibinfo {author} {\bibnamefont {{Miller}},
  \bibfnamefont {M.~C.}}} (\bibinfo {year} {2013}),\ \href@noop {} {\bibfield
  {journal} {\bibinfo  {journal} {ArXiv e-prints}\ }}\Eprint
  {http://arxiv.org/abs/1312.0029} {arXiv:1312.0029 [astro-ph.HE]} \BibitemShut
  {NoStop}%
\bibitem [{\citenamefont {{Miller}}\ \emph {et~al.}(2013)\citenamefont
  {{Miller}}, \citenamefont {{Boutloukos}}, \citenamefont {{Lo}},\ and\
  \citenamefont {{Lamb}}}]{Miller13b}%
  \BibitemOpen
  \bibfield  {author} {\bibinfo {author} {\bibnamefont {{Miller}},
  \bibfnamefont {M.~C.}}, \bibinfo {author} {\bibfnamefont {S.}~\bibnamefont
  {{Boutloukos}}}, \bibinfo {author} {\bibfnamefont {K.~H.}\ \bibnamefont
  {{Lo}}}, \ and\ \bibinfo {author} {\bibfnamefont {F.~K.}\ \bibnamefont
  {{Lamb}}}} (\bibinfo {year} {2013}),\ in\ \href {\doibase
  10.1017/S1743921312019308} {\emph {\bibinfo {booktitle} {IAU Symposium}}},\
  \bibinfo {series} {IAU Symposium}, Vol.\ \bibinfo {volume} {290},\ \bibinfo
  {editor} {edited by\ \bibinfo {editor} {\bibfnamefont {C.~M.}\ \bibnamefont
  {{Zhang}}}, \bibinfo {editor} {\bibfnamefont {T.}~\bibnamefont {{Belloni}}},
  \bibinfo {editor} {\bibfnamefont {M.}~\bibnamefont {{M{\'e}ndez}}}, \ and\
  \bibinfo {editor} {\bibfnamefont {S.~N.}\ \bibnamefont {{Zhang}}}},\ pp.\
  \bibinfo {pages} {101--108}\BibitemShut {NoStop}%
\bibitem [{\citenamefont {{Miller}}\ and\ \citenamefont
  {{Lamb}}(1998)}]{Miller98}%
  \BibitemOpen
  \bibfield  {author} {\bibinfo {author} {\bibnamefont {{Miller}},
  \bibfnamefont {M.~C.}}, \ and\ \bibinfo {author} {\bibfnamefont {F.~K.}\
  \bibnamefont {{Lamb}}}} (\bibinfo {year} {1998}),\ \href {\doibase
  10.1086/311335} {\bibfield  {journal} {\bibinfo  {journal} {\apjl}\ }\textbf
  {\bibinfo {volume} {499}},\ \bibinfo {pages} {L37}},\ \Eprint
  {http://arxiv.org/abs/astro-ph/9711325} {astro-ph/9711325} \BibitemShut
  {NoStop}%
\bibitem [{\citenamefont {{Miller}}\ and\ \citenamefont
  {{Lamb}}(2015)}]{Miller15}%
  \BibitemOpen
  \bibfield  {author} {\bibinfo {author} {\bibnamefont {{Miller}},
  \bibfnamefont {M.~C.}}, \ and\ \bibinfo {author} {\bibfnamefont {F.~K.}\
  \bibnamefont {{Lamb}}}} (\bibinfo {year} {2015}),\ \href {\doibase
  10.1088/0004-637X/808/1/31} {\bibfield  {journal} {\bibinfo  {journal}
  {\apj}\ }\textbf {\bibinfo {volume} {808}},\ \bibinfo {eid} {31}},\ \Eprint
  {http://arxiv.org/abs/1407.2579} {arXiv:1407.2579 [astro-ph.HE]} \BibitemShut
  {NoStop}%
\bibitem [{\citenamefont {{Morsink}}\ and\ \citenamefont
  {{Leahy}}(2011)}]{Morsink11}%
  \BibitemOpen
  \bibfield  {author} {\bibinfo {author} {\bibnamefont {{Morsink}},
  \bibfnamefont {S.~M.}}, \ and\ \bibinfo {author} {\bibfnamefont {D.~A.}\
  \bibnamefont {{Leahy}}}} (\bibinfo {year} {2011}),\ \href {\doibase
  10.1088/0004-637X/726/1/56} {\bibfield  {journal} {\bibinfo  {journal}
  {\apj}\ }\textbf {\bibinfo {volume} {726}},\ \bibinfo {eid} {56}},\ \Eprint
  {http://arxiv.org/abs/0911.0887} {arXiv:0911.0887 [astro-ph.HE]} \BibitemShut
  {NoStop}%
\bibitem [{\citenamefont {{Morsink}}\ \emph {et~al.}(2007)\citenamefont
  {{Morsink}}, \citenamefont {{Leahy}}, \citenamefont {{Cadeau}},\ and\
  \citenamefont {{Braga}}}]{Morsink07}%
  \BibitemOpen
  \bibfield  {author} {\bibinfo {author} {\bibnamefont {{Morsink}},
  \bibfnamefont {S.~M.}}, \bibinfo {author} {\bibfnamefont {D.~A.}\
  \bibnamefont {{Leahy}}}, \bibinfo {author} {\bibfnamefont {C.}~\bibnamefont
  {{Cadeau}}}, \ and\ \bibinfo {author} {\bibfnamefont {J.}~\bibnamefont
  {{Braga}}}} (\bibinfo {year} {2007}),\ \href {\doibase 10.1086/518648}
  {\bibfield  {journal} {\bibinfo  {journal} {\apj}\ }\textbf {\bibinfo
  {volume} {663}},\ \bibinfo {pages} {1244}},\ \Eprint
  {http://arxiv.org/abs/astro-ph/0703123} {astro-ph/0703123} \BibitemShut
  {NoStop}%
\bibitem [{\citenamefont {{Motch}}\ \emph {et~al.}(2013)\citenamefont
  {{Motch}}, \citenamefont {{Wilms}}, \citenamefont {{Barret}}, \citenamefont
  {{Becker}}, \citenamefont {{Bogdanov}}, \citenamefont {{Boirin}},
  \citenamefont {{Corbel}}, \citenamefont {{Cackett}}, \citenamefont
  {{Campana}}, \citenamefont {{de Martino}} \emph {et~al.}}]{Motch13}%
  \BibitemOpen
  \bibfield  {author} {\bibinfo {author} {\bibnamefont {{Motch}}, \bibfnamefont
  {C.}}, \bibinfo {author} {\bibfnamefont {J.}~\bibnamefont {{Wilms}}},
  \bibinfo {author} {\bibfnamefont {D.}~\bibnamefont {{Barret}}}, \bibinfo
  {author} {\bibfnamefont {W.}~\bibnamefont {{Becker}}}, \bibinfo {author}
  {\bibfnamefont {S.}~\bibnamefont {{Bogdanov}}}, \bibinfo {author}
  {\bibfnamefont {L.}~\bibnamefont {{Boirin}}}, \bibinfo {author}
  {\bibfnamefont {S.}~\bibnamefont {{Corbel}}}, \bibinfo {author}
  {\bibfnamefont {E.}~\bibnamefont {{Cackett}}}, \bibinfo {author}
  {\bibfnamefont {S.}~\bibnamefont {{Campana}}}, \bibinfo {author}
  {\bibfnamefont {D.}~\bibnamefont {{de Martino}}},  \emph {et~al.}} (\bibinfo
  {year} {2013}),\ \href@noop {} {\bibfield  {journal} {\bibinfo  {journal}
  {ArXiv e-prints}\ }}\Eprint {http://arxiv.org/abs/1306.2334} {arXiv:1306.2334
  [astro-ph.HE]} \BibitemShut {NoStop}%
\bibitem [{\citenamefont {{Moustakidis}}(2015)}]{Moustakidis15}%
  \BibitemOpen
  \bibfield  {author} {\bibinfo {author} {\bibnamefont {{Moustakidis}},
  \bibfnamefont {C.~C.}}} (\bibinfo {year} {2015}),\ \href {\doibase
  10.1103/PhysRevC.91.035804} {\bibfield  {journal} {\bibinfo  {journal}
  {\prc}\ }\textbf {\bibinfo {volume} {91}}~(\bibinfo {number} {3}),\ \bibinfo
  {eid} {035804}}\BibitemShut {NoStop}%
\bibitem [{\citenamefont {{Muno}}\ \emph {et~al.}(2004)\citenamefont {{Muno}},
  \citenamefont {{Galloway}},\ and\ \citenamefont {{Chakrabarty}}}]{Muno04}%
  \BibitemOpen
  \bibfield  {author} {\bibinfo {author} {\bibnamefont {{Muno}}, \bibfnamefont
  {M.~P.}}, \bibinfo {author} {\bibfnamefont {D.~K.}\ \bibnamefont
  {{Galloway}}}, \ and\ \bibinfo {author} {\bibfnamefont {D.}~\bibnamefont
  {{Chakrabarty}}}} (\bibinfo {year} {2004}),\ \href {\doibase 10.1086/420812}
  {\bibfield  {journal} {\bibinfo  {journal} {\apj}\ }\textbf {\bibinfo
  {volume} {608}},\ \bibinfo {pages} {930}},\ \Eprint
  {http://arxiv.org/abs/astro-ph/0310726} {astro-ph/0310726} \BibitemShut
  {NoStop}%
\bibitem [{\citenamefont {{Nakar}}(2007)}]{Nakar07}%
  \BibitemOpen
  \bibfield  {author} {\bibinfo {author} {\bibnamefont {{Nakar}}, \bibfnamefont
  {E.}}} (\bibinfo {year} {2007}),\ \href {\doibase
  10.1016/j.physrep.2007.02.005} {\bibfield  {journal} {\bibinfo  {journal}
  {\physrep}\ }\textbf {\bibinfo {volume} {442}},\ \bibinfo {pages} {166}},\
  \Eprint {http://arxiv.org/abs/astro-ph/0701748} {astro-ph/0701748}
  \BibitemShut {NoStop}%
\bibitem [{\citenamefont {{Nandra}}\ \emph {et~al.}(2013)\citenamefont
  {{Nandra}}, \citenamefont {{Barret}}, \citenamefont {{Barcons}},
  \citenamefont {{Fabian}}, \citenamefont {{den Herder}}, \citenamefont
  {{Piro}}, \citenamefont {{Watson}}, \citenamefont {{Adami}}, \citenamefont
  {{Aird}}, \citenamefont {{Afonso}},\ and\ \citenamefont {et~al.}}]{Nandra13}%
  \BibitemOpen
  \bibfield  {author} {\bibinfo {author} {\bibnamefont {{Nandra}},
  \bibfnamefont {K.}}, \bibinfo {author} {\bibfnamefont {D.}~\bibnamefont
  {{Barret}}}, \bibinfo {author} {\bibfnamefont {X.}~\bibnamefont {{Barcons}}},
  \bibinfo {author} {\bibfnamefont {A.}~\bibnamefont {{Fabian}}}, \bibinfo
  {author} {\bibfnamefont {J.-W.}\ \bibnamefont {{den Herder}}}, \bibinfo
  {author} {\bibfnamefont {L.}~\bibnamefont {{Piro}}}, \bibinfo {author}
  {\bibfnamefont {M.}~\bibnamefont {{Watson}}}, \bibinfo {author}
  {\bibfnamefont {C.}~\bibnamefont {{Adami}}}, \bibinfo {author} {\bibfnamefont
  {J.}~\bibnamefont {{Aird}}}, \bibinfo {author} {\bibfnamefont {J.~M.}\
  \bibnamefont {{Afonso}}}, \ and\ \bibinfo {author} {\bibnamefont {et~al.}}}
  (\bibinfo {year} {2013}),\ \href@noop {} {\bibfield  {journal} {\bibinfo
  {journal} {ArXiv e-prints}\ }}\Eprint {http://arxiv.org/abs/1306.2307}
  {arXiv:1306.2307 [astro-ph.HE]} \BibitemShut {NoStop}%
\bibitem [{\citenamefont {{Nik{\v s}i{\'c}}}\ \emph {et~al.}(2015)\citenamefont
  {{Nik{\v s}i{\'c}}}, \citenamefont {{Paar}}, \citenamefont {{Reinhard}},\
  and\ \citenamefont {{Vretenar}}}]{Niksic15}%
  \BibitemOpen
  \bibfield  {author} {\bibinfo {author} {\bibnamefont {{Nik{\v s}i{\'c}}},
  \bibfnamefont {T.}}, \bibinfo {author} {\bibfnamefont {N.}~\bibnamefont
  {{Paar}}}, \bibinfo {author} {\bibfnamefont {P.-G.}\ \bibnamefont
  {{Reinhard}}}, \ and\ \bibinfo {author} {\bibfnamefont {D.}~\bibnamefont
  {{Vretenar}}}} (\bibinfo {year} {2015}),\ \href {\doibase
  10.1088/0954-3899/42/3/034008} {\bibfield  {journal} {\bibinfo  {journal}
  {Journal of Physics G Nuclear Physics}\ }\textbf {\bibinfo {volume}
  {42}}~(\bibinfo {number} {3}),\ \bibinfo {eid} {034008}},\ \Eprint
  {http://arxiv.org/abs/1407.0530} {arXiv:1407.0530 [nucl-th]} \BibitemShut
  {NoStop}%
\bibitem [{\citenamefont {{Oppenheimer}}\ and\ \citenamefont
  {{Volkoff}}(1939)}]{Oppenheimer39}%
  \BibitemOpen
  \bibfield  {author} {\bibinfo {author} {\bibnamefont {{Oppenheimer}},
  \bibfnamefont {J.~R.}}, \ and\ \bibinfo {author} {\bibfnamefont {G.~M.}\
  \bibnamefont {{Volkoff}}}} (\bibinfo {year} {1939}),\ \href {\doibase
  10.1103/PhysRev.55.374} {\bibfield  {journal} {\bibinfo  {journal} {Physical
  Review}\ }\textbf {\bibinfo {volume} {55}},\ \bibinfo {pages}
  {374}}\BibitemShut {NoStop}%
\bibitem [{\citenamefont {{{\"O}zel}}(2006)}]{Ozel06}%
  \BibitemOpen
  \bibfield  {author} {\bibinfo {author} {\bibnamefont {{{\"O}zel}},
  \bibfnamefont {F.}}} (\bibinfo {year} {2006}),\ \href {\doibase
  10.1038/nature04858} {\bibfield  {journal} {\bibinfo  {journal} {\nat}\
  }\textbf {\bibinfo {volume} {441}},\ \bibinfo {pages} {1115}},\ \Eprint
  {http://arxiv.org/abs/astro-ph/0605106} {astro-ph/0605106} \BibitemShut
  {NoStop}%
\bibitem [{\citenamefont {{{\"O}zel}}\ \emph {et~al.}(2012)\citenamefont
  {{{\"O}zel}}, \citenamefont {{Gould}},\ and\ \citenamefont
  {{G{\"u}ver}}}]{Ozel12}%
  \BibitemOpen
  \bibfield  {author} {\bibinfo {author} {\bibnamefont {{{\"O}zel}},
  \bibfnamefont {F.}}, \bibinfo {author} {\bibfnamefont {A.}~\bibnamefont
  {{Gould}}}, \ and\ \bibinfo {author} {\bibfnamefont {T.}~\bibnamefont
  {{G{\"u}ver}}}} (\bibinfo {year} {2012}),\ \href {\doibase
  10.1088/0004-637X/748/1/5} {\bibfield  {journal} {\bibinfo  {journal} {\apj}\
  }\textbf {\bibinfo {volume} {748}},\ \bibinfo {eid} {5}},\ \Eprint
  {http://arxiv.org/abs/1104.5027} {arXiv:1104.5027} \BibitemShut {NoStop}%
\bibitem [{\citenamefont {{{\"O}zel}}\ \emph {et~al.}(2009)\citenamefont
  {{{\"O}zel}}, \citenamefont {{G{\"u}ver}},\ and\ \citenamefont
  {{Psaltis}}}]{Ozel09a}%
  \BibitemOpen
  \bibfield  {author} {\bibinfo {author} {\bibnamefont {{{\"O}zel}},
  \bibfnamefont {F.}}, \bibinfo {author} {\bibfnamefont {T.}~\bibnamefont
  {{G{\"u}ver}}}, \ and\ \bibinfo {author} {\bibfnamefont {D.}~\bibnamefont
  {{Psaltis}}}} (\bibinfo {year} {2009}),\ \href {\doibase
  10.1088/0004-637X/693/2/1775} {\bibfield  {journal} {\bibinfo  {journal}
  {\apj}\ }\textbf {\bibinfo {volume} {693}},\ \bibinfo {pages} {1775}},\
  \Eprint {http://arxiv.org/abs/0810.1521} {arXiv:0810.1521} \BibitemShut
  {NoStop}%
\bibitem [{\citenamefont {{{\"O}zel}}\ and\ \citenamefont
  {{Psaltis}}(2003)}]{Ozel03}%
  \BibitemOpen
  \bibfield  {author} {\bibinfo {author} {\bibnamefont {{{\"O}zel}},
  \bibfnamefont {F.}}, \ and\ \bibinfo {author} {\bibfnamefont
  {D.}~\bibnamefont {{Psaltis}}}} (\bibinfo {year} {2003}),\ \href {\doibase
  10.1086/346197} {\bibfield  {journal} {\bibinfo  {journal} {\apjl}\ }\textbf
  {\bibinfo {volume} {582}},\ \bibinfo {pages} {L31}},\ \Eprint
  {http://arxiv.org/abs/astro-ph/0209225} {astro-ph/0209225} \BibitemShut
  {NoStop}%
\bibitem [{\citenamefont {{{\"O}zel}}\ and\ \citenamefont
  {{Psaltis}}(2009)}]{Ozel09}%
  \BibitemOpen
  \bibfield  {author} {\bibinfo {author} {\bibnamefont {{{\"O}zel}},
  \bibfnamefont {F.}}, \ and\ \bibinfo {author} {\bibfnamefont
  {D.}~\bibnamefont {{Psaltis}}}} (\bibinfo {year} {2009}),\ \href {\doibase
  10.1103/PhysRevD.80.103003} {\bibfield  {journal} {\bibinfo  {journal}
  {\prd}\ }\textbf {\bibinfo {volume} {80}}~(\bibinfo {number} {10}),\ \bibinfo
  {eid} {103003}},\ \Eprint {http://arxiv.org/abs/0905.1959} {arXiv:0905.1959
  [astro-ph.HE]} \BibitemShut {NoStop}%
\bibitem [{\citenamefont {{Ozel}}\ \emph {et~al.}(2015)\citenamefont {{Ozel}},
  \citenamefont {{Psaltis}}, \citenamefont {{Guver}}, \citenamefont {{Baym}},
  \citenamefont {{Heinke}},\ and\ \citenamefont {{Guillot}}}]{Ozel15}%
  \BibitemOpen
  \bibfield  {author} {\bibinfo {author} {\bibnamefont {{Ozel}}, \bibfnamefont
  {F.}}, \bibinfo {author} {\bibfnamefont {D.}~\bibnamefont {{Psaltis}}},
  \bibinfo {author} {\bibfnamefont {T.}~\bibnamefont {{Guver}}}, \bibinfo
  {author} {\bibfnamefont {G.}~\bibnamefont {{Baym}}}, \bibinfo {author}
  {\bibfnamefont {C.}~\bibnamefont {{Heinke}}}, \ and\ \bibinfo {author}
  {\bibfnamefont {S.}~\bibnamefont {{Guillot}}}} (\bibinfo {year} {2015}),\
  \href@noop {} {\bibfield  {journal} {\bibinfo  {journal} {ArXiv e-prints}\
  }}\Eprint {http://arxiv.org/abs/1505.05155} {arXiv:1505.05155 [astro-ph.HE]}
  \BibitemShut {NoStop}%
\bibitem [{\citenamefont {{{\"O}zel}}\ \emph {et~al.}(2010)\citenamefont
  {{{\"O}zel}}, \citenamefont {{Psaltis}}, \citenamefont {{Ransom}},
  \citenamefont {{Demorest}},\ and\ \citenamefont {{Alford}}}]{Ozel10}%
  \BibitemOpen
  \bibfield  {author} {\bibinfo {author} {\bibnamefont {{{\"O}zel}},
  \bibfnamefont {F.}}, \bibinfo {author} {\bibfnamefont {D.}~\bibnamefont
  {{Psaltis}}}, \bibinfo {author} {\bibfnamefont {S.}~\bibnamefont {{Ransom}}},
  \bibinfo {author} {\bibfnamefont {P.}~\bibnamefont {{Demorest}}}, \ and\
  \bibinfo {author} {\bibfnamefont {M.}~\bibnamefont {{Alford}}}} (\bibinfo
  {year} {2010}),\ \href {\doibase 10.1088/2041-8205/724/2/L199} {\bibfield
  {journal} {\bibinfo  {journal} {\apjl}\ }\textbf {\bibinfo {volume} {724}},\
  \bibinfo {pages} {L199}},\ \Eprint {http://arxiv.org/abs/1010.5790}
  {arXiv:1010.5790 [astro-ph.HE]} \BibitemShut {NoStop}%
\bibitem [{\citenamefont {{Papitto}}\ \emph {et~al.}(2013)\citenamefont
  {{Papitto}}, \citenamefont {{Ferrigno}}, \citenamefont {{Bozzo}},
  \citenamefont {{Rea}}, \citenamefont {{Pavan}}, \citenamefont {{Burderi}},
  \citenamefont {{Burgay}}, \citenamefont {{Campana}}, \citenamefont {{di
  Salvo}}, \citenamefont {{Falanga}} \emph {et~al.}}]{Papitto13}%
  \BibitemOpen
  \bibfield  {author} {\bibinfo {author} {\bibnamefont {{Papitto}},
  \bibfnamefont {A.}}, \bibinfo {author} {\bibfnamefont {C.}~\bibnamefont
  {{Ferrigno}}}, \bibinfo {author} {\bibfnamefont {E.}~\bibnamefont {{Bozzo}}},
  \bibinfo {author} {\bibfnamefont {N.}~\bibnamefont {{Rea}}}, \bibinfo
  {author} {\bibfnamefont {L.}~\bibnamefont {{Pavan}}}, \bibinfo {author}
  {\bibfnamefont {L.}~\bibnamefont {{Burderi}}}, \bibinfo {author}
  {\bibfnamefont {M.}~\bibnamefont {{Burgay}}}, \bibinfo {author}
  {\bibfnamefont {S.}~\bibnamefont {{Campana}}}, \bibinfo {author}
  {\bibfnamefont {T.}~\bibnamefont {{di Salvo}}}, \bibinfo {author}
  {\bibfnamefont {M.}~\bibnamefont {{Falanga}}},  \emph {et~al.}} (\bibinfo
  {year} {2013}),\ \href {\doibase 10.1038/nature12470} {\bibfield  {journal}
  {\bibinfo  {journal} {\nat}\ }\textbf {\bibinfo {volume} {501}},\ \bibinfo
  {pages} {517}},\ \Eprint {http://arxiv.org/abs/1305.3884} {arXiv:1305.3884
  [astro-ph.HE]} \BibitemShut {NoStop}%
\bibitem [{\citenamefont {{Passamonti}}\ and\ \citenamefont
  {{Lander}}(2013)}]{Passamonti13}%
  \BibitemOpen
  \bibfield  {author} {\bibinfo {author} {\bibnamefont {{Passamonti}},
  \bibfnamefont {A.}}, \ and\ \bibinfo {author} {\bibfnamefont {S.~K.}\
  \bibnamefont {{Lander}}}} (\bibinfo {year} {2013}),\ \href {\doibase
  10.1093/mnras/sts372} {\bibfield  {journal} {\bibinfo  {journal} {\mnras}\
  }\textbf {\bibinfo {volume} {429}},\ \bibinfo {pages} {767}},\ \Eprint
  {http://arxiv.org/abs/1210.2969} {arXiv:1210.2969 [astro-ph.SR]} \BibitemShut
  {NoStop}%
\bibitem [{\citenamefont {{Passamonti}}\ and\ \citenamefont
  {{Lander}}(2014)}]{Passamonti14}%
  \BibitemOpen
  \bibfield  {author} {\bibinfo {author} {\bibnamefont {{Passamonti}},
  \bibfnamefont {A.}}, \ and\ \bibinfo {author} {\bibfnamefont {S.~K.}\
  \bibnamefont {{Lander}}}} (\bibinfo {year} {2014}),\ \href {\doibase
  10.1093/mnras/stt2134} {\bibfield  {journal} {\bibinfo  {journal} {\mnras}\
  }\textbf {\bibinfo {volume} {438}},\ \bibinfo {pages} {156}},\ \Eprint
  {http://arxiv.org/abs/1307.3210} {arXiv:1307.3210 [astro-ph.SR]} \BibitemShut
  {NoStop}%
\bibitem [{\citenamefont {{Patruno}}(2010)}]{Patruno10}%
  \BibitemOpen
  \bibfield  {author} {\bibinfo {author} {\bibnamefont {{Patruno}},
  \bibfnamefont {A.}}} (\bibinfo {year} {2010}),\ \href {\doibase
  10.1088/0004-637X/722/1/909} {\bibfield  {journal} {\bibinfo  {journal}
  {\apj}\ }\textbf {\bibinfo {volume} {722}},\ \bibinfo {pages} {909}},\
  \Eprint {http://arxiv.org/abs/1006.0815} {arXiv:1006.0815 [astro-ph.HE]}
  \BibitemShut {NoStop}%
\bibitem [{\citenamefont {{Patruno}}\ \emph {et~al.}(2014)\citenamefont
  {{Patruno}}, \citenamefont {{Archibald}}, \citenamefont {{Hessels}},
  \citenamefont {{Bogdanov}}, \citenamefont {{Stappers}}, \citenamefont
  {{Bassa}}, \citenamefont {{Janssen}}, \citenamefont {{Kaspi}}, \citenamefont
  {{Tendulkar}},\ and\ \citenamefont {{Lyne}}}]{Patruno14}%
  \BibitemOpen
  \bibfield  {author} {\bibinfo {author} {\bibnamefont {{Patruno}},
  \bibfnamefont {A.}}, \bibinfo {author} {\bibfnamefont {A.~M.}\ \bibnamefont
  {{Archibald}}}, \bibinfo {author} {\bibfnamefont {J.~W.~T.}\ \bibnamefont
  {{Hessels}}}, \bibinfo {author} {\bibfnamefont {S.}~\bibnamefont
  {{Bogdanov}}}, \bibinfo {author} {\bibfnamefont {B.~W.}\ \bibnamefont
  {{Stappers}}}, \bibinfo {author} {\bibfnamefont {C.~G.}\ \bibnamefont
  {{Bassa}}}, \bibinfo {author} {\bibfnamefont {G.~H.}\ \bibnamefont
  {{Janssen}}}, \bibinfo {author} {\bibfnamefont {V.~M.}\ \bibnamefont
  {{Kaspi}}}, \bibinfo {author} {\bibfnamefont {S.}~\bibnamefont
  {{Tendulkar}}}, \ and\ \bibinfo {author} {\bibfnamefont {A.~G.}\ \bibnamefont
  {{Lyne}}}} (\bibinfo {year} {2014}),\ \href {\doibase
  10.1088/2041-8205/781/1/L3} {\bibfield  {journal} {\bibinfo  {journal}
  {\apjl}\ }\textbf {\bibinfo {volume} {781}},\ \bibinfo {eid} {L3}},\ \Eprint
  {http://arxiv.org/abs/1310.7549} {arXiv:1310.7549 [astro-ph.HE]} \BibitemShut
  {NoStop}%
\bibitem [{\citenamefont {{Patruno}}\ and\ \citenamefont
  {{Watts}}(2012)}]{Patruno12}%
  \BibitemOpen
  \bibfield  {author} {\bibinfo {author} {\bibnamefont {{Patruno}},
  \bibfnamefont {A.}}, \ and\ \bibinfo {author} {\bibfnamefont {A.~L.}\
  \bibnamefont {{Watts}}}} (\bibinfo {year} {2012}),\ \href@noop {} {\bibfield
  {journal} {\bibinfo  {journal} {ArXiv e-prints}\ }}\Eprint
  {http://arxiv.org/abs/1206.2727} {arXiv:1206.2727 [astro-ph.HE]} \BibitemShut
  {NoStop}%
\bibitem [{\citenamefont {{Pechenick}}\ \emph {et~al.}(1983)\citenamefont
  {{Pechenick}}, \citenamefont {{Ftaclas}},\ and\ \citenamefont
  {{Cohen}}}]{Pechenick83}%
  \BibitemOpen
  \bibfield  {author} {\bibinfo {author} {\bibnamefont {{Pechenick}},
  \bibfnamefont {K.~R.}}, \bibinfo {author} {\bibfnamefont {C.}~\bibnamefont
  {{Ftaclas}}}, \ and\ \bibinfo {author} {\bibfnamefont {J.~M.}\ \bibnamefont
  {{Cohen}}}} (\bibinfo {year} {1983}),\ \href {\doibase 10.1086/161498}
  {\bibfield  {journal} {\bibinfo  {journal} {\apj}\ }\textbf {\bibinfo
  {volume} {274}},\ \bibinfo {pages} {846}}\BibitemShut {NoStop}%
\bibitem [{\citenamefont {{Philippov}}\ and\ \citenamefont
  {{Spitkovsky}}(2014)}]{Philippov14}%
  \BibitemOpen
  \bibfield  {author} {\bibinfo {author} {\bibnamefont {{Philippov}},
  \bibfnamefont {A.~A.}}, \ and\ \bibinfo {author} {\bibfnamefont
  {A.}~\bibnamefont {{Spitkovsky}}}} (\bibinfo {year} {2014}),\ \href {\doibase
  10.1088/2041-8205/785/2/L33} {\bibfield  {journal} {\bibinfo  {journal}
  {\apjl}\ }\textbf {\bibinfo {volume} {785}},\ \bibinfo {eid} {L33}},\ \Eprint
  {http://arxiv.org/abs/1312.4970} {arXiv:1312.4970 [astro-ph.HE]} \BibitemShut
  {NoStop}%
\bibitem [{\citenamefont {{Piekarewicz}}\ \emph {et~al.}(2012)\citenamefont
  {{Piekarewicz}}, \citenamefont {{Agrawal}}, \citenamefont {{Col{\`o}}},
  \citenamefont {{Nazarewicz}}, \citenamefont {{Paar}}, \citenamefont
  {{Reinhard}}, \citenamefont {{Roca-Maza}},\ and\ \citenamefont
  {{Vretenar}}}]{Piekarewicz12}%
  \BibitemOpen
  \bibfield  {author} {\bibinfo {author} {\bibnamefont {{Piekarewicz}},
  \bibfnamefont {J.}}, \bibinfo {author} {\bibfnamefont {B.~K.}\ \bibnamefont
  {{Agrawal}}}, \bibinfo {author} {\bibfnamefont {G.}~\bibnamefont
  {{Col{\`o}}}}, \bibinfo {author} {\bibfnamefont {W.}~\bibnamefont
  {{Nazarewicz}}}, \bibinfo {author} {\bibfnamefont {N.}~\bibnamefont
  {{Paar}}}, \bibinfo {author} {\bibfnamefont {P.-G.}\ \bibnamefont
  {{Reinhard}}}, \bibinfo {author} {\bibfnamefont {X.}~\bibnamefont
  {{Roca-Maza}}}, \ and\ \bibinfo {author} {\bibfnamefont {D.}~\bibnamefont
  {{Vretenar}}}} (\bibinfo {year} {2012}),\ \href {\doibase
  10.1103/PhysRevC.85.041302} {\bibfield  {journal} {\bibinfo  {journal}
  {\prc}\ }\textbf {\bibinfo {volume} {85}}~(\bibinfo {number} {4}),\ \bibinfo
  {eid} {041302}},\ \Eprint {http://arxiv.org/abs/1201.3807} {arXiv:1201.3807
  [nucl-th]} \BibitemShut {NoStop}%
\bibitem [{\citenamefont {{Pletsch}}\ \emph {et~al.}(2012)\citenamefont
  {{Pletsch}}, \citenamefont {{Guillemot}}, \citenamefont {{Allen}},
  \citenamefont {{Kramer}}, \citenamefont {{Aulbert}}, \citenamefont
  {{Fehrmann}}, \citenamefont {{Ray}}, \citenamefont {{Barr}}, \citenamefont
  {{Belfiore}}, \citenamefont {{Camilo}} \emph {et~al.}}]{Pletsch12}%
  \BibitemOpen
  \bibfield  {author} {\bibinfo {author} {\bibnamefont {{Pletsch}},
  \bibfnamefont {H.~J.}}, \bibinfo {author} {\bibfnamefont {L.}~\bibnamefont
  {{Guillemot}}}, \bibinfo {author} {\bibfnamefont {B.}~\bibnamefont
  {{Allen}}}, \bibinfo {author} {\bibfnamefont {M.}~\bibnamefont {{Kramer}}},
  \bibinfo {author} {\bibfnamefont {C.}~\bibnamefont {{Aulbert}}}, \bibinfo
  {author} {\bibfnamefont {H.}~\bibnamefont {{Fehrmann}}}, \bibinfo {author}
  {\bibfnamefont {P.~S.}\ \bibnamefont {{Ray}}}, \bibinfo {author}
  {\bibfnamefont {E.~D.}\ \bibnamefont {{Barr}}}, \bibinfo {author}
  {\bibfnamefont {A.}~\bibnamefont {{Belfiore}}}, \bibinfo {author}
  {\bibfnamefont {F.}~\bibnamefont {{Camilo}}},  \emph {et~al.}} (\bibinfo
  {year} {2012}),\ \href {\doibase 10.1088/0004-637X/744/2/105} {\bibfield
  {journal} {\bibinfo  {journal} {\apj}\ }\textbf {\bibinfo {volume} {744}},\
  \bibinfo {eid} {105}},\ \Eprint {http://arxiv.org/abs/1111.0523}
  {arXiv:1111.0523 [astro-ph.HE]} \BibitemShut {NoStop}%
\bibitem [{\citenamefont {{Poutanen}}(2008)}]{Poutanen08}%
  \BibitemOpen
  \bibfield  {author} {\bibinfo {author} {\bibnamefont {{Poutanen}},
  \bibfnamefont {J.}}} (\bibinfo {year} {2008}),\ in\ \href {\doibase
  10.1063/1.3031209} {\emph {\bibinfo {booktitle} {American Institute of
  Physics Conference Series}}},\ \bibinfo {series} {American Institute of
  Physics Conference Series}, Vol.\ \bibinfo {volume} {1068},\ \bibinfo
  {editor} {edited by\ \bibinfo {editor} {\bibfnamefont {R.}~\bibnamefont
  {{Wijnands}}}, \bibinfo {editor} {\bibfnamefont {D.}~\bibnamefont
  {{Altamirano}}}, \bibinfo {editor} {\bibfnamefont {P.}~\bibnamefont
  {{Soleri}}}, \bibinfo {editor} {\bibfnamefont {N.}~\bibnamefont
  {{Degenaar}}}, \bibinfo {editor} {\bibfnamefont {N.}~\bibnamefont {{Rea}}},
  \bibinfo {editor} {\bibfnamefont {P.}~\bibnamefont {{Casella}}}, \bibinfo
  {editor} {\bibfnamefont {A.}~\bibnamefont {{Patruno}}}, \ and\ \bibinfo
  {editor} {\bibfnamefont {M.}~\bibnamefont {{Linares}}}},\ pp.\ \bibinfo
  {pages} {77--86},\ \Eprint {http://arxiv.org/abs/0809.2400} {arXiv:0809.2400}
  \BibitemShut {NoStop}%
\bibitem [{\citenamefont {{Poutanen}}\ and\ \citenamefont
  {{Beloborodov}}(2006)}]{Poutanen06}%
  \BibitemOpen
  \bibfield  {author} {\bibinfo {author} {\bibnamefont {{Poutanen}},
  \bibfnamefont {J.}}, \ and\ \bibinfo {author} {\bibfnamefont {A.~M.}\
  \bibnamefont {{Beloborodov}}}} (\bibinfo {year} {2006}),\ \href {\doibase
  10.1111/j.1365-2966.2006.11088.x} {\bibfield  {journal} {\bibinfo  {journal}
  {\mnras}\ }\textbf {\bibinfo {volume} {373}},\ \bibinfo {pages} {836}},\
  \Eprint {http://arxiv.org/abs/astro-ph/0608663} {astro-ph/0608663}
  \BibitemShut {NoStop}%
\bibitem [{\citenamefont {{Poutanen}}\ and\ \citenamefont
  {{Gierli{\'n}ski}}(2003)}]{Poutanen03}%
  \BibitemOpen
  \bibfield  {author} {\bibinfo {author} {\bibnamefont {{Poutanen}},
  \bibfnamefont {J.}}, \ and\ \bibinfo {author} {\bibfnamefont
  {M.}~\bibnamefont {{Gierli{\'n}ski}}}} (\bibinfo {year} {2003}),\ \href
  {\doibase 10.1046/j.1365-8711.2003.06773.x} {\bibfield  {journal} {\bibinfo
  {journal} {\mnras}\ }\textbf {\bibinfo {volume} {343}},\ \bibinfo {pages}
  {1301}},\ \Eprint {http://arxiv.org/abs/astro-ph/0303084} {astro-ph/0303084}
  \BibitemShut {NoStop}%
\bibitem [{\citenamefont {{Poutanen}}\ \emph {et~al.}(2009)\citenamefont
  {{Poutanen}}, \citenamefont {{Ibragimov}},\ and\ \citenamefont
  {{Annala}}}]{Poutanen09}%
  \BibitemOpen
  \bibfield  {author} {\bibinfo {author} {\bibnamefont {{Poutanen}},
  \bibfnamefont {J.}}, \bibinfo {author} {\bibfnamefont {A.}~\bibnamefont
  {{Ibragimov}}}, \ and\ \bibinfo {author} {\bibfnamefont {M.}~\bibnamefont
  {{Annala}}}} (\bibinfo {year} {2009}),\ \href {\doibase
  10.1088/0004-637X/706/1/L129} {\bibfield  {journal} {\bibinfo  {journal}
  {\apjl}\ }\textbf {\bibinfo {volume} {706}},\ \bibinfo {pages} {L129}},\
  \Eprint {http://arxiv.org/abs/0910.5868} {arXiv:0910.5868 [astro-ph.HE]}
  \BibitemShut {NoStop}%
\bibitem [{\citenamefont {{Poutanen}}\ \emph {et~al.}(2014)\citenamefont
  {{Poutanen}}, \citenamefont {{N{\"a}ttil{\"a}}}, \citenamefont {{Kajava}},
  \citenamefont {{Latvala}}, \citenamefont {{Galloway}}, \citenamefont
  {{Kuulkers}},\ and\ \citenamefont {{Suleimanov}}}]{Poutanen14}%
  \BibitemOpen
  \bibfield  {author} {\bibinfo {author} {\bibnamefont {{Poutanen}},
  \bibfnamefont {J.}}, \bibinfo {author} {\bibfnamefont {J.}~\bibnamefont
  {{N{\"a}ttil{\"a}}}}, \bibinfo {author} {\bibfnamefont {J.~J.~E.}\
  \bibnamefont {{Kajava}}}, \bibinfo {author} {\bibfnamefont {O.-M.}\
  \bibnamefont {{Latvala}}}, \bibinfo {author} {\bibfnamefont {D.~K.}\
  \bibnamefont {{Galloway}}}, \bibinfo {author} {\bibfnamefont
  {E.}~\bibnamefont {{Kuulkers}}}, \ and\ \bibinfo {author} {\bibfnamefont
  {V.~F.}\ \bibnamefont {{Suleimanov}}}} (\bibinfo {year} {2014}),\ \href
  {\doibase 10.1093/mnras/stu1139} {\bibfield  {journal} {\bibinfo  {journal}
  {\mnras}\ }\textbf {\bibinfo {volume} {442}},\ \bibinfo {pages} {3777}},\
  \Eprint {http://arxiv.org/abs/1405.2663} {arXiv:1405.2663 [astro-ph.HE]}
  \BibitemShut {NoStop}%
\bibitem [{\citenamefont {{Psaltis}}\ and\ \citenamefont
  {{{\"O}zel}}(2014)}]{Psaltis14a}%
  \BibitemOpen
  \bibfield  {author} {\bibinfo {author} {\bibnamefont {{Psaltis}},
  \bibfnamefont {D.}}, \ and\ \bibinfo {author} {\bibfnamefont
  {F.}~\bibnamefont {{{\"O}zel}}}} (\bibinfo {year} {2014}),\ \href {\doibase
  10.1088/0004-637X/792/2/87} {\bibfield  {journal} {\bibinfo  {journal}
  {\apj}\ }\textbf {\bibinfo {volume} {792}},\ \bibinfo {eid} {87}},\ \Eprint
  {http://arxiv.org/abs/1305.6615} {arXiv:1305.6615 [astro-ph.HE]} \BibitemShut
  {NoStop}%
\bibitem [{\citenamefont {{Psaltis}}\ \emph {et~al.}(2014)\citenamefont
  {{Psaltis}}, \citenamefont {{{\"O}zel}},\ and\ \citenamefont
  {{Chakrabarty}}}]{Psaltis14b}%
  \BibitemOpen
  \bibfield  {author} {\bibinfo {author} {\bibnamefont {{Psaltis}},
  \bibfnamefont {D.}}, \bibinfo {author} {\bibfnamefont {F.}~\bibnamefont
  {{{\"O}zel}}}, \ and\ \bibinfo {author} {\bibfnamefont {D.}~\bibnamefont
  {{Chakrabarty}}}} (\bibinfo {year} {2014}),\ \href {\doibase
  10.1088/0004-637X/787/2/136} {\bibfield  {journal} {\bibinfo  {journal}
  {\apj}\ }\textbf {\bibinfo {volume} {787}},\ \bibinfo {eid} {136}},\ \Eprint
  {http://arxiv.org/abs/1311.1571} {arXiv:1311.1571 [astro-ph.HE]} \BibitemShut
  {NoStop}%
\bibitem [{\citenamefont {{Radhakrishnan}}\ and\ \citenamefont
  {{Srinivasan}}(1982)}]{Radhakrishnan82}%
  \BibitemOpen
  \bibfield  {author} {\bibinfo {author} {\bibnamefont {{Radhakrishnan}},
  \bibfnamefont {V.}}, \ and\ \bibinfo {author} {\bibfnamefont
  {G.}~\bibnamefont {{Srinivasan}}}} (\bibinfo {year} {1982}),\ \href@noop {}
  {\bibfield  {journal} {\bibinfo  {journal} {Current Science}\ }\textbf
  {\bibinfo {volume} {51}},\ \bibinfo {pages} {1096}}\BibitemShut {NoStop}%
\bibitem [{\citenamefont {{Rauch}}\ \emph {et~al.}(2008)\citenamefont
  {{Rauch}}, \citenamefont {{Suleimanov}},\ and\ \citenamefont
  {{Werner}}}]{Rauch08}%
  \BibitemOpen
  \bibfield  {author} {\bibinfo {author} {\bibnamefont {{Rauch}}, \bibfnamefont
  {T.}}, \bibinfo {author} {\bibfnamefont {V.}~\bibnamefont {{Suleimanov}}}, \
  and\ \bibinfo {author} {\bibfnamefont {K.}~\bibnamefont {{Werner}}}}
  (\bibinfo {year} {2008}),\ \href {\doibase 10.1051/0004-6361:200810129}
  {\bibfield  {journal} {\bibinfo  {journal} {\aap}\ }\textbf {\bibinfo
  {volume} {490}},\ \bibinfo {pages} {1127}},\ \Eprint
  {http://arxiv.org/abs/0809.2170} {arXiv:0809.2170} \BibitemShut {NoStop}%
\bibitem [{\citenamefont {{Ray}}\ \emph {et~al.}(2010)\citenamefont {{Ray}},
  \citenamefont {{Chakrabarty}}, \citenamefont {{Wilson-Hodge}}, \citenamefont
  {{Phlips}}, \citenamefont {{Remillard}}, \citenamefont {{Levine}},
  \citenamefont {{Wood}}, \citenamefont {{Wolff}}, \citenamefont {{Gwon}},
  \citenamefont {{Strohmayer}} \emph {et~al.}}]{Ray10}%
  \BibitemOpen
  \bibfield  {author} {\bibinfo {author} {\bibnamefont {{Ray}}, \bibfnamefont
  {P.~S.}}, \bibinfo {author} {\bibfnamefont {D.}~\bibnamefont
  {{Chakrabarty}}}, \bibinfo {author} {\bibfnamefont {C.~A.}\ \bibnamefont
  {{Wilson-Hodge}}}, \bibinfo {author} {\bibfnamefont {B.~F.}\ \bibnamefont
  {{Phlips}}}, \bibinfo {author} {\bibfnamefont {R.~A.}\ \bibnamefont
  {{Remillard}}}, \bibinfo {author} {\bibfnamefont {A.~M.}\ \bibnamefont
  {{Levine}}}, \bibinfo {author} {\bibfnamefont {K.~S.}\ \bibnamefont
  {{Wood}}}, \bibinfo {author} {\bibfnamefont {M.~T.}\ \bibnamefont {{Wolff}}},
  \bibinfo {author} {\bibfnamefont {C.~S.}\ \bibnamefont {{Gwon}}}, \bibinfo
  {author} {\bibfnamefont {T.~E.}\ \bibnamefont {{Strohmayer}}},  \emph
  {et~al.}} (\bibinfo {year} {2010}),\ in\ \href {\doibase 10.1117/12.857385}
  {\emph {\bibinfo {booktitle} {Society of Photo-Optical Instrumentation
  Engineers (SPIE) Conference Series}}},\ \bibinfo {series} {Society of
  Photo-Optical Instrumentation Engineers (SPIE) Conference Series}, Vol.\
  \bibinfo {volume} {7732},\ p.~\bibinfo {pages} {48},\ \Eprint
  {http://arxiv.org/abs/1007.0988} {arXiv:1007.0988 [astro-ph.IM]} \BibitemShut
  {NoStop}%
\bibitem [{\citenamefont {{Read}}\ \emph {et~al.}(2013)\citenamefont {{Read}},
  \citenamefont {{Baiotti}}, \citenamefont {{Creighton}}, \citenamefont
  {{Friedman}}, \citenamefont {{Giacomazzo}}, \citenamefont {{Kyutoku}},
  \citenamefont {{Markakis}}, \citenamefont {{Rezzolla}}, \citenamefont
  {{Shibata}},\ and\ \citenamefont {{Taniguchi}}}]{Read13}%
  \BibitemOpen
  \bibfield  {author} {\bibinfo {author} {\bibnamefont {{Read}}, \bibfnamefont
  {J.~S.}}, \bibinfo {author} {\bibfnamefont {L.}~\bibnamefont {{Baiotti}}},
  \bibinfo {author} {\bibfnamefont {J.~D.~E.}\ \bibnamefont {{Creighton}}},
  \bibinfo {author} {\bibfnamefont {J.~L.}\ \bibnamefont {{Friedman}}},
  \bibinfo {author} {\bibfnamefont {B.}~\bibnamefont {{Giacomazzo}}}, \bibinfo
  {author} {\bibfnamefont {K.}~\bibnamefont {{Kyutoku}}}, \bibinfo {author}
  {\bibfnamefont {C.}~\bibnamefont {{Markakis}}}, \bibinfo {author}
  {\bibfnamefont {L.}~\bibnamefont {{Rezzolla}}}, \bibinfo {author}
  {\bibfnamefont {M.}~\bibnamefont {{Shibata}}}, \ and\ \bibinfo {author}
  {\bibfnamefont {K.}~\bibnamefont {{Taniguchi}}}} (\bibinfo {year} {2013}),\
  \href {\doibase 10.1103/PhysRevD.88.044042} {\bibfield  {journal} {\bibinfo
  {journal} {\prd}\ }\textbf {\bibinfo {volume} {88}}~(\bibinfo {number} {4}),\
  \bibinfo {eid} {044042}},\ \Eprint {http://arxiv.org/abs/1306.4065}
  {arXiv:1306.4065 [gr-qc]} \BibitemShut {NoStop}%
\bibitem [{\citenamefont {{Read}}\ \emph {et~al.}(2009)\citenamefont {{Read}},
  \citenamefont {{Lackey}}, \citenamefont {{Owen}},\ and\ \citenamefont
  {{Friedman}}}]{Read09}%
  \BibitemOpen
  \bibfield  {author} {\bibinfo {author} {\bibnamefont {{Read}}, \bibfnamefont
  {J.~S.}}, \bibinfo {author} {\bibfnamefont {B.~D.}\ \bibnamefont {{Lackey}}},
  \bibinfo {author} {\bibfnamefont {B.~J.}\ \bibnamefont {{Owen}}}, \ and\
  \bibinfo {author} {\bibfnamefont {J.~L.}\ \bibnamefont {{Friedman}}}}
  (\bibinfo {year} {2009}),\ \href {\doibase 10.1103/PhysRevD.79.124032}
  {\bibfield  {journal} {\bibinfo  {journal} {\prd}\ }\textbf {\bibinfo
  {volume} {79}}~(\bibinfo {number} {12}),\ \bibinfo {eid} {124032}},\ \Eprint
  {http://arxiv.org/abs/0812.2163} {arXiv:0812.2163} \BibitemShut {NoStop}%
\bibitem [{\citenamefont {{Reardon}}\ \emph {et~al.}(2016)\citenamefont
  {{Reardon}}, \citenamefont {{Hobbs}}, \citenamefont {{Coles}}, \citenamefont
  {{Levin}}, \citenamefont {{Keith}}, \citenamefont {{Bailes}}, \citenamefont
  {{Bhat}}, \citenamefont {{Burke-Spolaor}}, \citenamefont {{Dai}},
  \citenamefont {{Kerr}}, \citenamefont {{Lasky}}, \citenamefont
  {{Manchester}}, \citenamefont {{Os{\l}owski}}, \citenamefont {{Ravi}},
  \citenamefont {{Shannon}}, \citenamefont {{van Straten}}, \citenamefont
  {{Toomey}}, \citenamefont {{Wang}}, \citenamefont {{Wen}}, \citenamefont
  {{You}},\ and\ \citenamefont {{Zhu}}}]{Reardon15}%
  \BibitemOpen
  \bibfield  {author} {\bibinfo {author} {\bibnamefont {{Reardon}},
  \bibfnamefont {D.~J.}}, \bibinfo {author} {\bibfnamefont {G.}~\bibnamefont
  {{Hobbs}}}, \bibinfo {author} {\bibfnamefont {W.}~\bibnamefont {{Coles}}},
  \bibinfo {author} {\bibfnamefont {Y.}~\bibnamefont {{Levin}}}, \bibinfo
  {author} {\bibfnamefont {M.~J.}\ \bibnamefont {{Keith}}}, \bibinfo {author}
  {\bibfnamefont {M.}~\bibnamefont {{Bailes}}}, \bibinfo {author}
  {\bibfnamefont {N.~D.~R.}\ \bibnamefont {{Bhat}}}, \bibinfo {author}
  {\bibfnamefont {S.}~\bibnamefont {{Burke-Spolaor}}}, \bibinfo {author}
  {\bibfnamefont {S.}~\bibnamefont {{Dai}}}, \bibinfo {author} {\bibfnamefont
  {M.}~\bibnamefont {{Kerr}}}, \bibinfo {author} {\bibfnamefont {P.~D.}\
  \bibnamefont {{Lasky}}}, \bibinfo {author} {\bibfnamefont {R.~N.}\
  \bibnamefont {{Manchester}}}, \bibinfo {author} {\bibfnamefont
  {S.}~\bibnamefont {{Os{\l}owski}}}, \bibinfo {author} {\bibfnamefont
  {V.}~\bibnamefont {{Ravi}}}, \bibinfo {author} {\bibfnamefont {R.~M.}\
  \bibnamefont {{Shannon}}}, \bibinfo {author} {\bibfnamefont {W.}~\bibnamefont
  {{van Straten}}}, \bibinfo {author} {\bibfnamefont {L.}~\bibnamefont
  {{Toomey}}}, \bibinfo {author} {\bibfnamefont {J.}~\bibnamefont {{Wang}}},
  \bibinfo {author} {\bibfnamefont {L.}~\bibnamefont {{Wen}}}, \bibinfo
  {author} {\bibfnamefont {X.~P.}\ \bibnamefont {{You}}}, \ and\ \bibinfo
  {author} {\bibfnamefont {X.-J.}\ \bibnamefont {{Zhu}}}} (\bibinfo {year}
  {2016}),\ \href {\doibase 10.1093/mnras/stv2395} {\bibfield  {journal}
  {\bibinfo  {journal} {\mnras}\ }\textbf {\bibinfo {volume} {455}},\ \bibinfo
  {pages} {1751}},\ \Eprint {http://arxiv.org/abs/1510.04434} {arXiv:1510.04434
  [astro-ph.HE]} \BibitemShut {NoStop}%
\bibitem [{\citenamefont {{Roca-Maza}}\ \emph {et~al.}(2011)\citenamefont
  {{Roca-Maza}}, \citenamefont {{Centelles}}, \citenamefont {{Vi{\~n}as}},\
  and\ \citenamefont {{Warda}}}]{RocaMaza11}%
  \BibitemOpen
  \bibfield  {author} {\bibinfo {author} {\bibnamefont {{Roca-Maza}},
  \bibfnamefont {X.}}, \bibinfo {author} {\bibfnamefont {M.}~\bibnamefont
  {{Centelles}}}, \bibinfo {author} {\bibfnamefont {X.}~\bibnamefont
  {{Vi{\~n}as}}}, \ and\ \bibinfo {author} {\bibfnamefont {M.}~\bibnamefont
  {{Warda}}}} (\bibinfo {year} {2011}),\ \href {\doibase
  10.1103/PhysRevLett.106.252501} {\bibfield  {journal} {\bibinfo  {journal}
  {Physical Review Letters}\ }\textbf {\bibinfo {volume} {106}}~(\bibinfo
  {number} {25}),\ \bibinfo {eid} {252501}},\ \Eprint
  {http://arxiv.org/abs/1103.1762} {arXiv:1103.1762 [nucl-th]} \BibitemShut
  {NoStop}%
\bibitem [{\citenamefont {{Romanova}}\ \emph {et~al.}(2008)\citenamefont
  {{Romanova}}, \citenamefont {{Kulkarni}},\ and\ \citenamefont
  {{Lovelace}}}]{Romanova08}%
  \BibitemOpen
  \bibfield  {author} {\bibinfo {author} {\bibnamefont {{Romanova}},
  \bibfnamefont {M.~M.}}, \bibinfo {author} {\bibfnamefont {A.~K.}\
  \bibnamefont {{Kulkarni}}}, \ and\ \bibinfo {author} {\bibfnamefont
  {R.~V.~E.}\ \bibnamefont {{Lovelace}}}} (\bibinfo {year} {2008}),\ \href
  {\doibase 10.1086/527298} {\bibfield  {journal} {\bibinfo  {journal} {\apjl}\
  }\textbf {\bibinfo {volume} {673}},\ \bibinfo {pages} {L171}}\BibitemShut
  {NoStop}%
\bibitem [{\citenamefont {{Rosswog}}(2015)}]{Rosswog15}%
  \BibitemOpen
  \bibfield  {author} {\bibinfo {author} {\bibnamefont {{Rosswog}},
  \bibfnamefont {S.}}} (\bibinfo {year} {2015}),\ \href {\doibase
  10.1142/S0218271815300128} {\bibfield  {journal} {\bibinfo  {journal}
  {International Journal of Modern Physics D}\ }\textbf {\bibinfo {volume}
  {24}},\ \bibinfo {eid} {1530012}},\ \Eprint {http://arxiv.org/abs/1501.02081}
  {arXiv:1501.02081 [astro-ph.HE]} \BibitemShut {NoStop}%
\bibitem [{\citenamefont {{Ruderman}}(1991)}]{Ruderman91}%
  \BibitemOpen
  \bibfield  {author} {\bibinfo {author} {\bibnamefont {{Ruderman}},
  \bibfnamefont {M.}}} (\bibinfo {year} {1991}),\ \href@noop {} {\bibfield
  {journal} {\bibinfo  {journal} {\apj}\ }\textbf {\bibinfo {volume} {366}},\
  \bibinfo {pages} {261}}\BibitemShut {NoStop}%
\bibitem [{\citenamefont {{Rutledge}}\ \emph {et~al.}(1999)\citenamefont
  {{Rutledge}}, \citenamefont {{Bildsten}}, \citenamefont {{Brown}},
  \citenamefont {{Pavlov}},\ and\ \citenamefont {{Zavlin}}}]{Rutledge99}%
  \BibitemOpen
  \bibfield  {author} {\bibinfo {author} {\bibnamefont {{Rutledge}},
  \bibfnamefont {R.~E.}}, \bibinfo {author} {\bibfnamefont {L.}~\bibnamefont
  {{Bildsten}}}, \bibinfo {author} {\bibfnamefont {E.~F.}\ \bibnamefont
  {{Brown}}}, \bibinfo {author} {\bibfnamefont {G.~G.}\ \bibnamefont
  {{Pavlov}}}, \ and\ \bibinfo {author} {\bibfnamefont {V.~E.}\ \bibnamefont
  {{Zavlin}}}} (\bibinfo {year} {1999}),\ \href {\doibase 10.1086/306990}
  {\bibfield  {journal} {\bibinfo  {journal} {\apj}\ }\textbf {\bibinfo
  {volume} {514}},\ \bibinfo {pages} {945}},\ \Eprint
  {http://arxiv.org/abs/astro-ph/9810288} {astro-ph/9810288} \BibitemShut
  {NoStop}%
\bibitem [{\citenamefont {Sagert}\ \emph {et~al.}(2012)\citenamefont {Sagert},
  \citenamefont {Tolos}, \citenamefont {Chatterjee}, \citenamefont
  {Schaffner-Bielich},\ and\ \citenamefont {Sturm}}]{Sagert12}%
  \BibitemOpen
  \bibfield  {author} {\bibinfo {author} {\bibnamefont {Sagert}, \bibfnamefont
  {I.}}, \bibinfo {author} {\bibfnamefont {L.}~\bibnamefont {Tolos}}, \bibinfo
  {author} {\bibfnamefont {D.}~\bibnamefont {Chatterjee}}, \bibinfo {author}
  {\bibfnamefont {J.}~\bibnamefont {Schaffner-Bielich}}, \ and\ \bibinfo
  {author} {\bibfnamefont {C.}~\bibnamefont {Sturm}}} (\bibinfo {year}
  {2012}),\ \href {\doibase 10.1103/PhysRevC.86.045802} {\bibfield  {journal}
  {\bibinfo  {journal} {Phys.Rev.}\ }\textbf {\bibinfo {volume} {C86}},\
  \bibinfo {pages} {045802}},\ \Eprint {http://arxiv.org/abs/1111.6058}
  {arXiv:1111.6058 [astro-ph.SR]} \BibitemShut {NoStop}%
\bibitem [{\citenamefont {{Samuelsson}}\ and\ \citenamefont
  {{Andersson}}(2007)}]{Samuelsson07}%
  \BibitemOpen
  \bibfield  {author} {\bibinfo {author} {\bibnamefont {{Samuelsson}},
  \bibfnamefont {L.}}, \ and\ \bibinfo {author} {\bibfnamefont
  {N.}~\bibnamefont {{Andersson}}}} (\bibinfo {year} {2007}),\ \href {\doibase
  10.1111/j.1365-2966.2006.11147.x} {\bibfield  {journal} {\bibinfo  {journal}
  {\mnras}\ }\textbf {\bibinfo {volume} {374}},\ \bibinfo {pages} {256}},\
  \Eprint {http://arxiv.org/abs/astro-ph/0609265} {astro-ph/0609265}
  \BibitemShut {NoStop}%
\bibitem [{\citenamefont {{Schinckel}}\ \emph {et~al.}(2012)\citenamefont
  {{Schinckel}}, \citenamefont {{Bunton}}, \citenamefont {{Cornwell}},
  \citenamefont {{Feain}},\ and\ \citenamefont {{Hay}}}]{Schinckel12}%
  \BibitemOpen
  \bibfield  {author} {\bibinfo {author} {\bibnamefont {{Schinckel}},
  \bibfnamefont {A.~E.}}, \bibinfo {author} {\bibfnamefont {J.~D.}\
  \bibnamefont {{Bunton}}}, \bibinfo {author} {\bibfnamefont {T.~J.}\
  \bibnamefont {{Cornwell}}}, \bibinfo {author} {\bibfnamefont
  {I.}~\bibnamefont {{Feain}}}, \ and\ \bibinfo {author} {\bibfnamefont
  {S.~G.}\ \bibnamefont {{Hay}}}} (\bibinfo {year} {2012}),\ in\ \href
  {\doibase 10.1117/12.926959} {\emph {\bibinfo {booktitle} {Society of
  Photo-Optical Instrumentation Engineers (SPIE) Conference Series}}},\
  \bibinfo {series} {Society of Photo-Optical Instrumentation Engineers (SPIE)
  Conference Series}, Vol.\ \bibinfo {volume} {8444},\ p.~\bibinfo {pages}
  {2}\BibitemShut {NoStop}%
\bibitem [{\citenamefont {{Servillat}}\ \emph {et~al.}(2012)\citenamefont
  {{Servillat}}, \citenamefont {{Heinke}}, \citenamefont {{Ho}}, \citenamefont
  {{Grindlay}}, \citenamefont {{Hong}}, \citenamefont {{van den Berg}},\ and\
  \citenamefont {{Bogdanov}}}]{Servillat12}%
  \BibitemOpen
  \bibfield  {author} {\bibinfo {author} {\bibnamefont {{Servillat}},
  \bibfnamefont {M.}}, \bibinfo {author} {\bibfnamefont {C.~O.}\ \bibnamefont
  {{Heinke}}}, \bibinfo {author} {\bibfnamefont {W.~C.~G.}\ \bibnamefont
  {{Ho}}}, \bibinfo {author} {\bibfnamefont {J.~E.}\ \bibnamefont
  {{Grindlay}}}, \bibinfo {author} {\bibfnamefont {J.}~\bibnamefont {{Hong}}},
  \bibinfo {author} {\bibfnamefont {M.}~\bibnamefont {{van den Berg}}}, \ and\
  \bibinfo {author} {\bibfnamefont {S.}~\bibnamefont {{Bogdanov}}}} (\bibinfo
  {year} {2012}),\ \href {\doibase 10.1111/j.1365-2966.2012.20976.x} {\bibfield
   {journal} {\bibinfo  {journal} {\mnras}\ }\textbf {\bibinfo {volume}
  {423}},\ \bibinfo {pages} {1556}},\ \Eprint {http://arxiv.org/abs/1203.5807}
  {arXiv:1203.5807 [astro-ph.HE]} \BibitemShut {NoStop}%
\bibitem [{\citenamefont {{Shibata}}\ and\ \citenamefont
  {{Taniguchi}}(2011)}]{Shibata11}%
  \BibitemOpen
  \bibfield  {author} {\bibinfo {author} {\bibnamefont {{Shibata}},
  \bibfnamefont {M.}}, \ and\ \bibinfo {author} {\bibfnamefont
  {K.}~\bibnamefont {{Taniguchi}}}} (\bibinfo {year} {2011}),\ \href {\doibase
  10.12942/lrr-2011-6} {\bibfield  {journal} {\bibinfo  {journal} {Living
  Reviews in Relativity}\ }\textbf {\bibinfo {volume} {14}},\ \bibinfo {pages}
  {6}}\BibitemShut {NoStop}%
\bibitem [{\citenamefont {{Singh}}\ \emph {et~al.}(2014)\citenamefont
  {{Singh}}, \citenamefont {{Tandon}}, \citenamefont {{Agrawal}}, \citenamefont
  {{Antia}}, \citenamefont {{Manchanda}}, \citenamefont {{Yadav}},
  \citenamefont {{Seetha}}, \citenamefont {{Ramadevi}}, \citenamefont {{Rao}},
  \citenamefont {{Bhattacharya}} \emph {et~al.}}]{Singh14}%
  \BibitemOpen
  \bibfield  {author} {\bibinfo {author} {\bibnamefont {{Singh}}, \bibfnamefont
  {K.~P.}}, \bibinfo {author} {\bibfnamefont {S.~N.}\ \bibnamefont {{Tandon}}},
  \bibinfo {author} {\bibfnamefont {P.~C.}\ \bibnamefont {{Agrawal}}}, \bibinfo
  {author} {\bibfnamefont {H.~M.}\ \bibnamefont {{Antia}}}, \bibinfo {author}
  {\bibfnamefont {R.~K.}\ \bibnamefont {{Manchanda}}}, \bibinfo {author}
  {\bibfnamefont {J.~S.}\ \bibnamefont {{Yadav}}}, \bibinfo {author}
  {\bibfnamefont {S.}~\bibnamefont {{Seetha}}}, \bibinfo {author}
  {\bibfnamefont {M.~C.}\ \bibnamefont {{Ramadevi}}}, \bibinfo {author}
  {\bibfnamefont {A.~R.}\ \bibnamefont {{Rao}}}, \bibinfo {author}
  {\bibfnamefont {D.}~\bibnamefont {{Bhattacharya}}},  \emph {et~al.}}
  (\bibinfo {year} {2014}),\ in\ \href {\doibase 10.1117/12.2062667} {\emph
  {\bibinfo {booktitle} {Society of Photo-Optical Instrumentation Engineers
  (SPIE) Conference Series}}},\ \bibinfo {series} {Society of Photo-Optical
  Instrumentation Engineers (SPIE) Conference Series}, Vol.\ \bibinfo {volume}
  {9144},\ p.~\bibinfo {pages} {1}\BibitemShut {NoStop}%
\bibitem [{\citenamefont {{Sotani}}\ \emph {et~al.}(2008)\citenamefont
  {{Sotani}}, \citenamefont {{Kokkotas}},\ and\ \citenamefont
  {{Stergioulas}}}]{Sotani08}%
  \BibitemOpen
  \bibfield  {author} {\bibinfo {author} {\bibnamefont {{Sotani}},
  \bibfnamefont {H.}}, \bibinfo {author} {\bibfnamefont {K.~D.}\ \bibnamefont
  {{Kokkotas}}}, \ and\ \bibinfo {author} {\bibfnamefont {N.}~\bibnamefont
  {{Stergioulas}}}} (\bibinfo {year} {2008}),\ \href {\doibase
  10.1111/j.1745-3933.2007.00420.x} {\bibfield  {journal} {\bibinfo  {journal}
  {\mnras}\ }\textbf {\bibinfo {volume} {385}},\ \bibinfo {pages} {L5}},\
  \Eprint {http://arxiv.org/abs/0710.1113} {arXiv:0710.1113} \BibitemShut
  {NoStop}%
\bibitem [{\citenamefont {{Spitkovsky}}\ \emph {et~al.}(2002)\citenamefont
  {{Spitkovsky}}, \citenamefont {{Levin}},\ and\ \citenamefont
  {{Ushomirsky}}}]{Spitkovsky02}%
  \BibitemOpen
  \bibfield  {author} {\bibinfo {author} {\bibnamefont {{Spitkovsky}},
  \bibfnamefont {A.}}, \bibinfo {author} {\bibfnamefont {Y.}~\bibnamefont
  {{Levin}}}, \ and\ \bibinfo {author} {\bibfnamefont {G.}~\bibnamefont
  {{Ushomirsky}}}} (\bibinfo {year} {2002}),\ \href {\doibase 10.1086/338040}
  {\bibfield  {journal} {\bibinfo  {journal} {\apj}\ }\textbf {\bibinfo
  {volume} {566}},\ \bibinfo {pages} {1018}},\ \Eprint
  {http://arxiv.org/abs/astro-ph/0108074} {astro-ph/0108074} \BibitemShut
  {NoStop}%
\bibitem [{\citenamefont {{Stappers}}\ \emph {et~al.}(2014)\citenamefont
  {{Stappers}}, \citenamefont {{Archibald}}, \citenamefont {{Hessels}},
  \citenamefont {{Bassa}}, \citenamefont {{Bogdanov}}, \citenamefont
  {{Janssen}}, \citenamefont {{Kaspi}}, \citenamefont {{Lyne}}, \citenamefont
  {{Patruno}}, \citenamefont {{Tendulkar}}, \citenamefont {{Hill}},\ and\
  \citenamefont {{Glanzman}}}]{Stappers14}%
  \BibitemOpen
  \bibfield  {author} {\bibinfo {author} {\bibnamefont {{Stappers}},
  \bibfnamefont {B.~W.}}, \bibinfo {author} {\bibfnamefont {A.~M.}\
  \bibnamefont {{Archibald}}}, \bibinfo {author} {\bibfnamefont {J.~W.~T.}\
  \bibnamefont {{Hessels}}}, \bibinfo {author} {\bibfnamefont {C.~G.}\
  \bibnamefont {{Bassa}}}, \bibinfo {author} {\bibfnamefont {S.}~\bibnamefont
  {{Bogdanov}}}, \bibinfo {author} {\bibfnamefont {G.~H.}\ \bibnamefont
  {{Janssen}}}, \bibinfo {author} {\bibfnamefont {V.~M.}\ \bibnamefont
  {{Kaspi}}}, \bibinfo {author} {\bibfnamefont {A.~G.}\ \bibnamefont {{Lyne}}},
  \bibinfo {author} {\bibfnamefont {A.}~\bibnamefont {{Patruno}}}, \bibinfo
  {author} {\bibfnamefont {S.}~\bibnamefont {{Tendulkar}}}, \bibinfo {author}
  {\bibfnamefont {A.~B.}\ \bibnamefont {{Hill}}}, \ and\ \bibinfo {author}
  {\bibfnamefont {T.}~\bibnamefont {{Glanzman}}}} (\bibinfo {year} {2014}),\
  \href {\doibase 10.1088/0004-637X/790/1/39} {\bibfield  {journal} {\bibinfo
  {journal} {\apj}\ }\textbf {\bibinfo {volume} {790}},\ \bibinfo {eid} {39}},\
  \Eprint {http://arxiv.org/abs/1311.7506} {arXiv:1311.7506 [astro-ph.HE]}
  \BibitemShut {NoStop}%
\bibitem [{\citenamefont {{Steiner}}\ \emph {et~al.}(2010)\citenamefont
  {{Steiner}}, \citenamefont {{Lattimer}},\ and\ \citenamefont
  {{Brown}}}]{Steiner10}%
  \BibitemOpen
  \bibfield  {author} {\bibinfo {author} {\bibnamefont {{Steiner}},
  \bibfnamefont {A.~W.}}, \bibinfo {author} {\bibfnamefont {J.~M.}\
  \bibnamefont {{Lattimer}}}, \ and\ \bibinfo {author} {\bibfnamefont {E.~F.}\
  \bibnamefont {{Brown}}}} (\bibinfo {year} {2010}),\ \href@noop {} {\bibfield
  {journal} {\bibinfo  {journal} {\apj}\ }\textbf {\bibinfo {volume} {722}},\
  \bibinfo {pages} {33}},\ \Eprint {http://arxiv.org/abs/1005.0811}
  {arXiv:1005.0811 [astro-ph.HE]} \BibitemShut {NoStop}%
\bibitem [{\citenamefont {{Steiner}}\ \emph {et~al.}(2013)\citenamefont
  {{Steiner}}, \citenamefont {{Lattimer}},\ and\ \citenamefont
  {{Brown}}}]{Steiner13}%
  \BibitemOpen
  \bibfield  {author} {\bibinfo {author} {\bibnamefont {{Steiner}},
  \bibfnamefont {A.~W.}}, \bibinfo {author} {\bibfnamefont {J.~M.}\
  \bibnamefont {{Lattimer}}}, \ and\ \bibinfo {author} {\bibfnamefont {E.~F.}\
  \bibnamefont {{Brown}}}} (\bibinfo {year} {2013}),\ \href {\doibase
  10.1088/2041-8205/765/1/L5} {\bibfield  {journal} {\bibinfo  {journal}
  {\apjl}\ }\textbf {\bibinfo {volume} {765}},\ \bibinfo {eid} {L5}},\ \Eprint
  {http://arxiv.org/abs/1205.6871} {arXiv:1205.6871 [nucl-th]} \BibitemShut
  {NoStop}%
\bibitem [{\citenamefont {{Steiner}}\ and\ \citenamefont
  {{Watts}}(2009)}]{Steiner09}%
  \BibitemOpen
  \bibfield  {author} {\bibinfo {author} {\bibnamefont {{Steiner}},
  \bibfnamefont {A.~W.}}, \ and\ \bibinfo {author} {\bibfnamefont {A.~L.}\
  \bibnamefont {{Watts}}}} (\bibinfo {year} {2009}),\ \href@noop {} {\bibfield
  {journal} {\bibinfo  {journal} {Physical Review Letters}\ }\textbf {\bibinfo
  {volume} {103}}~(\bibinfo {number} {18}),\ \bibinfo {eid} {181101}},\ \Eprint
  {http://arxiv.org/abs/0902.1683} {arXiv:0902.1683 [astro-ph.HE]} \BibitemShut
  {NoStop}%
\bibitem [{\citenamefont {{Stergioulas}}(2003)}]{Stergioulas03}%
  \BibitemOpen
  \bibfield  {author} {\bibinfo {author} {\bibnamefont {{Stergioulas}},
  \bibfnamefont {N.}}} (\bibinfo {year} {2003}),\ \href {\doibase
  10.12942/lrr-2003-3} {\bibfield  {journal} {\bibinfo  {journal} {Living
  Reviews in Relativity}\ }\textbf {\bibinfo {volume} {6}},\ \bibinfo {pages}
  {3}},\ \Eprint {http://arxiv.org/abs/gr-qc/0302034} {gr-qc/0302034}
  \BibitemShut {NoStop}%
\bibitem [{\citenamefont {{Stergioulas}}\ and\ \citenamefont
  {{Friedman}}(1995)}]{Stergioulas95}%
  \BibitemOpen
  \bibfield  {author} {\bibinfo {author} {\bibnamefont {{Stergioulas}},
  \bibfnamefont {N.}}, \ and\ \bibinfo {author} {\bibfnamefont {J.~L.}\
  \bibnamefont {{Friedman}}}} (\bibinfo {year} {1995}),\ \href {\doibase
  10.1086/175605} {\bibfield  {journal} {\bibinfo  {journal} {\apj}\ }\textbf
  {\bibinfo {volume} {444}},\ \bibinfo {pages} {306}},\ \Eprint
  {http://arxiv.org/abs/astro-ph/9411032} {astro-ph/9411032} \BibitemShut
  {NoStop}%
\bibitem [{\citenamefont {{Strohmayer}}\ and\ \citenamefont
  {{Markwardt}}(2002)}]{Strohmayer02}%
  \BibitemOpen
  \bibfield  {author} {\bibinfo {author} {\bibnamefont {{Strohmayer}},
  \bibfnamefont {T.~E.}}, \ and\ \bibinfo {author} {\bibfnamefont {C.~B.}\
  \bibnamefont {{Markwardt}}}} (\bibinfo {year} {2002}),\ \href@noop {}
  {\bibfield  {journal} {\bibinfo  {journal} {\apj}\ }\textbf {\bibinfo
  {volume} {577}},\ \bibinfo {pages} {337}},\ \Eprint
  {http://arxiv.org/abs/astro-ph/0205435} {astro-ph/0205435} \BibitemShut
  {NoStop}%
\bibitem [{\citenamefont {{Strohmayer}}\ and\ \citenamefont
  {{Watts}}(2005)}]{Strohmayer05}%
  \BibitemOpen
  \bibfield  {author} {\bibinfo {author} {\bibnamefont {{Strohmayer}},
  \bibfnamefont {T.~E.}}, \ and\ \bibinfo {author} {\bibfnamefont {A.~L.}\
  \bibnamefont {{Watts}}}} (\bibinfo {year} {2005}),\ \href {\doibase
  10.1086/497911} {\bibfield  {journal} {\bibinfo  {journal} {\apjl}\ }\textbf
  {\bibinfo {volume} {632}},\ \bibinfo {pages} {L111}},\ \Eprint
  {http://arxiv.org/abs/astro-ph/0508206} {astro-ph/0508206} \BibitemShut
  {NoStop}%
\bibitem [{\citenamefont {{Strohmayer}}\ and\ \citenamefont
  {{Watts}}(2006)}]{Strohmayer06a}%
  \BibitemOpen
  \bibfield  {author} {\bibinfo {author} {\bibnamefont {{Strohmayer}},
  \bibfnamefont {T.~E.}}, \ and\ \bibinfo {author} {\bibfnamefont {A.~L.}\
  \bibnamefont {{Watts}}}} (\bibinfo {year} {2006}),\ \href {\doibase
  10.1086/508703} {\bibfield  {journal} {\bibinfo  {journal} {\apj}\ }\textbf
  {\bibinfo {volume} {653}},\ \bibinfo {pages} {593}},\ \Eprint
  {http://arxiv.org/abs/astro-ph/0608463} {astro-ph/0608463} \BibitemShut
  {NoStop}%
\bibitem [{\citenamefont {{Sturm}}\ \emph {et~al.}(2001)\citenamefont
  {{Sturm}}, \citenamefont {{B{\"o}ttcher}}, \citenamefont {{D{\c e}bowski}},
  \citenamefont {{F{\"o}rster}}, \citenamefont {{Grosse}}, \citenamefont
  {{Koczo{\'n}}}, \citenamefont {{Kohlmeyer}}, \citenamefont {{Laue}},
  \citenamefont {{Mang}}, \citenamefont {{Naumann}} \emph {et~al.}}]{Sturm01}%
  \BibitemOpen
  \bibfield  {author} {\bibinfo {author} {\bibnamefont {{Sturm}}, \bibfnamefont
  {C.}}, \bibinfo {author} {\bibfnamefont {I.}~\bibnamefont {{B{\"o}ttcher}}},
  \bibinfo {author} {\bibfnamefont {M.}~\bibnamefont {{D{\c e}bowski}}},
  \bibinfo {author} {\bibfnamefont {A.}~\bibnamefont {{F{\"o}rster}}}, \bibinfo
  {author} {\bibfnamefont {E.}~\bibnamefont {{Grosse}}}, \bibinfo {author}
  {\bibfnamefont {P.}~\bibnamefont {{Koczo{\'n}}}}, \bibinfo {author}
  {\bibfnamefont {B.}~\bibnamefont {{Kohlmeyer}}}, \bibinfo {author}
  {\bibfnamefont {F.}~\bibnamefont {{Laue}}}, \bibinfo {author} {\bibfnamefont
  {M.}~\bibnamefont {{Mang}}}, \bibinfo {author} {\bibfnamefont
  {L.}~\bibnamefont {{Naumann}}},  \emph {et~al.}} (\bibinfo {year} {2001}),\
  \href@noop {} {\bibfield  {journal} {\bibinfo  {journal} {Physical Review
  Letters}\ }\textbf {\bibinfo {volume} {86}},\ \bibinfo {pages} {39}},\
  \Eprint {http://arxiv.org/abs/nucl-ex/0011001} {nucl-ex/0011001} \BibitemShut
  {NoStop}%
\bibitem [{\citenamefont {{Suleimanov}}\ \emph
  {et~al.}(2011{\natexlab{a}})\citenamefont {{Suleimanov}}, \citenamefont
  {{Poutanen}}, \citenamefont {{Revnivtsev}},\ and\ \citenamefont
  {{Werner}}}]{Suleimanov11}%
  \BibitemOpen
  \bibfield  {author} {\bibinfo {author} {\bibnamefont {{Suleimanov}},
  \bibfnamefont {V.}}, \bibinfo {author} {\bibfnamefont {J.}~\bibnamefont
  {{Poutanen}}}, \bibinfo {author} {\bibfnamefont {M.}~\bibnamefont
  {{Revnivtsev}}}, \ and\ \bibinfo {author} {\bibfnamefont {K.}~\bibnamefont
  {{Werner}}}} (\bibinfo {year} {2011}{\natexlab{a}}),\ \href {\doibase
  10.1088/0004-637X/742/2/122} {\bibfield  {journal} {\bibinfo  {journal}
  {\apj}\ }\textbf {\bibinfo {volume} {742}},\ \bibinfo {eid} {122}},\ \Eprint
  {http://arxiv.org/abs/1004.4871} {arXiv:1004.4871 [astro-ph.HE]} \BibitemShut
  {NoStop}%
\bibitem [{\citenamefont {{Suleimanov}}\ \emph
  {et~al.}(2011{\natexlab{b}})\citenamefont {{Suleimanov}}, \citenamefont
  {{Poutanen}},\ and\ \citenamefont {{Werner}}}]{Suleimanov11b}%
  \BibitemOpen
  \bibfield  {author} {\bibinfo {author} {\bibnamefont {{Suleimanov}},
  \bibfnamefont {V.}}, \bibinfo {author} {\bibfnamefont {J.}~\bibnamefont
  {{Poutanen}}}, \ and\ \bibinfo {author} {\bibfnamefont {K.}~\bibnamefont
  {{Werner}}}} (\bibinfo {year} {2011}{\natexlab{b}}),\ \href {\doibase
  10.1051/0004-6361/201015845} {\bibfield  {journal} {\bibinfo  {journal}
  {\aap}\ }\textbf {\bibinfo {volume} {527}},\ \bibinfo {eid} {A139}},\ \Eprint
  {http://arxiv.org/abs/1009.6147} {arXiv:1009.6147 [astro-ph.HE]} \BibitemShut
  {NoStop}%
\bibitem [{\citenamefont {{Suleimanov}}\ \emph {et~al.}(2012)\citenamefont
  {{Suleimanov}}, \citenamefont {{Poutanen}},\ and\ \citenamefont
  {{Werner}}}]{Suleimanov12}%
  \BibitemOpen
  \bibfield  {author} {\bibinfo {author} {\bibnamefont {{Suleimanov}},
  \bibfnamefont {V.}}, \bibinfo {author} {\bibfnamefont {J.}~\bibnamefont
  {{Poutanen}}}, \ and\ \bibinfo {author} {\bibfnamefont {K.}~\bibnamefont
  {{Werner}}}} (\bibinfo {year} {2012}),\ \href {\doibase
  10.1051/0004-6361/201219480} {\bibfield  {journal} {\bibinfo  {journal}
  {\aap}\ }\textbf {\bibinfo {volume} {545}},\ \bibinfo {eid} {A120}},\ \Eprint
  {http://arxiv.org/abs/1208.1467} {arXiv:1208.1467 [astro-ph.HE]} \BibitemShut
  {NoStop}%
\bibitem [{\citenamefont {{Takami}}\ \emph {et~al.}(2014)\citenamefont
  {{Takami}}, \citenamefont {{Rezzolla}},\ and\ \citenamefont
  {{Baiotti}}}]{Takami14}%
  \BibitemOpen
  \bibfield  {author} {\bibinfo {author} {\bibnamefont {{Takami}},
  \bibfnamefont {K.}}, \bibinfo {author} {\bibfnamefont {L.}~\bibnamefont
  {{Rezzolla}}}, \ and\ \bibinfo {author} {\bibfnamefont {L.}~\bibnamefont
  {{Baiotti}}}} (\bibinfo {year} {2014}),\ \href {\doibase
  10.1103/PhysRevLett.113.091104} {\bibfield  {journal} {\bibinfo  {journal}
  {Physical Review Letters}\ }\textbf {\bibinfo {volume} {113}}~(\bibinfo
  {number} {9}),\ \bibinfo {eid} {091104}},\ \Eprint
  {http://arxiv.org/abs/1403.5672} {arXiv:1403.5672 [gr-qc]} \BibitemShut
  {NoStop}%
\bibitem [{\citenamefont {{Takatsuka}}\ \emph {et~al.}(2008)\citenamefont
  {{Takatsuka}}, \citenamefont {{Nishizaki}},\ and\ \citenamefont
  {{Tamagaki}}}]{Takatsuka08}%
  \BibitemOpen
  \bibfield  {author} {\bibinfo {author} {\bibnamefont {{Takatsuka}},
  \bibfnamefont {T.}}, \bibinfo {author} {\bibfnamefont {S.}~\bibnamefont
  {{Nishizaki}}}, \ and\ \bibinfo {author} {\bibfnamefont {R.}~\bibnamefont
  {{Tamagaki}}}} (\bibinfo {year} {2008}),\ \href {\doibase
  10.1143/PTPS.174.80} {\bibfield  {journal} {\bibinfo  {journal} {Progress of
  Theoretical Physics Supplement}\ }\textbf {\bibinfo {volume} {174}},\
  \bibinfo {pages} {80}}\BibitemShut {NoStop}%
\bibitem [{\citenamefont {{Tamii}}\ \emph {et~al.}(2011)\citenamefont
  {{Tamii}}, \citenamefont {{Poltoratska}}, \citenamefont {{von
  Neumann-Cosel}}, \citenamefont {{Fujita}}, \citenamefont {{Adachi}},
  \citenamefont {{Bertulani}}, \citenamefont {{Carter}}, \citenamefont
  {{Dozono}}, \citenamefont {{Fujita}}, \citenamefont {{Fujita}} \emph
  {et~al.}}]{Tamii11}%
  \BibitemOpen
  \bibfield  {author} {\bibinfo {author} {\bibnamefont {{Tamii}}, \bibfnamefont
  {A.}}, \bibinfo {author} {\bibfnamefont {I.}~\bibnamefont {{Poltoratska}}},
  \bibinfo {author} {\bibfnamefont {P.}~\bibnamefont {{von Neumann-Cosel}}},
  \bibinfo {author} {\bibfnamefont {Y.}~\bibnamefont {{Fujita}}}, \bibinfo
  {author} {\bibfnamefont {T.}~\bibnamefont {{Adachi}}}, \bibinfo {author}
  {\bibfnamefont {C.~A.}\ \bibnamefont {{Bertulani}}}, \bibinfo {author}
  {\bibfnamefont {J.}~\bibnamefont {{Carter}}}, \bibinfo {author}
  {\bibfnamefont {M.}~\bibnamefont {{Dozono}}}, \bibinfo {author}
  {\bibfnamefont {H.}~\bibnamefont {{Fujita}}}, \bibinfo {author}
  {\bibfnamefont {K.}~\bibnamefont {{Fujita}}},  \emph {et~al.}} (\bibinfo
  {year} {2011}),\ \href {\doibase 10.1103/PhysRevLett.107.062502} {\bibfield
  {journal} {\bibinfo  {journal} {Physical Review Letters}\ }\textbf {\bibinfo
  {volume} {107}}~(\bibinfo {number} {6}),\ \bibinfo {eid} {062502}},\ \Eprint
  {http://arxiv.org/abs/1104.5431} {arXiv:1104.5431 [nucl-ex]} \BibitemShut
  {NoStop}%
\bibitem [{\citenamefont {{Tauris}}(2012)}]{Tauris12}%
  \BibitemOpen
  \bibfield  {author} {\bibinfo {author} {\bibnamefont {{Tauris}},
  \bibfnamefont {T.~M.}}} (\bibinfo {year} {2012}),\ \href {\doibase
  10.1126/science.1216355} {\bibfield  {journal} {\bibinfo  {journal}
  {Science}\ }\textbf {\bibinfo {volume} {335}},\ \bibinfo {pages} {561}},\
  \Eprint {http://arxiv.org/abs/1202.0551} {arXiv:1202.0551 [astro-ph.SR]}
  \BibitemShut {NoStop}%
\bibitem [{\citenamefont {{The LIGO Scientific Collaboration}}\ \emph
  {et~al.}(2015)\citenamefont {{The LIGO Scientific Collaboration}},
  \citenamefont {{Aasi}}, \citenamefont {{Abbott}}, \citenamefont {{Abbott}},
  \citenamefont {{Abbott}}, \citenamefont {{Abernathy}}, \citenamefont
  {{Ackley}}, \citenamefont {{Adams}}, \citenamefont {{Adams}}, \citenamefont
  {{Addesso}},\ and\ \citenamefont {et~al.}}]{LIGO15}%
  \BibitemOpen
  \bibfield  {author} {\bibinfo {author} {\bibnamefont {{The LIGO Scientific
  Collaboration}},}, \bibinfo {author} {\bibfnamefont {J.}~\bibnamefont
  {{Aasi}}}, \bibinfo {author} {\bibfnamefont {B.~P.}\ \bibnamefont
  {{Abbott}}}, \bibinfo {author} {\bibfnamefont {R.}~\bibnamefont {{Abbott}}},
  \bibinfo {author} {\bibfnamefont {T.}~\bibnamefont {{Abbott}}}, \bibinfo
  {author} {\bibfnamefont {M.~R.}\ \bibnamefont {{Abernathy}}}, \bibinfo
  {author} {\bibfnamefont {K.}~\bibnamefont {{Ackley}}}, \bibinfo {author}
  {\bibfnamefont {C.}~\bibnamefont {{Adams}}}, \bibinfo {author} {\bibfnamefont
  {T.}~\bibnamefont {{Adams}}}, \bibinfo {author} {\bibfnamefont
  {P.}~\bibnamefont {{Addesso}}}, \ and\ \bibinfo {author} {\bibnamefont
  {et~al.}}} (\bibinfo {year} {2015}),\ \href {\doibase
  10.1088/0264-9381/32/7/074001} {\bibfield  {journal} {\bibinfo  {journal}
  {Classical and Quantum Gravity}\ }\textbf {\bibinfo {volume} {32}}~(\bibinfo
  {number} {7}),\ \bibinfo {eid} {074001}},\ \Eprint
  {http://arxiv.org/abs/1411.4547} {arXiv:1411.4547 [gr-qc]} \BibitemShut
  {NoStop}%
\bibitem [{\citenamefont {{Timokhin}}\ \emph {et~al.}(2008)\citenamefont
  {{Timokhin}}, \citenamefont {{Eichler}},\ and\ \citenamefont
  {{Lyubarsky}}}]{Timokhin08}%
  \BibitemOpen
  \bibfield  {author} {\bibinfo {author} {\bibnamefont {{Timokhin}},
  \bibfnamefont {A.~N.}}, \bibinfo {author} {\bibfnamefont {D.}~\bibnamefont
  {{Eichler}}}, \ and\ \bibinfo {author} {\bibfnamefont {Y.}~\bibnamefont
  {{Lyubarsky}}}} (\bibinfo {year} {2008}),\ \href {\doibase 10.1086/587925}
  {\bibfield  {journal} {\bibinfo  {journal} {\apj}\ }\textbf {\bibinfo
  {volume} {680}},\ \bibinfo {pages} {1398}},\ \Eprint
  {http://arxiv.org/abs/0706.3698} {arXiv:0706.3698} \BibitemShut {NoStop}%
\bibitem [{\citenamefont {{Tolman}}(1939)}]{Tolman39}%
  \BibitemOpen
  \bibfield  {author} {\bibinfo {author} {\bibnamefont {{Tolman}},
  \bibfnamefont {R.~C.}}} (\bibinfo {year} {1939}),\ \href {\doibase
  10.1103/PhysRev.55.364} {\bibfield  {journal} {\bibinfo  {journal} {Physical
  Review}\ }\textbf {\bibinfo {volume} {55}},\ \bibinfo {pages}
  {364}}\BibitemShut {NoStop}%
\bibitem [{\citenamefont {Tolos}\ \emph {et~al.}(2008)\citenamefont {Tolos},
  \citenamefont {Friman},\ and\ \citenamefont {Schwenk}}]{Tolos08}%
  \BibitemOpen
  \bibfield  {author} {\bibinfo {author} {\bibnamefont {Tolos}, \bibfnamefont
  {L.}}, \bibinfo {author} {\bibfnamefont {B.}~\bibnamefont {Friman}}, \ and\
  \bibinfo {author} {\bibfnamefont {A.}~\bibnamefont {Schwenk}}} (\bibinfo
  {year} {2008}),\ \href {\doibase 10.1016/j.nuclphysa.2008.02.309} {\bibfield
  {journal} {\bibinfo  {journal} {Nucl.Phys.}\ }\textbf {\bibinfo {volume}
  {A806}},\ \bibinfo {pages} {105}},\ \Eprint {http://arxiv.org/abs/0711.3613}
  {arXiv:0711.3613 [nucl-th]} \BibitemShut {NoStop}%
\bibitem [{\citenamefont {{Trippa}}\ \emph {et~al.}(2008)\citenamefont
  {{Trippa}}, \citenamefont {{Col{\`o}}},\ and\ \citenamefont
  {{Vigezzi}}}]{Trippa08}%
  \BibitemOpen
  \bibfield  {author} {\bibinfo {author} {\bibnamefont {{Trippa}},
  \bibfnamefont {L.}}, \bibinfo {author} {\bibfnamefont {G.}~\bibnamefont
  {{Col{\`o}}}}, \ and\ \bibinfo {author} {\bibfnamefont {E.}~\bibnamefont
  {{Vigezzi}}}} (\bibinfo {year} {2008}),\ \href {\doibase
  10.1103/PhysRevC.77.061304} {\bibfield  {journal} {\bibinfo  {journal}
  {\prc}\ }\textbf {\bibinfo {volume} {77}}~(\bibinfo {number} {6}),\ \bibinfo
  {eid} {061304}},\ \Eprint {http://arxiv.org/abs/0802.3658} {arXiv:0802.3658
  [nucl-th]} \BibitemShut {NoStop}%
\bibitem [{\citenamefont {{Tsang}}\ \emph {et~al.}(2012)\citenamefont
  {{Tsang}}, \citenamefont {{Stone}}, \citenamefont {{Camera}}, \citenamefont
  {{Danielewicz}}, \citenamefont {{Gandolfi}}, \citenamefont {{Hebeler}},
  \citenamefont {{Horowitz}}, \citenamefont {{Lee}}, \citenamefont {{Lynch}},
  \citenamefont {{Kohley}} \emph {et~al.}}]{Tsang12}%
  \BibitemOpen
  \bibfield  {author} {\bibinfo {author} {\bibnamefont {{Tsang}}, \bibfnamefont
  {M.~B.}}, \bibinfo {author} {\bibfnamefont {J.~R.}\ \bibnamefont {{Stone}}},
  \bibinfo {author} {\bibfnamefont {F.}~\bibnamefont {{Camera}}}, \bibinfo
  {author} {\bibfnamefont {P.}~\bibnamefont {{Danielewicz}}}, \bibinfo {author}
  {\bibfnamefont {S.}~\bibnamefont {{Gandolfi}}}, \bibinfo {author}
  {\bibfnamefont {K.}~\bibnamefont {{Hebeler}}}, \bibinfo {author}
  {\bibfnamefont {C.~J.}\ \bibnamefont {{Horowitz}}}, \bibinfo {author}
  {\bibfnamefont {J.}~\bibnamefont {{Lee}}}, \bibinfo {author} {\bibfnamefont
  {W.~G.}\ \bibnamefont {{Lynch}}}, \bibinfo {author} {\bibfnamefont
  {Z.}~\bibnamefont {{Kohley}}},  \emph {et~al.}} (\bibinfo {year} {2012}),\
  \href {\doibase 10.1103/PhysRevC.86.015803} {\bibfield  {journal} {\bibinfo
  {journal} {\prc}\ }\textbf {\bibinfo {volume} {86}}~(\bibinfo {number} {1}),\
  \bibinfo {eid} {015803}},\ \Eprint {http://arxiv.org/abs/1204.0466}
  {arXiv:1204.0466 [nucl-ex]} \BibitemShut {NoStop}%
\bibitem [{\citenamefont {{Tsang}}\ \emph {et~al.}(2009)\citenamefont
  {{Tsang}}, \citenamefont {{Zhang}}, \citenamefont {{Danielewicz}},
  \citenamefont {{Famiano}}, \citenamefont {{Li}}, \citenamefont {{Lynch}},\
  and\ \citenamefont {{Steiner}}}]{Tsang09}%
  \BibitemOpen
  \bibfield  {author} {\bibinfo {author} {\bibnamefont {{Tsang}}, \bibfnamefont
  {M.~B.}}, \bibinfo {author} {\bibfnamefont {Y.}~\bibnamefont {{Zhang}}},
  \bibinfo {author} {\bibfnamefont {P.}~\bibnamefont {{Danielewicz}}}, \bibinfo
  {author} {\bibfnamefont {M.}~\bibnamefont {{Famiano}}}, \bibinfo {author}
  {\bibfnamefont {Z.}~\bibnamefont {{Li}}}, \bibinfo {author} {\bibfnamefont
  {W.~G.}\ \bibnamefont {{Lynch}}}, \ and\ \bibinfo {author} {\bibfnamefont
  {A.~W.}\ \bibnamefont {{Steiner}}}} (\bibinfo {year} {2009}),\ \href
  {\doibase 10.1103/PhysRevLett.102.122701} {\bibfield  {journal} {\bibinfo
  {journal} {Physical Review Letters}\ }\textbf {\bibinfo {volume}
  {102}}~(\bibinfo {number} {12}),\ \bibinfo {eid} {122701}},\ \Eprint
  {http://arxiv.org/abs/0811.3107} {arXiv:0811.3107 [nucl-ex]} \BibitemShut
  {NoStop}%
\bibitem [{\citenamefont {{van Haarlem}}\ \emph {et~al.}(2013)\citenamefont
  {{van Haarlem}}, \citenamefont {{Wise}}, \citenamefont {{Gunst}},
  \citenamefont {{Heald}}, \citenamefont {{McKean}}, \citenamefont {{Hessels}},
  \citenamefont {{de Bruyn}}, \citenamefont {{Nijboer}}, \citenamefont
  {{Swinbank}}, \citenamefont {{Fallows}},\ and\ \citenamefont
  {et~al.}}]{vanHaarlem13}%
  \BibitemOpen
  \bibfield  {author} {\bibinfo {author} {\bibnamefont {{van Haarlem}},
  \bibfnamefont {M.~P.}}, \bibinfo {author} {\bibfnamefont {M.~W.}\
  \bibnamefont {{Wise}}}, \bibinfo {author} {\bibfnamefont {A.~W.}\
  \bibnamefont {{Gunst}}}, \bibinfo {author} {\bibfnamefont {G.}~\bibnamefont
  {{Heald}}}, \bibinfo {author} {\bibfnamefont {J.~P.}\ \bibnamefont
  {{McKean}}}, \bibinfo {author} {\bibfnamefont {J.~W.~T.}\ \bibnamefont
  {{Hessels}}}, \bibinfo {author} {\bibfnamefont {A.~G.}\ \bibnamefont {{de
  Bruyn}}}, \bibinfo {author} {\bibfnamefont {R.}~\bibnamefont {{Nijboer}}},
  \bibinfo {author} {\bibfnamefont {J.}~\bibnamefont {{Swinbank}}}, \bibinfo
  {author} {\bibfnamefont {R.}~\bibnamefont {{Fallows}}}, \ and\ \bibinfo
  {author} {\bibnamefont {et~al.}}} (\bibinfo {year} {2013}),\ \href {\doibase
  10.1051/0004-6361/201220873} {\bibfield  {journal} {\bibinfo  {journal}
  {\aap}\ }\textbf {\bibinfo {volume} {556}},\ \bibinfo {eid} {A2}},\ \Eprint
  {http://arxiv.org/abs/1305.3550} {arXiv:1305.3550 [astro-ph.IM]} \BibitemShut
  {NoStop}%
\bibitem [{\citenamefont {{van Hoven}}\ and\ \citenamefont
  {{Levin}}(2011)}]{vanHoven11}%
  \BibitemOpen
  \bibfield  {author} {\bibinfo {author} {\bibnamefont {{van Hoven}},
  \bibfnamefont {M.}}, \ and\ \bibinfo {author} {\bibfnamefont
  {Y.}~\bibnamefont {{Levin}}}} (\bibinfo {year} {2011}),\ \href {\doibase
  10.1111/j.1365-2966.2010.17499.x} {\bibfield  {journal} {\bibinfo  {journal}
  {\mnras}\ }\textbf {\bibinfo {volume} {410}},\ \bibinfo {pages} {1036}},\
  \Eprint {http://arxiv.org/abs/1006.0348} {arXiv:1006.0348 [astro-ph.HE]}
  \BibitemShut {NoStop}%
\bibitem [{\citenamefont {{van Hoven}}\ and\ \citenamefont
  {{Levin}}(2012)}]{vanHoven12}%
  \BibitemOpen
  \bibfield  {author} {\bibinfo {author} {\bibnamefont {{van Hoven}},
  \bibfnamefont {M.}}, \ and\ \bibinfo {author} {\bibfnamefont
  {Y.}~\bibnamefont {{Levin}}}} (\bibinfo {year} {2012}),\ \href {\doibase
  10.1111/j.1365-2966.2011.20177.x} {\bibfield  {journal} {\bibinfo  {journal}
  {\mnras}\ }\textbf {\bibinfo {volume} {420}},\ \bibinfo {pages} {3035}},\
  \Eprint {http://arxiv.org/abs/1110.2107} {arXiv:1110.2107 [astro-ph.HE]}
  \BibitemShut {NoStop}%
\bibitem [{\citenamefont {{van Paradijs}}(1979)}]{vanParadijs79}%
  \BibitemOpen
  \bibfield  {author} {\bibinfo {author} {\bibnamefont {{van Paradijs}},
  \bibfnamefont {J.}}} (\bibinfo {year} {1979}),\ \href {\doibase
  10.1086/157535} {\bibfield  {journal} {\bibinfo  {journal} {\apj}\ }\textbf
  {\bibinfo {volume} {234}},\ \bibinfo {pages} {609}}\BibitemShut {NoStop}%
\bibitem [{\citenamefont {{van Paradijs}}\ and\ \citenamefont
  {{Lewin}}(1987)}]{vanParadijs87}%
  \BibitemOpen
  \bibfield  {author} {\bibinfo {author} {\bibnamefont {{van Paradijs}},
  \bibfnamefont {J.}}, \ and\ \bibinfo {author} {\bibfnamefont {W.~H.~G.}\
  \bibnamefont {{Lewin}}}} (\bibinfo {year} {1987}),\ \href@noop {} {\bibfield
  {journal} {\bibinfo  {journal} {\aap}\ }\textbf {\bibinfo {volume} {172}},\
  \bibinfo {pages} {L20}}\BibitemShut {NoStop}%
\bibitem [{\citenamefont {{Vida{\~n}a}}(2015)}]{Vidana15}%
  \BibitemOpen
  \bibfield  {author} {\bibinfo {author} {\bibnamefont {{Vida{\~n}a}},
  \bibfnamefont {I.}}} (\bibinfo {year} {2015}),\ in\ \href {\doibase
  10.1063/1.4909561} {\emph {\bibinfo {booktitle} {American Institute of
  Physics Conference Series}}},\ \bibinfo {series} {American Institute of
  Physics Conference Series}, Vol.\ \bibinfo {volume} {1645},\ pp.\ \bibinfo
  {pages} {79--85}\BibitemShut {NoStop}%
\bibitem [{\citenamefont {{Viironen}}\ and\ \citenamefont
  {{Poutanen}}(2004)}]{Viironen04}%
  \BibitemOpen
  \bibfield  {author} {\bibinfo {author} {\bibnamefont {{Viironen}},
  \bibfnamefont {K.}}, \ and\ \bibinfo {author} {\bibfnamefont
  {J.}~\bibnamefont {{Poutanen}}}} (\bibinfo {year} {2004}),\ \href {\doibase
  10.1051/0004-6361:20041084} {\bibfield  {journal} {\bibinfo  {journal}
  {\aap}\ }\textbf {\bibinfo {volume} {426}},\ \bibinfo {pages} {985}},\
  \Eprint {http://arxiv.org/abs/astro-ph/0408250} {astro-ph/0408250}
  \BibitemShut {NoStop}%
\bibitem [{\citenamefont {{Watts}}\ \emph {et~al.}(2015)\citenamefont
  {{Watts}}, \citenamefont {{Espinoza}}, \citenamefont {{Xu}}, \citenamefont
  {{Andersson}}, \citenamefont {{Antoniadis}}, \citenamefont {{Antonopoulou}},
  \citenamefont {{Buchner}}, \citenamefont {{Datta}}, \citenamefont
  {{Demorest}}, \citenamefont {{Freire}} \emph {et~al.}}]{Watts14}%
  \BibitemOpen
  \bibfield  {author} {\bibinfo {author} {\bibnamefont {{Watts}}, \bibfnamefont
  {A.}}, \bibinfo {author} {\bibfnamefont {C.~M.}\ \bibnamefont {{Espinoza}}},
  \bibinfo {author} {\bibfnamefont {R.}~\bibnamefont {{Xu}}}, \bibinfo {author}
  {\bibfnamefont {N.}~\bibnamefont {{Andersson}}}, \bibinfo {author}
  {\bibfnamefont {J.}~\bibnamefont {{Antoniadis}}}, \bibinfo {author}
  {\bibfnamefont {D.}~\bibnamefont {{Antonopoulou}}}, \bibinfo {author}
  {\bibfnamefont {S.}~\bibnamefont {{Buchner}}}, \bibinfo {author}
  {\bibfnamefont {S.}~\bibnamefont {{Datta}}}, \bibinfo {author} {\bibfnamefont
  {P.}~\bibnamefont {{Demorest}}}, \bibinfo {author} {\bibfnamefont
  {P.}~\bibnamefont {{Freire}}},  \emph {et~al.}} (\bibinfo {year} {2015}),\
  in\ \href@noop {} {\emph {\bibinfo {booktitle} {Advancing Astrophysics with
  the Square Kilometre Array (AASKA14)}}},\ p.~\bibinfo {pages} {43},\ \Eprint
  {http://arxiv.org/abs/1501.00042} {arXiv:1501.00042 [astro-ph.SR]}
  \BibitemShut {NoStop}%
\bibitem [{\citenamefont {{Watts}}(2012)}]{Watts12}%
  \BibitemOpen
  \bibfield  {author} {\bibinfo {author} {\bibnamefont {{Watts}}, \bibfnamefont
  {A.~L.}}} (\bibinfo {year} {2012}),\ \href {\doibase
  10.1146/annurev-astro-040312-132617} {\bibfield  {journal} {\bibinfo
  {journal} {Ann. Rev. Astron. Astrophys.}\ }\textbf {\bibinfo {volume} {50}},\
  \bibinfo {pages} {609}},\ \Eprint {http://arxiv.org/abs/1203.2065}
  {arXiv:1203.2065 [astro-ph.HE]} \BibitemShut {NoStop}%
\bibitem [{\citenamefont {{Watts}}\ \emph {et~al.}(2008)\citenamefont
  {{Watts}}, \citenamefont {{Krishnan}}, \citenamefont {{Bildsten}},\ and\
  \citenamefont {{Schutz}}}]{Watts08b}%
  \BibitemOpen
  \bibfield  {author} {\bibinfo {author} {\bibnamefont {{Watts}}, \bibfnamefont
  {A.~L.}}, \bibinfo {author} {\bibfnamefont {B.}~\bibnamefont {{Krishnan}}},
  \bibinfo {author} {\bibfnamefont {L.}~\bibnamefont {{Bildsten}}}, \ and\
  \bibinfo {author} {\bibfnamefont {B.~F.}\ \bibnamefont {{Schutz}}}} (\bibinfo
  {year} {2008}),\ \href {\doibase 10.1111/j.1365-2966.2008.13594.x} {\bibfield
   {journal} {\bibinfo  {journal} {\mnras}\ }\textbf {\bibinfo {volume}
  {389}},\ \bibinfo {pages} {839}},\ \Eprint {http://arxiv.org/abs/0803.4097}
  {arXiv:0803.4097} \BibitemShut {NoStop}%
\bibitem [{\citenamefont {{Watts}}\ and\ \citenamefont
  {{Reddy}}(2007)}]{Watts07}%
  \BibitemOpen
  \bibfield  {author} {\bibinfo {author} {\bibnamefont {{Watts}}, \bibfnamefont
  {A.~L.}}, \ and\ \bibinfo {author} {\bibfnamefont {S.}~\bibnamefont
  {{Reddy}}}} (\bibinfo {year} {2007}),\ \href {\doibase
  10.1111/j.1745-3933.2007.00336.x} {\bibfield  {journal} {\bibinfo  {journal}
  {\mnras}\ }\textbf {\bibinfo {volume} {379}},\ \bibinfo {pages} {L63}},\
  \Eprint {http://arxiv.org/abs/astro-ph/0609364} {astro-ph/0609364}
  \BibitemShut {NoStop}%
\bibitem [{\citenamefont {{Watts}}\ and\ \citenamefont
  {{Strohmayer}}(2006)}]{Watts06}%
  \BibitemOpen
  \bibfield  {author} {\bibinfo {author} {\bibnamefont {{Watts}}, \bibfnamefont
  {A.~L.}}, \ and\ \bibinfo {author} {\bibfnamefont {T.~E.}\ \bibnamefont
  {{Strohmayer}}}} (\bibinfo {year} {2006}),\ \href {\doibase 10.1086/500735}
  {\bibfield  {journal} {\bibinfo  {journal} {\apjl}\ }\textbf {\bibinfo
  {volume} {637}},\ \bibinfo {pages} {L117}},\ \Eprint
  {http://arxiv.org/abs/astro-ph/0512630} {astro-ph/0512630} \BibitemShut
  {NoStop}%
\bibitem [{\citenamefont {{Webb}}\ and\ \citenamefont
  {{Barret}}(2007)}]{Webb07}%
  \BibitemOpen
  \bibfield  {author} {\bibinfo {author} {\bibnamefont {{Webb}}, \bibfnamefont
  {N.~A.}}, \ and\ \bibinfo {author} {\bibfnamefont {D.}~\bibnamefont
  {{Barret}}}} (\bibinfo {year} {2007}),\ \href {\doibase 10.1086/522877}
  {\bibfield  {journal} {\bibinfo  {journal} {\apj}\ }\textbf {\bibinfo
  {volume} {671}},\ \bibinfo {pages} {727}},\ \Eprint
  {http://arxiv.org/abs/0708.3816} {arXiv:0708.3816} \BibitemShut {NoStop}%
\bibitem [{\citenamefont {{Weissenborn}}\ \emph {et~al.}(2012)\citenamefont
  {{Weissenborn}}, \citenamefont {{Chatterjee}},\ and\ \citenamefont
  {{Schaffner-Bielich}}}]{Weissenborn12}%
  \BibitemOpen
  \bibfield  {author} {\bibinfo {author} {\bibnamefont {{Weissenborn}},
  \bibfnamefont {S.}}, \bibinfo {author} {\bibfnamefont {D.}~\bibnamefont
  {{Chatterjee}}}, \ and\ \bibinfo {author} {\bibfnamefont {J.}~\bibnamefont
  {{Schaffner-Bielich}}}} (\bibinfo {year} {2012}),\ \href {\doibase
  10.1103/PhysRevC.85.065802} {\bibfield  {journal} {\bibinfo  {journal}
  {\prc}\ }\textbf {\bibinfo {volume} {85}}~(\bibinfo {number} {6}),\ \bibinfo
  {eid} {065802}},\ \Eprint {http://arxiv.org/abs/1112.0234} {arXiv:1112.0234
  [astro-ph.HE]} \BibitemShut {NoStop}%
\bibitem [{\citenamefont {{Wienholtz}}\ \emph {et~al.}(2013)\citenamefont
  {{Wienholtz}}, \citenamefont {{Beck}}, \citenamefont {{Blaum}}, \citenamefont
  {{Borgmann}}, \citenamefont {{Breitenfeldt}}, \citenamefont {{Cakirli}},
  \citenamefont {{George}}, \citenamefont {{Herfurth}}, \citenamefont {{Holt}},
  \citenamefont {{Kowalska}} \emph {et~al.}}]{Wienholtz13}%
  \BibitemOpen
  \bibfield  {author} {\bibinfo {author} {\bibnamefont {{Wienholtz}},
  \bibfnamefont {F.}}, \bibinfo {author} {\bibfnamefont {D.}~\bibnamefont
  {{Beck}}}, \bibinfo {author} {\bibfnamefont {K.}~\bibnamefont {{Blaum}}},
  \bibinfo {author} {\bibfnamefont {C.}~\bibnamefont {{Borgmann}}}, \bibinfo
  {author} {\bibfnamefont {M.}~\bibnamefont {{Breitenfeldt}}}, \bibinfo
  {author} {\bibfnamefont {R.}~\bibnamefont {{Cakirli}}}, \bibinfo {author}
  {\bibfnamefont {S.}~\bibnamefont {{George}}}, \bibinfo {author}
  {\bibfnamefont {F.}~\bibnamefont {{Herfurth}}}, \bibinfo {author}
  {\bibfnamefont {J.}~\bibnamefont {{Holt}}}, \bibinfo {author} {\bibfnamefont
  {M.}~\bibnamefont {{Kowalska}}},  \emph {et~al.}} (\bibinfo {year} {2013}),\
  \href@noop {} {\bibfield  {journal} {\bibinfo  {journal} {\nat}\ }\textbf
  {\bibinfo {volume} {498}},\ \bibinfo {pages} {346}}\BibitemShut {NoStop}%
\bibitem [{\citenamefont {{Wijnands}}\ and\ \citenamefont {{van der
  Klis}}(1998)}]{Wijnands98}%
  \BibitemOpen
  \bibfield  {author} {\bibinfo {author} {\bibnamefont {{Wijnands}},
  \bibfnamefont {R.}}, \ and\ \bibinfo {author} {\bibfnamefont
  {M.}~\bibnamefont {{van der Klis}}}} (\bibinfo {year} {1998}),\ \href
  {\doibase 10.1038/28557} {\bibfield  {journal} {\bibinfo  {journal} {\nat}\
  }\textbf {\bibinfo {volume} {394}},\ \bibinfo {pages} {344}}\BibitemShut
  {NoStop}%
\bibitem [{\citenamefont {{Witten}}(1984)}]{Witten84}%
  \BibitemOpen
  \bibfield  {author} {\bibinfo {author} {\bibnamefont {{Witten}},
  \bibfnamefont {E.}}} (\bibinfo {year} {1984}),\ \href@noop {} {\bibfield
  {journal} {\bibinfo  {journal} {\prd}\ }\textbf {\bibinfo {volume} {30}},\
  \bibinfo {pages} {272}}\BibitemShut {NoStop}%
\bibitem [{\citenamefont {{Woods}}\ and\ \citenamefont
  {{Thompson}}(2006)}]{Woods06}%
  \BibitemOpen
  \bibfield  {author} {\bibinfo {author} {\bibnamefont {{Woods}}, \bibfnamefont
  {P.~M.}}, \ and\ \bibinfo {author} {\bibfnamefont {C.}~\bibnamefont
  {{Thompson}}}} (\bibinfo {year} {2006}),\ \enquote {\bibinfo {title} {{Soft
  gamma repeaters and anomalous X-ray pulsars: magnetar candidates}},}\ in\
  \href@noop {} {\emph {\bibinfo {booktitle} {Compact stellar X-ray
  sources}}},\ \bibinfo {editor} {edited by\ \bibinfo {editor} {\bibfnamefont
  {W.~H.~G.}\ \bibnamefont {{Lewin}}}\ and\ \bibinfo {editor} {\bibfnamefont
  {M.}~\bibnamefont {{van der Klis}}}},\ pp.\ \bibinfo {pages}
  {547--586}\BibitemShut {NoStop}%
\bibitem [{\citenamefont {{Zavlin}}\ \emph {et~al.}(1996)\citenamefont
  {{Zavlin}}, \citenamefont {{Pavlov}},\ and\ \citenamefont
  {{Shibanov}}}]{Zavlin96}%
  \BibitemOpen
  \bibfield  {author} {\bibinfo {author} {\bibnamefont {{Zavlin}},
  \bibfnamefont {V.~E.}}, \bibinfo {author} {\bibfnamefont {G.~G.}\
  \bibnamefont {{Pavlov}}}, \ and\ \bibinfo {author} {\bibfnamefont {Y.~A.}\
  \bibnamefont {{Shibanov}}}} (\bibinfo {year} {1996}),\ \href@noop {}
  {\bibfield  {journal} {\bibinfo  {journal} {\aap}\ }\textbf {\bibinfo
  {volume} {315}},\ \bibinfo {pages} {141}},\ \Eprint
  {http://arxiv.org/abs/astro-ph/9604072} {astro-ph/9604072} \BibitemShut
  {NoStop}%
\bibitem [{\citenamefont {{Zdunik}}\ and\ \citenamefont
  {{Haensel}}(2013)}]{Zdunik13}%
  \BibitemOpen
  \bibfield  {author} {\bibinfo {author} {\bibnamefont {{Zdunik}},
  \bibfnamefont {J.~L.}}, \ and\ \bibinfo {author} {\bibfnamefont
  {P.}~\bibnamefont {{Haensel}}}} (\bibinfo {year} {2013}),\ \href {\doibase
  10.1051/0004-6361/201220697} {\bibfield  {journal} {\bibinfo  {journal}
  {\aap}\ }\textbf {\bibinfo {volume} {551}},\ \bibinfo {eid} {A61}},\ \Eprint
  {http://arxiv.org/abs/1211.1231} {arXiv:1211.1231 [astro-ph.SR]} \BibitemShut
  {NoStop}%
\end{thebibliography}%

\end{document}